\documentclass[12pt]{iopart}
\pdfoutput=1
\usepackage{float}
\usepackage{morefloats} 
\usepackage{datetime}
\usepackage{iopams}
\usepackage{epsf}
\usepackage{epsfig}
\usepackage{color}
\usepackage{psfrag}
\usepackage{verbatim}
\usepackage{setstack}
\usepackage{amsopn}
\usepackage{appendix}
\usepackage{graphicx}
\usepackage{graphics}
\usepackage{subfig}
\usepackage[margin=3.3cm]{geometry}

\usepackage{float}
\usepackage{tikz}

\usepackage{array}
\usepackage{tabularx}
\usepackage{multirow}
\usepackage{booktabs}
\usepackage{ctable}

\newcolumntype{?}{!{\vrule width 2pt}}

\newcommand{\be}{\begin{eqnarray}}	
\newcommand{\ee}{\end{eqnarray}}

\newcommand{\comments}[1]{}   %%%%%%%%%%%%%%% comments

\begin{document}

%\unitlength = 1mm
\eqnobysec

\title{Critical percolation in the dynamics of the $2d$ ferromagnetic Ising model}
\author{Thibault Blanchard$^1$,  Leticia F. Cugliandolo$^2$, \\
Marco Picco$^2$ and Alessandro Tartaglia$^2$
}

\address{
$^1$ Lyc\'ee d'Arsonval,\\
65, rue du Pont de Cr\'eteil, 
94100 Saint Maur des Foss\'es
\\
$^2$ Sorbonne Universit\'es, Universit\'e Pierre et Marie Curie -- Paris VI, \\
Laboratoire de Physique Th\'eorique et Hautes Energies,\\
4 Place Jussieu, %Tour 13-14, 5\'eme \'etage, 
75252 Paris Cedex 05, France
}

\begin{abstract}
We study the early time dynamics of the $2d$
ferromagnetic Ising model  instantaneously quenched from  the disordered to
the ordered, low temperature, phase.
We evolve the system with kinetic Monte Carlo rules that do not  conserve the  order parameter. 
We confirm the rapid approach to random critical percolation in a time-scale that 
diverges with the system size but is much shorter than the equilibration time.  We study the scaling properties of the evolution towards critical percolation and we 
identify an associated growing length, different from the curvature driven one. By working with the model defined on 
square, triangular and honeycomb microscopic geometries we establish the
dependence of this growing length on the lattice coordination. We discuss the interplay with 
the usual coarsening mechanism and the eventual fall into and escape from metastability.\\
\end{abstract}

%\today
\newpage
\tableofcontents
\newpage

\section{Introduction}
\label{sec:intro}

Coarsening is the process whereby a system, initiated in a homogeneous and disordered configuration,
progressively achieves local order in two or more competing equilibrium or absorbing states. It is 
a problem with manifold technological implications that, still after many decades of research, continues
to pose interesting theoretical challenges.

The theory of coarsening or 
phase ordering kinetics~\cite{Bray94,Puri09-article,CorberiPoliti} is based on the dynamic scaling hypothesis.
It states that 
at long times  the system enters a scaling 
regime regulated by a single growing length, $\ell_d(t)$, such that the 
structure is statistically invariant when distances are 
measured with respect to it.  For this hypothesis to apply, measuring times are asked to be longer than a microscopic 
time-scale, $t_0$, and observation distances $r$ are required to be such that 
$a \ll r\ll L$ with $a$ a microscopic length-scale
and $L$ the linear size of the system. The way in which the length $\ell_d$ grows is 
determined by mesoscopic mechanisms and defines dynamic universality classes.
In the absence of frustration and/or quenched disorder
$\ell_d$ typically grows algebraically $\ell_d(t) \simeq t^{1/z_d}$, and 
the best known cases are the curvature driven class or model A 
with $z_d=2$, and the locally conserved order parameter class or model B with $z_d=3$, in the classification introduced in 
Ref.~\cite{Hohenberg-Halperin}. The actual time needed to reach this scaling regime had not been the object of
detailed studies until recently.

Most of the analyses of coarsening phenomena are based on investigations of the space-time 
correlation function or, equivalently, the dynamic structure factor. The time-evolving domain structure,
that has not been as much studied so far, 
should contain additional information and be of interest
from both practical and theoretical viewpoints.

From the existence of a {\it single} growing length $\ell_d$ implied by the dynamic scaling hypothesis 
one may conclude that, on the one side, the instantaneous distribution of domain sizes is peaked at the value 
$\ell_d^d(t)$ with the power $d$ being the space dimension and, on the other side, the systems attain equilibrium when this growing length 
$\ell_d$ reaches the systems size $L$, i.e., after times of the order of 
$t_L \simeq L^{z_d}$.  None of these conclusions are totally valid, as was recently 
shown in a series of papers.

Let us  focus on the $2d$ cases from now on.
The dynamic number density of cluster areas in the $2d$ Ising model 
evolving with non-conserved order parameter dynamics and quenched from infinite to a subcritical temperature
was studied in~\cite{ArBrCuSi07,SiArBrCu07}. It was shown in these papers that after a 
short time scale the number density takes a form with two distinct regimes 
separated by $\ell_d^2(t)$: at short length scales the area dynamics is determined
by the coarsening mechanism while at long length scales the number density 
decays algebraically, with a power law that is numerically equal to the one 
of random critical percolation. The geometric properties of clusters and interfaces of various kinds also show this 
crossover. Similar results were found under weak disorder~\cite{SiArBrCu08} and 
for conserved order parameter dynamics~\cite{SiSaArBrCu09}.
Very generally and quite surprisingly, the systems first approach the morphology of 
critical percolation, with one (or more) percolating cluster(s), to later evolve 
following their coarsening dynamics. The number density of areas (also interfaces) 
satisfies dynamic scaling and the ``typical'' area $\ell_d^2(t)$ appears as a shoulder 
in the number density for curvature driven dynamics~\cite{ArBrCuSi07,SiArBrCu07,SiArBrCu08} and as a maximum for 
phase separation~\cite{SiSaArBrCu09}. The role played by an early approach to critical percolation
was stressed in these studies.

Metastability in the zero temperature quenches of the $2d$ Ising model with non-conserved order parameter dynamics 
was studied in a series of works~\cite{SpKrRe01,SpKrRe02,BaKrRe09,OlKrRe12}. 
The existence of metastable states under these conditions was first signalled in~\cite{SpKrRe01,SpKrRe02}
and the passage to a critical percolating state was exploited in~\cite{BaKrRe09,OlKrRe12}
to predict their probability of occurrence. These states are, typically, configurations with stripes and 
flat interfaces that are stable with respect to the zero-temperature dynamics. At finite 
though sub-critical temperature 
these states trap the dynamics for very long time scales, indeed longer than the naively
expected $L^{z_d}$ ones, and the actual equilibration time becomes much longer than these.

The careful analysis of the time scale needed to reach a critical percolating state that will not be destroyed by the 
stochastic dynamics, with a percolating cluster that will simply grow ever after,
unveiled that it actually scales with the system size. Numerically, an algebraic dependence was found~\cite{BlCoCuPi14}
\begin{equation}
t_p \simeq L^{z_p}
\end{equation}
with an exponent $z_p$ that depends on the coordination of the lattice,  $n_c$, and the microscopic 
dynamics.
In~\cite{BlCoCuPi14} the following conjecture on its dependence  
on $n_c$ and the conventional dynamic exponent, $z_d=2$,
\begin{equation}
z_p = z_d/n_c
\label{eq:zp-zdnc}
\end{equation}
was given (for lattices that do not allow for early freezing, as the honeycomb one, and in the absence of quenched 
disorder). This dependence
was verified with relatively good numerical accuracy on the triangular ($n_c=6$), bow-tie (on average $n_c=5$), square ($n_c=4$), 
and Kagom\'e ($n_c=4$) lattices using kinetic Monte Carlo with non-conserved order parameter updates. The approach to critical percolation is preserved under 
weak quenched disorder although the time-scales involved are different~\cite{InCoCuPi16}. The study of local spin-exchange Kawasaki dynamics confirmed the 
passage by critical percolation although the analysis of the dependence of $t_p$ with $L$ proved to be much harder~\cite{TaCuPi16}. The same applies
to the voter rules~\cite{TaCuPi15}.

In this paper we extend the analysis of the early approach to random critical percolation 
in the $2d$ Ising model with non-conserved order parameter dynamics at zero and finite temperature.
We distinguish the first time at which the system attains a critical percolation structure, called $t_{p_1}$ in the text,
from the time after which the critical percolating structure becomes stable, in the sense that it is not broken by the subsequent dynamics,
and that we call $t_p$. The role played by the fact that there are {\it two} large clusters in competition in the magnetic models 
compared to the single leading cluster of the percolation problem is also discussed. Moreover, we elaborate upon 
the understanding of the problem as one with an effective lattice spacing $\ell_d(t)$~\cite{InCoCuPi16}.
 
We demonstrate that the time regime in which the system approaches a critical percolation pattern that will not be 
destroyed by the ensuing dynamics is characterised by dynamic scaling with respect to 
the growing length 
\begin{equation}
\ell_p(t) \simeq \ell_d(t) \ t^{1/\zeta}
\end{equation} 
that, for an algebraic $\ell_d(t) \simeq t^{1/z_d}$, implies
\begin{equation}
\ell_p(t) \simeq t^{1/z_p} \qquad\qquad
\mbox{or}
\qquad\qquad
\ell_p(t) \simeq \ell_d^n(t)  
\; .
\end{equation} 
We thoroughly investigate  the dependence of $z_p$ (and $n$) on the coordination of the lattice.
We anticipate that we found a small
change in the dependence of $z_p$ on $n_c$ and $z_d$ 
with respect to the one given in Eq.~(\ref{eq:zp-zdnc})~\cite{BlCoCuPi14}.

In order to give strong support to our statements
we show results for quantities that have not been considered in previous 
works and we set the stage for the discussion of other microscopic dynamics that we will treat in a future publication. 
We also set the problem in two situations 
not considered so far. On the one hand, we use a  
honeycomb lattice that is known to have  peculiar coarsening dynamics~\cite{TakanoMiyashita} due to the stability of 
some finite-size clusters at zero temperature. On the other hand we study the effects of thermal fluctuations.
We finalise the analysis of this problem with the study of the finite-size scaling of the last time regime in which diagonal stripes
turn since they are not fully stable at zero temperature on particular lattices, or the system approaches equilibrium
helped to leave the metastable states by thermal fluctuations.

Concretely, we simulate the $2d$ Ising model dynamics with 
the single spin flip Monte Carlo (MC) updates defined in App.~\ref{app:MC}. We implement the 
continuous time Monte Carlo approach (CTMC), also referred to as Kinetic Monte Carlo (KMC) in 
the literature~\cite{Bortz-etal74,Barkema},
 to gain computer time. In~App.~\ref{app:MC}
we discuss the relation between this algorithm and the master-equation approach with 
Glauber transition probabilities, putting special emphasis on the distinction between the 
blocked states with respect to one and the other rules.

We organised the manuscript in 
six more Sections. In the next one, Sec.~\ref{subsec:cases}, we 
define the model and the lattices on which it is defined in our study.
In App.~\ref{app:MC} we explain 
the  implementation of the Monte Carlo algorithm with the 
Continuous Time setting, that we used in the simulations (CTMC).
In Sec.~\ref{sec:observables},  we define a large number of observables apt to study this problem;
in this paper we show results for some of them only. In the following Section,
Sec.~\ref{sec:phenomenon} we present the phenomenon by using just two observables, in the simplest and 
hopefully clearest possible way. We give an extensive description of the behaviour of
many other observables that complete our understanding of the phenomenon in Sec.~\ref{sec:results}. 
The theme of Sec.~\ref{sec:metastability} is the analysis of thermal fluctuations and the study of the 
final time-regime in which the system approaches equilibrium escaping from eventual metastable states.
Finally, we close the paper in the concluding Section~\ref{sec:conclusions}.

\section{The model}
\label{subsec:cases}

In the series of studies of the geometry of coarsening systems that we are currently carrying out~\cite{BlCoCuPi14,InCoCuPi16,TaCuPi16,TaCuPi15} 
we focus on models with bimodal variables, $s_i = \pm 1$, placed on the 
vertices of $2d$ lattices with  linear size $L$. 

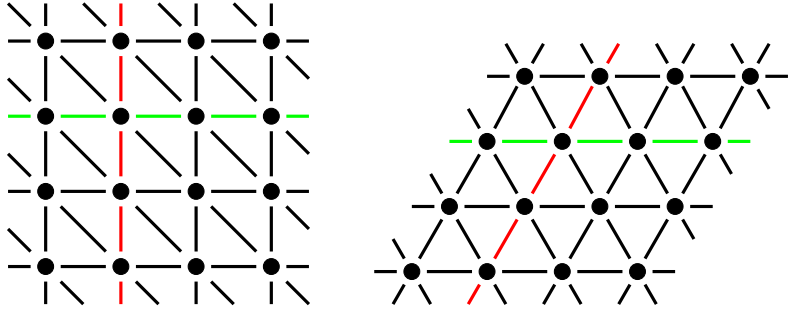
\begin{figure}[h!]
\centering
\begin{tikzpicture}

% First row, starting from bottom with the edges
\draw[very thick]  (0.5,1) -- (0.8,1);
\draw[black,fill] (1.,1.) circle (3pt);
\draw[very thick]  (1.2,1) -- (1.8,1);
\draw[black,fill] (2.,1.) circle (3pt);
\draw[black,very thick]  (2.2,1) -- (2.8,1);
\draw[black,fill] (3.,1.) circle (3pt);
\draw[very thick]  (3.2,1) -- (3.8,1);
\draw[black,fill] (4.,1.) circle (3pt);
\draw[very thick]  (4.2,1) -- (4.5,1);

% Second row
\draw[black,very thick]  (0.5,2) -- (0.8,2);
\draw[black,fill] (1.,2.) circle (3pt);
\draw[black,very thick]  (1.2,2) -- (1.8,2);
\draw[black,fill] (2.,2.) circle (3pt);
\draw[black,very thick]  (2.2,2) -- (2.8,2);
\draw[black,fill] (3.,2.) circle (3pt);
\draw[black,very thick]  (3.2,2) -- (3.8,2);
\draw[black,fill] (4.,2.) circle (3pt);
\draw[black,very thick]  (4.2,2) -- (4.5,2);

% Third row
\draw[color=green,very thick]  (0.5,3) -- (0.8,3);
\draw[black,fill] (1.,3.) circle (3pt);
\draw[color=green,very thick]  (1.2,3) -- (1.8,3);
\draw[black,fill] (2.,3.) circle (3pt);
\draw[color=green,,very thick]  (2.2,3) -- (2.8,3);
\draw[black,fill] (3.,3.) circle (3pt);
\draw[color=green,very thick]  (3.2,3) -- (3.8,3);
\draw[black,fill] (4.,3.) circle (3pt);
\draw[color=green,very thick]  (4.2,3) -- (4.5,3);

% Fourth row
\draw[very thick]  (0.5,4) -- (0.8,4);
\draw[black,fill] (1.,4.) circle (3pt);
\draw[very thick]  (1.2,4) -- (1.8,4);
\draw[black,fill] (2.,4.) circle (3pt);
\draw[black,very thick]  (2.2,4) -- (2.8,4);
\draw[black,fill] (3.,4.) circle (3pt);
\draw[very thick]  (3.2,4) -- (3.8,4);
\draw[black,fill] (4.,4.) circle (3pt);
\draw[very thick]  (4.2,4) -- (4.5,4);

%%%
% 1st to 2nd row
\draw[very thick]  (1.,1.2) -- (1.,1.8);
\draw[color=red,very thick]  (2.,1.2) -- (2.,1.8);
\draw[very thick]  (3.,1.2) -- (3.,1.8);
\draw[very thick]  (4.,1.2) -- (4.,1.8);
% diagonals
\draw[very thick]  (0.8,1.2) -- (0.5,1.5);
\draw[very thick]  (1.8,1.2) -- (1.2,1.8);
\draw[very thick]  (2.8,1.2) -- (2.2,1.8);
\draw[color=black,very thick]  (3.8,1.2) -- (3.2,1.8);

%%%
% 2nd to 3nd row
\draw[very thick]  (1.,2.2) -- (1.,2.8);
\draw[color=red,very thick]  (2.,2.2) -- (2.,2.8);
\draw[very thick]  (3.,2.2) -- (3.,2.8);
\draw[very thick]  (4.,2.2) -- (4.,2.8);
% diagonals
\draw[very thick]  (0.8,2.2) -- (0.5,2.5);
\draw[very thick]  (1.8,2.2) -- (1.2,2.8);
\draw[color=black,very thick]  (2.8,2.2) -- (2.2,2.8);
\draw[very thick]  (3.8,2.2) -- (3.2,2.8);

%%%
% 3nd row to 4th row
\draw[very thick]  (1.,3.2) -- (1.,3.8);
\draw[color=red,very thick]  (2.,3.2) -- (2.,3.8);
\draw[very thick]  (3.,3.2) -- (3.,3.8);
\draw[very thick]  (4.,3.2) -- (4.,3.8);        
% diagonals
\draw[very thick]  (0.8,3.2) -- (0.5,3.5);
\draw[color=black,very thick]  (1.8,3.2) -- (1.2,3.8);
\draw[very thick]  (2.8,3.2) -- (2.2,3.8);
\draw[very thick]  (3.8,3.2) -- (3.2,3.8);

%%%
% Above 4th row
\draw[very thick]  (1.,4.2) -- (1.,4.5);
\draw[color=red,very thick]  (2.,4.2) -- (2.,4.5);
\draw[very thick]  (3.,4.2) -- (3.,4.5);
\draw[very thick]  (4.,4.2) -- (4.,4.5);
% diagonals
\draw[color=black,very thick]  (0.8,4.2) -- (0.5,4.5);
\draw[very thick]  (1.8,4.2) -- (1.5,4.5);
\draw[very thick]  (2.8,4.2) -- (2.5,4.5);
\draw[very thick]  (3.8,4.2) -- (3.5,4.5);

%%%
% Below 1st row
\draw[very thick]  (1.,0.5) -- (1.,0.8);
\draw[color=red,very thick]  (2.,0.5) -- (2.,0.8);
\draw[very thick]  (3.,0.5) -- (3.,0.8);
\draw[very thick]  (4.,0.5) -- (4.,0.8);
% diagonals
\draw[very thick]  (1.2,0.8) -- (1.5,0.5);
\draw[very thick]  (2.2,0.8) -- (2.5,0.5);
\draw[very thick]  (3.2,0.8) -- (3.5,0.5);
\draw[color=black,very thick]  (4.2,0.8) -- (4.5,0.5);

%%%
% On the right of the 4th column
\draw[very thick]  (4.2,1.8) -- (4.5,1.5);
\draw[very thick]  (4.2,2.8) -- (4.5,2.5);
\draw[very thick]  (4.2,3.8) -- (4.5,3.5);

\end{tikzpicture}\quad\quad%
\begin{tikzpicture}
%%%%%%
% Triangular lattice

% First row, starting from bottom with the edges
\draw[very thick]  (0.5,1) -- (0.8,1);
\draw[black,fill] (1.,1.) circle (3pt);
\draw[very thick]  (1.2,1) -- (1.8,1);
\draw[black,fill] (2.,1.) circle (3pt);
\draw[black,very thick]  (2.2,1) -- (2.8,1);
\draw[black,fill] (3.,1.) circle (3pt);
\draw[very thick]  (3.2,1) -- (3.8,1);
\draw[black,fill] (4.,1.) circle (3pt);
\draw[very thick]  (4.2,1) -- (4.5,1);

% 2nd row
\draw[very thick]  (1.,1.866) -- (1.3,1.866);
\draw[black,fill] (1.5,1.866) circle (3pt);
\draw[very thick]  (1.7,1.866) -- (2.3,1.866);
\draw[black,fill] (2.5,1.866) circle (3pt);
\draw[black,very thick]  (2.7,1.866) -- (3.3,1.866);
\draw[black,fill] (3.5,1.866) circle (3pt);
\draw[very thick]  (3.7,1.866) -- (4.3,1.866);
\draw[black,fill] (4.5,1.866) circle (3pt);
\draw[very thick]  (4.7,1.866) -- (5.0,1.866);

% 3rd row
\draw[color=green,very thick]  (1.5,2.732) -- (1.8,2.732);
\draw[black,fill] (2,2.732) circle (3pt);
\draw[color=green,very thick]  (2.2,2.732) -- (2.8,2.732);
\draw[black,fill] (3,2.732) circle (3pt);
\draw[color=green,very thick]  (3.2,2.732) -- (3.8,2.732);
\draw[black,fill] (4,2.732) circle (3pt);
\draw[color=green,very thick]  (4.2,2.732) -- (4.8,2.732);
\draw[black,fill] (5,2.732) circle (3pt);
\draw[color=green,very thick]  (5.2,2.732) -- (5.5,2.732);

% 4th row
\draw[very thick]  (2.,3.598) -- (2.3,3.598);
\draw[black,fill] (2.5,3.598) circle (3pt);
\draw[very thick]  (2.7,3.598) -- (3.3,3.598);
\draw[black,fill] (3.5,3.598) circle (3pt);
\draw[very thick]  (3.7,3.598) -- (4.3,3.598);
\draw[black,fill] (4.5,3.598) circle (3pt);
\draw[very thick]  (4.7,3.598) -- (5.3,3.598);
\draw[black,fill] (5.5,3.598) circle (3pt);
\draw[very thick]  (5.7,3.598) -- (6.,3.598);

% links from 1st to 2nd row
\draw[very thick]  (1.1,1.173) -- (1.4,1.693);
\draw[color=red,very thick]  (2.1,1.173) -- (2.4,1.693);
\draw[very thick]  (3.1,1.173) -- (3.4,1.693);
\draw[very thick]  (4.1,1.173) -- (4.4,1.693);

\draw[very thick]  (0.9,1.173) -- (0.75,1.433);
\draw[very thick]  (1.9,1.173) -- (1.6,1.693);
\draw[very thick]  (2.9,1.173) -- (2.6,1.693);
\draw[color=black,very thick]  (3.9,1.173) -- (3.6,1.693);

% links from 2nd to 3rd row
\draw[very thick]  (1.6,2.039) -- (1.9,2.559);
\draw[color=red,very thick]  (2.6,2.039) -- (2.9,2.559);
\draw[very thick]  (3.6,2.039) -- (3.9,2.559);
\draw[very thick]  (4.6,2.039) -- (4.9,2.559);

\draw[very thick]  (1.4,2.039) -- (1.25,2.299);
\draw[very thick]  (2.4,2.039) -- (2.1,2.559);
\draw[color=black,very thick]  (3.4,2.039) -- (3.1,2.559);
\draw[very thick]  (4.4,2.039) -- (4.1,2.559);

% links from 3rd to 4th row

\draw[very thick]  (2.1,2.905) -- (2.4,3.465);
\draw[color=red,very thick]  (3.1,2.905) -- (3.4,3.465);
\draw[very thick]  (4.1,2.905) -- (4.4,3.465);
\draw[very thick]  (5.1,2.905) -- (5.4,3.465);

\draw[very thick]  (1.9,2.905) -- (1.75,3.165);
\draw[color=black,very thick]  (2.9,2.905) -- (2.6,3.465);
\draw[very thick]  (3.9,2.905) -- (3.6,3.465);
\draw[very thick]  (4.9,2.905) -- (4.6,3.465);

% Above 4th row
\draw[very thick]  (2.6,3.771) -- (2.75,4.031);
\draw[color=red,very thick]  (3.6,3.771) -- (3.75,4.031);
\draw[very thick]  (4.6,3.771) -- (4.75,4.031);
\draw[very thick]  (5.6,3.771) -- (5.75,4.031);

\draw[color=black,very thick]  (2.4,3.771) -- (2.25,4.031);
\draw[very thick]  (3.4,3.771) -- (3.25,4.031);
\draw[very thick]  (4.4,3.771) -- (4.25,4.031);
\draw[very thick]  (5.4,3.771) -- (5.25,4.031);

% Below 1st row
\draw[very thick]  (1.1,0.827) -- (1.25,0.567);
\draw[very thick]  (2.1,0.827) -- (2.25,0.567);
\draw[very thick]  (3.1,0.827) -- (3.25,0.567);
\draw[color=black,very thick]  (4.1,0.827) -- (4.25,0.567);

\draw[very thick]  (0.9,0.827) -- (0.75,0.567);
\draw[color=red,very thick]  (1.9,0.827) -- (1.75,0.567);
\draw[very thick]  (2.9,0.827) -- (2.75,0.567);
\draw[very thick]  (3.9,0.827) -- (3.75,0.567);

% On the right border
\draw[very thick]  (4.6,1.693) -- (4.75,1.433);
\draw[very thick]  (5.1,2.559) -- (5.25,2.299);
\draw[very thick]  (5.6,3.425) -- (5.75,3.165);

\end{tikzpicture}
\caption{\small On the left we show a $4 \times 4$
triangular lattice with PBC constructed from a square lattice by adding diagonal bonds.
This is the graphical way in which we portrayed a triangular lattice in our numerical simulations.
On the right we present the standard representation, with the same lattice spacing.
We also show how an horizontal cycle (depicted in green) and a vertical cycle (in red) 
transform when going from one representation to the other one.
}
\label{fig:triangular}
\end{figure}

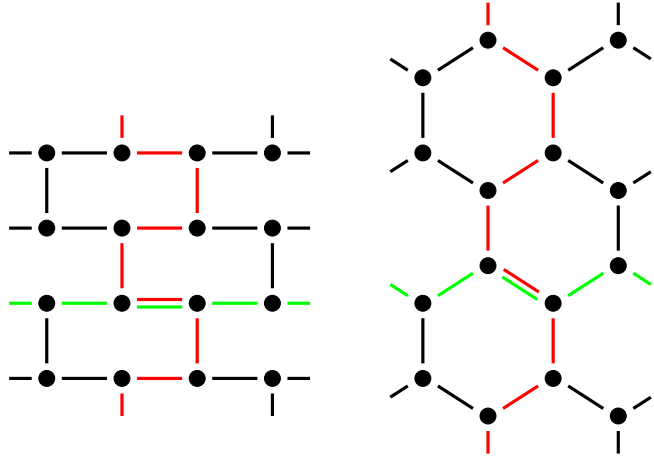
\begin{figure}
\centering
\begin{tikzpicture}
% Square lattice with some links missing
% First raw, starting from bottom with the edges
\draw[very thick]  (0.5,1) -- (0.8,1);
\draw[black,fill] (1.,1.) circle (3pt);
\draw[very thick]  (1.2,1) -- (1.8,1);
\draw[black,fill] (2.,1.) circle (3pt);
\draw[color=red,very thick]  (2.2,1) -- (2.8,1);
\draw[black,fill] (3.,1.) circle (3pt);
\draw[very thick]  (3.2,1) -- (3.8,1);
\draw[black,fill] (4.,1.) circle (3pt);
\draw[very thick]  (4.2,1) -- (4.5,1);

% Second raw
\draw[color=green,very thick]  (0.5,2) -- (0.8,2);
\draw[black,fill] (1.,2.) circle (3pt);
\draw[color=green,very thick]  (1.2,2) -- (1.8,2);
\draw[black,fill] (2.,2.) circle (3pt);
\draw[color=green,very thick]  (2.2,1.95) -- (2.8,1.95);
\draw[color=red,very thick]  (2.2,2.05) -- (2.8,2.05);
\draw[black,fill] (3.,2.) circle (3pt);
\draw[color=green,very thick]  (3.2,2) -- (3.8,2);
\draw[black,fill] (4.,2.) circle (3pt);
\draw[color=green,very thick]  (4.2,2) -- (4.5,2);

% Third raw
\draw[very thick]  (0.5,3) -- (0.8,3);
\draw[black,fill] (1.,3.) circle (3pt);
\draw[very thick]  (1.2,3) -- (1.8,3);
\draw[black,fill] (2.,3.) circle (3pt);
\draw[color=red,very thick]  (2.2,3) -- (2.8,3);
\draw[black,fill] (3.,3.) circle (3pt);
\draw[very thick]  (3.2,3) -- (3.8,3);
\draw[black,fill] (4.,3.) circle (3pt);
\draw[very thick]  (4.2,3) -- (4.5,3);

% Fourth raw
\draw[very thick]  (0.5,4) -- (0.8,4);
\draw[black,fill] (1.,4.) circle (3pt);
\draw[very thick]  (1.2,4) -- (1.8,4);
\draw[black,fill] (2.,4.) circle (3pt);
\draw[color=red,very thick]  (2.2,4) -- (2.8,4);
\draw[black,fill] (3.,4.) circle (3pt);
\draw[very thick]  (3.2,4) -- (3.8,4);
\draw[black,fill] (4.,4.) circle (3pt);
\draw[very thick]  (4.2,4) -- (4.5,4);

% 1st to 2nd raw
\draw[very thick]  (1.,1.2) -- (1.,1.8);
\draw[color=red,very thick]  (3.,1.2) -- (3.,1.8);

% 2nd to 3nd raw
\draw[color=red,very thick]  (2.,2.2) -- (2.,2.8);
\draw[very thick]  (4.,2.2) -- (4.,2.8);

% 3nd raw to 4th raw
\draw[very thick]  (1.,3.2) -- (1.,3.8);
\draw[color=red,very thick]  (3.,3.2) -- (3.,3.8);

% Above 4th raw
\draw[color=red,very thick]  (2.,4.2) -- (2.,4.5);
\draw[very thick]  (4.,4.2) -- (4.,4.5);

% Bellow 1st raw
\draw[color=red,very thick]  (2.,0.5) -- (2.,0.8);
\draw[very thick]  (4.,0.5) -- (4.,0.8);

%  Honeycomb lattice 

% 1st raw, starting from bottom
\draw[very thick]  (5.567,0.75) -- (5.8,0.9);
\draw[black,fill] (6.,1.) circle (3pt);
\draw[very thick]  (6.2,0.9) -- (6.666,0.6);
\draw[black,fill] (6.866,0.5) circle (3pt);
\draw[color=red,very thick]  (7.066,0.6) -- (7.532,0.9);
\draw[black,fill] (7.732,1.) circle (3pt);
\draw[very thick]  (7.932,0.9) -- (8.398,0.6);
\draw[black,fill] (8.598,0.5) circle (3pt);
\draw[very thick]  (8.798,0.6) -- (9.031,0.75);

\draw[very thick]  (6.,1.2) -- (6.,1.8);
\draw[color=red,very thick]  (7.732,1.2) -- (7.732,1.8);

\draw[color=green,very thick]  (5.567,2.25) -- (5.8,2.1);
\draw[black,fill] (6.,2.) circle (3pt);
\draw[color=green,very thick]  (6.2,2.1) -- (6.666,2.4);
\draw[black,fill] (6.866,2.5) circle (3pt);
\draw[color=green,very thick]  (7.056,2.35) -- (7.522,2.05);
\draw[color=red,very thick]  (7.076,2.45) -- (7.542,2.15);
%\draw[very thick]  (7.066,2.4) -- (7.532,2.1);
\draw[black,fill] (7.732,2.) circle (3pt);
\draw[color=green,very thick]  (7.932,2.1) -- (8.398,2.4);
\draw[black,fill] (8.598,2.5) circle (3pt);
\draw[color=green,very thick]  (8.798,2.4) -- (9.031,2.25);

\draw[color=red,very thick]  (6.866,2.7) -- (6.866,3.3);
\draw[very thick]  (8.598,2.7) -- (8.598,3.3);

\draw[very thick]  (5.567,3.75) -- (5.8,3.9);
\draw[black,fill] (6.,4.) circle (3pt);
\draw[very thick]  (6.2,3.9) -- (6.666,3.6);
\draw[black,fill] (6.866,3.5) circle (3pt);
\draw[color=red,very thick]  (7.066,3.6) -- (7.532,3.9);
\draw[black,fill] (7.732,4.) circle (3pt);
\draw[very thick]  (7.932,3.9) -- (8.398,3.6);
\draw[black,fill] (8.598,3.5) circle (3pt);
\draw[very thick]  (8.798,3.6) -- (9.031,3.75);

\draw[very thick]  (6.,4.2) -- (6.,4.8);
\draw[color=red,very thick]  (7.732,4.2) -- (7.732,4.8);

\draw[very thick]  (5.567,5.25) -- (5.8,5.1);
\draw[black,fill] (6.,5.) circle (3pt);
\draw[very thick]  (6.2,5.1) -- (6.666,5.4);
\draw[black,fill] (6.866,5.5) circle (3pt);
\draw[color=red,very thick]  (7.066,5.4) -- (7.532,5.1);
\draw[black,fill] (7.732,5.) circle (3pt);
\draw[very thick]  (7.932,5.1) -- (8.398,5.4);
\draw[black,fill] (8.598,5.5) circle (3pt);
\draw[very thick]  (8.798,5.4) -- (9.031,5.25);

\draw[color=red,very thick]  (6.866,5.7) -- (6.866,6.);
\draw[very thick]  (8.598,5.7) -- (8.598,6.);

\draw[color=red,very thick]  (6.866,0.0) -- (6.866,0.3);
\draw[very thick]  (8.598,0.0) -- (8.598,0.3);

\end{tikzpicture}
\caption{\small
On the left we show a $4 \times 4$ honeycomb lattice with PBC constructed from the square lattice  
by removing some of the vertical bonds, as described in the main text. On the right we present the standard image, 
with the same lattice spacing.
Here again we show how an horizontal cycle (depicted in green) and a vertical cycle (in red) 
transform when going from one representation to the other one.
In the standard presentation, the width of the lattice along the horizontal direction is $W_{x} = \sqrt{3}/{2} \; L $, while
the width along the vertical direction is $W_{y} = 3/2 \; L $, with $L=4$ in this particular case.
This corresponds to an aspect ratio $W_{y}/W_{x} = \sqrt{3}$.
}
\label{fig:honeycomb}
\end{figure}

The ferromagnetic $2d$ Ising model Hamiltonian is defined by the Hamiltonian
\begin{equation}
H_J[\{s_i\}] = -J \sum_{\langle ij\rangle} s_i s_j
\label{eq:2dIM}
\end{equation}
with $J>0$ and the sum running over nearest-neighbours on the lattice (each pair counted once).
We consider three types of lattices: triangular, square and honeycomb. The former has connectivity $n_c=6$, the 
intermediate one $n_c=4$, and the latter $n_c=3$. 
In our numerical simulations, we constructed the triangular and honeycomb lattices from a square lattice in the following manner. 
We built the triangular lattice  by adding a diagonal bond between 
the position $(i,j)$ and $(i+1,j+1)$, see Fig.~\ref{fig:triangular}. To create the
honeycomb lattice we removed the bond between each site $(i,j)$ and its neighbour $(i,j+1)$ if $i+j$ is an even number, 
and the bond between each site $(i,j)$ and its neighbour $(i,j-1)$ otherwise, see Fig.~\ref{fig:honeycomb}.
The number of vertices is always $N=L\times L$, and we take either
free boundary conditions (FBC) or periodic boundary conditions (PBC). 
This model undergoes a second order phase transition at
a critical  temperature, $T_c$, and, for $J=1$,
$\beta^{\rm sq}_c = 1/(k_BT^{\rm sq}_c) = {1\over 2} \ln{(1+\sqrt{2})} \simeq 0.44$ on the square
lattice, $\beta^{\rm tr}_c = 1/(k_BT^{\rm tr}_c) = \frac{1}{4} \ln{3} \simeq 0.28$ on the triangular lattice, and
$\beta^{\rm honey}_c = 1/(k_BT^{\rm honey}_c) = \frac{1}{2} \ln{(2+\sqrt{3})} \simeq 0.66$ on the honeycomb lattice.
The initial condition is always taken to be a random state with no correlations, obtained by choosing $s_i =+ 1$
or $s_i = -1$ with probability $1/2$ on each lattice site (long-range correlated initial conditions, as the ones of the critical 
Ising point fall in a different class~\cite{ArBrCuSi07,BrayHumayunNewman,ChakrabortyDas,Corberi16}). 
Under a mapping to occupation numbers,
{\large $\frac{1+s_i}{2}$}, this state corresponds to a realisation of site percolation with $p=1/2$. It is therefore right at
the critical percolation point for the triangular lattice and below the critical percolation points in the other 
two cases since $p_c^{\rm tr}=1/2$, $p_c^{\rm sq} \approx 0.59$ and $p_c^{\rm honey} \approx 0.69$.

We consider kinetic local Monte Carlo (similar to local Glauber dynamics) for the spin updates. These rules
satisfy detailed-balance and do not conserve the order parameter.
We also study the effect of a non-vanishing working temperature. More details on the 
implementation of the numerical algorithm, and its comparison to the Glauber transition probabilities in the 
master equation formalism, are given in App.~\ref{app:MC}.

\section{Observables}
\label{sec:observables}

We now list all the observables that we will use in this study. We will choose some quantities among this list to define and 
characterise three growing lengths that control (i) the approach to stable percolation, (ii) the curvature driven 
coarsening processes with usual dynamic scaling, and (iii) the approach to equilibrium that includes, in certain cases, 
a escape from metastability. 
 
 The averaged {\it magnetisation} density in absolute value is defined as 
 \begin{equation}
 m(t,L) = \frac{1}{L^2} \left| \sum_{i=1}^{L^2} \langle s_i(t) \rangle \right|
 \end{equation}
 and the averaged total magnetisation is $M(t,L) = L^2 \, m(t,L)$. 
 Here, and in what follows, $\langle \dots \rangle$ represents an average 
 over initial conditions and/or stochastic dynamic paths.
 
 In the case of the ferromagnetic Ising model, 
 we define a {\it growing length} as the inverse of the excess energy,
\begin{equation}
\ell_{G}(t) = \frac{E_{\mathrm{eq}}(T) }{ E_{\mathrm{eq}}(T) - E(t)}
\label{eq:growing-length-excess-energy}
\; , 
\end{equation}
with $E(t)$ the energy of the dynamic configuration 
evaluated from the Hamiltonian~(\ref{eq:2dIM}), and 
$E_{\mathrm{eq}}(T)$ the equilibrium energy of the Ising model at temperature $T$.
(We did not write explicitly the dependence on $L$ due to finite size corrections here.)
The excess energy is concentrated on the broken bonds. 
For example, for the ground state of the Ising model on a square lattice with linear size $L$,
$E_0 = E_{\mathrm{eq}}(T=0) = -2 \, L^2$ (since we add each bond over nearest neighbours once and we set $J=1$).
In all the cases that we are going to present in this article, the temperature $T$ at which the system evolves
under the Monte Carlo heat bath rule is at most
$T_c/2$, with $T_c$ the critical temperature, and $E_{\mathrm{eq}}(T_c/2)$ is very close to the ground state energy.
For example, in the case of the square lattice, $E_{\mathrm{eq}}(T_c/2)/L^{2} \simeq -1.99$.

In the paramagnetic initial state, $E(0) \simeq 0$ and $\ell_{G}(0) \simeq 1$.
As the system approaches thermal equilibrium at the target temperature after the quench,  the growing length increases and 
approaches the system size. We will take $\ell_G(t)$ as 
our estimate for the usual dynamic growing length $\ell_d(t)$. The distinction between $\ell_G(t)$ and the theoretically expected 
$\ell_d(t) \simeq t^{1/2}$ is especially important at very early times, 
when the system is approaching critical percolation.

The average {\it overlap} between two replicas is defined as
\begin{equation}
Q(t,t_w;L) = \frac{1}{L^2} \sum_{i=1}^{L^2} \langle s_i (t) \sigma^{(t_{w})}_i(t)\rangle 
\label{eq:overlap}
\end{equation}
where $\{ \sigma^{(t_{w})}_i \}$ is a replica of the system $\{ s_i \}$ ``created'' at the time $t_w$, that evolves with an independent thermal noise
for $t>t_{w}$. More precisely, $\sigma^{(t_{w})}_i(t) = s_i(t)$ for $t\leq t_w$, while for $t>t_w$ the two spin configurations $\{ s_i \}$
and$\{ \sigma^{(t_{w})}_i \}$ evolve with two completely independent realizations of the spin-flip dynamics.
%where $s_i(t_w) = \sigma_i (t_w)$ are two copies of the system created at time $t_w$ that evolve independently 
%after this instant. 

This quantity should approach 
\begin{eqnarray}
Q(t,t_w; L) \underset{t\gg t_w}{\longrightarrow} 
\left\{ 
\begin{array}{ll}
\mbox{const} >0 \qquad & t_w > t_p(L)
\\
0 \qquad & t_w < t_p(L) 
\end{array} 
\right.
\label{eq:def-Q}
\end{eqnarray}
and it was used in~\cite{BlCoCuPi14} to estimate $t_p(L)$, the time after which the percolating structure no longer changes,
in the Ising model with kinetic Monte Carlo dynamics with non-conserved order parameter.

We will not spend much time discussing {\it persistence}, but we will just measure the exponent that characterises 
its decay in time to refute claims in the literature for its identity with the one of the vanishing 
waiting time overlap, $Q(t,0;L)$.
As a reminder, persistence is a measure of the ``resilience'' of a reference state, in this case the initial one.  For spin models 
it is defined as the probability that a spin chosen at random has never flipped during the interval
that goes from the reference time, say, the initial time $t=0$,  to a measuring time $t$~\cite{DBG0,BMS}.

A {\it cluster} or {\it domain} is a set of spins with the same sign that are connected by nearest-neighbour bonds.
Its area $A$ is, simply, the number of sites that belong to it.
The {\it interface} between two domains of opposite order is defined on the lattice by following the 
nearest-neighbour broken bonds, that is to say, the links between sites with anti-parallel spins. Its 
length $l$ is also an interesting observable.

\begin{figure}
\begin{center}
\subfloat[]{\includegraphics[scale=0.25]{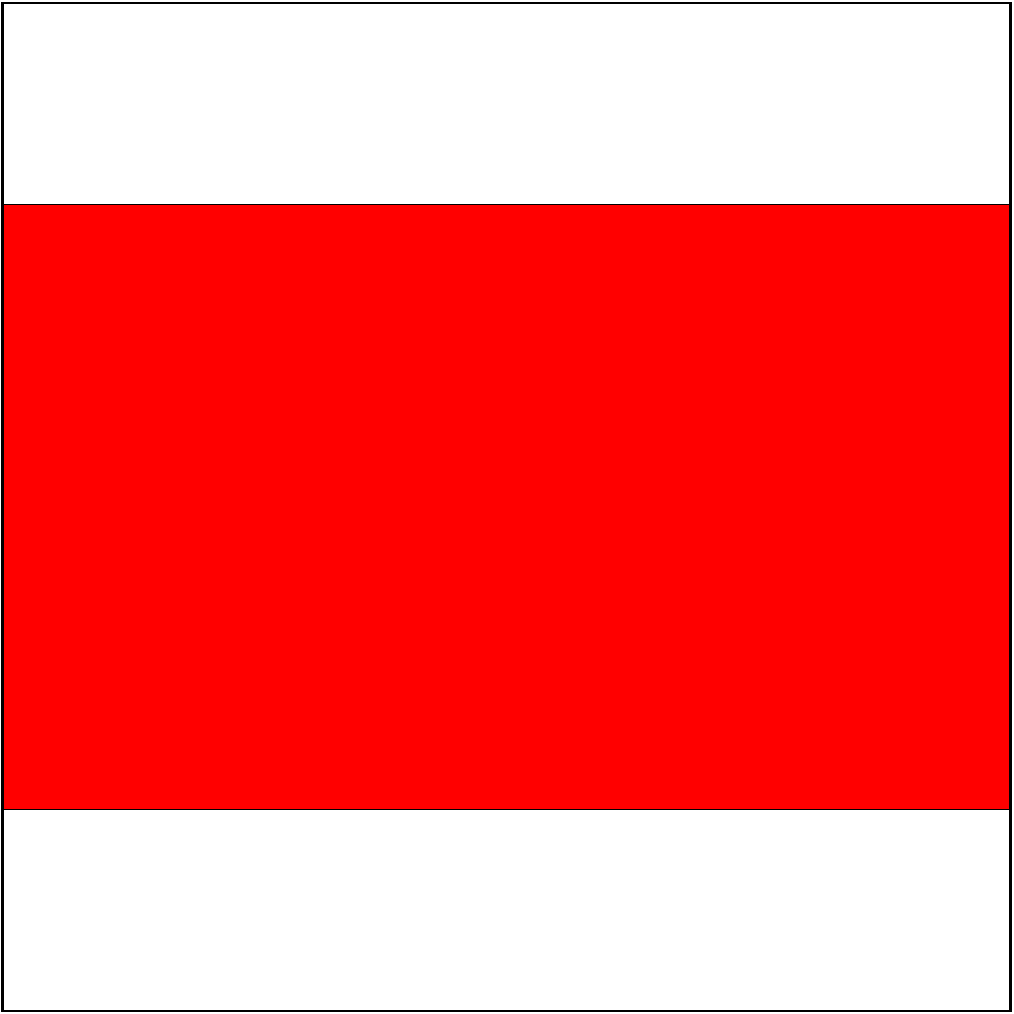}}\quad%
\subfloat[]{\includegraphics[scale=0.25]{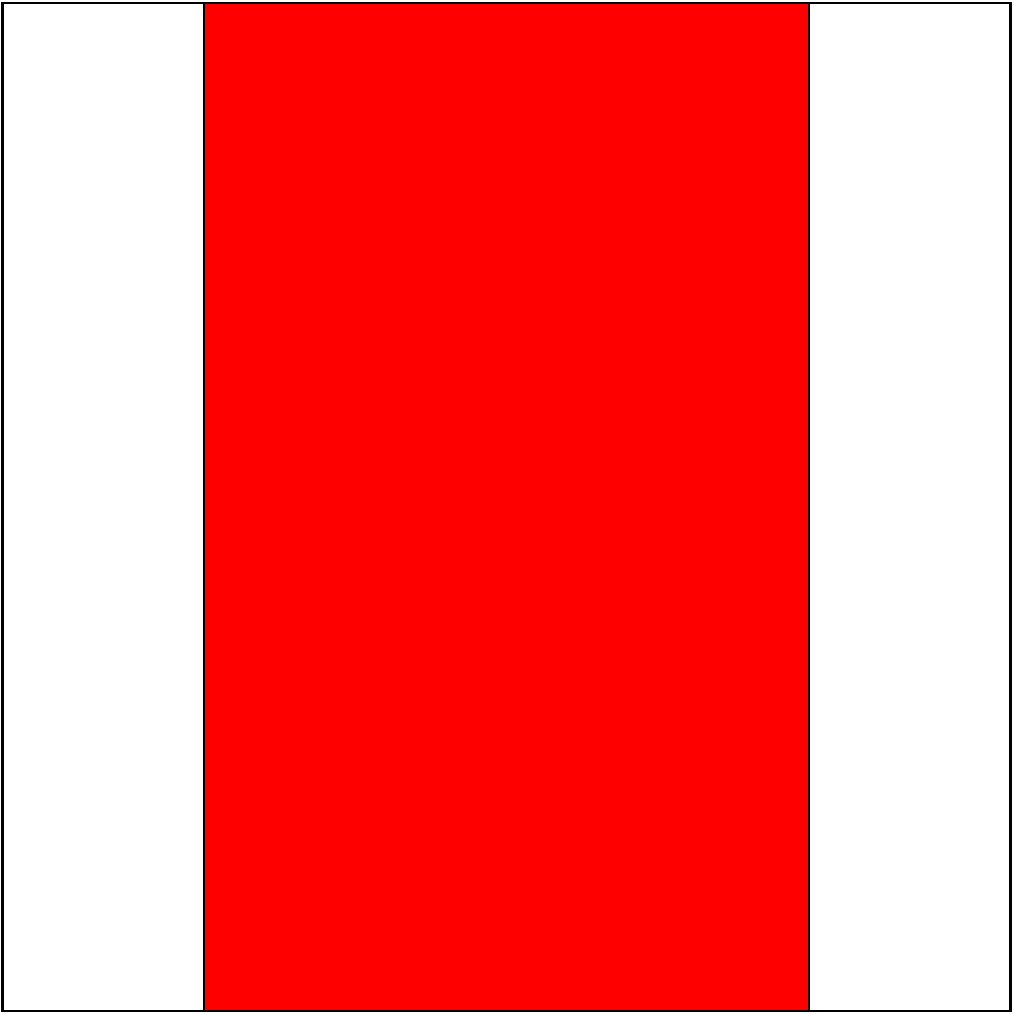}}\quad%
\subfloat[]{\includegraphics[scale=0.25]{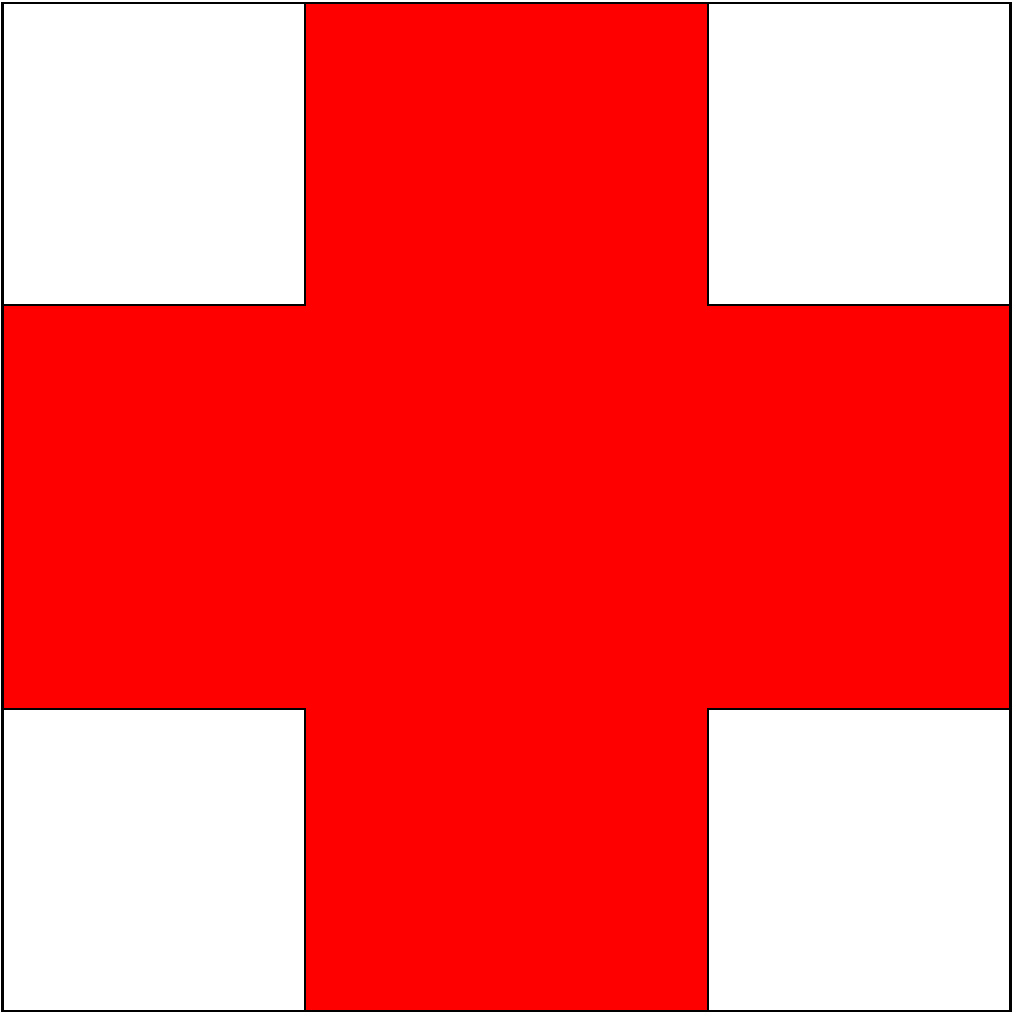}}\quad%
\subfloat[]{\includegraphics[scale=0.25]{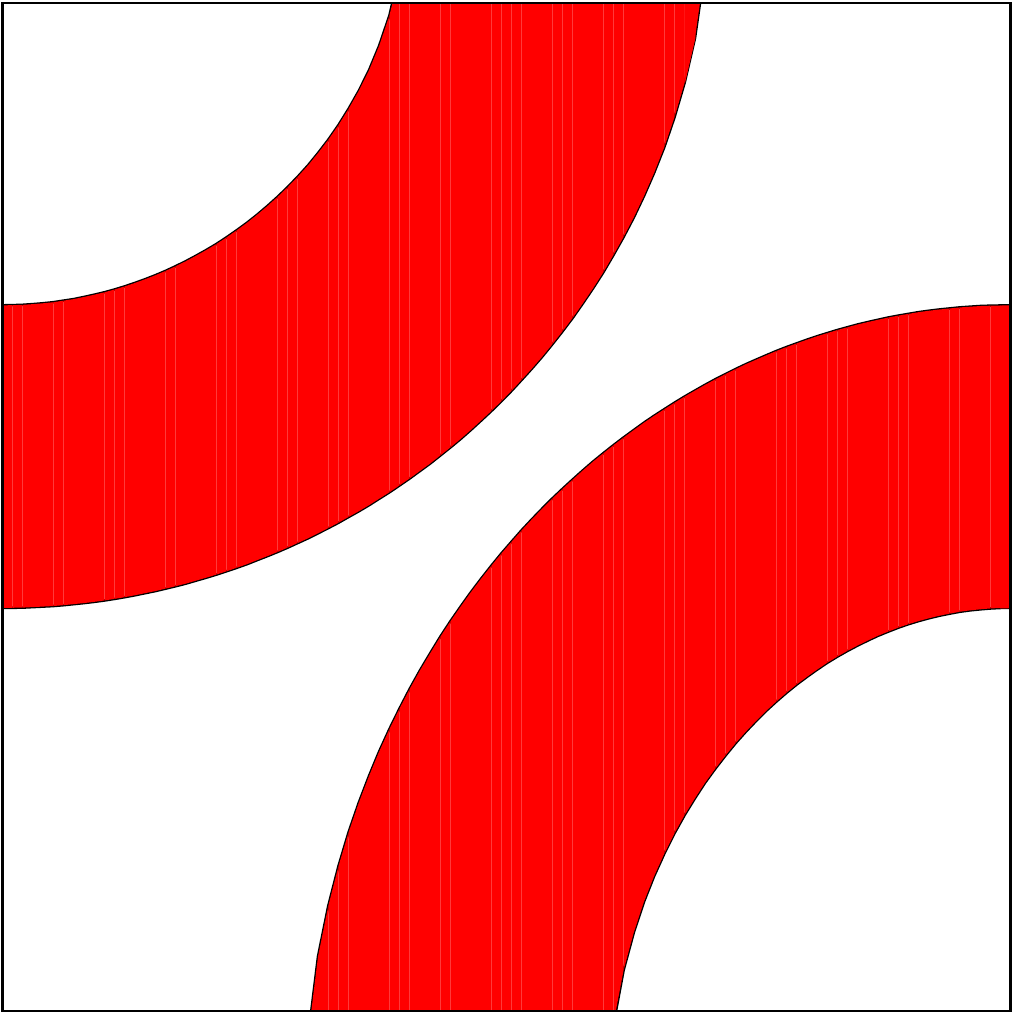}}\quad%
\end{center}
\caption{\small Sketches of wrapping clusters on a lattice with unit aspect ratio and PBC (i.e.,~on a torus).
The panels show in red clusters spanning the system horizontally, vertically, both horizontally and vertically, and 
diagonally. In the first, second and fourth cases, the topology of the red clusters implies 
the existence of a white percolating cluster next to them. On the contrary, in the third case the red cluster percolating in both directions
forbids the existence of other spanning clusters. 
%The dashed lines are examples of curves on the red cluster that wind around the system.
}
\label{fig:sketch}
\end{figure}

We must now give a proper definition of {\it percolating configurations} on a finite-size system 
and distinguish different possibilities. Let us first focus on PBC, i.e., a model defined on a torus, with a toroidal and a poloidal direction depicted as
horizontal and vertical directions when picturing the torus as a $2d$ sheet, see Fig.~\ref{fig:sketch}. 
A spin configuration percolates if there is at least one spin cluster that \textit{wraps} around the system, that is to say,
that winds around at least one of the two directions of the torus. 
%The dashed lines in Fig.~\ref{fig:sketch} represent winding curves inside the red clusters.
The wrapping cluster is separated by one or more interfaces from one or more clusters of the opposite phase and,
on a torus, all interfaces are closed. 
The interfaces can be homotopic to a point, as in panel ($\mbox{c}$) in  Fig.~\ref{fig:sketch}, 
or they can wind around the torus as in panels (a), (b) and (d) in the same figure. 
In general an interface can wind $a$ times across the toroidal direction and $b$ times around the poloidal direction:
$a$, $b$ take integer values, with the sign indicating if the curve is winding in the clockwise or anti-clockwise direction 
around the torus (only for cases in which it winds along both directions).
One can easily check that
$|a|$ and $|b|$ cannot be simultaneously larger than $2$. If one of the two is zero the other one is at most $1$.
To each of these configurations is associated a probability in $2d$ continuum critical percolation which we denote
by $\pi^{p}(a,b)$, following the notation in Ref.~\cite{Pi94}.

Thus, we can distinguish four different situations (see also Fig.~\ref{fig:sketch}):

\begin{itemize}
 \item A configuration with no wrapping cluster, with  probability denoted by $\pi^{p}(0)=\pi^{p}(0,0)$.
 \item A configuration that contains a cluster wrapping in both directions (which we also refer to as cross topology), 
with probability  denoted by $\pi^{p}(Z \times Z)$: starting from a point on the cluster, one can go around the torus as many times as
desired along both cycles and come back to the starting point.
 \item A configuration that contains a cluster wrapping only along one direction, meaning either $a=1$ and $b=0$ or $a=0$ and $b=1$
(that is to say, horizontal or vertical stripes).
 \item A configuration that contains a cluster wrapping in both directions but that does not self-intersect, i.e., for example $a=1$ and $|b|\ge1$:
in many cases we will refer to this situation as a diagonally striped configuration.
\end{itemize}

In our spin problems in which plus and minus spins are equivalent 
 $\pi^p(0) = \pi^{p}(Z \times Z)$ since a configuration that contains only non-percolating clusters of up spins necessarily contains 
a cluster of down spins percolating in both directions with a cross geometry, see Fig.~\ref{fig:sketch}~($\mbox{c}$).
For a lattice with unit aspect ratio, $\pi^p(a,b) = \pi^p(b,a) = \pi^p(a, -b)$.

The $\pi^p$'s have been calculated and checked numerically by Pinson~\cite{Pi94} for site critical percolation on lattices with unit aspect ratio and 
PBC. We report here the values of the $\pi^p$'s in the case
of a rectangular sheet of aspect ratio $1$: $\pi^{p}(0) = \pi^{p}(Z \times Z) \simeq 0.3095 $, $\pi^p(1,0) = \pi^p(0,1) \simeq 0.1694 $ and
$\pi^p(1,1) = \pi^p(1,-1) \simeq 0.0209$.
Since we will present data relative to the honeycomb lattice, we also mention here that, because of the way in which we constructed this lattice 
(see Fig.~\ref{fig:honeycomb}),
its aspect ratio is equal to $\sqrt{3}$ and the probabilities are \cite{BaKrRe09,PruMol04}
$\pi^p(0) =\pi^p( Z \times Z) \simeq 0.2560, \pi^p(1,0) \simeq 0.4221, \pi^p(0,1) \simeq 0.0408$,  and $\pi^p(1,1) = \pi^p(1,-1) \simeq 0.0125$
(for us the vertical direction is the longer one in our convention).

As far as our study is concerned, we expect that after a sufficiently long time after the quench 
the system takes one of the percolating configurations
above described. Accordingly, we introduce time-dependent probabilities that we will compute along the evolution. These are: 
the probability of having a cluster percolating in both directions with a cross topology, $\pi_{\mathrm{hv}}$, the probabilities
of having a cluster wrapping only horizontally or only vertically, $\pi_{\mathrm{h}}$ and  $\pi_{\mathrm{v}}$ respectively, and
the probability of having a cluster wrapping in both directions in what we call a diagonally striped configuration, $\pi_{\mathrm{diag}}$.
When the system enters the percolation regime, these time-dependent quantities should become constant and equal to the values 
at  $2d$ critical percolation which, in the case of a lattice with unit aspect ratio, are given by
\begin{eqnarray}
 \quad\;\; \pi_{\mathrm{hv}}  &=& \; \pi^p(0) + \pi^{p}(Z \times Z)  \; \simeq \; 0.6190 \; , \\
  \pi_{\mathrm{h}} + \pi_{\mathrm{v}} &=& \; \pi^p(1,0) + \pi^{p}(0,1) \;\;\, \simeq \; 0.3388 \; , 
  \label{eq:wrapping-probabilities-critical-perc}
  \\
  \quad \pi_{\mathrm{diag}} &=& \sum_{a=1}^{\infty} \sum_{b=1}^{\infty} \pi^p(a,b) \; +  \sum_{a=1}^{\infty} \sum_{b=-\infty}^{-1}  \pi^p(a,b) \\
  &\simeq& \; \pi^p(1,1) + \pi^p(1,-1)  \; \simeq \; 0.0418 \; . 
\end{eqnarray}
Note that, since we are dealing with Ising spin clusters, both $\pi^p(0)$ and $\pi^{p}(Z \times Z)$ contribute to $\pi_{\mathrm{hv}}$ by complementarity of the two
phases. Moreover, the probability of diagonal stripes, $\pi_{\mathrm{diag}}$, is rather small, of the order of $10^{-2}$, and the main
contributions to the series  come from the first two terms
indicated above. The remaining part is, in fact, of order $10^{-4}$. On lattices with aspect ratio different from one 
the wrapping probabilities $\pi_{\rm h}, \ \pi_{\rm v}, \ \pi_{\rm hv}$ and $\pi_{\rm diag}$ take different values and  
we will recall them in later Sections when necessary.

One can introduce similar probabilities in the case of FBC.
In this case, a spin cluster  percolates if there is a path of connected sites belonging to the cluster that crosses the system
from one border to the opposite one. The distinction between different geometries still applies.
The spanning probabilities have been computed by Cardy~\cite{Cardy92}
and Watts~\cite{Watts96}. In particular, later we will need the value on a square lattice with unit aspect ratio : 
$ \pi^{\rm FBC}_{\rm hv}  =  1/2 + \sqrt{3}/(2 \,\pi) \, \ln {(27/16)} \simeq 0.6442$. 
 
%%%%%%%%%%%%%%%%%%%%%%%%%%%%%%%%%%%%%%%%%%%%%%%%%%%%%%%%%%%%%%%%%%

Other interesting observables are the {\it area of the largest cluster} and the {\it length of its interface}.
Actually, while the definition of the area of a cluster is unequivocal, the interface of the cluster admits
several nonequivalent definitions. In this paper we will use two choices. One is the definition of the {\it hull}, that is to say, the  external 
boundary of the cluster constructed by joining the centres of the dual lattice by links that cut broken bonds between the cluster in question 
and its neighbour. Another definition will include the internal boundaries between the chosen 
cluster and clusters of the opposite phase that lie within it. As we will explain in the text we 
found that the length that better characterises the approach to critical percolation is the one of the 
hull of the largest cluster and we therefore focused on it.

In $2d$ critical percolation the largest cluster is a fractal object, thus both its area, $A_c$, and interface hull length, $l_c$,
are related to its linear size $l$ by fractal dimensions:
\begin{equation}
A_c \simeq l^{D_A} \; , 
\qquad\qquad
l_c \simeq l^{D_\ell}
\; , 
\end{equation}
with $D_A$ the surface fractal dimension and $D_\ell$ the interface fractal dimension. 
These dimensions can be exactly computed for the critical points of the $q$-state Potts model in two dimensions
 for $0 < q \le 4$ (where $q=2$ for the Ising model and $q\rightarrow 1$ for percolation)
through a Coulomb gas formulation~\cite{SaDu87}. 
The  parameter $\kappa$, related to $q$ through $\sqrt{q} = - 2 \cos{\left( 4 \pi/\kappa\right)}$,  
determines the universality class of the model near criticality. The  above-mentioned fractal dimensions are then expressed 
in the following form
\begin{equation}
 D_A = 2 - \frac{\beta}{\nu} = 1 + \frac{3\kappa}{32} + \frac{2}{\kappa}
 \; , 
 \qquad\qquad
 D_\ell = 1+ \frac{\kappa}{8}
 \; , 
 \end{equation}
where $\beta$ is the critical exponent of the order parameter and $\nu$ the one of the
equilibrium correlation length. $D_\ell$ is the hull fractal dimension.
For critical percolation $\kappa=6$~\cite{Smirnov01} and thus 
\begin{equation}
D_A = \frac{91}{48} \simeq 1.8958 \; , 
\qquad \qquad 
D_\ell = \frac{14}{8} = 1.75
\; . 
\end{equation}
We will show the evolution in time of the observables
$A_c$ and $l_c$ for the different types of coarsening dynamics described in Sec.~\ref{subsec:cases}
and compare the geometric properties of the dynamic largest cluster to the ones of the largest cluster at critical percolation.

We will also focus our attention on the statistics of domain areas. In particular we will show results regarding
the {\it number density of cluster areas} (also referred to as the distribution of cluster sizes), which
we denote by ${\mathcal N}(A,t, L)$. In general, for a finite-size system ${\mathcal N}(A,t, L)$ is given by the sum of two contributions
\begin{equation}
{\mathcal N}(A,t, L) \simeq N(A,t) + N_p(A,t,L)
\; , 
\label{eq:area-number-density}
\end{equation}
with the first term describing the weight of the finite areas and the second one the 
weight of the areas that span the sample.
At $t_p$ the last term should scale with $A/L^{D_A}$ with 
$D_A$ the fractal dimension of the percolating cluster, 
and keep a weak time-dependence, due to coarsening, that
essentially drives the system towards the equilibrium final state.
Concomitantly,  the number density of finite size clusters should have an algebraic decay similar to the one at critical percolation
\begin{equation}
N(A) \, = \, 2 c_d \ A^{-\tau_A}
\; , 
 \label{eq:na_eq}
\end{equation}
with $\tau_A $ a characteristic exponent related to $D_A$ by \cite{Stauffer94} 
\begin{equation}
\tau_A = 1+\frac{d}{D_A} = \frac{187}{91} \approx 2.0549 \ .
\end{equation} 
The normalisation constant has been computed exactly for 
hull-enclosed areas with the result $c_h=1/(8\pi\sqrt{3}) \approx 0.0229$~\cite{CaZi03} and the same factor $2$ in Eq.~(\ref{eq:na_eq}) 
due to the fact that there are two types of hull-enclosed areas (spins up and down) in the magnetic problem while there 
is only one kind (occupied sites) in the percolation problem. For the normalisation of the domain area distribution, there is no exact 
result. In~\cite{SiArBrCu07} the notation $2c_d$ for the pre-factor in the numerator was used.
The use of two sum rules, the facts that the total domain area should equal $L^2$, and that the total number of
domains is necessarily equal to the total number of hull-enclosed areas, yields $c_d = (\tau_A-2) (\tau_A-1)/2\approx 0.0289$ at first 
order in an expansion in $c_h$~\cite{SiArBrCu07}. Therefore, 
\begin{equation}
2c_d \simeq 0.0579 
\label{eq:cd-value}
\; .
\end{equation}
In this paper we will pay special attention to the way in which the finite-size area regime matches the one 
for the percolating clusters.

The percolation  hulls are,  in
the continuum limit, conformally invariant curves described by a stochastic Loewner
evolution SLE$_\kappa$, where the parameter $\kappa$ is the same as in the Coulomb gas representation mentioned above.
It can be determined numerically by computing the {\it variance of the winding angle}, $\langle \theta^2(x)\rangle$.
The winding angle $\theta(x)$, for two points chosen at random at a curvilinear distance $x$
along a curve, is defined as the angle between the lines that are tangent to the curve at those two points. 
(On a lattice the local tangent to a hull that separates two domains of opposite spin orientations
is a vector perpendicular to the {\it broken} bond at every point of the hull. 
Consequently, only a finite number of tangent directions are possible.
For instance, on a square lattice there are four directions. Of course, after averaging,  $\theta^{2}(x)$,
at any curvilinear distance $x$, becomes a real-valued function of $x$.)
For critical systems in two dimensions, this quantity is related to the fractal dimension of the curve
and to the parameter $\kappa$ associated to the universality
class~\cite{DuSa88,WiWi03} through
\begin{equation}
\langle \theta^2(x) \rangle = \mbox{cst} +{4 \kappa \over 8+\kappa } \ln{x} \; .
\label{ws}
\end{equation}
For critical percolation hulls, one should recover $\kappa = 6$ from these measurements. 
For comparison, for the critical Ising model, $\kappa = 3$, a very different value. This quantity should then be a good test to distinguish 
critical percolation from other types of criticality. 
In the case of PBC,
the average square winding angle can be computed for domain walls that wrap around the lattice (with zero average curvature) or for 
non-wrapping domain walls (with non-zero average curvature): both cases yield the same result
for sufficiently long domain walls and large system sizes. Moreover, we are interested in
the time-evolution of $\langle \theta^2(x,t)\rangle$ and its scaling behaviour.

Another interesting quantity is the two-time correlation function of what we call the {\it crossing number}.
We define the crossing number ${\rm n}_c(t)$ at a time $t$ as follows: 
if there exists $a$ horizontal crossing clusters and no vertical crossing cluster, ${\rm n}_c(t) = a$.
If there exists $a$ vertical crossing clusters and no horizontal crossing cluster, then ${\rm n}_c(t) = -a$. 
At sufficiently late times these two cases have $|a| \geq 2$.
For a configuration with a (unique) cluster crossing in both directions ${\rm n}_c(t)=1$, 
while for a configuration with no crossing cluster  ${\rm n}_c(t)=0$. 
We then define the correlation function of ${\rm n}_c$ as 
\begin{equation}
 {\cal O}_c(t,t') = \langle \delta_{{\rm n}_c(t),{\rm n}_c(t')} \rangle \;
 \label{eq:cross_correl}
\end{equation}
where $ \delta_{n,m}$ is the Kronecker delta.
We are particularly interested in the correlation between the 
crossing number at a given time $t$ and the one in the final
state of the system, i.e., in  the 
limit $t' \rightarrow \infty$ of ${\cal O}_c(t,t')$.
Thus we define ${\cal O}^{\infty}_c(t)=\lim_{t'\to\infty} {\cal O}_c(t,t')$. 
In the case of the relaxation dynamics following a quench from $T_0\to\infty$ to $T=0$ of the Ising model on the square lattice,
this function interpolates between $0$ and $1$ since at $t=0$ 
all the spin configurations are such that ${\rm n}_c(0)=0$ (for not too small lattice size), 
while in the final state $\lim_{t\rightarrow \infty} {\rm n}_c(t)  \neq 0$. This 
quantity is sensitive to $t_p$ 
since for all $t, t' > t_p$, ${\rm n}_c(t)={\rm n}_c(t')$.

In the framework of percolation theory, a useful tool to study the geometrical properties of clusters of occupied sites is the 
{\it pair connectedness function}, $g(r)$. 
This quantity is defined as the probability that
two lattice sites separated by a distance $r$ belong to the same cluster. At critical percolation in two dimensions, the behaviour of $g(r)$ for
large $r$ ($r \gg r_0$, with $r_0$ the lattice spacing) is known~\cite{Stauffer94,ChristensenMoloney,Saberi15}
\begin{equation}
 g(r) \sim r^{-2\Delta_\sigma}, \quad	r\gg r_0
\end{equation}
where $\Delta_\sigma=2-D_A$, $D_A$ being the fractal dimension of critical percolation clusters.

In order to assess the presence of a critical-percolation-like regime in the coarsening process occurrying in the quench dynamics of the Ising model,
we introduce an analogous quantity for a spin system. On a square lattice:
\begin{equation}
 g(r,t) = \frac{1}{4 L^{2}} \sum_{i} \sum_{i_{r}} \langle \gamma_{i,i_{r}}(t) \rangle 
\end{equation}
where the first summation is taken over all the lattice sites, the second over the four sites $i_{r}$ that are located at distance $r$ from site $i$
along the horizontal and vertical directions, and $\gamma_{i,j}(t)=1$ if the sites $i$ and $j$
belong to the same spin cluster (are occupied by the same spin and there is a path of sites with the same spin connecting the two sites) 
at time $t$, and equals $0$ otherwise.

\vspace{0.5cm}

The Monte Carlo step is the implicit time-unit in all our presentation. 

\section{The phenomenon}
\label{sec:phenomenon}

In this Section we illustrate, with the discussion of the snapshots and the 
presentation of just two observables, the phenomenon. 
The largest cluster area and the 
pair connectedness correlation function are the observables that provide the clearest evidence for critical 
percolation. The asymptotic dynamic growing length in this problem is
$\ell_d(t) =[\lambda(T) t]^{1/2}$ with $\lambda(T=0)\approx 2$ and a very slowly decreasing function of 
temperature~\cite{SiArBrCu07,Arenzon-etal15}. In the numerical analysis of the very early epochs we 
will use the evaluation of the growing length from the excess energy explained in Sec.~\ref{sec:observables}, that is
\begin{equation}
\ell_d(t) = \ell_G(t)
\end{equation}
unless otherwise stated.

\subsection{Snapshots}
\label{subsec:snapshots}
\begin{figure}[b]
\begin{center}
\includegraphics[scale=0.32]{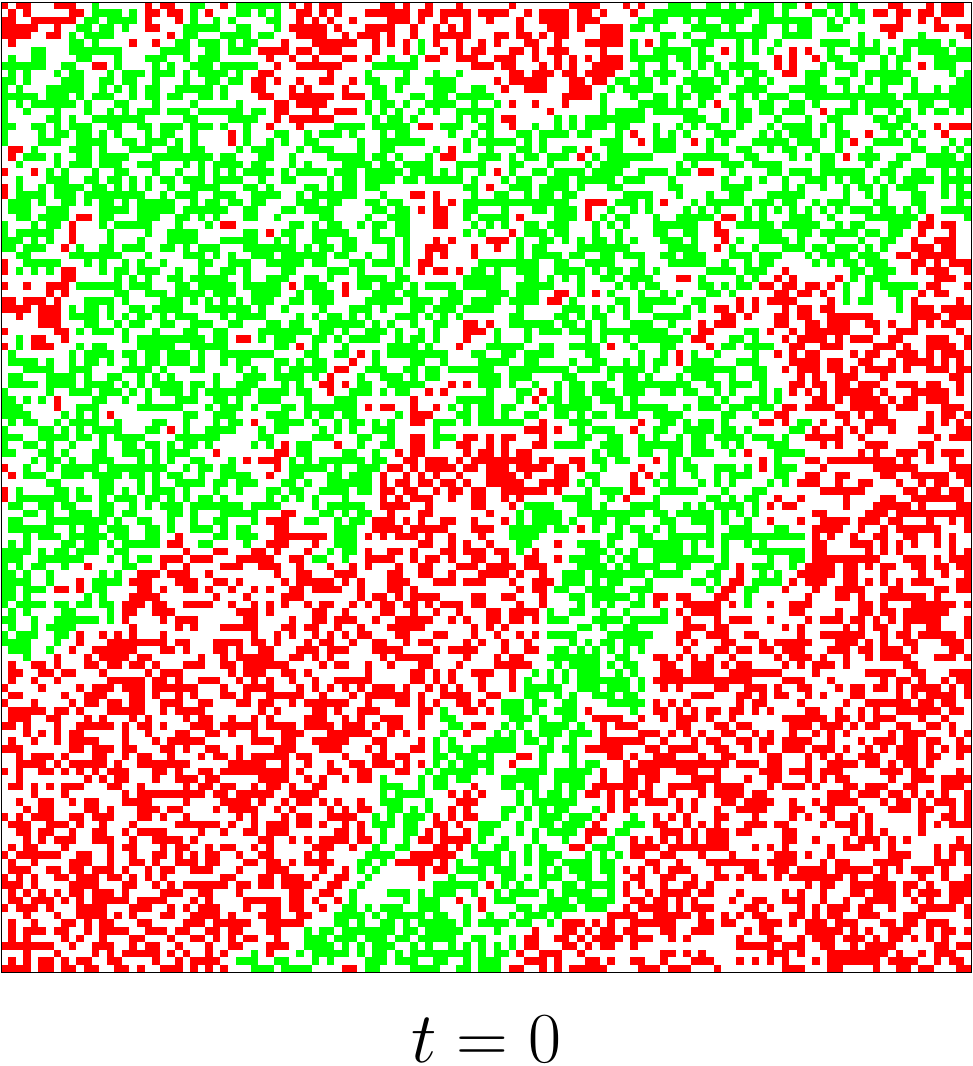}\quad%
\includegraphics[scale=0.32]{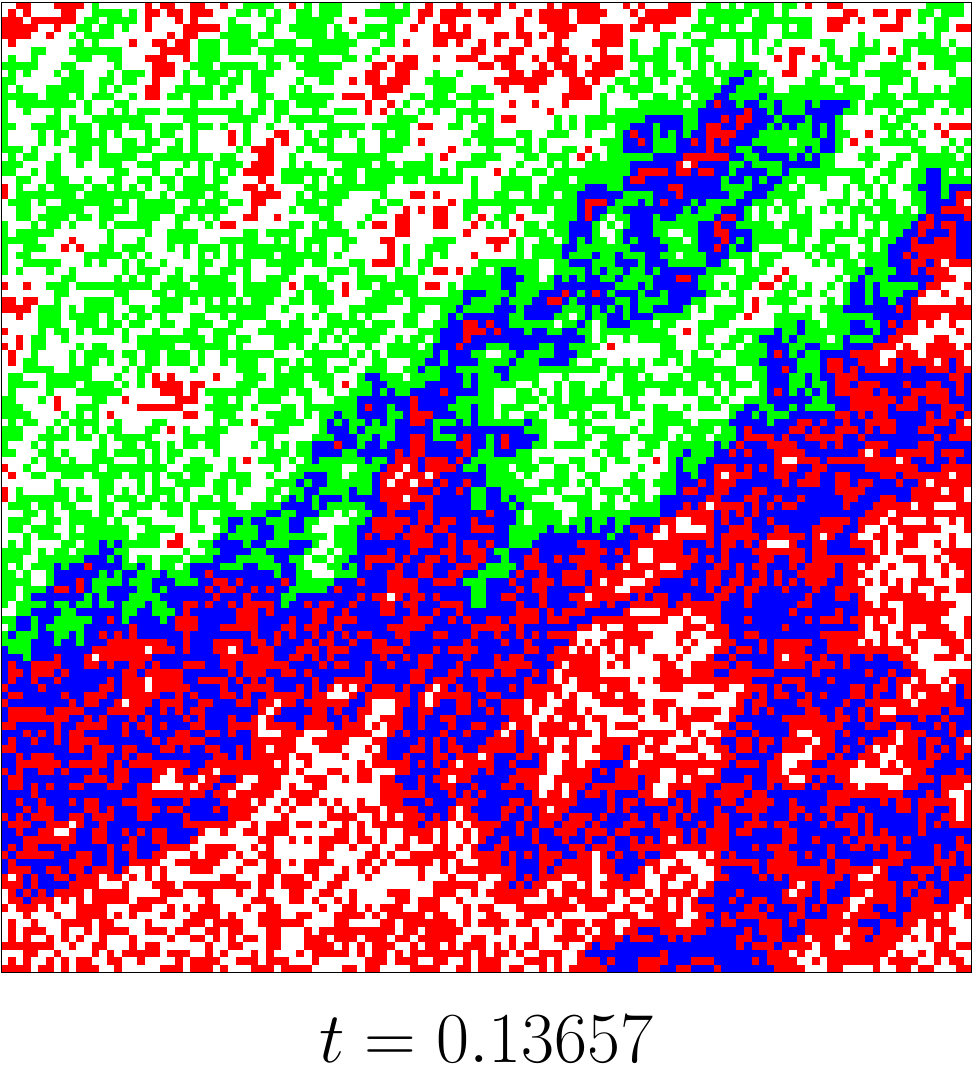}\quad%
\includegraphics[scale=0.32]{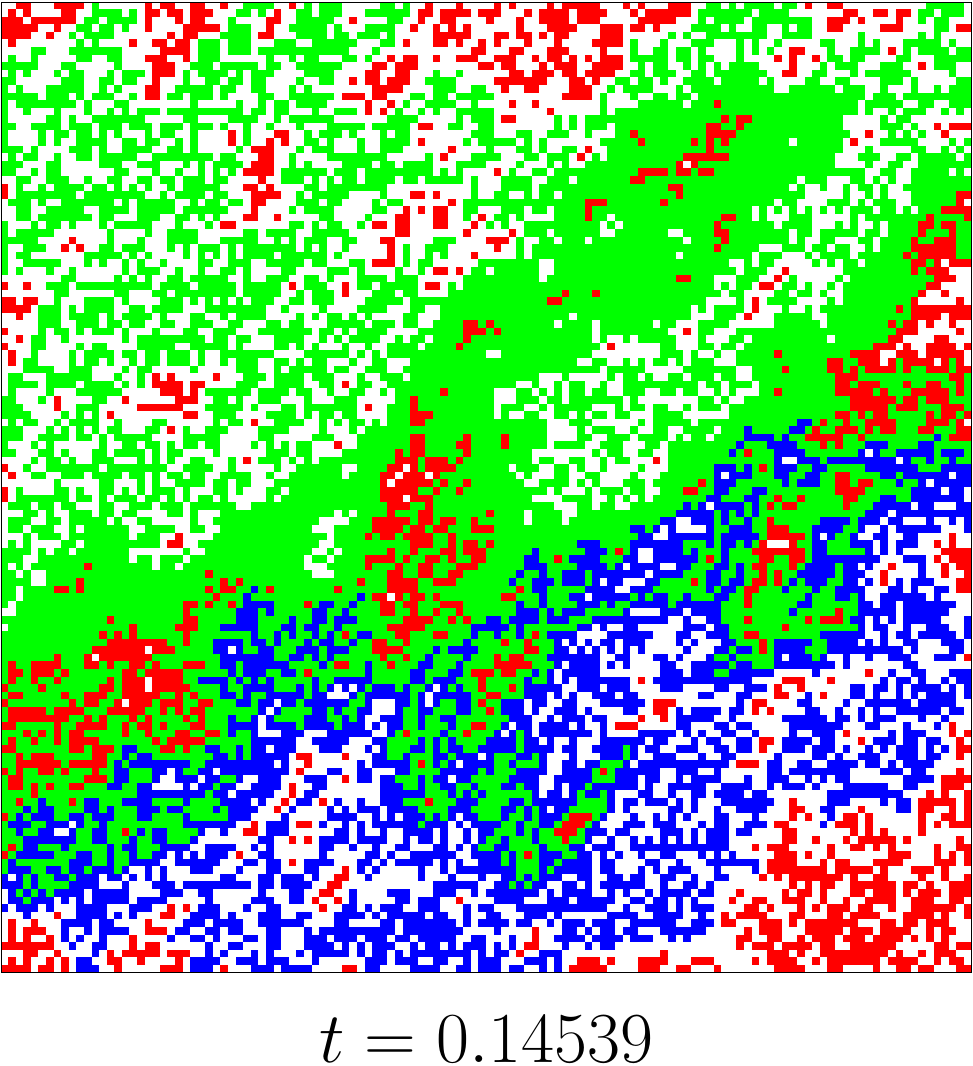}\quad%       
\includegraphics[scale=0.32]{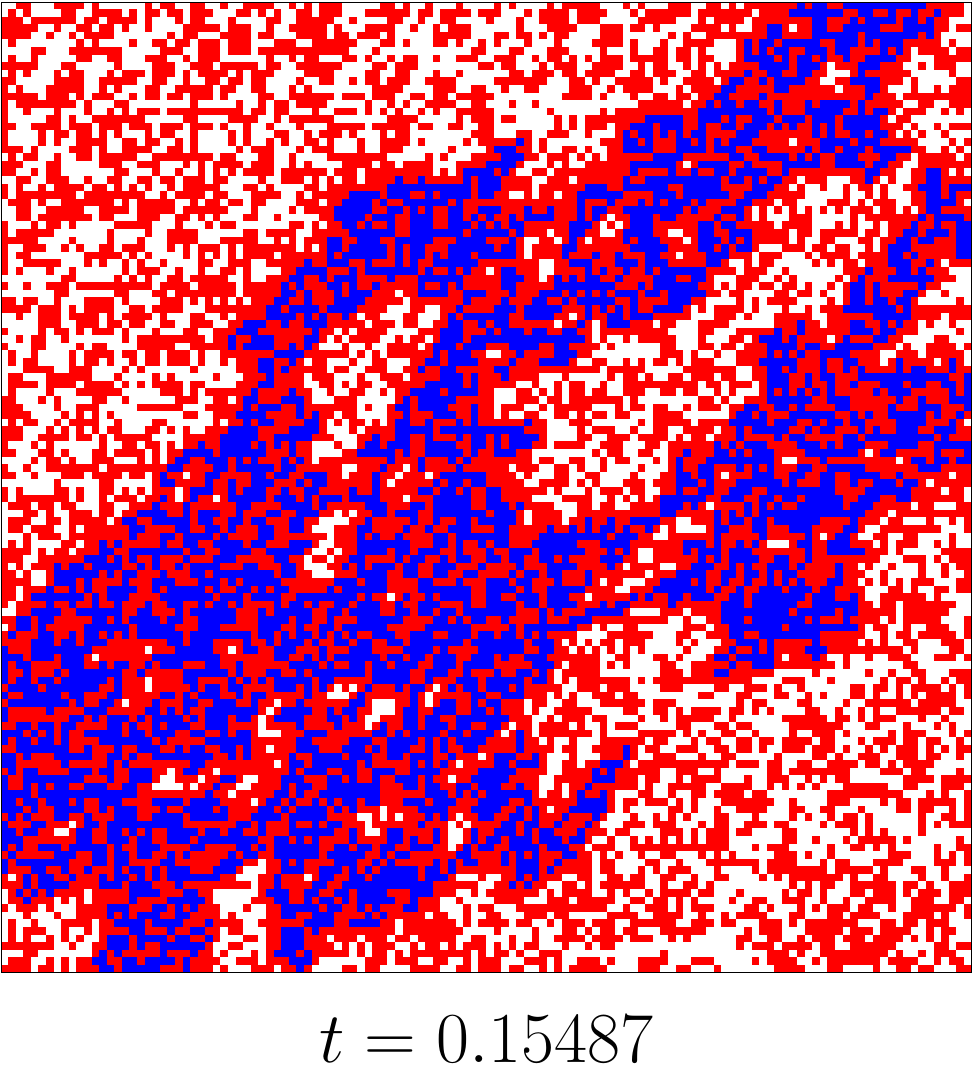}\quad%               
\vspace{10pt}%

\includegraphics[scale=0.32]{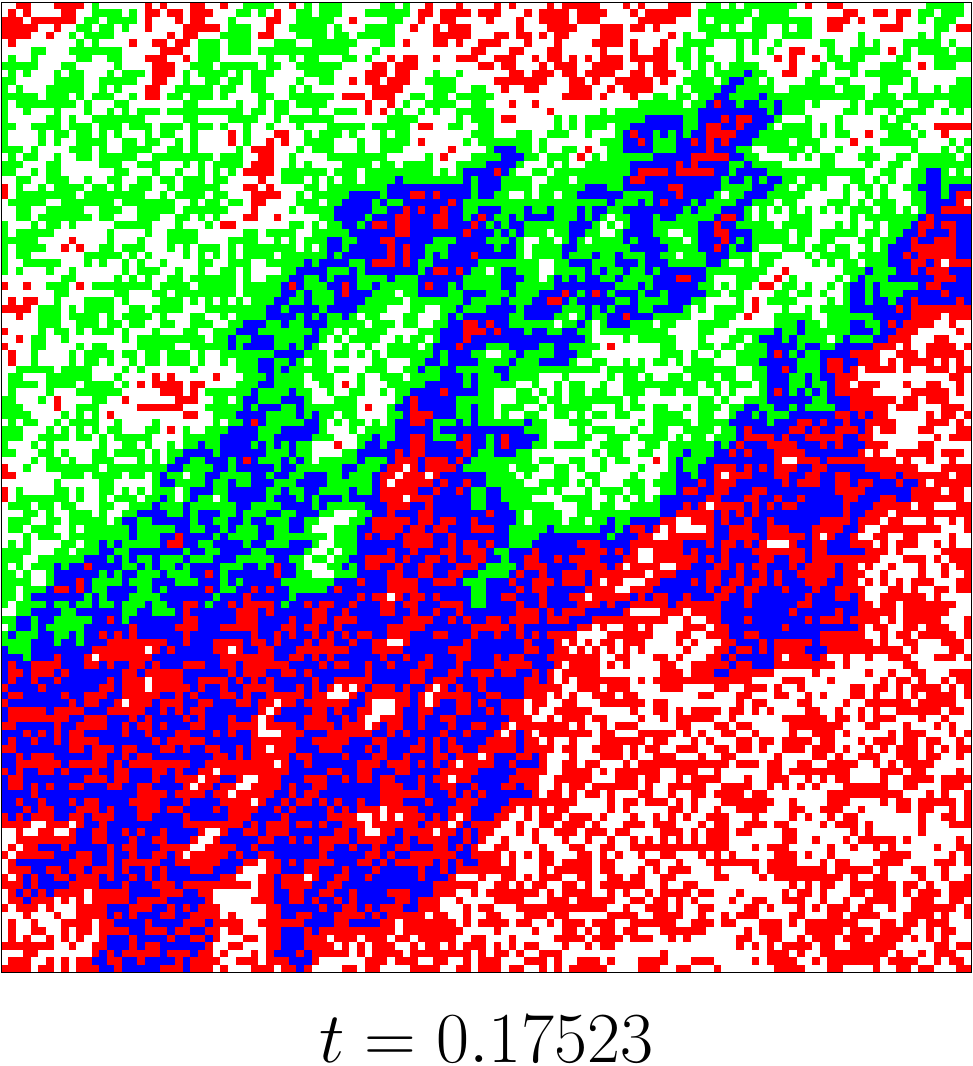}\quad%
\includegraphics[scale=0.32]{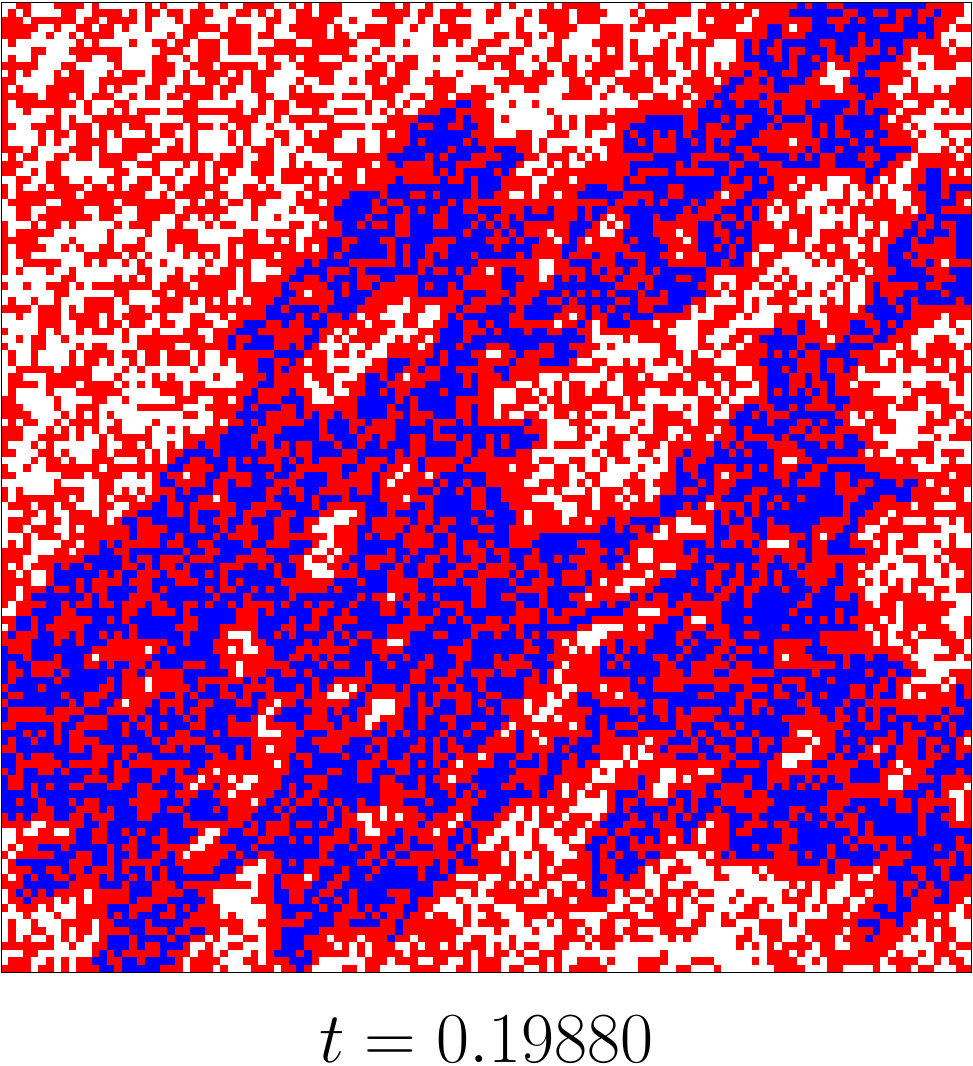}\quad%
\includegraphics[scale=0.32]{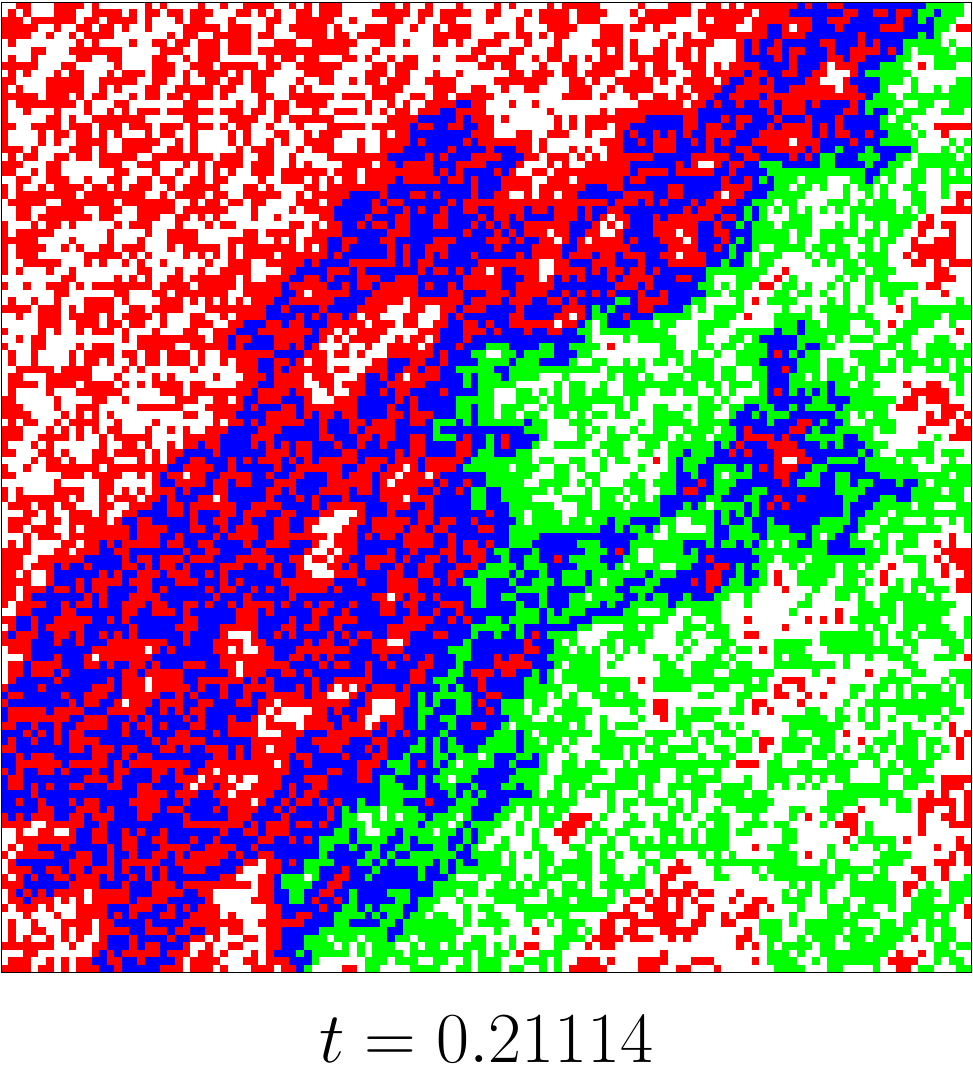}\quad%
\includegraphics[scale=0.32]{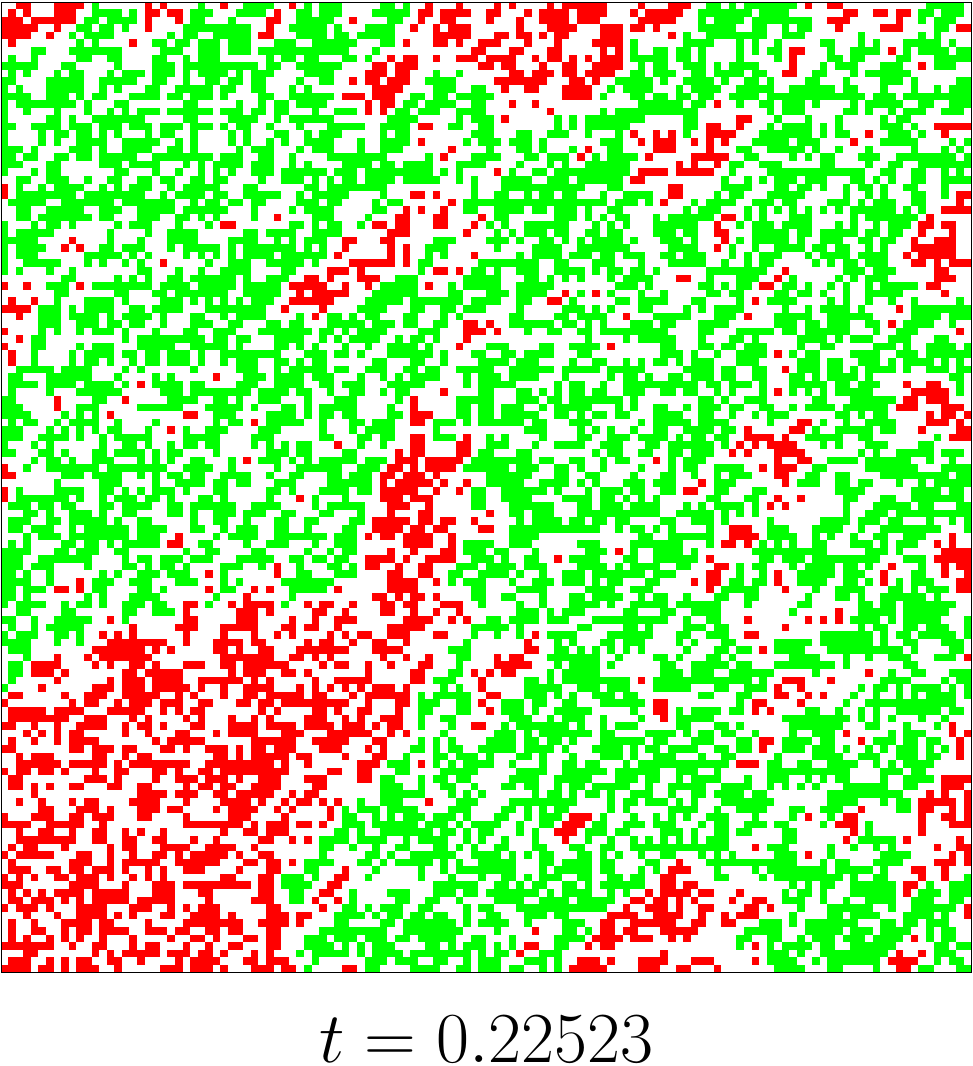}\quad%
\vspace{10pt}%

\includegraphics[scale=0.32]{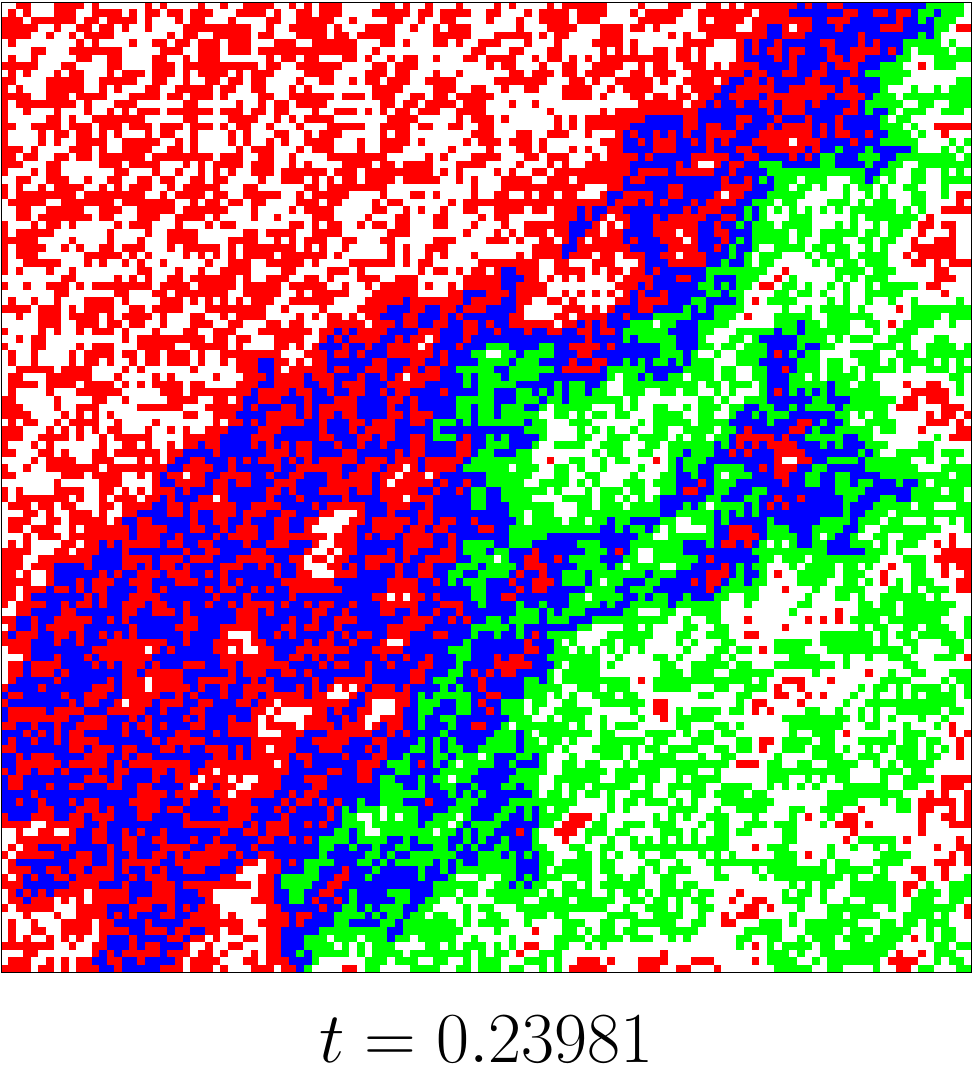}\quad%
\includegraphics[scale=0.32]{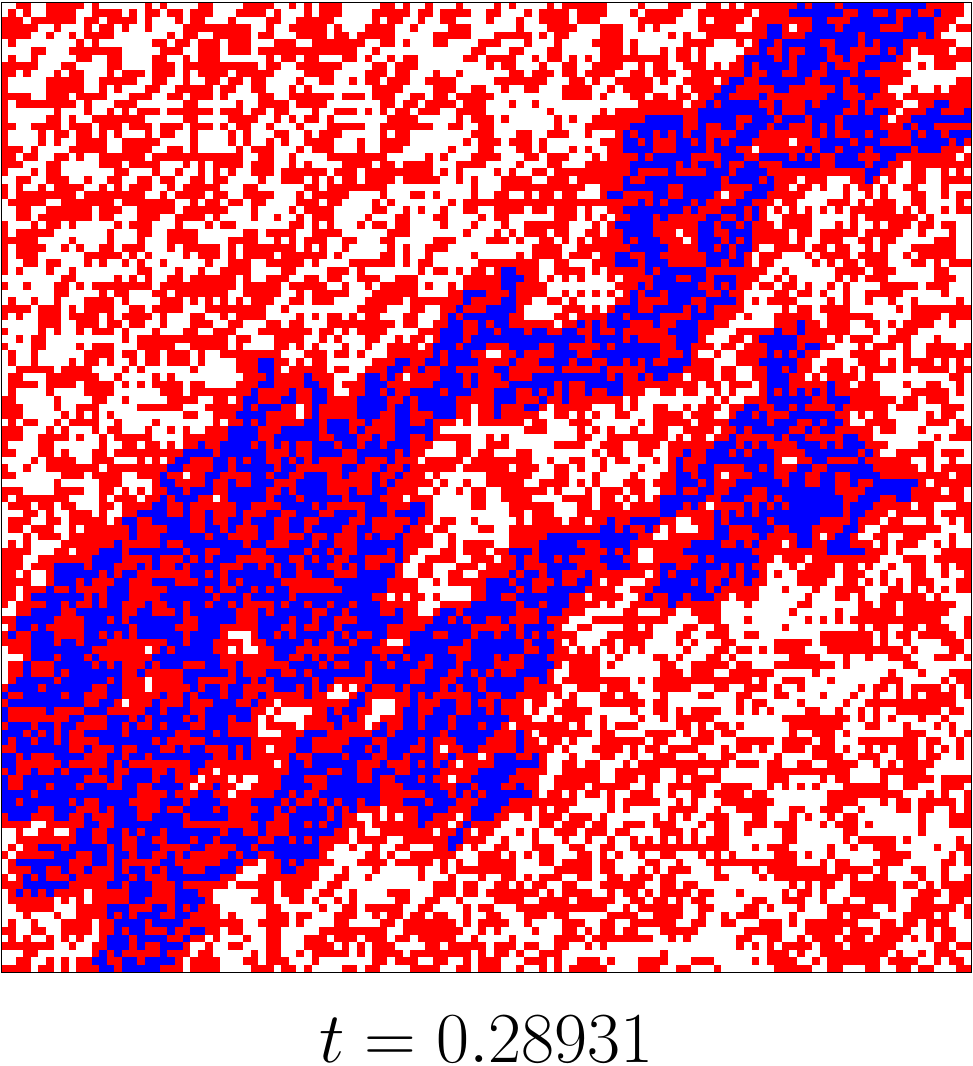}\quad%
\includegraphics[scale=0.32]{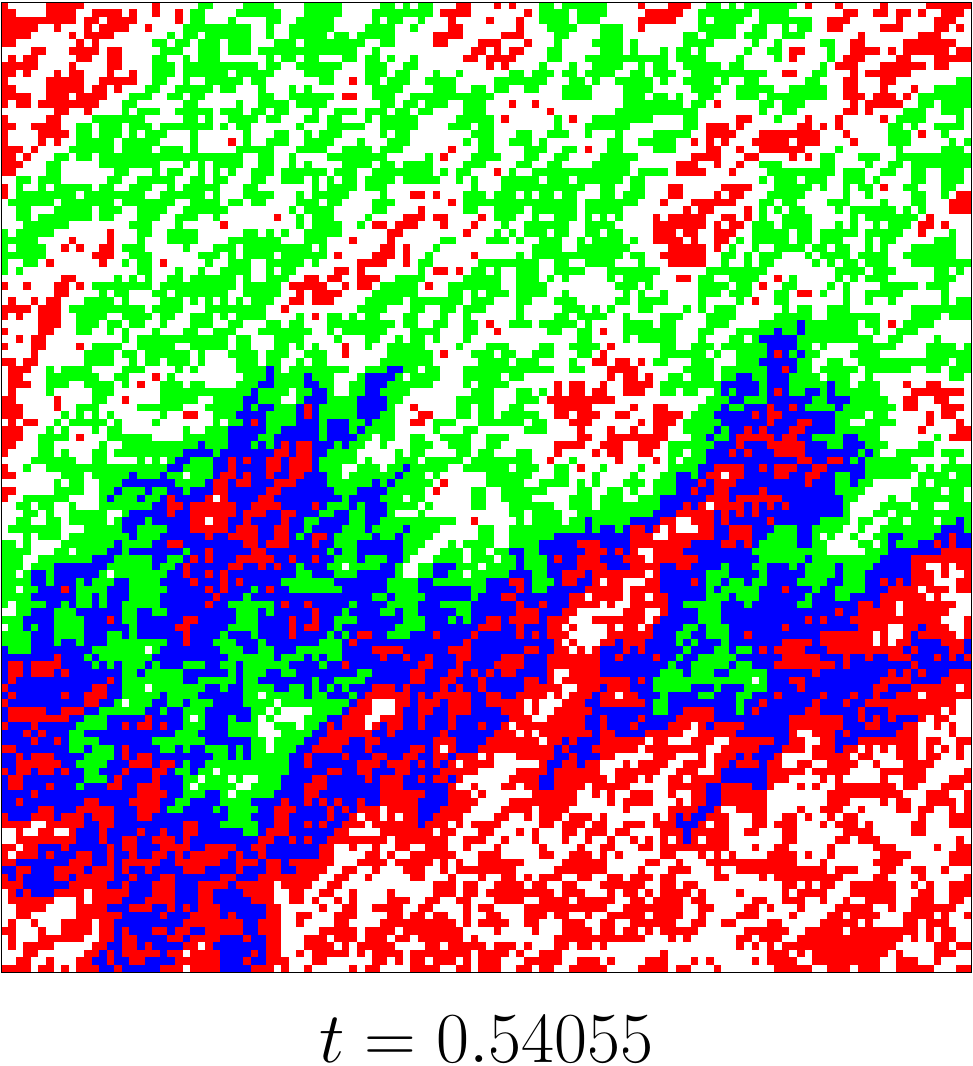}\quad%
\includegraphics[scale=0.32]{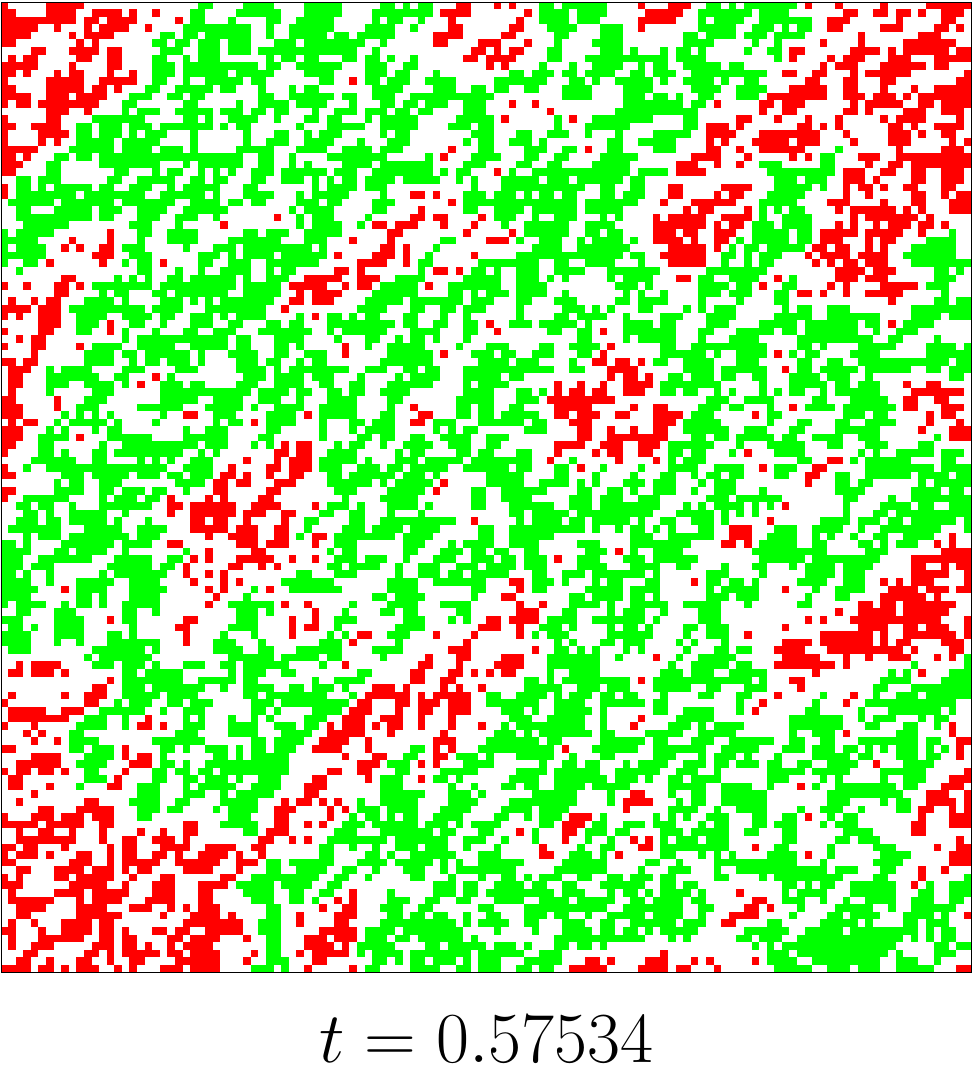}\quad%
\vspace{10pt}%

\includegraphics[scale=0.32]{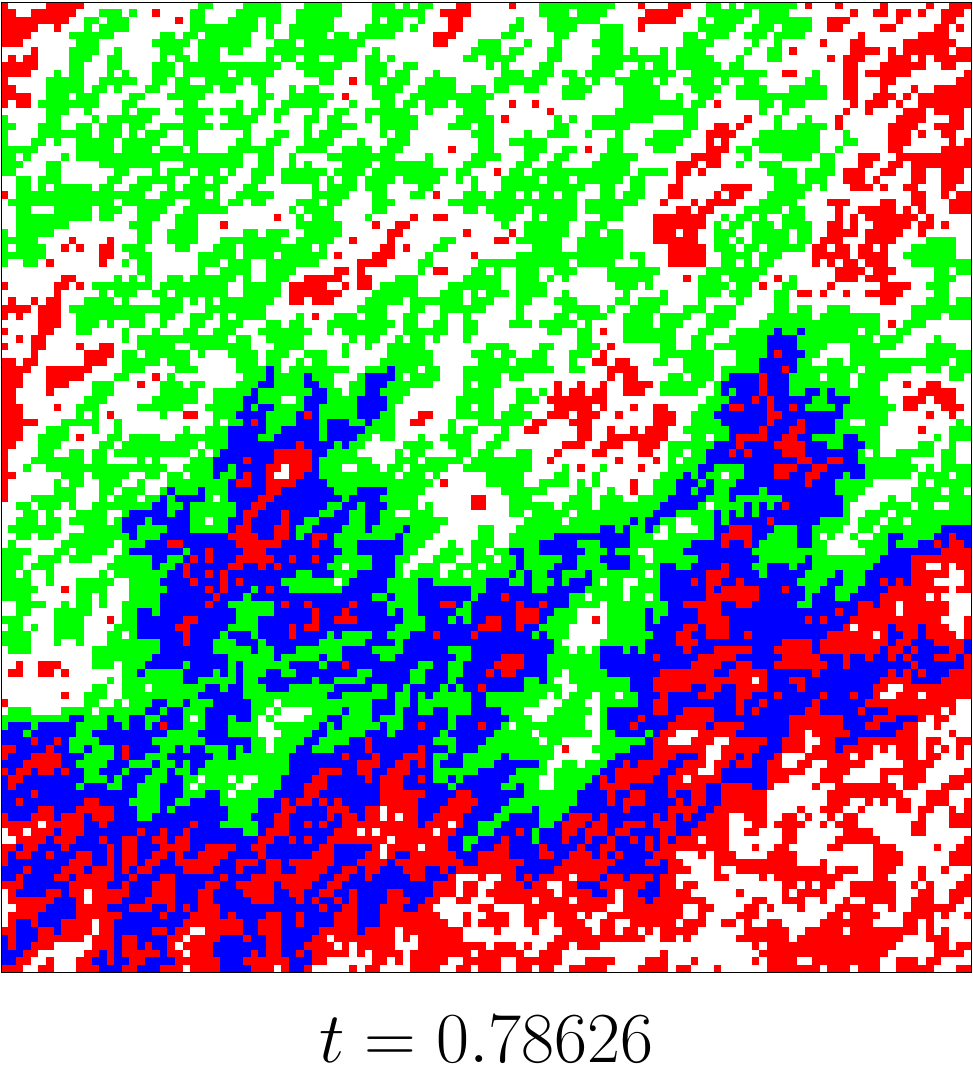}\quad%
\includegraphics[scale=0.32]{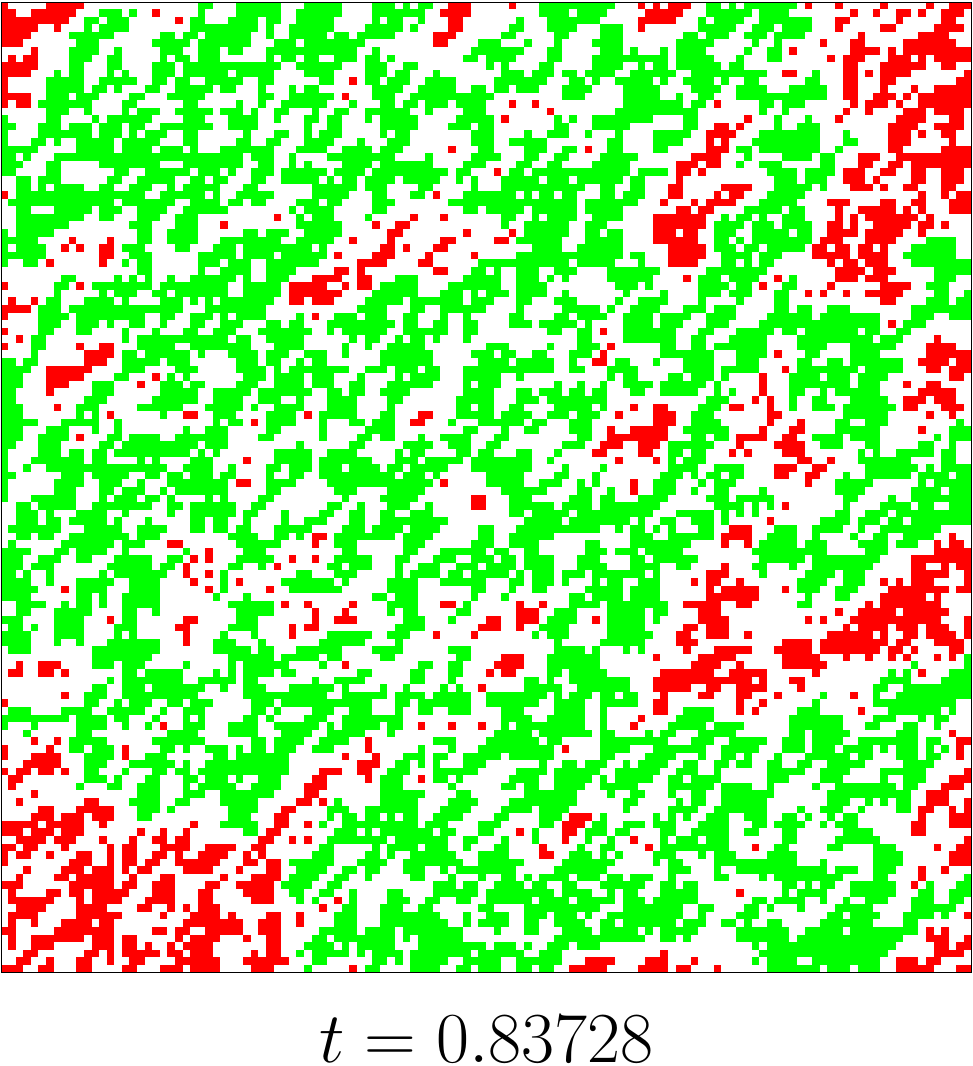}\quad%
\includegraphics[scale=0.32]{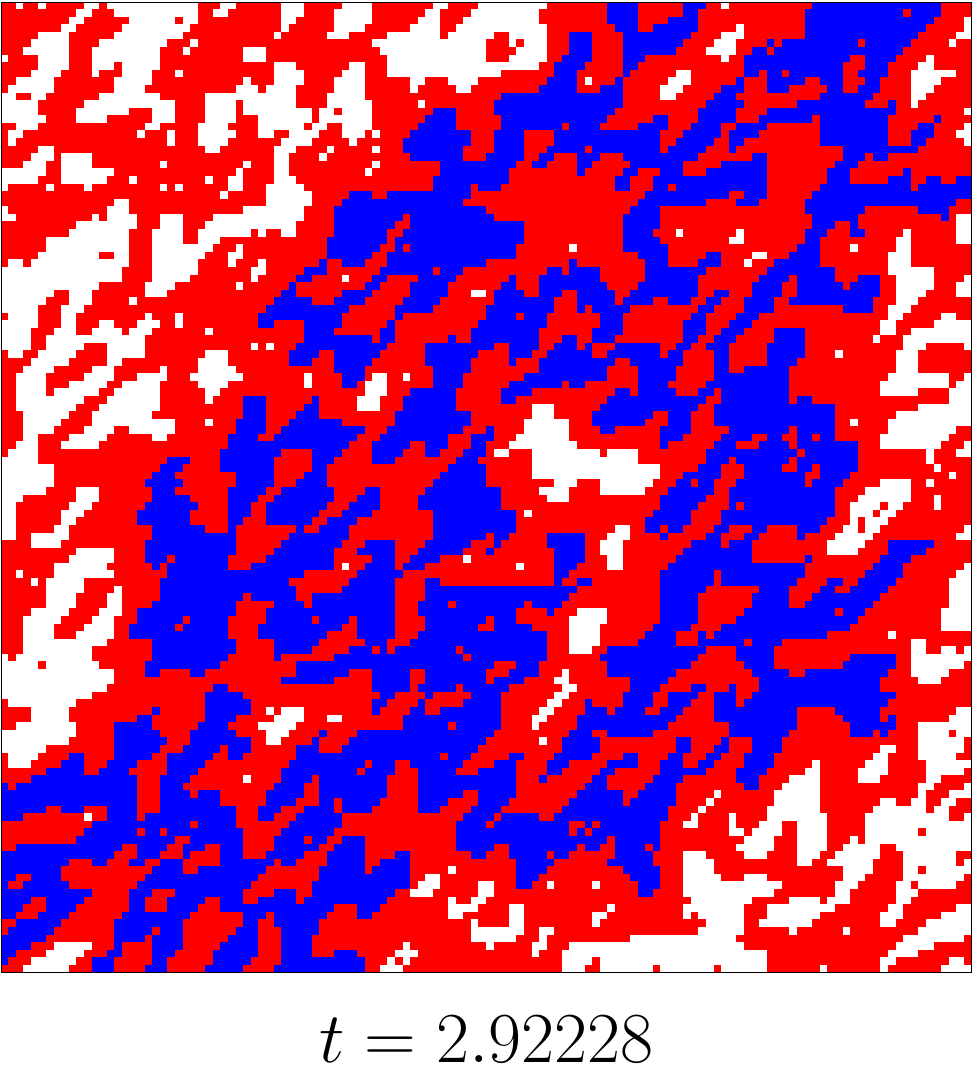}\quad%
\includegraphics[scale=0.32]{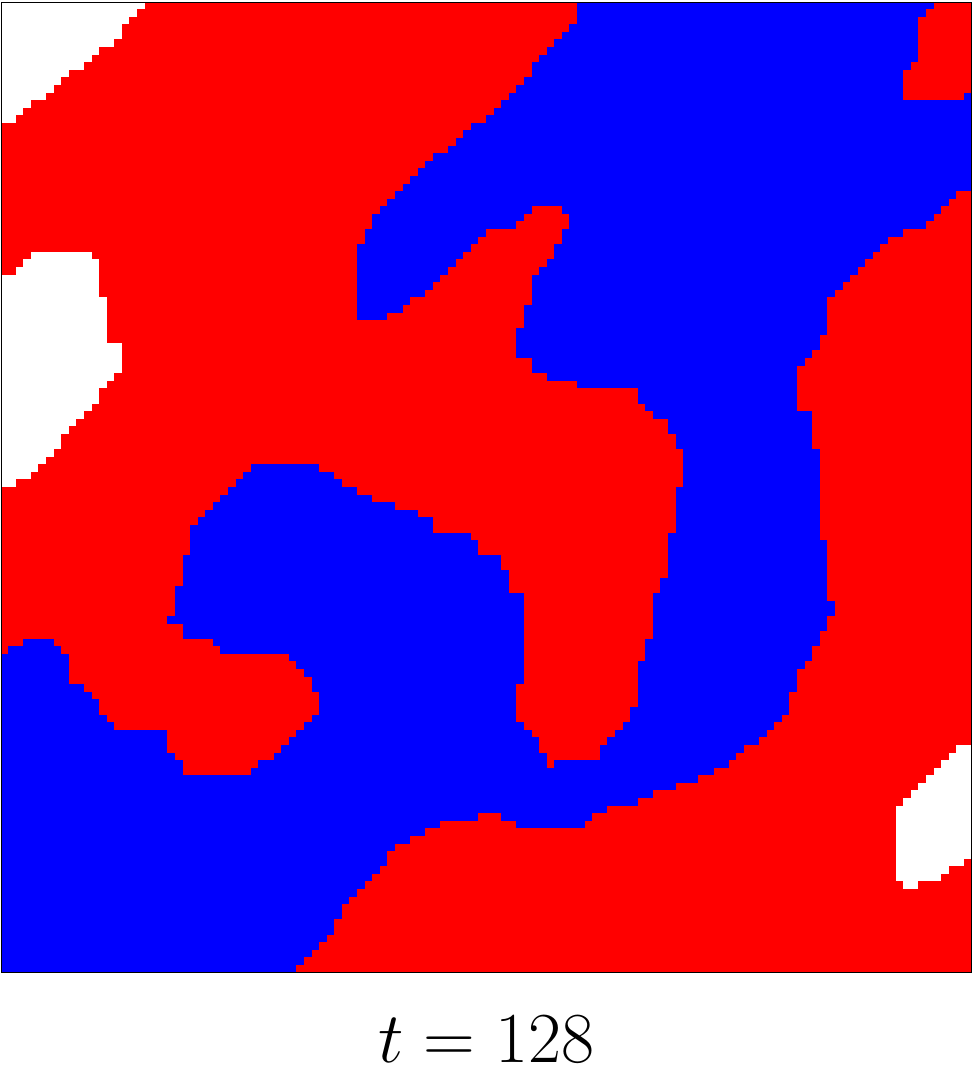}\quad%
\vspace{10pt}%

\end{center}
\caption{\small
Snapshots of the $2d$ Ising model on a triangular lattice with  $L=128$ 
and FBC, evolving with local spin flips at $T=0$ from  an infinite temperature initial condition. 
Spins $s_i=+1$ are shown as red points while spins $s_i=-1$ are shown as white points. 
A percolating cluster of spins $s_i=+1$ 
is shown in green and a percolating cluster of spins $s_i = -1$ is in blue.}
\label{fig:snapshots-triangular}
\end{figure}

We exhibit the presence of percolating clusters in time-evolving snapshots by highlighting them
on the figures with different colours. 

The triangular lattice, see Fig.~\ref{fig:triangular}, is particularly interesting since the initial state is right at the percolation
threshold and there is a percolating cluster at the start, that is $t_{p_1}=0$. A naive guess would be that this state survives after the
quench, implying $t_p=0$. However, this is not the case, as demonstrated by the 
series of snapshots in Fig.~\ref{fig:snapshots-triangular}, taken at 
different times after the quench. (The preferred diagonal inclination of the clusters is due to the 
way in which the triangular lattice was constructed, see Fig.~\ref{fig:triangular}, but does not influence the statistical 
properties of the structure.) While there is a percolating cluster in the initial configuration, this one  
disappears and is replaced by other percolating clusters until one of these eventually persists. In other words, 
the number of interfaces crossing the sample changes many times before reaching the final value. In 
consequence, $t_p$ is not zero and it actually scales with the system size~\cite{BlCoCuPi14}, even on this lattice.

Two other features in these plots merit some discussion.
First, in the next-to-last snapshot, at time $t\simeq 2.92$, the cluster that percolates in both horizontal and vertical directions 
 and remains at all subsequent times (highlighted in blue), is ``fatter'' than the ones that were present in the initial 
condition and at previous times. Some correlations have been built by the dynamics.
Second, one not only sees this percolating cluster but next to it, there is another one with the opposite spin orientation
that does not percolate but has an area of the same order of magnitude as the percolating one.

\subsection{Largest cluster}
\label{subsec:largest_cluster_intro}

In ordinary percolation, the area of the largest cluster of occupied sites
(divided by the size of the system, $L^2$) 
is the order parameter of the transition. As we mentioned in Sec~\ref{sec:observables}, right at the critical percolation point,
the size of the largest cluster, $A_c$, scales as $L^{D_A}$ with $D_A=91/48$,
where $L$ is the linear size of the system.

\begin{figure}[h]
\begin{center}
\includegraphics[scale=0.8]{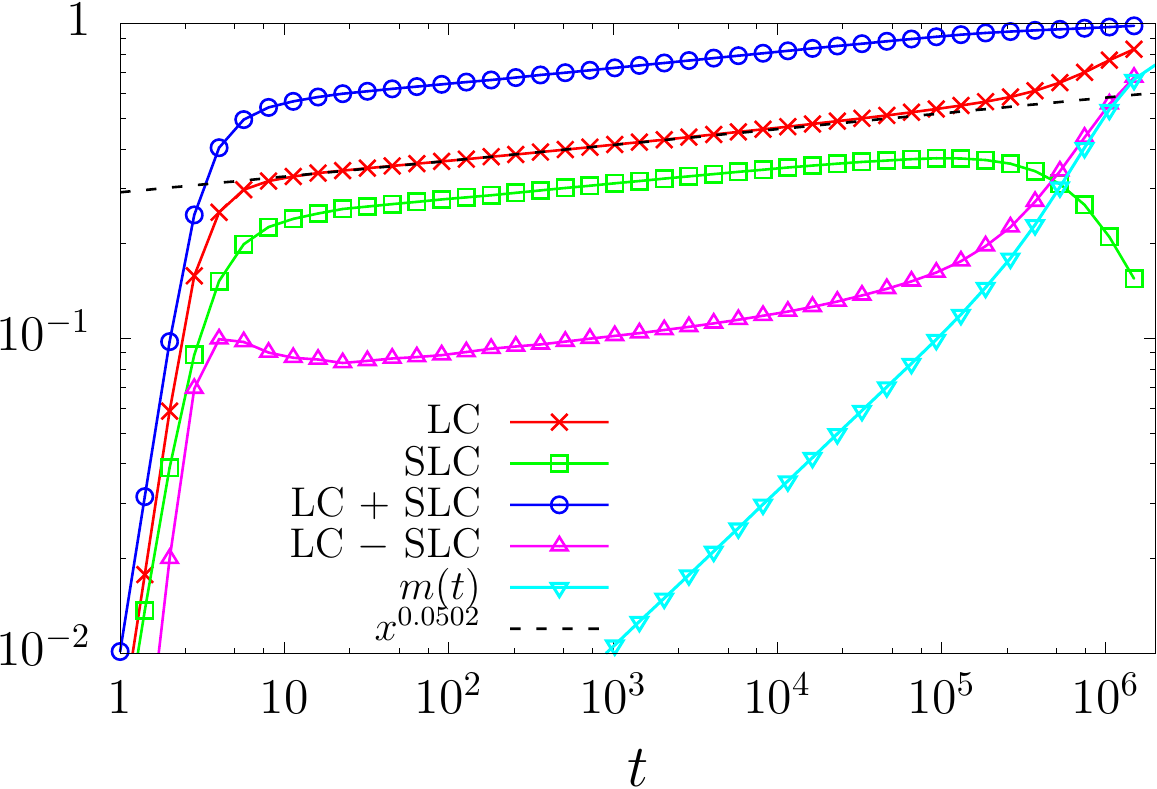}
\end{center}
\caption{\small Evolution of the (averaged) area of the largest (LC) and second largest (SLC) clusters, their sum (LC $+$ SLC) and their difference (LC $-$ SLC)
all normalised by the system area, $L^2$, and the magnetisation density $m(t)$, for the $T=0$  dynamics on a square lattice
with linear size $L=4096$ and PBC. The dashed line is a  
power law fit to the data for the largest cluster.
}
\label{F1}
\end{figure}

In the case of the  $T<T_c$ dynamics starting from a random initial condition,
we know that at a short time $t_{p_1}$ (zero for the triangular lattice and just a few steps on other lattices with finite size)
a first cluster that percolates appears. The critical-percolation-like clusters become stable after a still short time 
that scales with the system size as
$t_p \simeq L^{z_p}$~\cite{BlCoCuPi14,TaCuPi16,InCoCuPi16}.
The magnetisation density is very small at this $t_p$ since 
under the coarsening process it is characterised by $m(t) \simeq t/L^{z_d}$. 
The small magnetisation density is explained by the fact that in the spin problem at the same time that the largest cluster percolates, 
the second largest cluster with opposite magnetisation  
surrounds the largest one, although it does not necessarily percolate. 
These features are quantified in Fig.~\ref{F1}, that shows measurements on a square lattice with $L=4096$ and PBC,
averaged over a few thousands samples.
The size $A_c$ of the largest cluster
(LC) divided by the total size of the system $L^2$ is plotted as a function of time. 
We observe that after a short time $t \simeq 10$ the largest cluster occupies an important fraction of the system size, 
with $A_c/L^2 \simeq 0.32$.
The area of the second largest cluster (SLC) also occupies a sizeable part of 
space at this time, say $25\%$. 
In the same figure we display the sum of the largest and second largest cluster sizes (LC $+$ SLC) and their difference (LC $-$ SLC), 
still normalised by $L^2$. We will discuss their meaning below. 

Next, we observe that as time elapses, and due to the coarsening process,
the areas of both the largest and the second largest clusters increase as a power of time.
A fit of the function $f(t)= C \, t^{\alpha}$ to the data
$A_c(t)/L^2$ gives as a result the exponent $\alpha \simeq 0.0502$, and the fitting function is shown in the plot
with a dashed line. This algebraic evolution is observed up to a late time, $t_2 \simeq 10^5$. 
As one can see, the second largest cluster grows in the same way. 

This exponent can be easily understood using the following arguments.
In the static percolation problem the linear length of the system is measured in units of the 
lattice spacing $r_0$ and the area of the clusters is measured in units of the 
elementary area $r_0^2$.  At the static percolation transition, the size of the largest cluster  should scale
with the system linear size $L$ as $A_c/r_0^2 \propto (L/r_0)^{D_A}$, with $D_A$ the critical percolation clusters fractal dimension
introduced in Sec.~\ref{sec:observables}. Thus, the fraction of sites belonging to the largest cluster,
$(A_c/r_0^2)/(L/r_0)^2$, should scale with the linear length of the system
as $L^{D_A-2} = L^{-\beta/\nu}$, where $\beta$ and $\nu$ are the percolation critical exponents associated to the
order-parameter (fraction of sites belonging to the incipient percolating cluster) and the correlation length, respectively.

Because of coarsening,  the area of the largest (and second 
largest) cluster continues to grow at the expense of the smaller clusters that disappear. Therefore, the fact that  
$(A_c/r_0^2)/(L/r_0)^{D_A} = {\mathcal O}(1)$ at the critical threshold has to be extended to include the time-dependence. 
Arguing that under coarsening lengths are rescaled by the growing length $\ell_d(t)$, we introduce an effective dynamical 
block number $N(t) = L/\ell_d(t)$ or, equivalently, an effective ``dynamical lattice'' spacing
\begin{equation}
r_0 \to \ell_d(t)
\; . 
\end{equation}
The dynamic percolation problem is now set on a dynamic lattice and the
natural extension of the fractal scaling of the largest cluster is 
\begin{eqnarray}
\frac{A_c/r_0^2}{(L/r_0)^{D_A}} \to \frac{A_c(t)/\ell_d^2(t)}{(L/\ell_d(t))^{D_A}}
\label{eq:rescaling-vertical-axis}
\end{eqnarray}
that is equivalent to
$ A_c(t)/ N^2(t) \propto N(t)^{-\beta/\nu} =
(\ell_d(t)/L)^{\beta/\nu} \simeq (t^{1/z_d}/L)^{\beta/ \nu }$ 
with $\beta/{(z_d \nu)} = 5/96 \simeq 0.0521$ in excellent agreement with the measured power $\alpha$ in Fig.~\ref{F1}. 

Figure~\ref{F1} also shows the magnetisation density $m(t) = M(t)/L^2$. At $t\simeq 10$, 
this quantity is so small that it does not appear in the scale of the plot. 
It then increases following the power law $t/L^2$ and remains small ($< 0.1$) up to the 
time $t_2$. At longer times,  the size of the largest cluster increases faster, while the size of the second largest cluster decreases. 
The latter will remain finite until very late times since there is a finite probability that the final state for the zero-temperature dynamics
contains two clusters in a stripe configuration~\cite{SpKrRe01, SpKrRe02}.
The sum of the areas 
of the largest and second largest clusters (LC $+$ SLC) becomes very close to the total area $L^2$ for $t>t_2$. 
This means that most of the smaller clusters have disappeared. 
As a consequence, the magnetisation density increases very quickly and at $t > t_2$ it  is 
very close to the difference between the densities of the largest cluster and second largest clusters (indicated as LC $-$ SLC in Fig.~\ref{F1}).

The discussion above implies that during a very long period of time $10 < t < t_2$, the dynamics are characterised by the coexistence of two 
very large clusters, one percolating, the other one not necessarily, of different spin orientation (magnetisation) which grow as a 
consequence of the domain growth. For $t > t_2$, the dynamics are characterised by the evolution of only these two large clusters 
since most of the small ones have already disappeared. 
For the linear size considered here, $L=4096$, $t_2$ is close to $10^5$. We will 
show in Sec.~\ref{sec:metastability} that this value scales as $L^2$.

We now focus on the scaling properties associated with the {\it approach to} critical percolation.
As we have already stated, a time $t_p$ is needed to reach the stable critical percolation state in the dynamic problem. 
This characteristic time is a function of the system size,  $t_p\simeq L^{z_p}$, 
on all lattices including the triangular one, as it was shown in~\cite{BlCoCuPi14}, where the
exponent $z_p$ was measured from the asymptotic behaviour of the
two-copies-overlap, $Q$, 
and the correlation between the ``crossing number'' at a given time and  in the final (equilibrated) state,
${\mathcal O}_c$.
However, $t_p$ is not the time at which a percolating cluster  first appears in the system. 
In fact, a percolating cluster of positive or negative magnetisation first appears at an earlier time, $t_{p_1}<t_p$,
that does not necessarily scale like $t_p$.
On the triangular lattice, for example, the critical value of the occupation probability is $p_c=1/2$ and thus
$t_{p_1}=0$, since there is already one percolating spin cluster in the initial fully-disordered spin configuration.
The largest cluster present at $t=0$  
is surrounded by another very large one with the opposite spin orientation so as to ensure that 
the magnetisation density vanishes. But the largest cluster is not stable and it is broken in pieces by the dynamics 
until another stable one is created at time $t_p$.

\begin{figure}[h]
\begin{center}
 \includegraphics[scale=0.52]{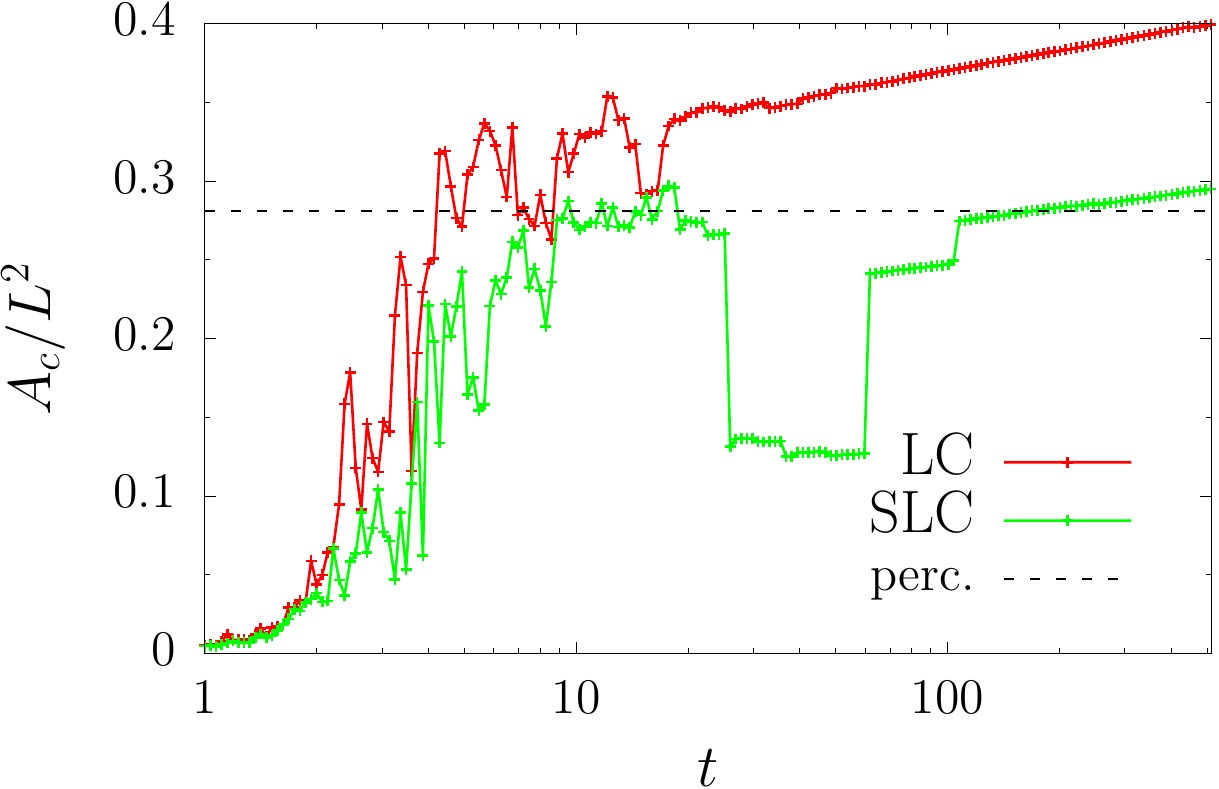}\quad%
 \includegraphics[scale=0.52]{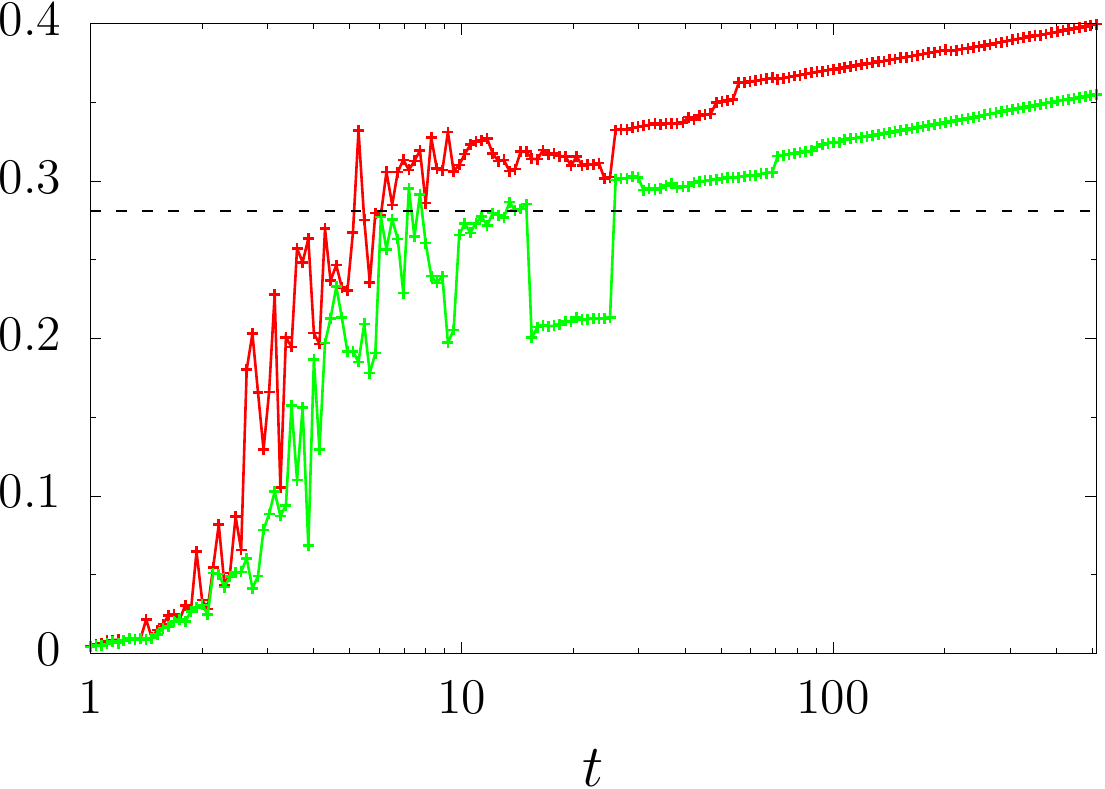}

 \includegraphics[scale=0.52]{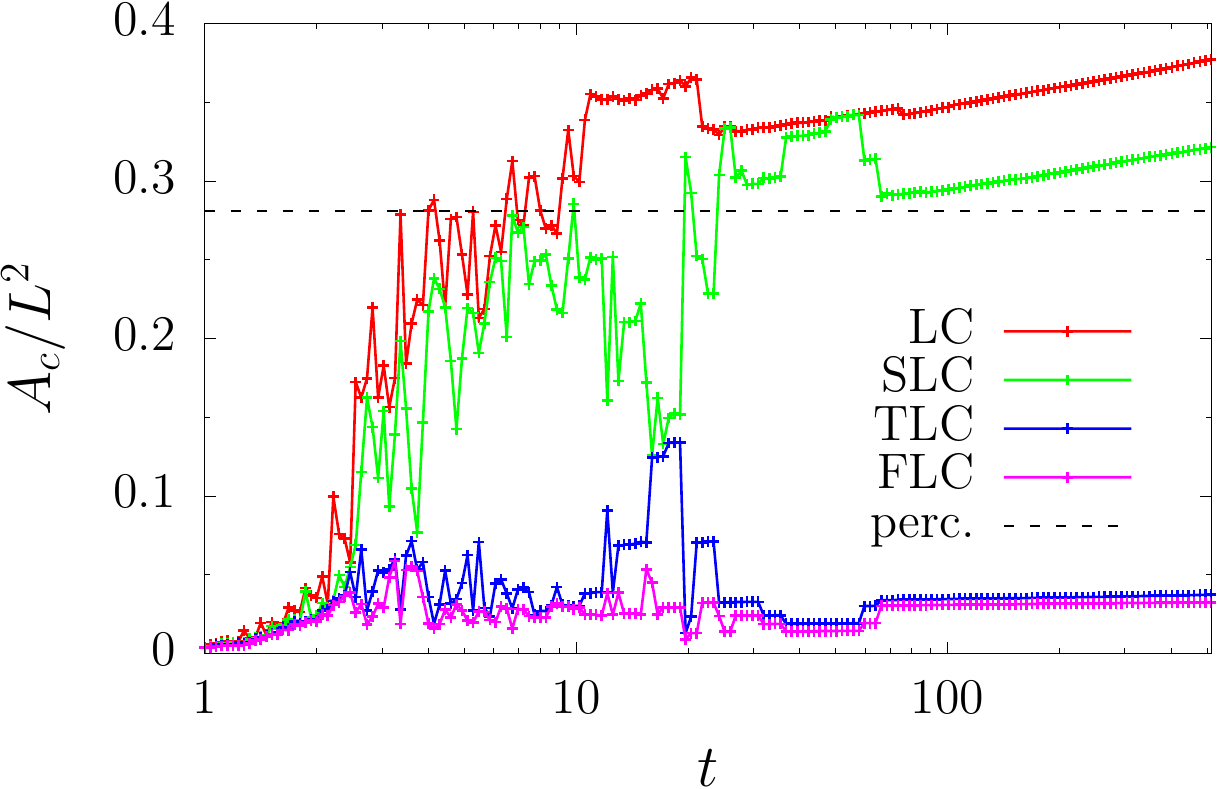}\quad%
 \includegraphics[scale=0.52]{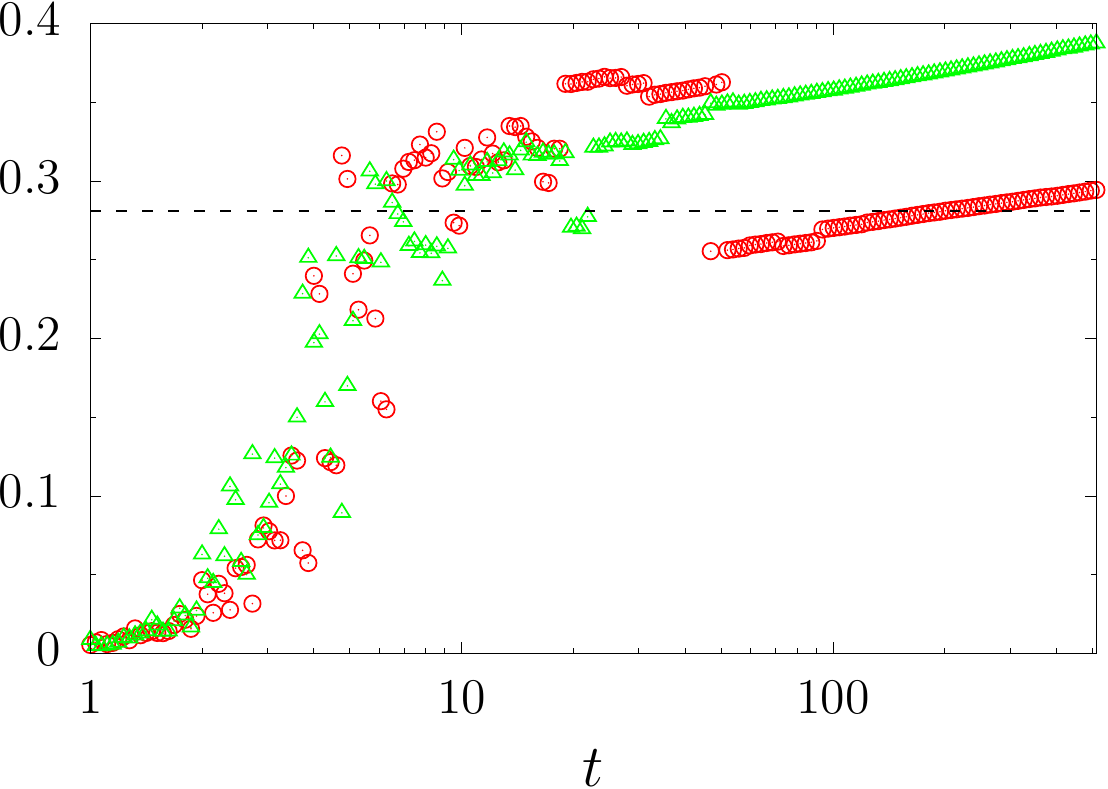}
\end{center}
\caption{\small
The two figures in the upper row show the fraction of sites, $A_c/L^2$ , belonging to the largest (LC) and second largest (SLC)
spin clusters as a function of time, for the $T=0$  dynamics on a square lattice with linear size $L=4096$.
Each figure represents an independent realisation of the dynamics.
In the panel on the lower left corner, for another realisation of the dynamics, 
we also show the time evolution of the fraction of sites belonging to the third largest (TL) and fourth largest (FL) clusters.
Finally, on the lower right corner we show the time evolution of
the fraction of sites belonging to the largest cluster of spin $+1$, indicated with red circles, and
the one of the largest cluster of spin $-1$, indicated with green triangles, 
for another numerical run.
The horizontal dashed line in each plot indicates the fraction of sites belonging to the
largest cluster in site percolation on a square lattice with occupation probability at the threshold value $p_c$.
} 
\label{2cl_sq} 
\end{figure}

In order to provide a better qualitative understanding of what is happening during the dynamics following a quench to zero temperature, 
we show in Fig.~\ref{2cl_sq}, in the top panels, the time evolution of the fraction of sites (area divided by $L^2$)
belonging to the largest cluster (LC) and the ones belonging to the second largest cluster (SLC), 
for two different and independent realisations of the $T=0$  dynamics, on a square lattice with PBC and linear size $L=4096$.
In the figure on the lower left corner, for an other independent realisation of the dynamics, 
we also show the fraction of sites belonging to the third largest cluster (TLC) and the fourth largest cluster (FLC).
Finally, in the figure on the lower right corner, for yet an other realisation, we show the time evolution of
the fraction of sites belonging to the largest cluster of spin $+1$, indicated with red circles, and
the one for the largest cluster of spin $-1$, indicated with green triangles.

The sizes of the LC and SLC grow rapidly in all cases. 
After a time $\simeq 10$ they have a size that is comparable to 
the one of a percolating cluster in site percolation on a square lattice of same linear size $L=4096$
(indicated as an horizontal dashed line), and, most importantly, they have opposite spin orientation.
On the contrary, the TLC and the FLC remain very small (shown only for the third sample). 
The LC and SLC sometimes exchange as one can see clearly from the fourth plot (lower right corner).
These exchanges stop after a time that is much longer than the one at which
these two clusters have reached the area of a typical percolating cluster in ordinary site percolation,
around $t \simeq 50$ (for the dynamics on a lattice of linear size $L=4096$).
Thus, the time when the cluster starts percolating is $t_{p_1} \simeq 10$ for $L=4096$, but
it is only after a longer time, e.g. $ t\simeq 50 >t_{p_1}$ in the last panel, that the two largest clusters 
stop exchanging themselves and become somehow ``stable''.
The time $t_p$ at which the very few big clusters (the LC, SLC and eventually
other few clusters) become stable is the time measured in~\cite{BlCoCuPi14} 
by analysing the two-copies-overlap $Q$ and the two-time-correlation of the crossing number ${\mathcal O}_c$.  

The value of $t_p$ will stem from an average over many realisations and it will turn
out to be in between the $t_{p_1}$ and the long time observed in the last panel. 
We also note that the same quantities averaged over many samples in Fig.~\ref{F1} do not show any distinction between $t_{p_1}$ and $t_p$.

\begin{figure}
\begin{center}
 \includegraphics[scale=0.55]{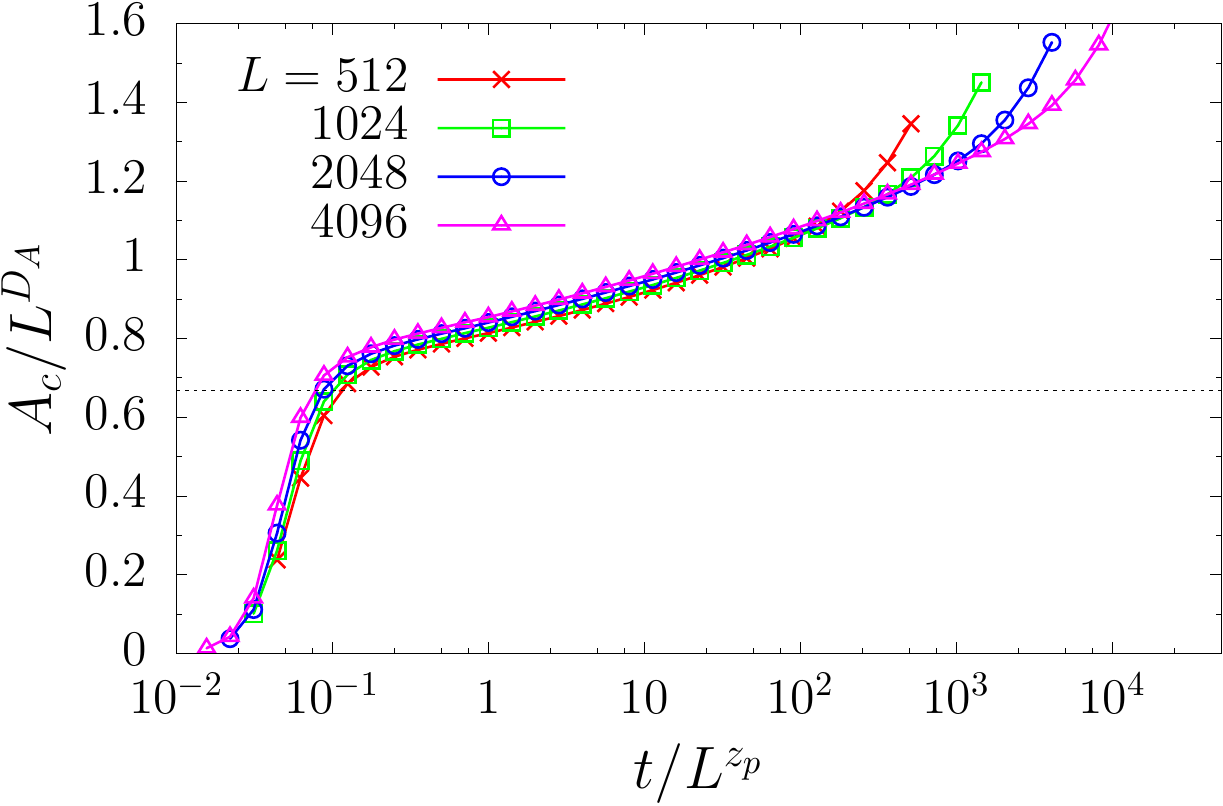}\quad%
 \includegraphics[scale=0.55]{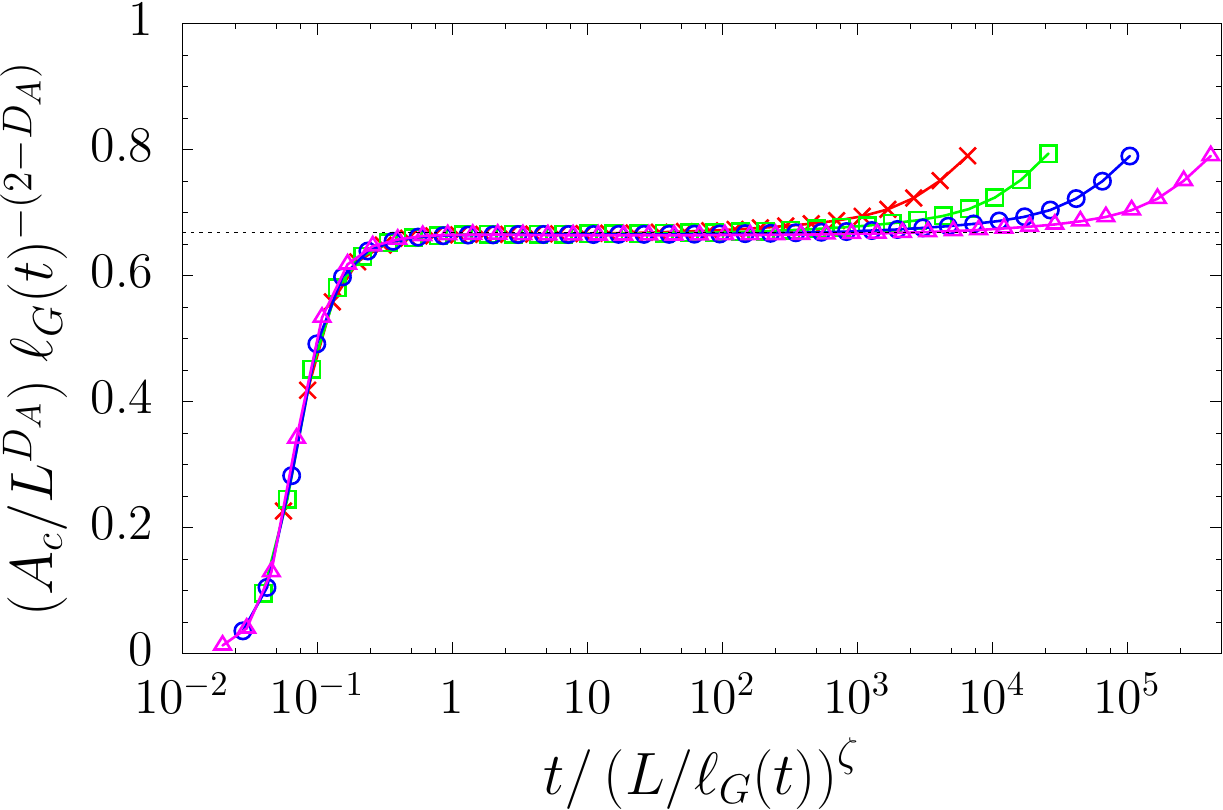}\quad%
\end{center}
\caption{\small Measurement of the (averaged) size $A_c$ of the largest cluster for the $T=0$  dynamics on a square lattice,
with different values of $L$, the linear size of the system.
On the left, we show $A_c/ L^{D_A}$ as a function of $t/L^{z_p}$, with $z_p=1/2$. 
On the right, we show $\ell_G(t)^{-(2-D_A)} A_c/L^{D_A} $ as a function of $t/ (L/ \ell_G(t))^{\zeta}$, where
$\ell_G(t)$, the characteristic length obtained from the excess energy, is taken as a measure of $\ell_d(t)$ and
$\zeta = 1/2$. $D_A=91/48$ is the fractal dimension of critical percolating clusters.
In both panels, the dashed horizontal line corresponds to the critical site percolation value on a square lattice,
$A_c/L^{D_A} \simeq 0.6683$.}
\label{F2}
\end{figure}

With this fact in mind, the most natural time-size scaling would be $t/t_p(L) \sim  t/L^{z_p}$, as done in the left panel of Fig.~\ref{F2} 
where we show the area of the largest cluster, $A_c$, divided by $L^{D_A}$ with
$D_A$ the fractal dimension of the percolating cluster in critical $2d$ site percolation,
as a function of $t/L^{z_p}$ for various system sizes 
$L=512, \ldots , 4096$, and the value of the exponent $z_p=1/2$ estimated in~\cite{BlCoCuPi14}.
The data roughly fall on a master curve but there are still rather strong finite-size corrections. These corrections correspond to the mixing 
of the two dynamic processes: approach to critical percolation occurring at $t < t_p \simeq L^{z_p}$ and usual coarsening 
arising afterwards. We will now disentangle the two contributions.

In the right panel in Fig.~\ref{F2} we attempt to take the coarsening phenomenon into account.
The rescaling in the vertical axis is motivated by the explanation around Eq.~(\ref{eq:rescaling-vertical-axis}), that 
suggests to focus on $A_c/L^{D_A} \times \ell^{D_A-2}_d(t)$, instead of just $A_c/L^{D_A}$.
The new proposal is to scale the data as a function of $\ell_p(t)/L$, a number that counts the number of
critical percolation ``blocks"  in a system with linear size $L$. If we further suggest  
\begin{equation}
\ell_p(t) \simeq \ell_d(t) \, (t/t_0)^{1/\zeta}
\label{eq:ell_p}
\end{equation}
where we used $\ell_d(t)$ as the dynamic lattice spacing, 
the scaling variable can also be written as
\begin{eqnarray}
\frac{\ell_p(t)}{L} &=& \frac{\ell_d(t) (t/t_0)^{1/\zeta}}{L} = \left(\frac{t/t_0}{(L/\ell_d(t))^\zeta} \right)^{1/\zeta}
 = \left(\frac{(t/t_0)^{1+\zeta/z_d}}{(L/r_0)^\zeta} \right)^{1/\zeta} 
 \nonumber\\
 &\equiv & \frac{(t/t_0)^{1/z_p}}{(L/r_0)}
\end{eqnarray}
(and we ignored a pre factor that measures the temperature dependence of the dynamic growing length and 
does not influence this argument).
The third member (without the irrelevant overall power $1/\zeta$) is the scaling variable used in the right panel in Fig.~\ref{F2}
with $\zeta=1/2$.

Shortly after the quench the dynamical characteristic length $\ell_d(t)$
can still be far from the asymptotic 
law $\ell_d(t) \simeq t^{1/2}$. It is, however, in this time regime that the approach to percolation 
occurs. For this reason we use a numerical estimate of $\ell_d(t)$, which is given by 
$\ell_G(t)$ defined by Eq.~\ref{eq:growing-length-excess-energy}. Then every time $\ell_d(t)$ is involved 
in the scaling analysis of the largest cluster size and 
other observables explored later in the paper, we will assume that $\ell_G(t)$ is a measure
of $\ell_d(t)$.
In Sec.~\ref{subsec:growing-length} we will give an insight on $\ell_G(t)$.

In the second line we made the following identification
\begin{equation}
\frac{1}{z_p} = \frac{1}{\zeta} + \frac{1}{z_d} 
\; . 
\label{eq:zp_and_zeta}
\end{equation}
With the knowledge that the best data collapse is found using $\zeta=1/2$ and that $z_d=2$, then
\begin{equation}
z_p = 2/5 \; . 
\end{equation}
This value is slightly different, $0.4$ vs. $0.5$, from the one that we estimated in~\cite{BlCoCuPi14}. We 
find, however, that it represents the numerical data more precisely and we stick to this way of reasoning in the 
remainder of this paper.

In both plots in Fig.~\ref{F2}, we also show the measured value
$A_c/L^{D_A}\simeq 0.6683$ for critical site percolation on the square lattice
\footnote{Note that for bond percolation on the square lattice, the same quantity $A_c/L^{D_A}\simeq 0.98....$ 
and for the site percolation on the triangular lattice, it is $A_c/L^{D_A}\simeq 0.655$.}.
It is in excellent agreement with the plateau in the rescaled value in the right panel.

\subsection{Pair connectedness function}
\label{subsec:pair-connectedness}

The correlation function used to characterise critical percolation is the 
pair connectedness $g(r)$ which measures 
the probability that two spins at a distance $r$ are in the same cluster.
In Sec.~\ref{sec:observables} we introduced the definition of $g(r,t)$
for a spin system undergoing quench dynamics, which is the one that we used for its 
practical computation in the Monte Carlo simulations.

\begin{figure}[h]
\begin{center}
 \includegraphics[scale=0.7]{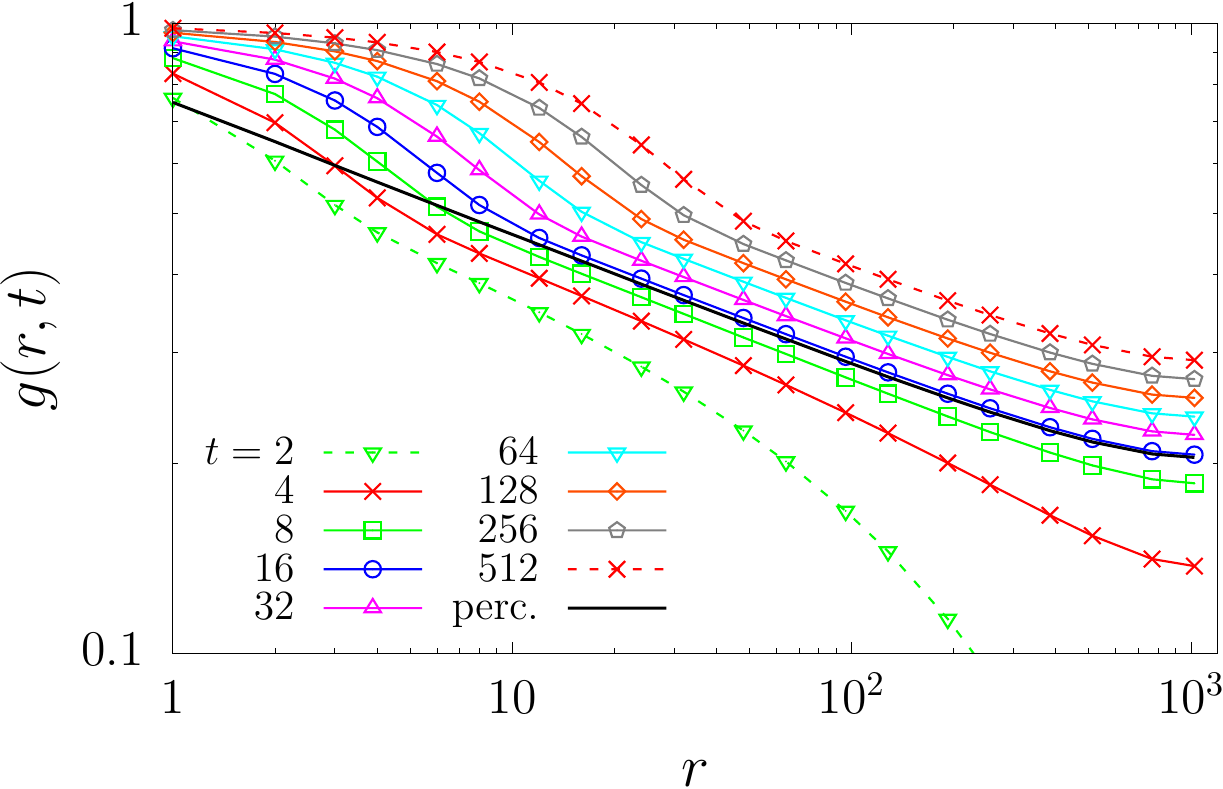}
\end{center}
\caption{\small Pair connectedness function $g(r,t)$ vs. $r$ for the $T=0$  dynamics 
on a square lattice with $L=2048$ and PBC, at the times shown in the key.
We also indicate with a continuous black line the pair connectedness for critical percolation on a square lattice with same size
and boundary conditions.}
\label{F3}
\end{figure}

In Fig.~\ref{F3} we show this ``two-point''  function at several times after the quench. The 
pair connectedness function in critical percolation
is also shown (with a black solid line).
The large distance behaviour at times longer than $16$ 
is very close to the one at the critical percolation point.

In Ref.~\cite{InCoCuPi16} the scaling properties of the pair connectedness function
were studied for random and clean Ising models evolving with kinetic Monte Carlo dynamics
with non-conserved order parameter. 
It was shown in this article that the data for $g(r,t)$
can be collapsed onto the same master curve in the percolation regime by rescaling the distance $r$ by the 
characteristic length $\ell_G(t)$ obtained from the excess energy, Eq.~(\ref{eq:growing-length-excess-energy}). 
This fact was explained in terms of a random site percolation problem at criticality with an effective
lattice spacing $\ell_d(t)$, evaluated with $\ell_G(t)$ in the models treated in~\cite{InCoCuPi16},
similarly to what we explained above. 

Specifically, the pair connectedness function is a function of $r$, $t$ and $L$, expected to scale as  
\begin{equation}
g(r,t,L) = g\left(\frac{r}{\ell_d(t)}, \frac{L}{\ell_p(t)}\right)
\end{equation}
before equilibration effects become important, that is to say, for $\ell_d(t) \ll L$.
For times $t$ beyond the characteristic time $t_p$ at which stable percolating clusters appear,
or equivalently for $t$ such that $\ell_p(t) \gg L$, the second argument vanishes. 
In this limit, the short  and long distance behaviour with respect to $\ell_d(t)$ can be distinguished.
The expectation is then that
\begin{equation}
 g(r,t) \sim \left( \frac{r}{\ell_d(t)} \right)^{-2\Delta_\sigma} \!\!\!\!\!  \qquad\qquad\quad r \gg \ell_d(t) \; ,
\end{equation}
while a correction will be needed at distances $r\ll \ell_d(t)$, with a crossover between the two extremes.

\begin{figure}[h]
\begin{center}
  \includegraphics[scale=0.55]{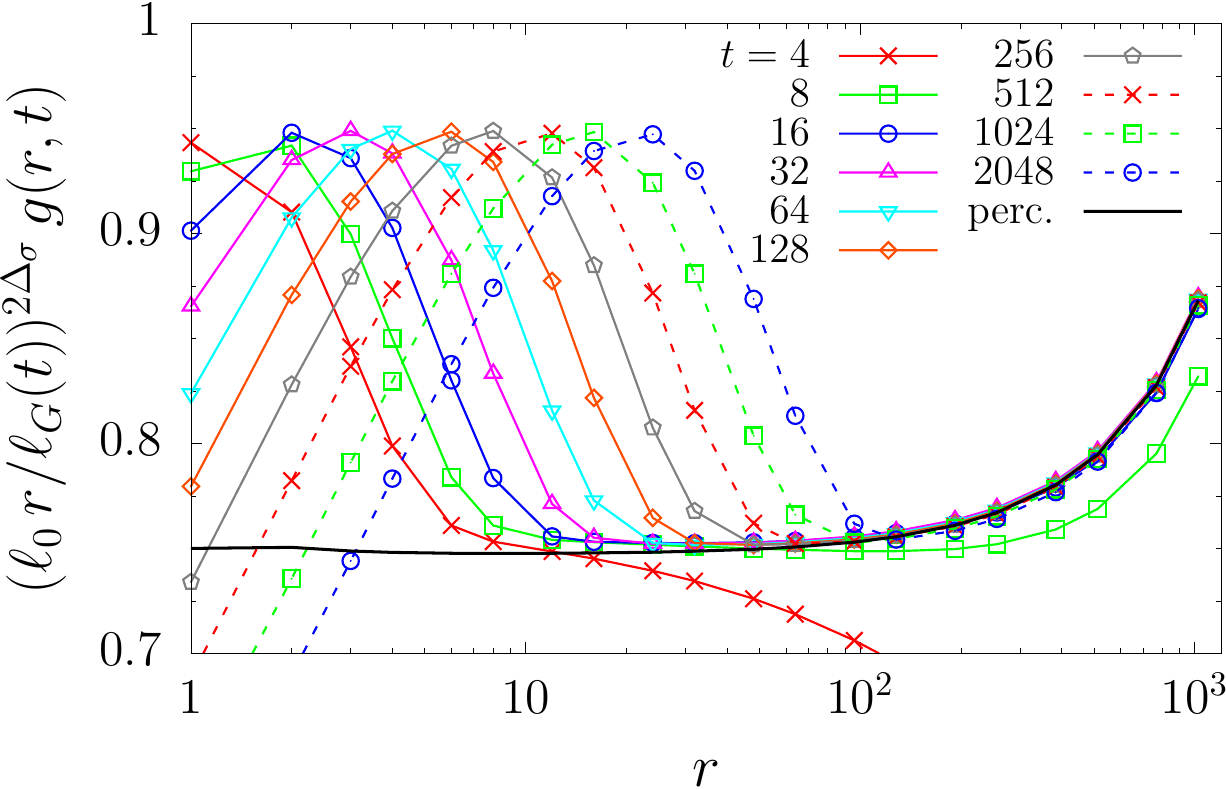}\quad%
  \includegraphics[scale=0.545]{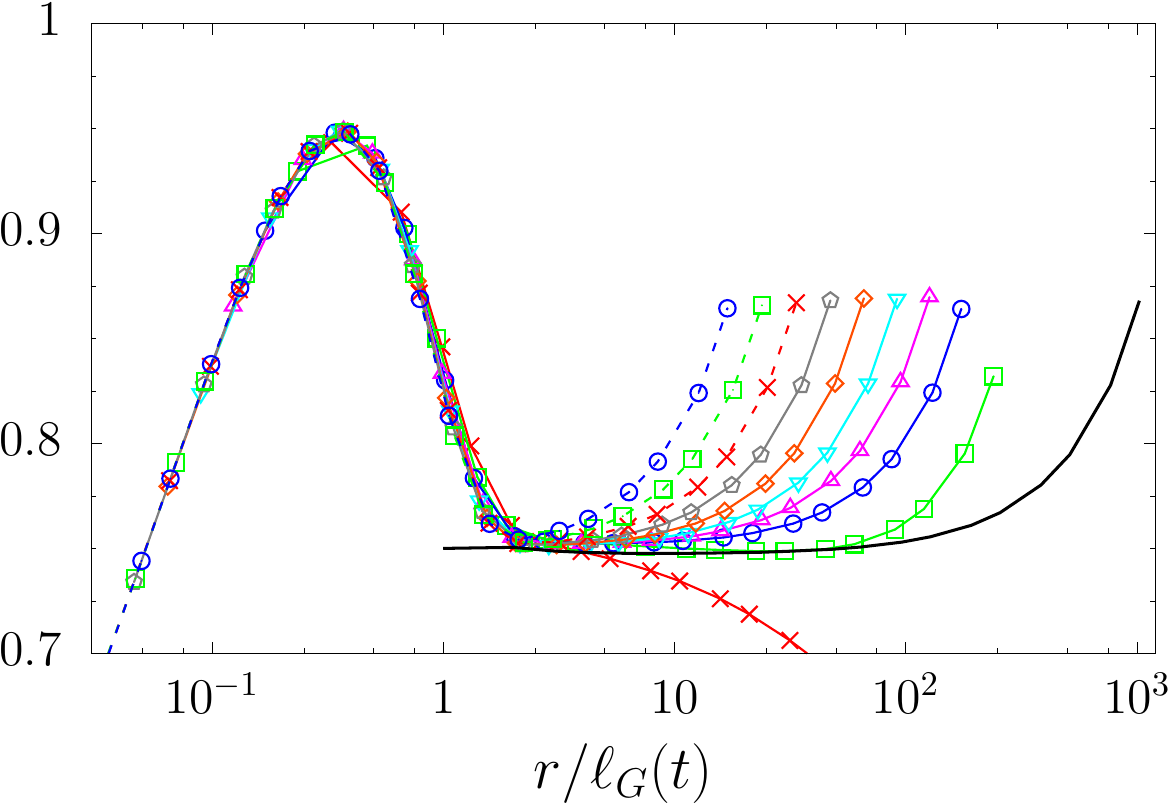}
\end{center}
\caption{\small 
Left panel: rescaled pair connectedness $(\ell_0 r/\ell_G(t))^{2 \Delta_\sigma}  g(r,t)$~vs.~$r$ 
for the $T=0$ dynamics on a square lattice with $L=2048$ and PBC, at the times shown in the key. 
Right panel: rescaled pair connectedness $(\ell_0 r/\ell_G(t))^{2 \Delta_\sigma}  g(r,t)$ vs. $r/\ell_G(t)$ 
at the same times. $\ell_G(t)$ is the characteristic length obtained as the inverse of the excess energy, defined
in Eq.~(\ref{eq:growing-length-excess-energy}), and it has been taken as a measure of $\ell_d(t)$.
In both figures, we also show $r^{2 \Delta_\sigma}  g(r,t)$ 
for critical percolation with a black curve. The value of the constant $\ell_0$ was chosen so that 
the data relative to the dynamical problem collapsed onto the data for critical percolation
in the region of the tail. The best result is given by $\ell_0 \simeq 5.5=\ell_G(t_p)$, see the text 
for an explanation.
}
\label{F4}
\end{figure}

Again, as explained in the previous section, we will take the numerical estimate of the excess energy characteristic length,
$\ell_G(t)$, as a measure of $\ell_d(t)$.
In the left panel of Fig.~\ref{F4} we display $(\ell_0 r/\ell_G(t))^{2 \Delta_\sigma}  g(r,t)$ {\it vs.} $r$
using $\Delta_\sigma=5/48$, the exponent of the critical percolation point. 
$\ell_0 = 5.5$ is a constant that we need to add to obtain the collapse of the dynamic data 
onto the critical percolation ones for $r>10^2$.
It corresponds to the numerical value of  $\ell_G(t)$  at $t\simeq 15$ 
and compatible with the expected behaviour $\ell_G(t) \simeq  \sqrt{2 t}$ since $5.5 \simeq \sqrt{2 \times 15}$.
This time is very close to the one at which we start observing percolation behaviour, see Fig.~\ref{F3}. 
It is then reasonable  to assume $\ell_0 = \ell_G(t_p)$. 
Furthermore, the correct way of scaling the distance $r$ on the horizontal axis
so that $g(r,t)$ matches the static counterpart in critical percolation at $t=t_p$ is 
$r \mapsto r / \ell_G(t)$, as it is done in the right panel of Fig.~\ref{F4} 
where  $\left[ \ell_0 r / \ell_G(t) \right]^{2 \Delta_\sigma}  \, g(r,t)$ is plotted against $r / \ell_G(t)$.
All data sets collapse with great precision, including the upturn of the curves at very long distances that is due to the PBC, 
and that is also present in the static data. 

The right panel in Fig.~\ref{F4} displays a more complete scaling of data, valid for 
long and short distances.

\subsection{Summary}

With the concise analysis of the behaviour of the largest cluster and pair connectedness 
correlation given in this Section, we illustrated the phenomenon that we will study in 
greater detail in the rest of the paper. 

The main conclusion so far is that at a characteristic time $t_p$ a stable critical percolation 
structure establishes and later grows, losing its critical properties. 
As the coarsening process starts right after the 
quench, the effective lattice spacing in the percolation problem is given by the 
dynamic growing length
\begin{equation}
r_0 \mapsto \ell_d(t) \; . 
\end{equation}

One of the questions that we will address below is how is the approach to critical 
percolation realised and which are the scaling laws that describe it. In particular, 
we will revisit the numerical determination of $z_p$.

\section{Detailed numerical analysis}
\label{sec:results}

In this Section we develop our analysis of the short time dynamics after the quench.
The logic of the order of presentation in this Section, and the main results obtained in each 
Subsection, are the following: 
\begin{itemize}
\item
Subsec.~\ref{subsec:order-parameter}.
The early scaling of the magnetisation density demonstrates that the approach to critical percolation 
is not due to the magnetisation of the sample. 
\item
Subsec.~\ref{subsec:overlap}.
The overlap gives a first proof of $t_p \simeq L^{z_p}$. 
\item
Subsec.~\ref{subsec:growing-length}.
We measure the excess energy growing length since the initial time.
The numerical values obtained provide the estimate of the dynamic growing 
length $\ell_d(t)$ used in our study.
\item
Subsec.~\ref{subsec:crossings}.
At $t_p$ the wrapping probabilities take the values of the critical percolation point.
\item
Subsec.~\ref{subsec:local-Glauber-angles}
The averaged square winding angles confirm the critical percolation phenomenon with $\kappa \approx 6$ 
and satisfy dynamic scaling with $\ell_d(t)$.
\item
Subsec.~\ref{subsec:largest-cluster}.
At $t_p$ the area and interface of the largest cluster have the fractal dimensions
of the critical percolation ones. 
 \item
 From the study of the number density of cluster areas in Subsec.~\ref{subsec:number-density-cluster}
we complete the understanding of the approach to critical percolation. 
\end{itemize}

\subsection{Order parameter}
\label{subsec:order-parameter}

The usual order parameter 
of the Ising model, the magnetisation, {\it is not} an adequate observable to detect the dynamic approach to critical percolation. 
Indeed, the magnetisation density scales as $m(t,L) \simeq ({t/ L^2})^{1/2}$ for 
small values of ${t/ L^2}$ (see the first panel in Fig.~\ref{FTmag}). 
Then, at $t \simeq L^{z_p}$, the magnetisation density is given by $m(L^{z_p},L) \simeq L^{{z_p/2} -1}$. 
Therefore, for any $z_p<2$ this quantity vanishes as a power of $L$. 
Taking $z_p=1/2$ for the square lattice, as measured in~\cite{BlCoCuPi14}, one has $m(L^{z_p},L) \simeq L^{-3/4}$ a very 
small value for large $L$, that vanishes in the thermodynamic limit. The value $z_p=2/5$ measured in Sec.~\ref{sec:phenomenon} implies $m(L^{z_p},L) 
\simeq L^{-4/5}$ also vanishing.
A vanishing magnetisation density is also found
on the other lattices. Accordingly, the percolation phenomenon that we observe is not due to the magnetisation of the 
sample.

\subsection{Overlap}
\label{subsec:overlap}

The two-replica overlap $Q(t,t_w,L)$ defined in Eq.~(\ref{eq:overlap}) 
was used in~\cite{BlCoCuPi14} to estimate the dependence of $t_p$ with $L$, for zero temperature quenches,
and the result $t_p \simeq L^{1/z_p}$ with 
$z_p = z_d/n$, $z_d=2$ and $n$ the coordination of the lattice, $n=n_c$, was thus found on the square, bow-tie,  Kagom\'e and  triangular lattices.

\begin{figure}[h!]
\begin{center}
\includegraphics[scale=0.58]{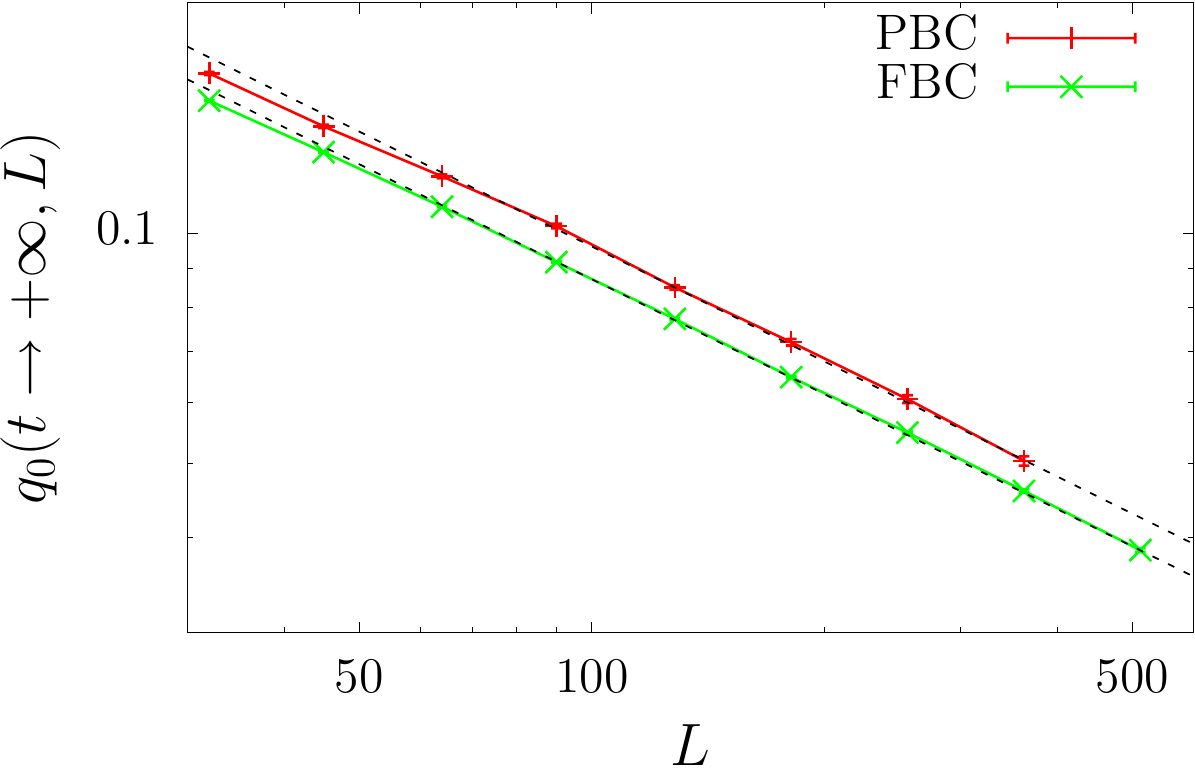}\quad%
\includegraphics[scale=0.58]{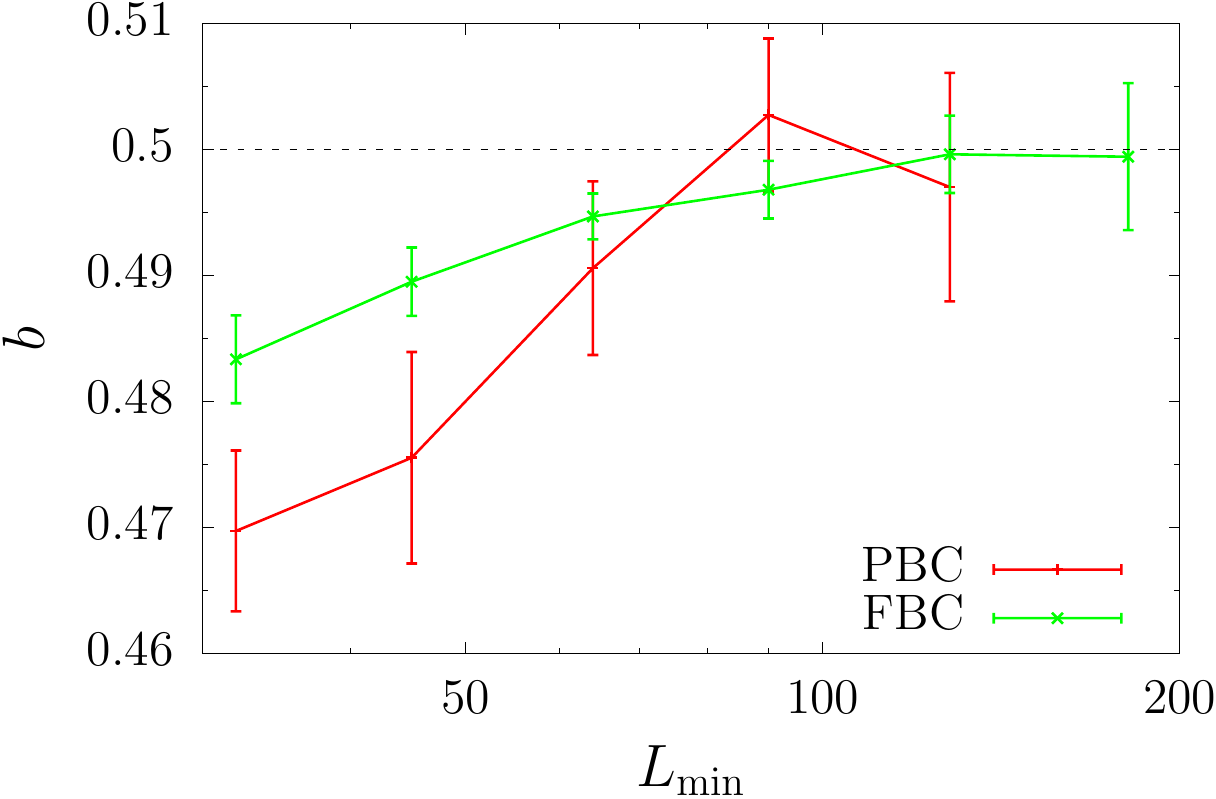}
\end{center}
\caption{\small 
Finite-size dependence of the two-replica overlap
between the initial configuration and the final frozen configuration,
$\lim_{t \rightarrow \infty}q_0 (t, L)$,
for the zero-temperature  dynamics on a
square lattice with PBC and FBC (the initial configuration is at infinite temperature).
In the left panel, we show the overlap as a function of the system size $L$. The dotted 
lines are proportional to $L^{-0.5}$.
In the right panel, we plot the exponent $b$ obtained from fitting the data as a function of $L_{\rm min}$. 
The horizontal dotted line is at $0.5$.
}
\label{Overlap12}
\end{figure}

In this paper, instead, we are going to show how a more precise analysis of the $L$-dependence of the
characteristic time $t_p$ can be made through other observables.
Nevertheless, in this Section we want to mention some other results regarding the two-replica overlap $Q(t,t_w,L)$
which are not strictly related to our problem of finding $t_p$, but still quite useful
to understand what is happening in the course of the relaxation dynamics following a sudden quench.

In a recent work~\cite{Machta13}, the overlap (\ref{eq:def-Q}) at $t_w=0$, $q_0(t,L)=Q(t,t_w=0;L)$ was measured numerically 
in the ferromagnetic $2d$ Ising model with kinetic Monte Carlo dynamics with non-conserved order parameter. It was shown in this paper 
that $q_0$ decreases algebraically, $q_0(t,L) \simeq t^{-\theta_h}$, 
with an exponent $\theta_h=0.22 (2)$  being very close to the one of the persistence probability, that is to say, the 
fraction of spins that have never flipped since the quench~\cite{DBG0,BMS}. Moreover, it was argued  that in equilibrium this quantity 
scales with the system size as  
$q_0 (t\gg t_{\rm eq}, L) \simeq L^{-b }$, with $t_{\rm eq}$ the equilibration time, and $b = 0.46 (2)$. Then, by an argument similar 
to the one for persistence, it can be argued that $b = z_d \theta_h = 2 \theta_h$, a relation roughly satisfied by the numerical 
data in~\cite{Machta13}. 

If one accepts $2\theta_h=b$, then 
the value  $b=0.46 (2)$ reported in~\cite{Machta13}
is not compatible with our previous measurement of 
$2 \theta_{\rm eff} = \, 2 \times 0.199(2)$~\cite{BlCuPi14} (see also~\cite{ChakrabortyDas16}). 
In order to settle this issue, we made our own measurement of $q_0$ but using much better statistics than in~\cite{Machta13}. 
In Fig.~\ref{Overlap12} we show our results 
for the square lattice with FBC averaged over $10^7$ samples and for PBC averaged over $10^6$ samples
(compared to only $3\times 10^4$ samples in \cite{Machta13}). 
In both cases, for large system sizes, the data are compatible with $b = 0.5$, shown as a dotted 
line in the left part of the figure that displays the measured values of the overlap between the initial configuration and
the final \textit{frozen} configuration (on a square lattice, it is either the fully magnetised configuration or a striped configuration),
as a function of the linear size $L$.
We denote it by $q_0 (t \rightarrow \infty , L)$.

In the right part of the figure, we show the exponent $b$ obtained from a fit of the data in the range $L 
\in [L_{\rm min},L_{\rm max}]$ with $L_{\rm max}=362$ for PBC and $L_{\rm max}=512$ for FBC. The exponents shown in this figure 
are functions of the smallest size $L_{\rm min}$ used in the fit. We observe that $b$ converges very quickly to $0.5$
for both boundary conditions. Moreover, the quality of the fit is always excellent if we remove the data with $L < 64$ while it 
deteriorates if we 
include data with smaller values of $L$. (The quality of the fit is measured by the value of the $\chi^2$ per number of degrees of freedom. 
This quantity is close or  smaller than one for $L_{\rm min} \geq 64$.) This result is at odds with the claim in~\cite{Machta13} 
that using poorer statistics obtained a value for the exponent $b$ that, the authors claim, is compatible with their estimate
of $2 \theta_h$. Instead, with much better statistics, we found $2\theta_{\rm eff} \simeq 0.40(2)$~\cite{BlCuPi14} and 
 here we measure $b \simeq 0.50(1)$, two different and well-distinguishable values. Therefore, $2\theta_h \neq 2\theta_{\rm eff}$,
 and with $q_0$ we cannot access the persistence exponent.

\subsection{Growing length}
\label{subsec:growing-length}

The Ising model on a square lattice evolving with single spin flip 
dynamics is the simplest coarsening system complying with the 
dynamic scaling hypothesis. At sufficiently long times, the  curvature driven mechanism for scalar non-conserved order
parameter~\cite{AlCa79} yields the growing length
\begin{equation}
\ell_d(t) =  \left[ \lambda_d(T) \, t \right]^{1/z_d} \qquad \mbox{with}
\qquad z_d=2 \; .
\label{eq:lambdadt}
\end{equation}
(This length is measured in units of the lattice spacing and time is measured in units of a microscopic 
time-scale that we did not write in this equation.)
The pre-factor $\lambda_d$ depends on temperature. It is very close to $2$ at $T=0$ and it monotonically decreases until reaching 
zero at $T_c$~\cite{SiArBrCu07,Arenzon-etal15}. Its variation at low temperature is very slow. 

\begin{figure}[h]
\begin{center}
\includegraphics[scale=0.75]{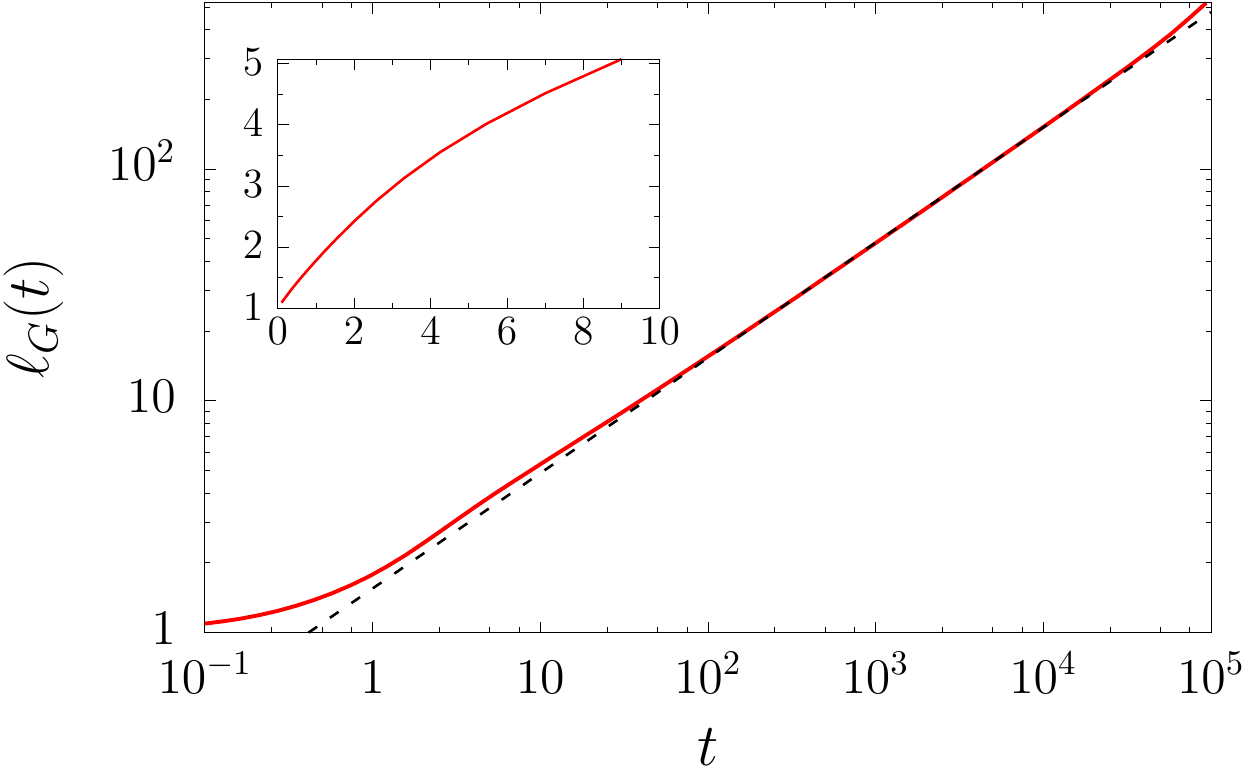}
\end{center}
\caption{\small Zero temperature dynamics on a square lattice with linear size $L=1280$ and PBC.
The plot shows the time evolution of the characteristic length $\ell_G$ (indicated with a red solid line)
obtained as the inverse of the excess energy, defined
in Eq.~(\ref{eq:growing-length-excess-energy}). The black dashed line represents the best fit of the function 
$f(t) = a \, t^b$ to the data in the interval $[10,10^4]$, 
yielding $a\simeq 1.54(1)$ and $b\simeq 0.4974(9)$. In the inset we show (in double linear scale) the same quantity
in the short time interval $[0,10]$.
}
\label{fig:lG_T0}
\end{figure}

The growing length (\ref{eq:lambdadt})
is easily recovered in numerical measurements of different observables.
For example, $\ell_G(t)$ evaluated from the excess energy in Eq.~(\ref{eq:growing-length-excess-energy})
is shown in  Fig.~\ref{fig:lG_T0}. The exponent $z_d=2$ establishes early after the quench, say 
at $t\simeq 10$, and is found over a wide time-interval,
before finite-size effects force saturation. However, the early time dependence of the growing length is 
especially important for our study of the approach to critical percolation. This is shown in a zoom
included as an inset in the figure. In the following, in all scaling analysis we will therefore
use the numerical evaluation of the growing length $\ell_G(t)$.  

(The zero-temperature dynamics on lattices that allow for 
finite-size blocked clusters, are peculiar. For instance, on the
honeycomb lattice, the excess energy growing length saturates at $\ell_G \simeq 4$ independently of the lattice size, 
see Fig.~\ref{GL_H}. This 
length corresponds, in this case, to the average distance between finite-size stable clusters.)

\subsection{Wrapping probabilities}
\label{subsec:crossings}
\begin{figure}[h]
\begin{center}
\includegraphics[scale=0.75]{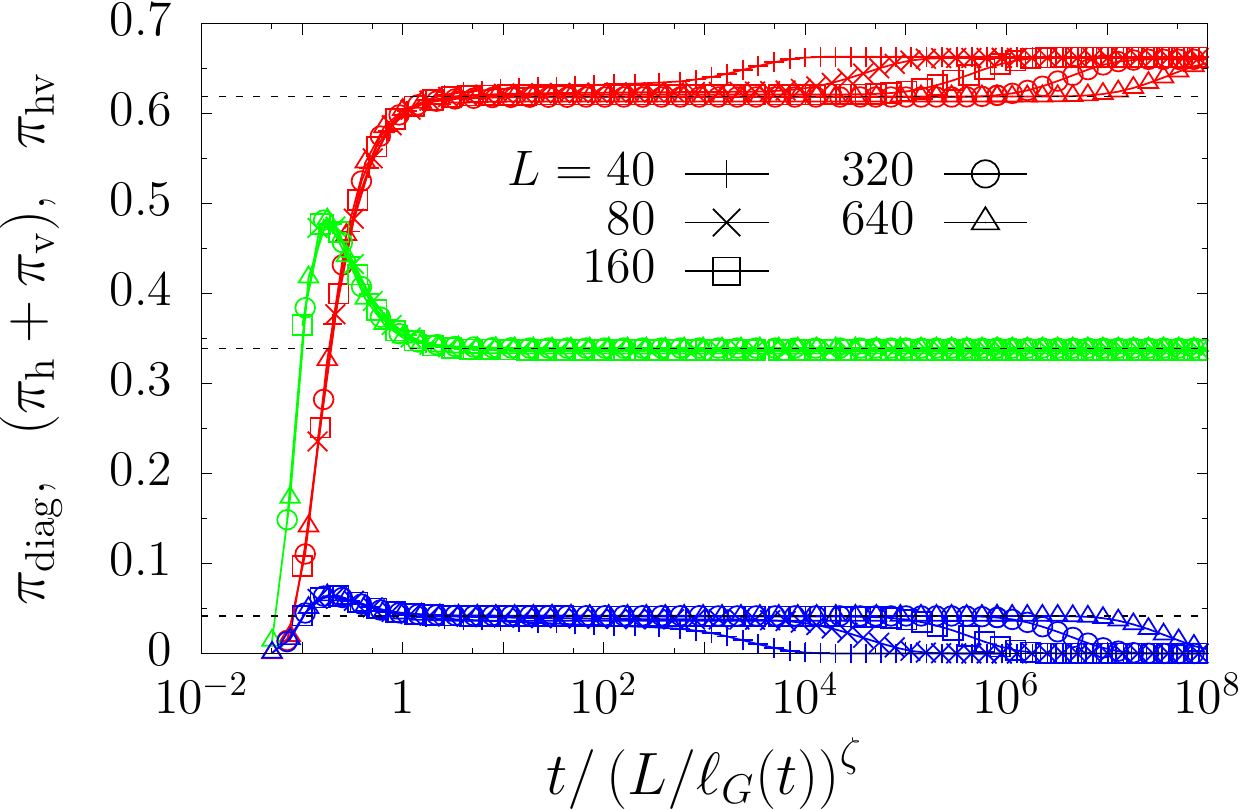}\vspace{2em}%

\includegraphics[scale=0.5]{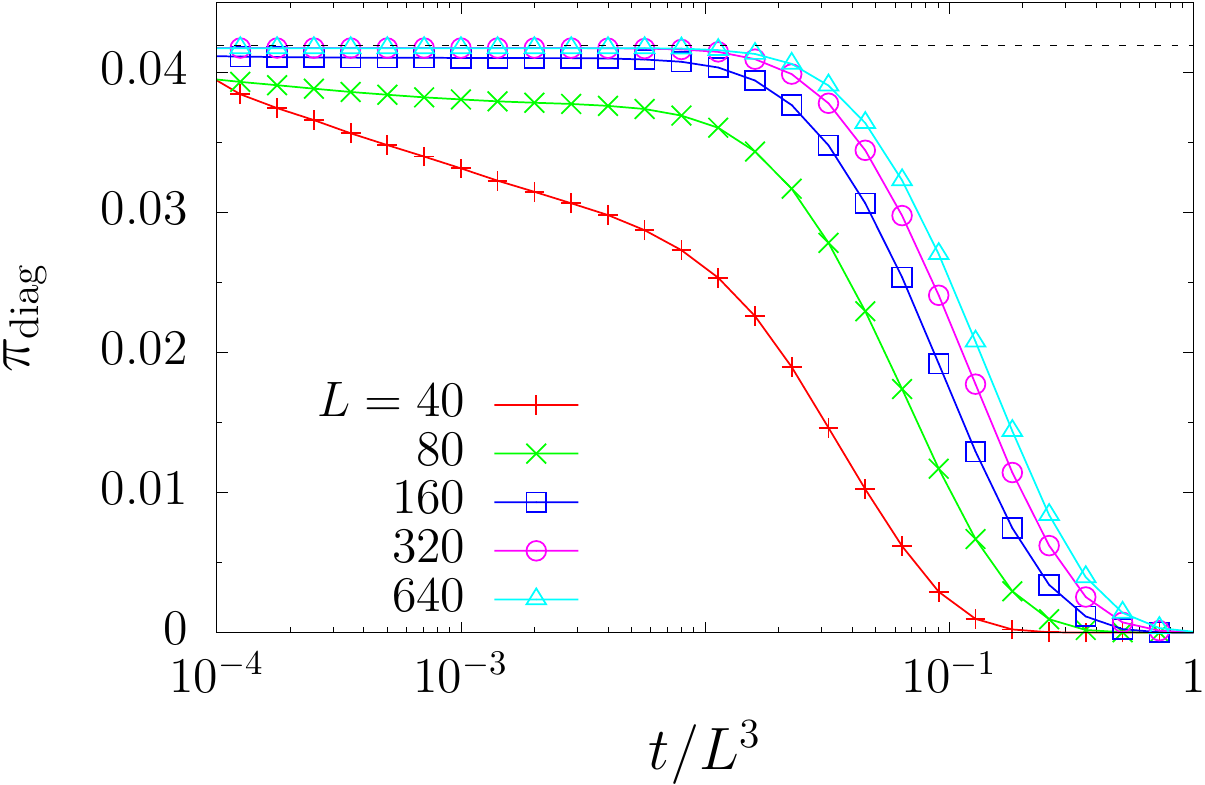}
\includegraphics[scale=0.5]{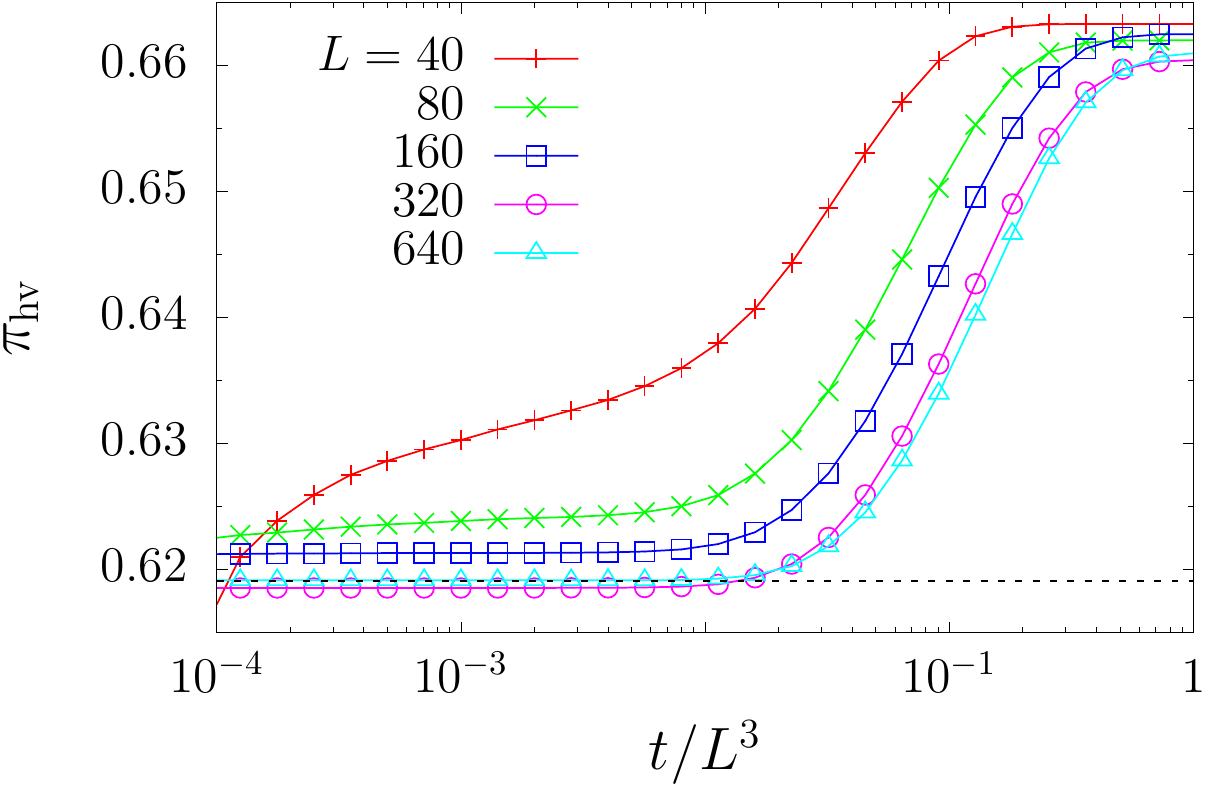}
\end{center}
\caption{\small Square lattice Ising model with PBC  evolving with dynamics at $T=0$. 
Upper panel: number of clusters that percolate
both horizontally and vertically (data above); either horizontally or vertically (data in the middle);
diagonally (data below), in all cases normalised by the total number of clusters that percolate in some direction.
Data are shown as a function of the rescaled time $t \, \left( L / \ell_G(t) \right)^{\zeta}$,
where $\ell_G(t)$, the characteristic length obtained as the inverse of the excess energy, is taken as a measure of $\ell_d(t)$, 
and $\zeta = 0.5$.
Lower panels: number of clusters that percolate diagonally (left) and horizontally and vertically (right) as a function of 
time scaled as $t/L^3$.
The horizontal dotted lines represent the associated probabilities in critical percolation.
}
\label{number-cluster-percolate}
\end{figure}

In Fig.~\ref{number-cluster-percolate} (upper panel) we show the probability of presence of clusters that 
wrap around the sample either horizontally and vertically, horizontally or vertically, or diagonally,
on a square lattice with PBC and for different values of the lattice linear size $L$.
Data are shown as a function of the rescaled time $t \, \left( L / \ell_G(t) \right)^{\zeta}$,
where $\ell_G(t)$, the characteristic length obtained as the inverse of the excess energy, is taken as a measure of
$\ell_d(t)$ the usual coarsening dynamic length scale.
The value of the exponent $\zeta$ was chosen so that the curves corresponding to different $L$ collapse one onto the other:
the best collapse is found by using $\zeta=0.50(1)$ implying 
\begin{equation}
\ell_p(t) \simeq \ell_d(t) (t/t_0)^{1/\zeta} \propto r_0 (t/t_0)^{1/z_p} \qquad
\mbox{and} \qquad z_p \approx 2/5 \; .
\end{equation}
This analysis confirms the value of $z_p$ found with the study of the largest cluster area scaling, a
value that is different from, but rather close to, the $z_p=1/2$ given in Ref.~\cite{BlCoCuPi14}.

These three probabilities, which are exclusive, add up to one at late times. For very early times, there can also be no wrappings, 
but this probability goes to zero very quickly (in units of $(L/\ell_d(t))^{\zeta}$).
The  curves for different system sizes scale well at small values of the scaling variable and until $10^4$ for the largest 
system size. The asymptotic values reached coincide with the predictions from critical percolation that are 
shown with dashed horizontal lines \cite{Pi94}. They correspond, from top to bottom to the probabilities 
$\pi_{\mathrm{hv}} = \pi^p(0)+\pi^p(Z\times Z)$, $\pi_{\mathrm{h}} + \pi_{\mathrm{v}} = \pi^p(1,0)+\pi^p(0,1)$ 
or $\pi_{\rm diag} \simeq \pi^p(1,1) + \pi^p(1,-1)$ in Eq.~(\ref{eq:wrapping-probabilities-critical-perc}).
Note that other situations 
can also exist, like a cluster winding in the $(2,1)$ direction, which would be a cluster wrapping twice in the horizontal direction and once in the vertical direction.
However, as we have already mentioned, the probabilities of such configurations are at least two orders of magnitude smaller than the ones
of the $(1,1)$ and $(1,-1)$ configurations, so we neglect them. 

From the data shown in Fig.~\ref{number-cluster-percolate} we can deduce that the characteristic time $t_p$ corresponds approximately to
the time at which the probabilities $\pi$s reach the plateaus set by the values of ordinary critical percolation.
This occur when $t / (L / \ell_d(t) )^{\zeta} \simeq 1$, approximately. Then

\begin{equation}
1 \simeq t_p/(L/\ell_d(t_p))^{\zeta} 
\qquad\qquad
\Rightarrow 
\qquad\qquad
t_p \simeq L^{\zeta z_d/(\zeta+z_d)}
\end{equation}
if one assumes $\ell_d(t) \sim t^{1/z_d}$ and then
\begin{equation}
t_p \simeq L^{2/5} \; . 
\end{equation}
given that $z_d=2$ and $\zeta \approx 1/2$.

If we use the relation $ t_p/(L/\ell_d(t_p))^{\zeta} \simeq 1$, using 
the numerical estimate for $\ell_d(t)$ given by $\ell_G(t)$, we find
$t_p \simeq 3.7(1), \, 4.9(1), \, 6.4(1), \, 8.5(1), \, 11.3(1)$  for $L=40, \, 80, \, 160, \, 320, \, 640$, respectively.
These are relatively short times implying
that most, if not all, numerical data in the literature lie in a regime in which the percolation structure is already
present. 

\vspace{0.5cm}

\begin{figure}[h]
\captionsetup[subfigure]{labelformat=empty}
\begin{center}
 \begin{tabular}{m{0.6\linewidth} m{0.5\linewidth}}
  \includegraphics[scale=0.65]{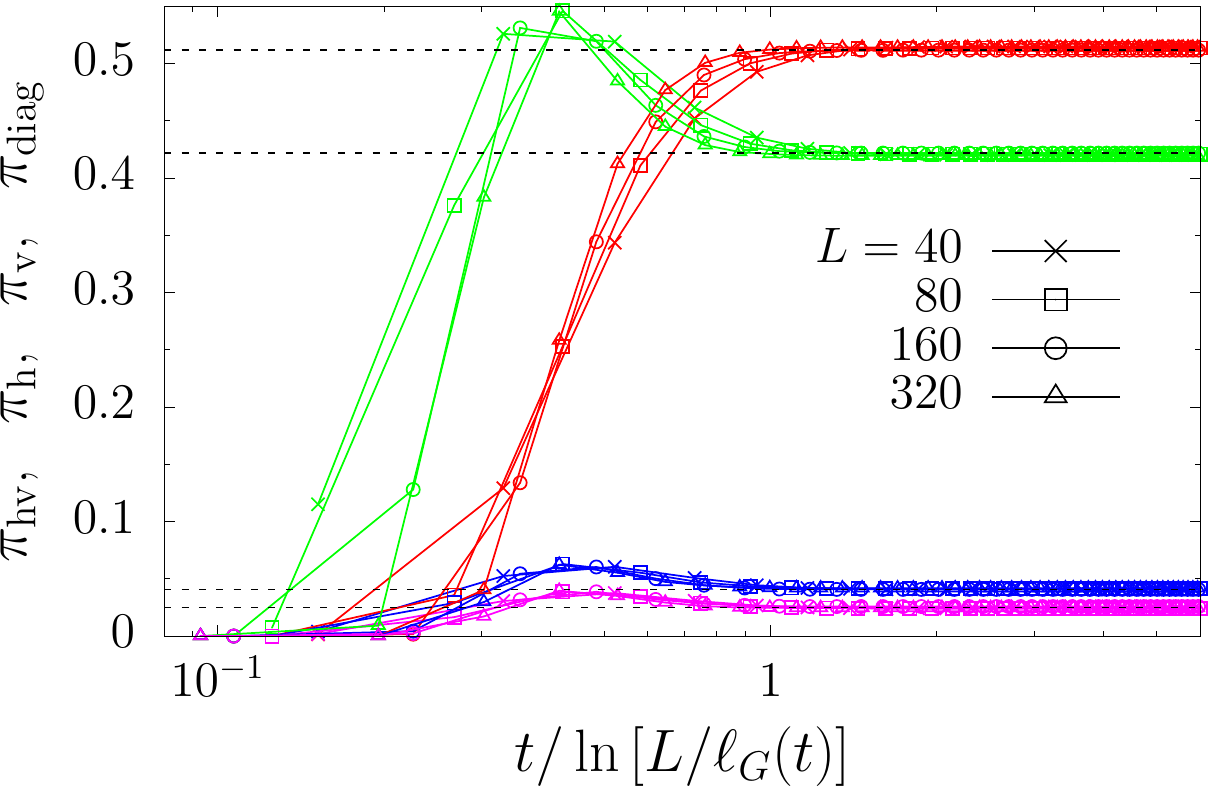}&
  \begin{minipage}[t]{0.5\textwidth}
  \subfloat[$t=1.75$]{\includegraphics[scale=0.5,angle=90]{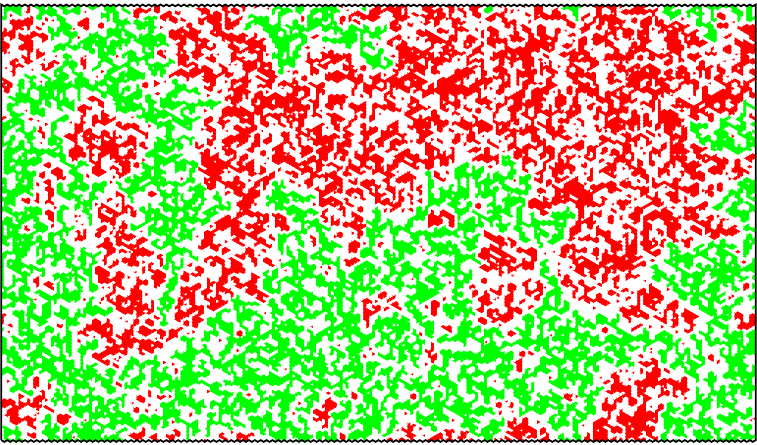}}\\
  \subfloat[$t=2$]{\includegraphics[scale=0.5,angle=90]{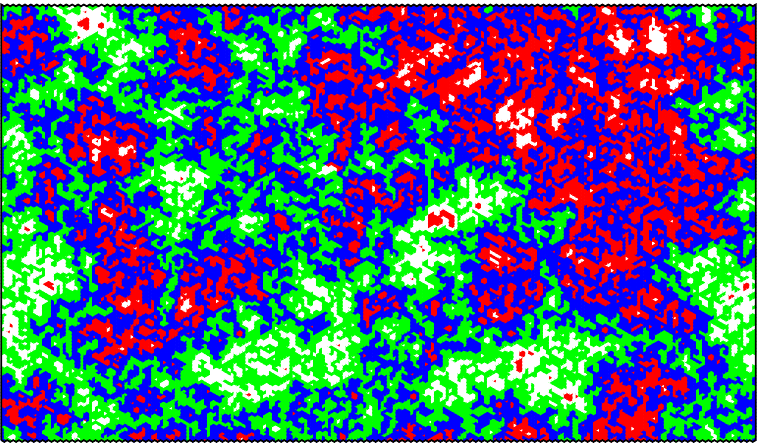}}%
  \end{minipage}
 \end{tabular}
\end{center}
\caption{\small 
$T=0$ dynamics on a honeycomb lattice with PBC.
On the left, we show the probabilities $\pi_{\rm hv}$ (red), $\pi_{\rm h}$ (green), $\pi_{\rm v}$ (blue), $\pi_{\rm diag}$ (purple),
for different lattice sizes $L$. The horizontal dashed lines correspond to the expected values at critical percolation
for a rectangular sheet with aspect ratio $\sqrt{3}$: $0.5120$, $0.4221$, $0.0408$ and $0.0250$, respectively.
Data are plotted against the scaling variable $t/\ln{ \left[ L/\ell_G(t) \right] }$, 
with $\ell_G(t)$ the excess energy growing length, taken as a measure of $\ell_d(t)$.
On the right, we show two snapshots of the evolution of a spin configuration on a honeycomb lattice with $L=160$ and PBC,
under zero temperature dynamics. The colour code is the same as in Fig.~\ref{fig:snapshots-triangular} and the times 
at which they were taken are indicated below the snapshots.
}
\label{Gl-Hon_pwrap}
\end{figure}

A cross-over to a  longer time-scale regime next appears and it corresponds to the 
disappearance of configurations with clusters percolating along a diagonal direction and the consequent increase in number 
of the
clusters that percolate along both Cartesian directions (cross topology). In fact, the interfaces winding in a diagonal direction
are not stable under zero-temperature dynamics on the square lattice, even though 
we showed that they can appear very early and
last for very long. This is due to the fact that, once they have established a \textit{ladder} shape (see, 
for example,~\cite{SpKrRe01,OlKrRe12}),
they can move in the perpendicular direction by means of spin flips with no energy cost, and they can thus 
wander for a very long time before 
disappearing by annihilating with another interface.
 
This last regime scales with a different power of 
the linear system size $L$, as shown in the lower panels in the same figure that display the proportion of clusters
percolating diagonally (left) and horizontally and vertically (right).
In fact, we could collapse the curves $\pi_{\rm diag}$ and $\pi_{\rm hv}$ corresponding to different system sizes, by choosing a
scaling variable $t/L^{u}$, with $u\simeq 3$ giving the best result. The exponent $u$ is then the numerical 
estimate of the exponent $z_{\rm eq}$ that controls the final approach to equilibrium in this case.
Note that the collapse works better for large system sizes: 
finite-size effects may render difficult the exact determination of this last scaling
regime by using the collapse method.

In Fig.~\ref{Gl-Hon_pwrap} we show similar probabilities computed on the honeycomb lattice.
As explained in the introduction, see Fig.~\ref{fig:honeycomb}, 
we built this lattice from the square one removing some vertical bonds. Then, wrapping 
around the lattice vertically is $\sqrt{3}$ longer than doing it horizontally.
The corresponding probabilities are~\cite{BaKrRe09,PruMol04}
$\pi_{\rm hv} = \pi^p(0) + \pi^p( Z \times Z) \simeq 0.5120$, \, 
$\pi_{\rm h} = \pi^p(1,0) \simeq 0.4221 $, \,
$\pi_{\rm v} = \pi^p(0,1) \simeq 0.0408 $ and 
$\pi_{\rm diag} = 1 - \pi_{\rm hv} - \pi_{\rm h} - \pi_{\rm v} \simeq \pi^p(1,1) + \pi^p(1,-1) \simeq 0.0250 $.
In the case of the zero-temperature dynamics on the honeycomb lattice, as we have already mentioned in the 
introduction to this Section,
the system gets blocked in a very short time in a spin configuration with a highly complex domain pattern, 
see the snapshots in Fig.~\ref{fig:Gl-Hon-snapshots}. This is due to the fact that the lattice has odd coordination number.
The domain pattern of these so-called \textit{frozen} configurations are richer in structure than the long-lived stripe states
occurring in the late stages of the coarsening dynamics on the square lattice.
Nevertheless we can still observe a transition from the initial fully disordered spin configuration to a 
critical-percolation-like state.

Since the time required by the system to freeze depends logarithmically on the system linear size $L$~\cite{TakanoMiyashita}, 
precisely $t_{\mathrm{freeze}}(L) \simeq \mathrm{const.} + 4.95 \, \ln{L}$,
the time $t_p$ to reach the percolation regime cannot be a power law $L^{z_p}$, 
as conjectured in the case of dynamics on the square lattice.
Instead, we expect $t_p(L) \propto \ln{L} $.
 In Fig.~\ref{Gl-Hon_pwrap}, we show that the wrapping probabilities for various sizes 
collapse with a rescaling of time by $\ln{ \left[ L/\ell_G(t) \right]}$, 
thus giving a first indication that for this lattice $t_p \sim \ln L$.

\subsection{Averaged squared winding angle}
\label{subsec:local-Glauber-angles}

We now consider the variance of the winding angle $\langle \theta^2(x,t)\rangle$, defined in Sec.~\ref{sec:observables}, 
on various lattices.

\begin{figure}[h]
\begin{center}
\includegraphics[scale=0.45]{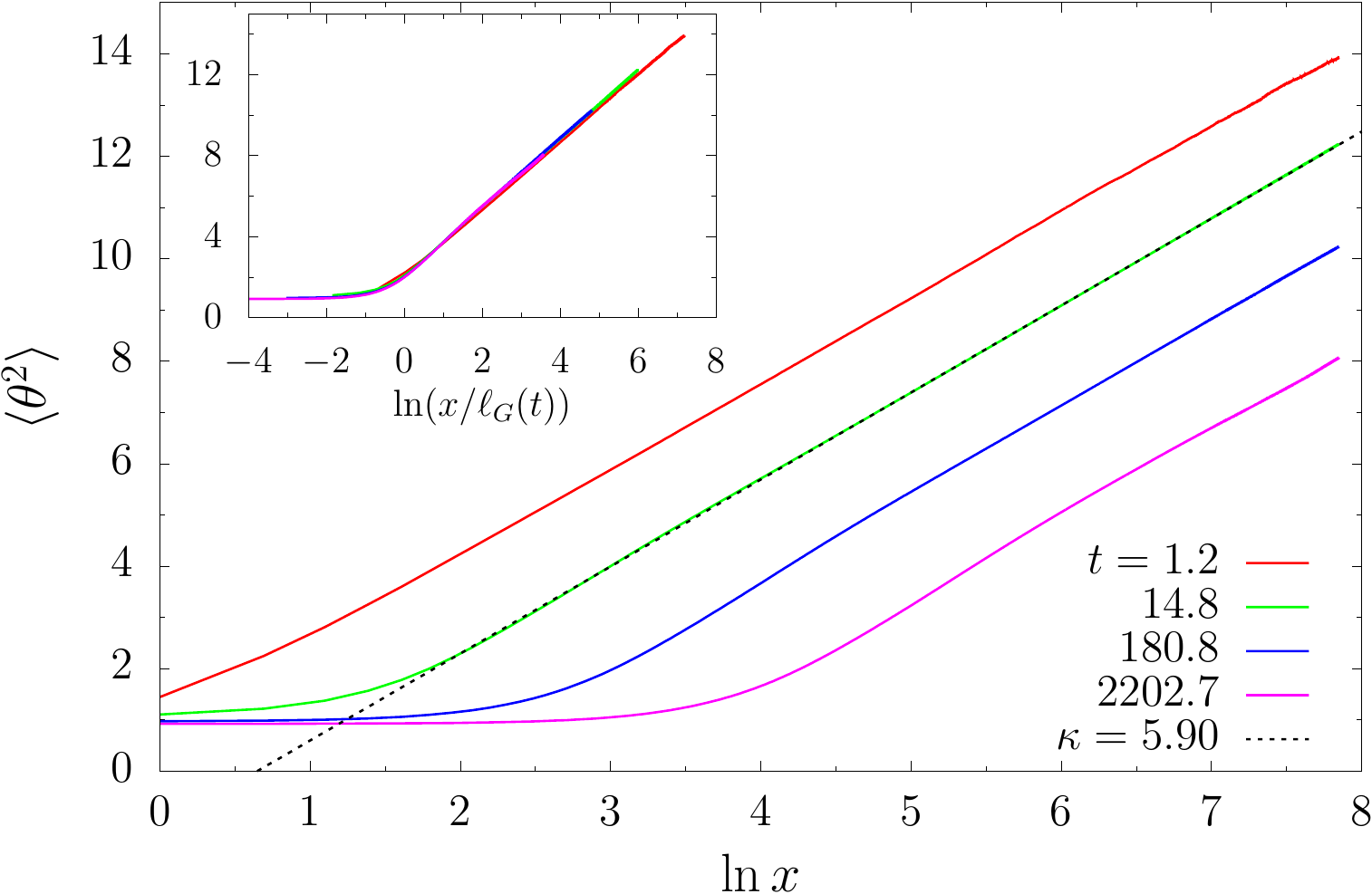}\quad%
\includegraphics[scale=0.45]{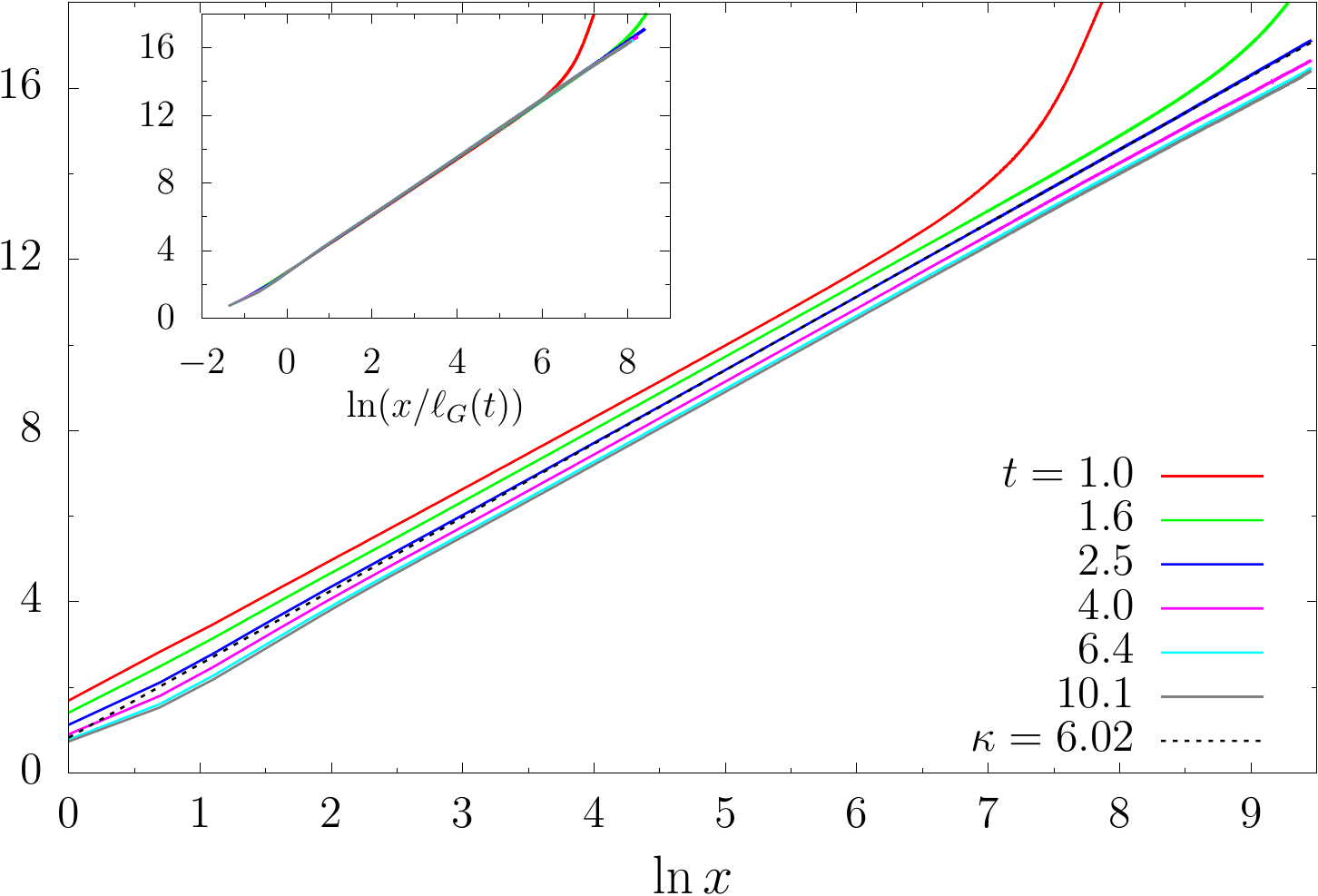}
\end{center}
\caption{\small Dynamics of the Ising model with $L=1280$
quenched to zero temperature. Averaged square winding angle, $\langle \theta^2 (x,t) \rangle$, against $\ln{x}$ with $x$ being the
curvilinear distance along a cluster interface. In the left panel the data are relative to the model on the square lattice and 
the  interfaces considered wrap around the system. The right panel is relative to closed interfaces on
the honeycomb lattice. The insets contain the same quantities plotted against $\ln[x/\ell_G(t)]$ with $\ell_G(t)$ 
the characteristic length obtained as the inverse of the excess energy.
}
\label{FigWGlauber}
\end{figure}

In the left panel of Fig.~\ref{FigWGlauber}, we show $\langle\theta^2(x,t)\rangle$ 
for domain walls that wrap around the lattice in one direction, plotted against $\ln{x}$ with
$x$ the curvilinear distance along the domain wall, in the case of the $T=0$  dynamics on a square lattice with PBC.
A fit of the function $f(x)=a + \frac{4 \kappa}{(8 +\kappa)} \ln{x}$ to the data at $t \simeq14.84$ (beyond $t_p$) is also shown.
The fit yields $\kappa \simeq 5.90(1)$, that is rather close to the expected $\kappa=6$ of critical percolation cluster hulls. 

\vspace{0.25cm}

In the right panel of Fig.~\ref{FigWGlauber},  we show $\langle \theta^2(x,t)\rangle$ for domain walls with a positive total winding angle, 
in the case of the $T=0$ dynamics on a honeycomb lattice with PBC.
These are domain walls that do not wrap around the lattice. Note that such interfaces exist for any short time considered. For the 
earliest time shown, the curve bends upwards at the longest length $x$,  indicating that these short-time/long-length domain walls
do not have the statistics of critical percolation. At the next time shown, 
$t=2.5$, the curve is nearly straight proving the $\ln x$ dependence, and a fit of the function $f(x)=a + \frac{4 \kappa}{(8 +\kappa)} \ln{x}$ yields
$\kappa \simeq 6.02(1)$, again very close to the value expected for critical percolation cluster hulls. 

We also note that for short curvilinear distance $x$ along a domain wall, the curves are nearly flat suggesting 
$\kappa =0$. This corresponds to the equilibration of the interfaces that become regular over a distance proportional to the dynamic growing length
$\ell_d(t)$. This remark allows us to rescale $\langle \theta^2(x,t)\rangle$ as a function of $x/\ell_d(t)$. 
This is done in the insets in the two panels, where we plot $\langle \theta^2(x,t)\rangle$ against $\ln{(x/\ell_G(t))}$,
taking again the characteristic length $\ell_G(t)$, obtained as the inverse of the excess energy, as a measure of $\ell_d(t)$.
As one can see, the measures corresponding to different times collapse one onto the other when performing this scaling.

\subsection{Largest cluster}
\label{subsec:largest-cluster}

In Sec.~\ref{sec:phenomenon} we exposed the main features of the approach to percolation 
phenomenon showing the time-dependent behaviour of the area of the largest
cluster and its scaling properties. Here we complete the analysis of this observable 
by working with different lattices. We also analyse the behaviour of the length of
its interface.

\begin{figure}[h]
\begin{center}%
 \includegraphics[scale=0.52]{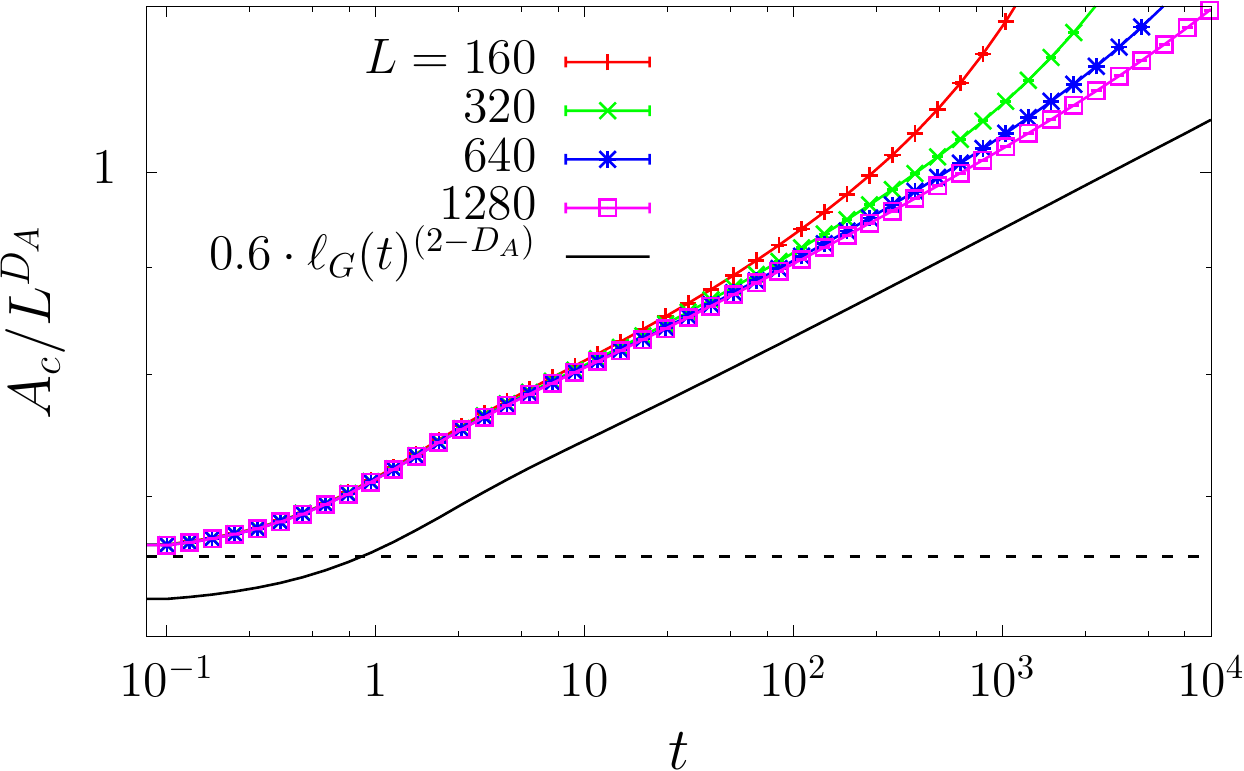}
 \includegraphics[scale=0.52]{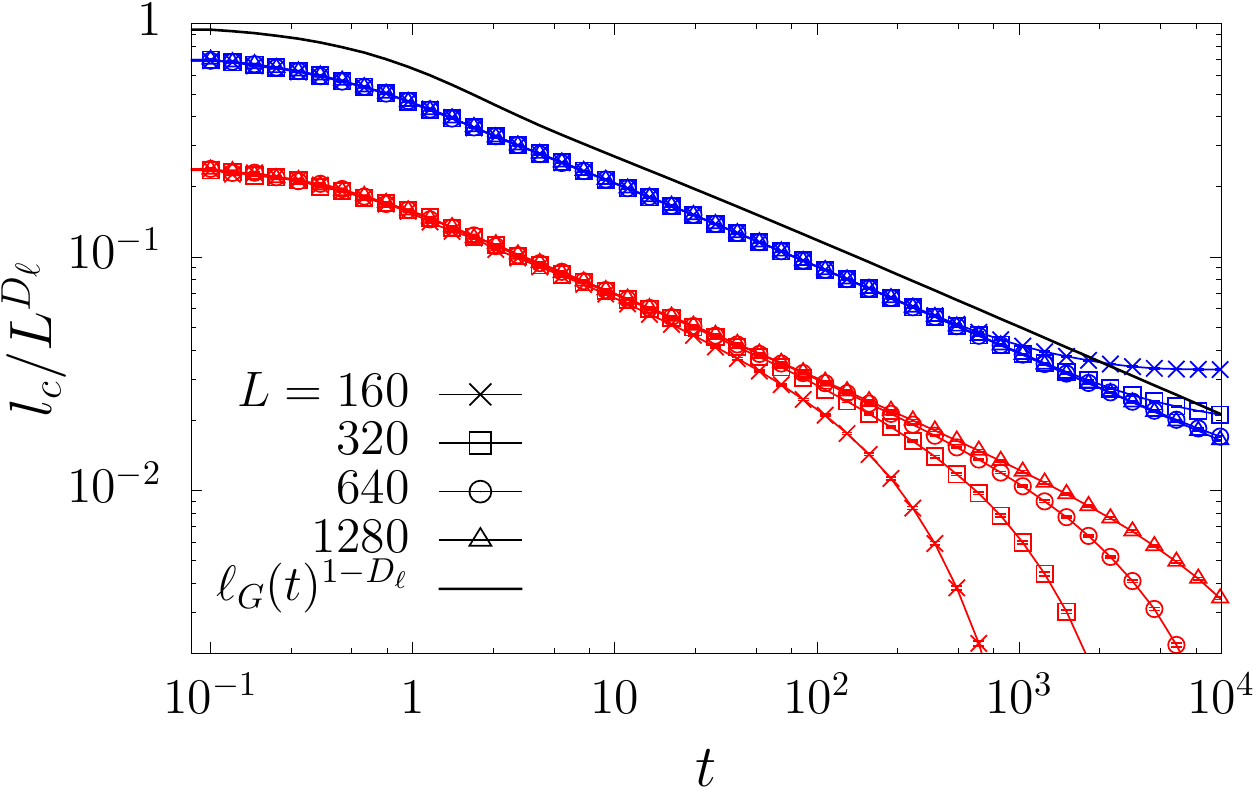}
\end{center}
\caption{\small Analysis of the geometric properties of the largest cluster in the case of the $T=0$ dynamics on a
triangular lattice with PBC, for different values of $L$. 
On the left $A_c/L^{D_A}$ vs. $t$, and on the right $l_c/L^{D_l}$ vs. $t$,
with $A_c$ the size of the largest cluster and $l_c$ the average length of its walls.
$D_A$ and $D_{\ell}$ are the fractal dimensions of the size and the interface of the percolating cluster
in $2d$ critical percolation. The interface of the largest cluster has two contributions: from the
wrapping hulls with zero total winding angle (shown in blue) and from the non-wrapping hulls with nonzero total winding angle 
(shown in red).
The horizontal dashed line in the left panel corresponds to the ratio $A_c/L^{D_A}$ for site percolation on a triangular lattice
at the critical occupation probability, that is approximately $0.655$.
}
\label{newftr}
\end{figure}

We now analyse the largest cluster geometric properties on the triangular lattice.
In  Fig.~\ref{newftr}, we show $A_c/L^{D_A}$ vs. $t$ (left) and $l_c/L^{D_l}$ vs. $t$ (right). In both cases, we also show a convenient power of the 
growing length $l_d(t)$. Concerning the areas, in the right panel
we plot $\ell_G(t)^{2-D_A}$ (multiplied by an arbitrary constant $0.6$) and we observe that, up to a constant, it behaves as 
$A_c/L^{D_A}$, apart from finite size corrections. The numerical value obtained at the earliest time (it  corresponds 
to $t=0$ but we show it at $t=0.1$ in order to remain on a logarithmic scale) is in good agreement with 
the corresponding value for site percolation on a triangular lattice at the critical point, 
$A/L^{D_A} \simeq 0.655$, that is shown as a horizontal dashed line.

Concerning the interface of the largest cluster, we show here the time
evolution of its length, denoted by $l_c$.
It is important to clarify that the interface of a cluster can be made of many hulls, also known as domain walls
(see Sec.~\ref{sec:observables} for the definition of domain wall on a lattice). In the case shown here we are considering 
the contribution to $l_c$ coming from wrapping hulls (having zero total winding angle) 
and the one coming from non-wrapping hulls (having nonzero total winding angle) separately, since, in general, they
may scale differently with $L$. Moreover, because of the coarsening process, non-wrapping hulls are destined to disappear,
while wrapping hulls can last forever.
We also show $\ell_G(t)^{1-D_l}$ which, apart from a proportionality constant and finite size corrections for large times, 
seems to behave like the contribution $l_c/L^{D_{\ell}}$ coming from wrapping hulls (the upper group of curves).
The data for non-wrapping hulls is similar but with much stronger finite size corrections appearing at shorter times.

\begin{figure}[h]
\begin{center}%
 \includegraphics[scale=0.52]{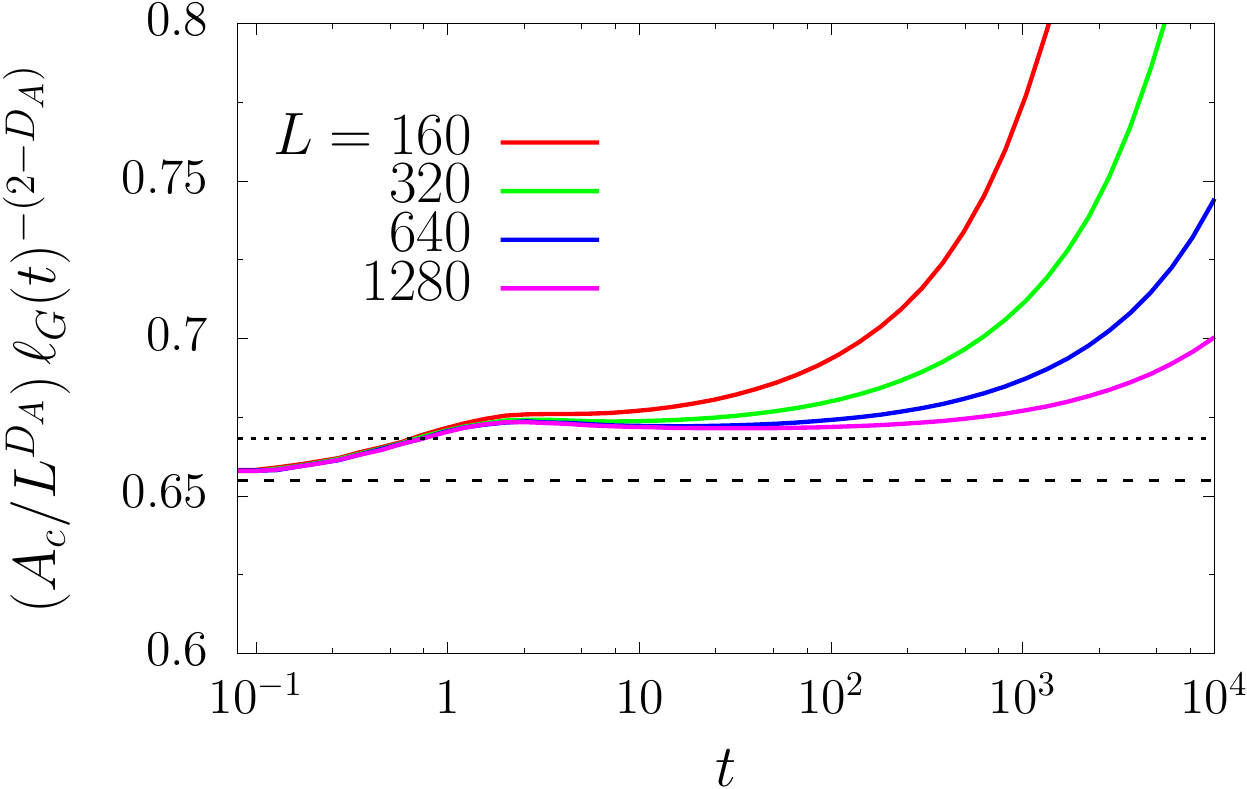}
 \includegraphics[scale=0.52]{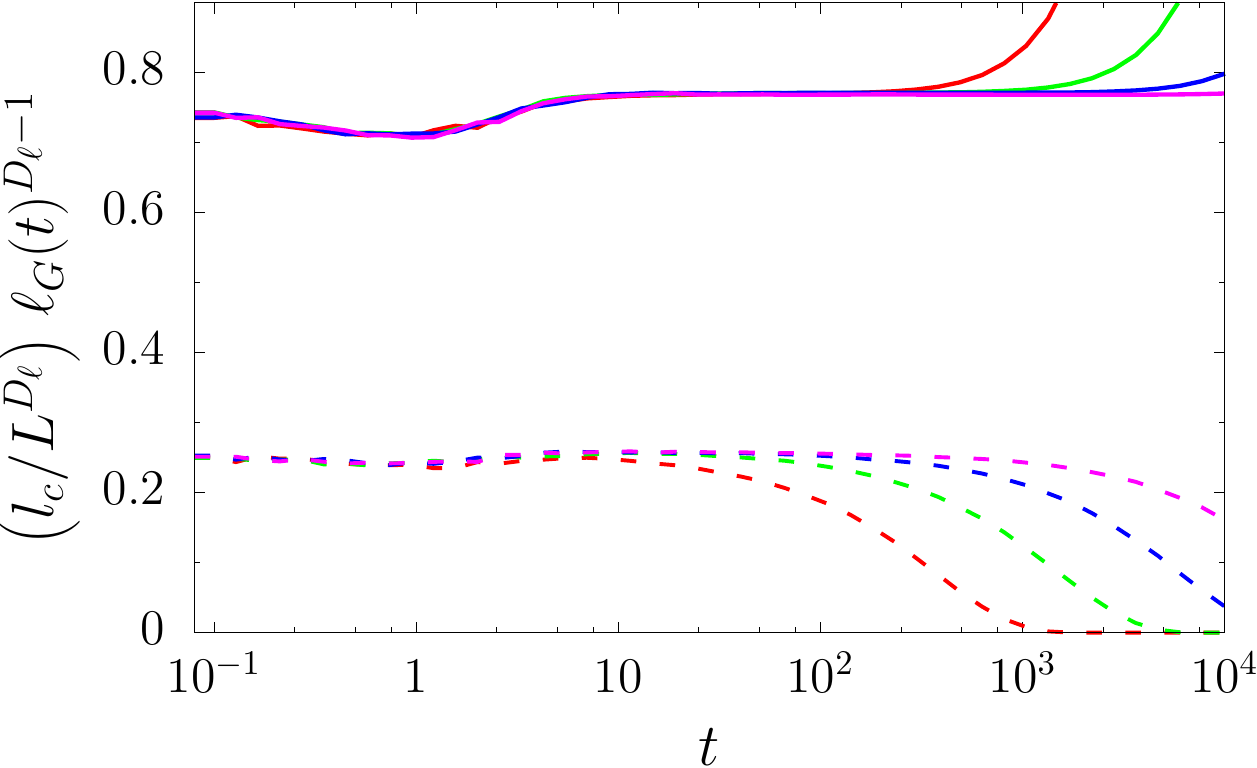}
\end{center}
\caption{\small Analysis of the geometric properties of the largest cluster in the case of the \,
$T=0$ dynamics on a triangular lattice with PBC, for different values of $L$. 
On the left $ ( A_c/L^{D_A} ) \; \ell_G(t)^{-(2-D_A)} $ vs. $t$, 
and on the right  $ (l_c/L^{D_{\ell}}) \; \ell_G(t)^{(D_{\ell}-1)} $ vs. $t$, where $\ell_G(t)$ is the characteristic 
length obtained as the inverse of the excess energy. As done in Fig.~\ref{newftr}, the contributions to the largest cluster
interface coming from wrapping (continuous lines) and non-wrapping hulls (dashed lines) have been separated.
The colour code is the same in both panels.
}
\label{newftr2}
\end{figure}

From the plots in Fig.~\ref{newftr} we can conclude that a better analysis of data is achieved 
by plotting ${ (A_c/ L^{D_A} )} \; \ell_G(t)^{-(2-D_A)}$ vs. $t$ and $ (l_c/L^{D_{\ell}}) \; \ell_G(t)^{D_{\ell}-1} $ vs. $t$. 
The two cases are shown in Fig.~\ref{newftr2}. 
We note that, apart from finite size corrections, $ ( A_c/L^{D_A} ) \, \ell_G(t)^{-(2-D_A)} $ is constant after a short time $\simeq 1$ 
which does not depend on the system size. We can interpret this value as the time it takes for the growing length to be in the asymptotic regime. 
After $t\simeq 1$, the rescaled quantity remains constant with a value that is very close to the one 
for the square lattice shown in the right panel of Fig.~\ref{F2}. To make this claim clearer,
in the same plot we also show the expected value of the ratio $A_c/L^{D_A}$ for site percolation 
at the critical occupation probability on the triangular lattice ($\sim 0.655$), indicated by a dashed horizontal line, 
and on the square lattice ($\sim 0.668$), indicated by a dotted horizontal line.
Similar results are obtained for the hull length: 
$ (l_c/L^{D_{\ell}}) \, \ell_G(t)^{(D_{\ell}-1)} $ is also constant after $ t \simeq 5$ and this  does not depend on the system 
size either, see the right panel in Fig.~\ref{newftr2}. Again, the two contributions to $l_c$ coming from wrapping hulls and non-wrapping ones
have been separated, with the former ones being represented by continuous lines, the latter by dashed lines.

\begin{figure}[h]
\begin{center}%
 \includegraphics[scale=0.52]{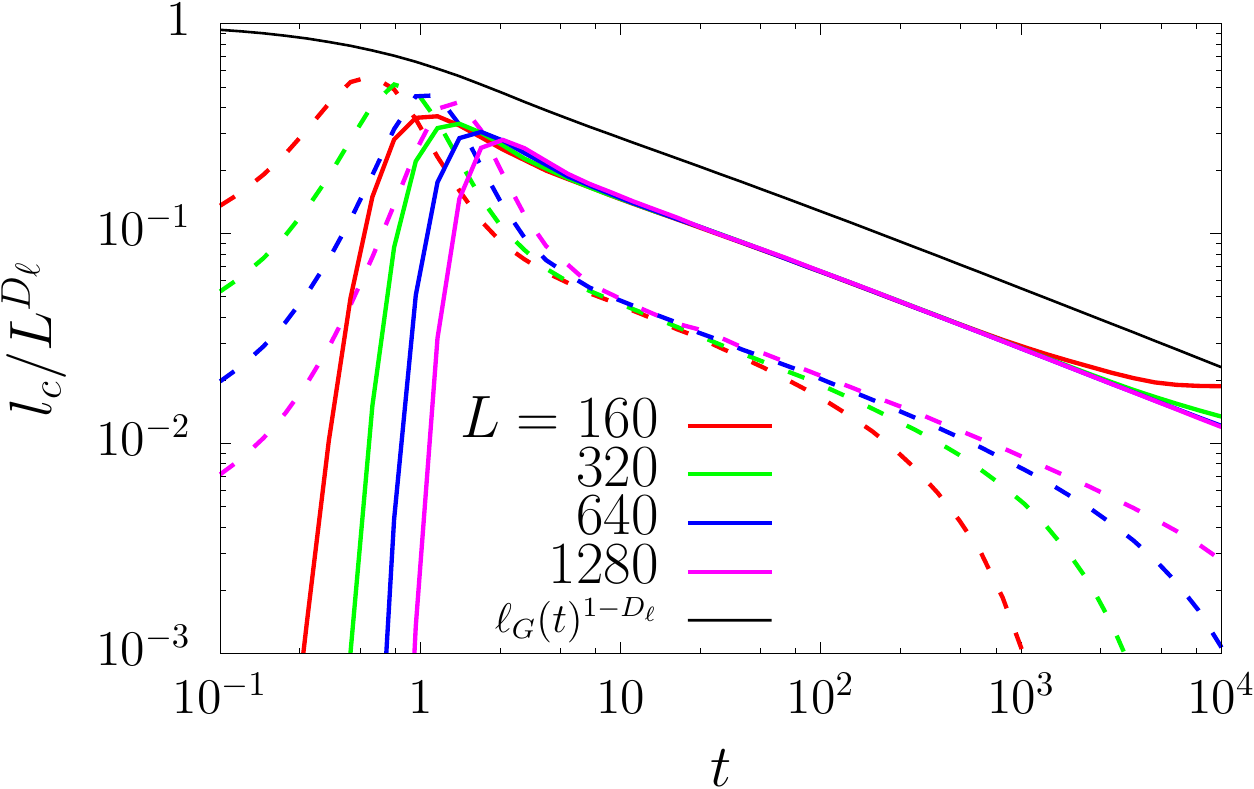}\quad%
 \includegraphics[scale=0.52]{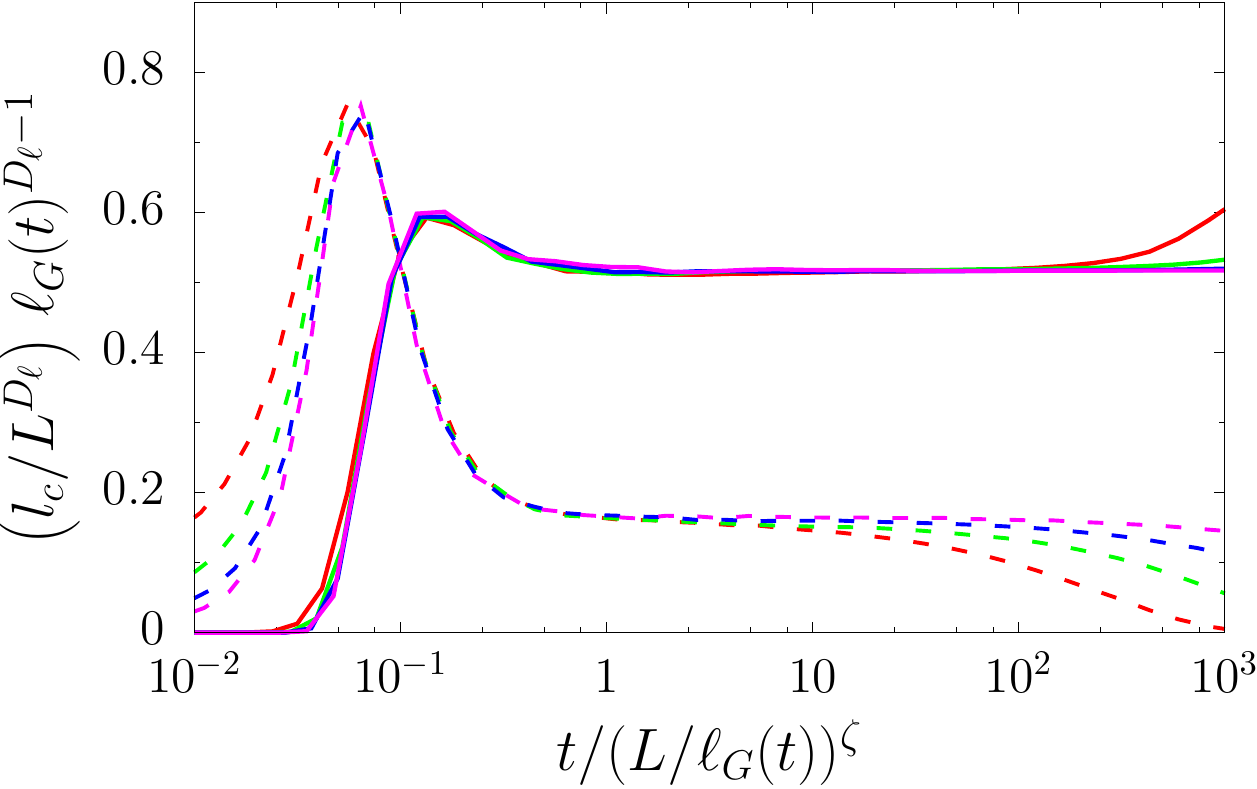}
\end{center}
\caption{\small 
Scaling analysis of the length of the largest cluster interface in the case of the $T=0$ dynamics
on the square lattice, for different values of $L$, the lattice linear size.
On the left  we show $l_c/L^{D_l}$ vs. $t$, while on the right we show 
$(l_c/L^{D_{\ell}}) \, \ell_d(t)^{(D_{\ell}-1)} $ against the rescaled time $t/ ( L / \ell_G(t))^{\zeta}$, where
$\ell_G(t)$ is the characteristic time obtained as the inverse of the excess energy, and $\zeta=0.5$. 
As in Figs.~\ref{newftr} and \ref{newftr2}, the contributions from the wrapping and non-wrapping hulls 
have been separated, with the former
indicated by continuous lines and the latter by dashed lines. In the left panel we also show
$\ell_G(t)^{1-D_{\ell}}$ (black solid line). The colour code is the same in both panels.
}
\label{newfsq}
\end{figure}

We have already shown the time evolution of the largest cluster size and its scaling properties in the case of the dynamics on the
square lattice in Sec.~\ref{subsec:largest_cluster_intro}. Here we complete the analysis by showing the scaling properties of the length of
its interface, $l_c$.
In the left panel of Fig.~\ref{newfsq}, we show $l_c/L^{D_l}$ vs. $t$, for systems with different linear size.
Again, we separate the contribution coming from wrapping domain walls, indicated by continuous lines,
from the one coming from non-wrapping ones, indicated by dashed lines.
We also show $\ell_G(t)^{1-D_{\ell}}$ to make a comparison, as was already done in the case of the dynamics on the triangular lattice.
After a crossover time that is system size dependent, both types of hulls have a similar behaviour and they seem to be just proportional to
$\ell_G(t)^{1-D_{\ell}}$ up to a second characteristic time (also dependent on $L$) where deviations caused by finite-size effects occur.
As it was done for  $A_c/L^{D_A}$ in Sec.~\ref{subsec:largest_cluster_intro}, it is possible to collapse the datasets
corresponding to different $L$ one onto the other in the small-$t$ region by plotting 
$ (l_c/L^{D_l} ) \, \ell_G(t)^{-(1-D_{\ell})}$ against the rescaled time $t / (L / \ell_G(t))^{\zeta}$.
As it was already explained, this is done to take into account the coarsening process occuring during the time regime
in which the system is approaching the critical-percolation-like state, something that is not present instead in the case
of the dynamics on the triangular lattice.
The value of the exponent $\zeta$ that gives us the best collapse is $\zeta\simeq 0.50(1)$ as in the case of the scaling of the largest
cluster size, see Fig.~\ref{F2}.

\vspace{0.25cm}
 
\begin{figure}[h]
\begin{center}%
\includegraphics[scale=0.52]{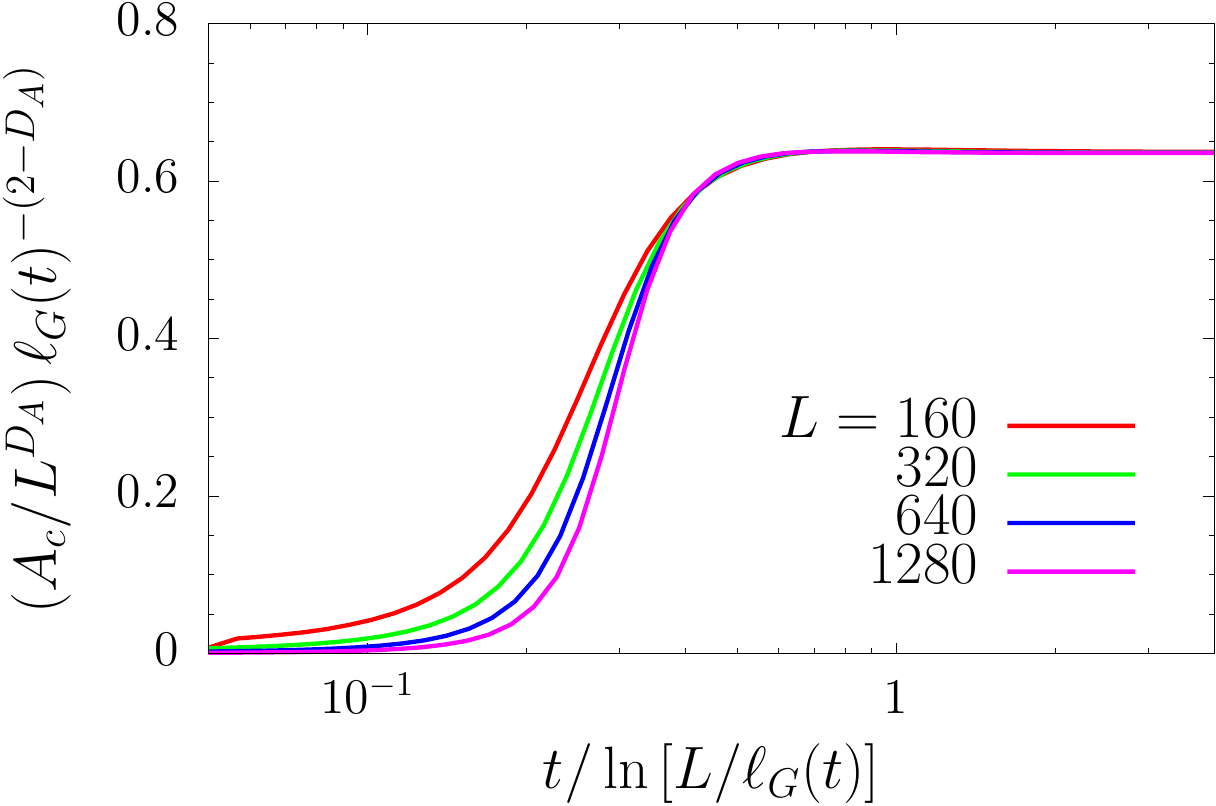}\quad%
\includegraphics[scale=0.52]{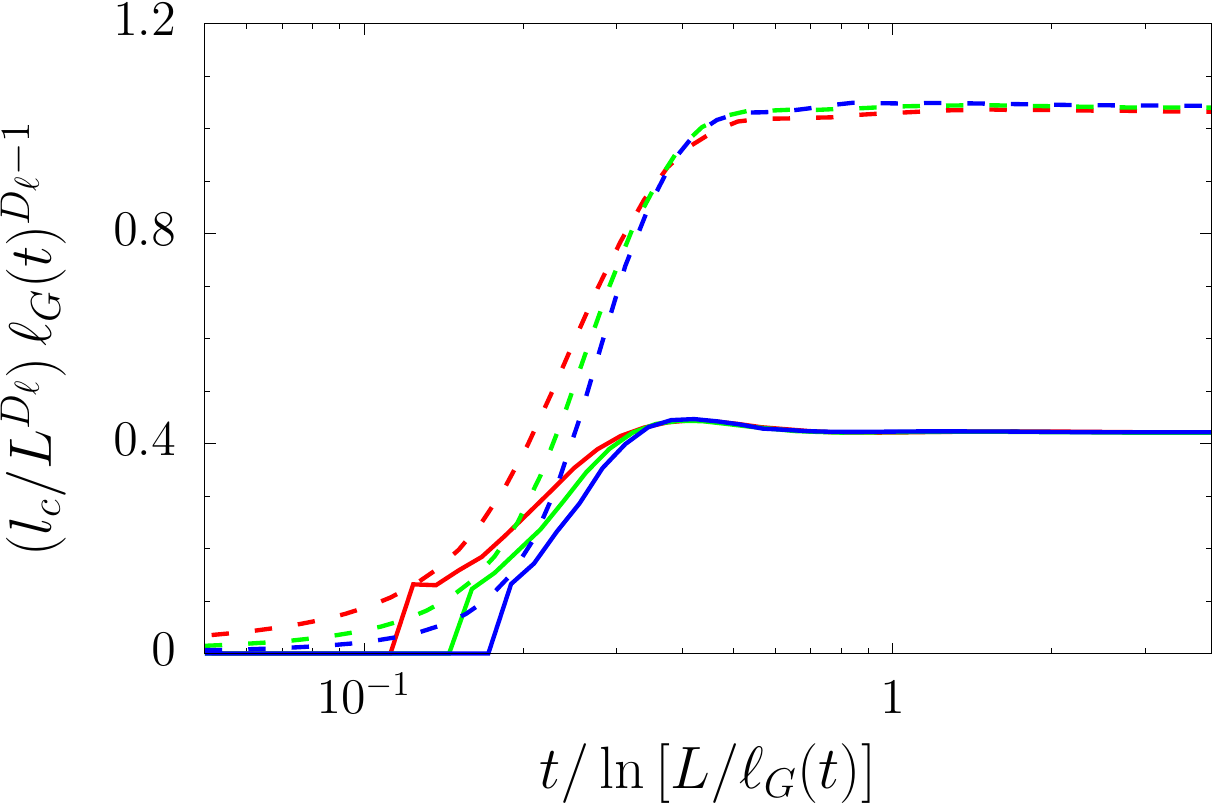}%
\end{center}
\caption{\small
Analysis of the geometric properties of the largest cluster in the case of the \,
$T=0$ dynamics on a honeycomb lattice with PBC, for different values of $L$. 
On the left we show  $A_c/L^{D_A} \; \ell_G(t)^{-(2-D_A)} $, while
on the right $ (l_c/L^{D_{\ell}}) \; \ell_G(t)^{D_{\ell}-1} $ both plotted against the rescaled time
$t / \ln{\left[ L / \ell_G(t) \right]} $
where $\ell_G(t)$ is the characteristic 
length obtained as the inverse of the excess energy. 
The continuous lines show the contribution of the wrapping clusters and dashed lines the 
ones of non-wrapping clusters. The colour code is the same in both panels.
}
\label{newf_hon}
\end{figure}

\begin{figure}[h!]
\begin{center}
\includegraphics[scale=0.52]{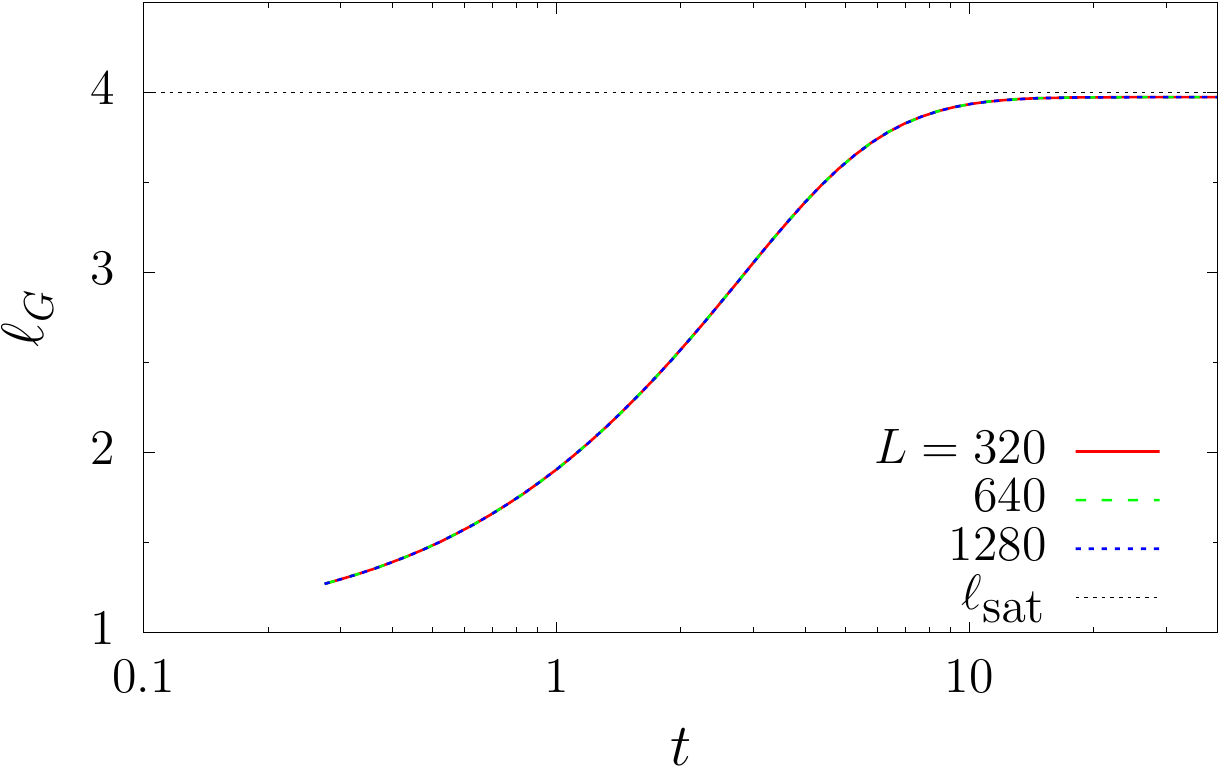}\quad%s
\includegraphics[scale=0.52]{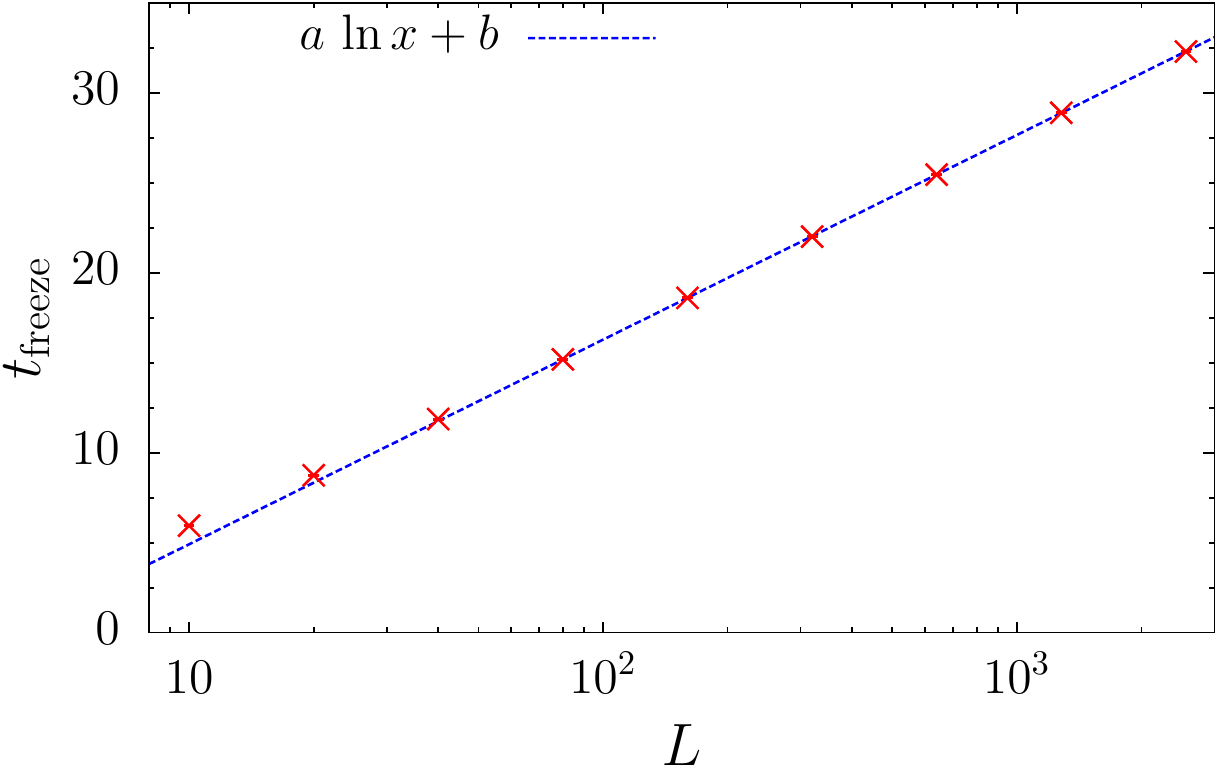}
\end{center}
\caption{\small In the left panel, the excess energy growing length, $\ell_G$, for the zero-temperature  dynamics
on the honeycomb lattice as a function of time, for lattice sizes $L=320, \ 640$ and $1280$.
$\ell_G$ saturates at $\ell_{\rm sat} \simeq 4$ at a time $\simeq 10$  independently of $L$.
In the right panel, the average freezing time,
$t_{\rm freeze}$, as a function of size $L$ (red data points) and the best fit, $t_{\rm freeze} \simeq
4.95 \ln{L} - 6.47$ (keeping only data with $L > 40$), indicated by the dashed blue line.  See 
Sec.~\ref{subsubsec:Hon-freeze} for a discussion. 
}
\label{GL_H}
\end{figure}

A similar analysis to what has been done for the dynamics on the square lattice is now performed on the 
honeycomb lattice in Fig.~\ref{newf_hon}. 
We show the two quantities $ (A_c/L^{D_A}) \, \ell_G(t)^{-(2-D_A)}$ (left panel) and 
$ (l_c/L^{D_{\ell}}) \, \ell_G(t)^{-(1-D_{\ell})}$ (right panel)
against $t/\ln{ \left[ L/\ell_d(t) \right]}$, where again
we used $\ell_G(t)$, the characteristic length obtained from the excess energy, as a measure of $\ell_d(t)$.
The reason for the peculiar scaling of time 
in the case of the zero-temperature dynamics on the honeycomb lattice is that, 
on this lattice, it freezes at a time  $t_{\rm freeze} %\simeq \mathrm{const.} +4.95 \, \ln L$, 
\simeq b +a\, \ln L$, with $a\simeq 4.940(5)$ and $b\simeq -6.46(3)$,
see Fig.~\ref{GL_H}-right, due to the fact that there are finite size stable clusters~\cite{TakanoMiyashita},
as discussed also in Sec.~\ref{subsec:crossings}.
In agreement with this fact, we find that the percolation time $t_p$ also 
scales as $ \ln L $ (and it ignores the fact that $\ell_G$ saturates at a finite value, see Fig.~\ref{GL_H}-left).  
We will discuss the behaviour of $\ell_G(t)$ and $t_{\rm freeze}$ on the honeycomb lattice in Sec.~\ref{subsubsec:Hon-freeze}.

\vspace{0.25cm}

\subsection{Number density of cluster areas}
\label{subsec:number-density-cluster}

The time-dependent distribution of domain areas was measured numerically in~\cite{ArBrCuSi07,SiArBrCu07}, after a quench from $T_0\to \infty$ to $T<T_c$, using square lattices. Three 
area regimes were identified in the functional form of ${\cal N}$. Thermal fluctuations 
generate very small domains and their  distribution falls-off exponentially just as in thermal equilibrium. 
The two remaining parts of the distribution are similar to the ones 
found at $T=0$.  A first regime in which areas are finite and the number density is affected by the 
coarsening process, and a second regime in which areas percolate across the sample and the 
number density presents a small bump. These 
two regimes are represented by the two terms in  
Eq.~(\ref{eq:area-number-density}). After a sufficiently long time, the fate of the finite size clusters 
is dictated by curvature-driven coarsening dynamics~\cite{AlCa79}
and an approximate expression for the time-dependence of the finite cluster size distribution was 
derived~\cite{ArBrCuSi07,SiArBrCu07}
\begin{equation}
N(A,t) \, \simeq \, \frac{2c_d \, [ \lambda_d(t-t_p+t_0) ]^{\tau_A-2}}{[ A + \lambda_d(t-t_p+t_0) ]^{\tau_A}}
\qquad\qquad 
t \geq  t_p 
 \; ,
  \label{eq:na_dynamics}
\end{equation}
where $\lambda_d $ is a material constant related to the diffusion coefficient of the hulls (closed curves separating domains of different phases),  see Eq.~(\ref{eq:lambdadt}),
and $t_0$ is a characteristic cutoff time, such that $A_0=\lambda_d t_0$ is a microscopic area that we set to be 1. This result was obtained assuming an 
initial state for the curvature-driven dynamics such that the distribution of domain areas
is the one in (\ref{eq:na_eq}) with a critical power law tail, that is to say, after the 
percolating time $t_p$.
A direct fit of the algebraic decay provides  a value of $\tau_A$
that is close to the expected one for critical percolation, $\tau_A \simeq 2.0549$, but  it is also close to the one for the $2d$ 
critical Ising model, $\tau_A \simeq 2.0267$. It is therefore difficult to distinguish 
between these two cases from the analysis of the algebraic piece. The $t$-dependent factor in the numerator ensures that the total number
of domain areas decays as $t^{-1}$, as expected from dynamic scaling. The following two limits can be read from Eq.~(\ref{eq:na_dynamics})
\begin{eqnarray}
N(A,t) \, 
\simeq 
\left\{
\begin{array}{ll}
\displaystyle{\frac{2c_d}{(\lambda_d t)^2}}
\qquad
& 
A \ll \lambda_d (t-t_p) \simeq \lambda_d t 
\vspace{0.25cm}
\\
\displaystyle{\frac{2c^{\rm eff}_d(t)}{A^{\tau_A}}} 
\qquad 
&  
A \gg \lambda_d (t-t_p) \simeq \lambda_d t 
\end{array}
\right.
\label{eq:NA-Sicilia}
\end{eqnarray}
where we took $t\gg t_p \gg t_0$ and we defined
\begin{equation}
c_d^{\rm eff}(t) \equiv c_d \, (\lambda_dt)^{\tau_A-2} = c_d \, [\, \ell_d(t) \, ]^{2(\tau_A-2)}
\; . 
\label{eq:eff-cd}
\end{equation}
Note that Eq.~(\ref{eq:na_dynamics}) can also be written as 
\begin{equation}
N(A,t) \simeq \frac{2c_d \; [\, \ell_d(t) \, ]^{2(\tau_A-2)} }{[A+\ell^2_d(t)]^{\tau_A}}
\label{NA-scaling}
\end{equation}
for $t\gg t_p-t_0$.

\begin{figure}[h!]
\begin{center}
 \includegraphics[scale=0.7]{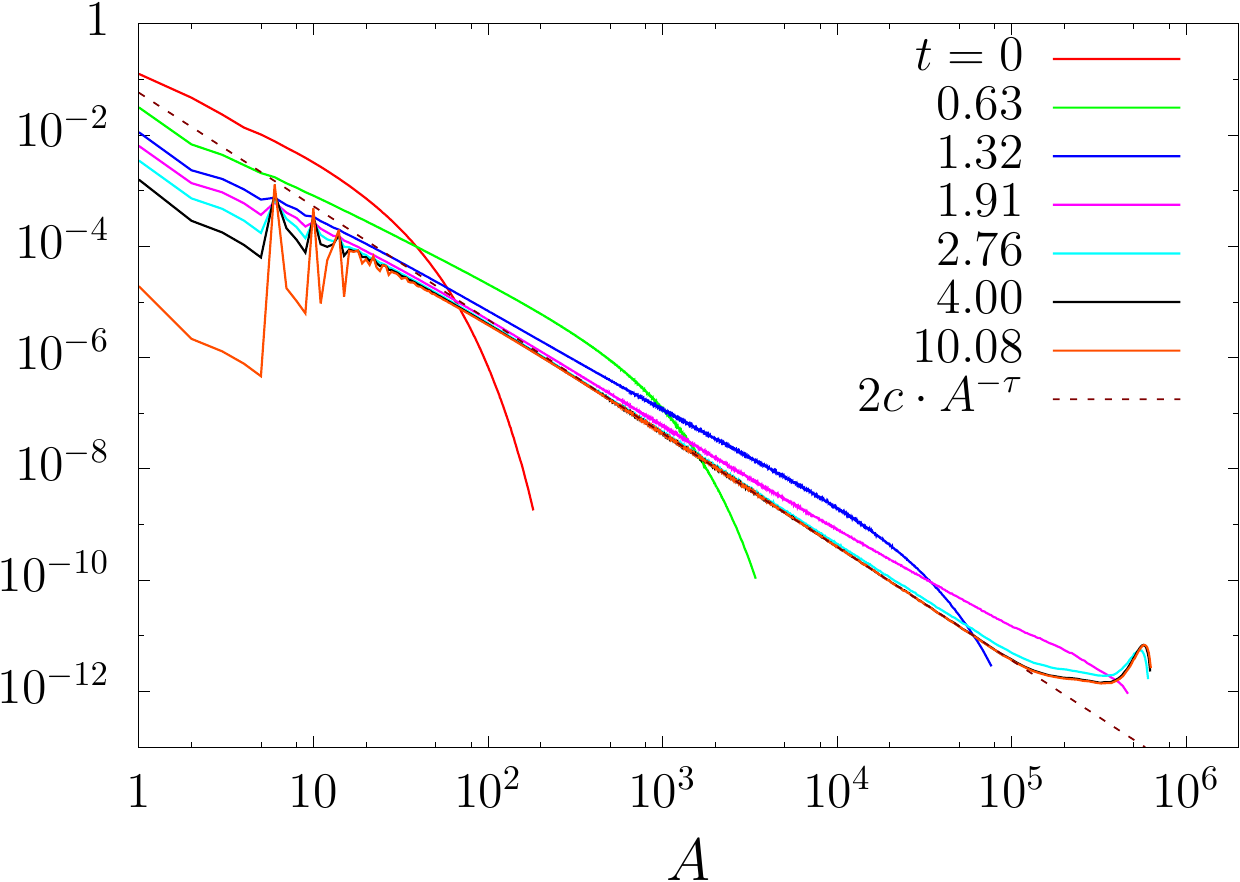}
\end{center}
\caption{\small
Time evolution of the number density of domain areas for the zero-temperature dynamics on
a honeycomb lattice of linear size $L=1280$. 
We show the domain area distribution ${\mathcal N}(A,t)$ vs. 
$A$ at various times given in the key. The function $f(A) = 2 c \, A^{-\tau}$ has been fitted to the data corresponding to time $t=10.08$ 
in the range $ [10^3, 5 \times 10^4] $ (the curve is represented by a dashed line). 
The fit  yields the estimates $c = 0.028(1) $ and 
$\tau = 2.035(5)$ that are close to the expected $c_d \simeq 0.0289$ and $ \tau_A = 187/91 \simeq 2.0549$.
}
\label{NA_HoIM}
\end{figure}

We investigate here the full ${\mathcal N}(A, t, L)$ in more detail focusing on its short-time behaviour for finite system sizes.
We emphasise that the global form of ${\mathcal N}(A, t, L)$ should be the same on all lattices. 
In Fig.~\ref{NA_HoIM} we present the complete domain area distribution, ${\mathcal N}(A,t)$, for the zero-temperature  dynamics
on a honeycomb lattice with linear size $L=1280$ and PBC, at various times after the quench. 
The initial fully disordered state with equal probability of up or down down spins on each site is not critical for the honeycomb lattice.
Thus, initially, the distribution of domain areas is expected to have an exponential cut-off at a relatively small area compared to the total size of the system,
as one can see from Fig.~\ref{NA_HoIM} (red curve).
But soon the distribution develops a power law behaviour $\mathcal{N}(A) \sim A^{-\tau}$ extending over many decades of domain sizes $A$.
This corresponds to the system having reached the critical-percolation-like state.
At the same time there is the appearance of the small \textit{bump} at very large values of $A$ appears due to the presence of domains that
percolate across the sample. 
Overall, the time evolution of ${\mathcal N}(A,t)$ strongly resembles the ones already 
found for the $T=0$  dynamics on the square lattice apart from the peaks at
relatively small areas, $A\simeq 10$, established at long times when the system is getting blocked in a spin configuration with lots of small stable domains
with definite number of sites, $6, 10, 14, \ldots$,  
a feature which is peculiar to the honeycomb lattice geometry.

In the following we focus our analysis on the scaling properties of the domain area distribution ${\cal N}$ by considering:
\begin{itemize}
\item
The triangular lattice separately.
\item
The contribution of the percolating clusters to the number density, that is to say, $N_p$, for the three lattices.
\item 
The dynamic approach to the percolation point, that is to say, the relatively short time-scales such that the 
bump $N_p$ in Eq.~(\ref{eq:area-number-density}) has not stabilised yet, and its scaling analysis, in 
the square and honeycomb lattices.
\end{itemize}

We first present the analysis of the first scaling regime after zero temperature quenches.
The study of the finite-size and time-dependence scaling properties of the bump $N_p$ is done under
finite temperature effects.

\subsubsection{The triangular lattice.}

\begin{figure}[h!]
\begin{center}
 \includegraphics[scale=0.7]{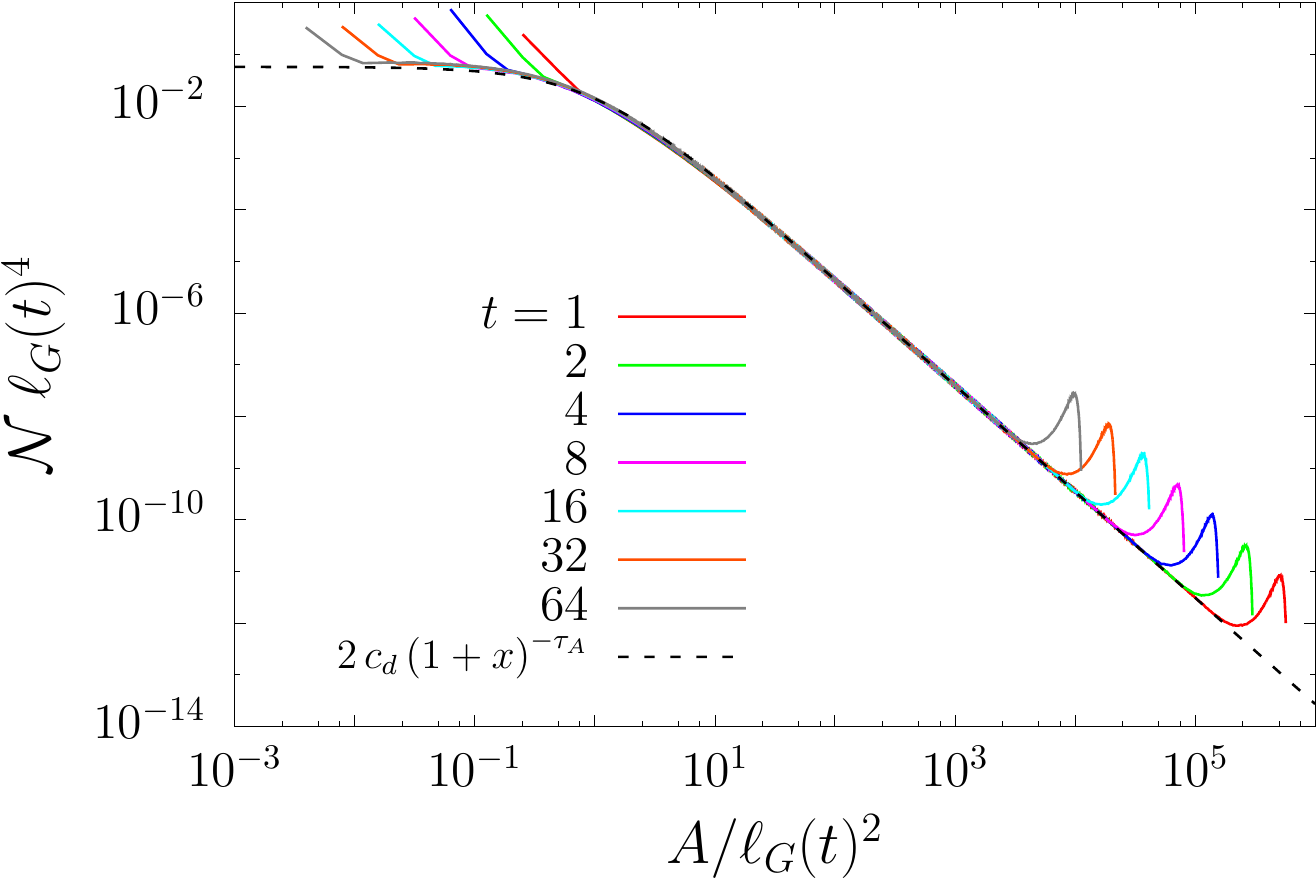}
\end{center}
\caption{\small $T=0$ dynamics on
a triangular lattice of linear size $L=2560$.
We show the scaling of the number density of areas $\mathcal{N}(A,t)$ implied by Eq.~(\ref{NA-scaling}) for finite areas. 
The quantity $\mathcal{N}(A,t) \, \ell_G(t)^4$ is plotted against the rescaled area $A/\ell_G(t)^2$
where $\ell_G(t)$ is the characteristic length scale obtained as the inverse of the excess energy.
The datasets corresponding to different times (indicated in the key) collapse on the master curve 
$ f(x) = 2 \, c_d \, \left(1 + x \right)^{-\tau_A} $ (dashed line), which is the expected analytic expression for the scaling function.}
\label{NATrIM-1}
\end{figure}

\vspace{0.25cm}
\begin{figure}[h!]
\begin{center}
\includegraphics[scale=0.54]{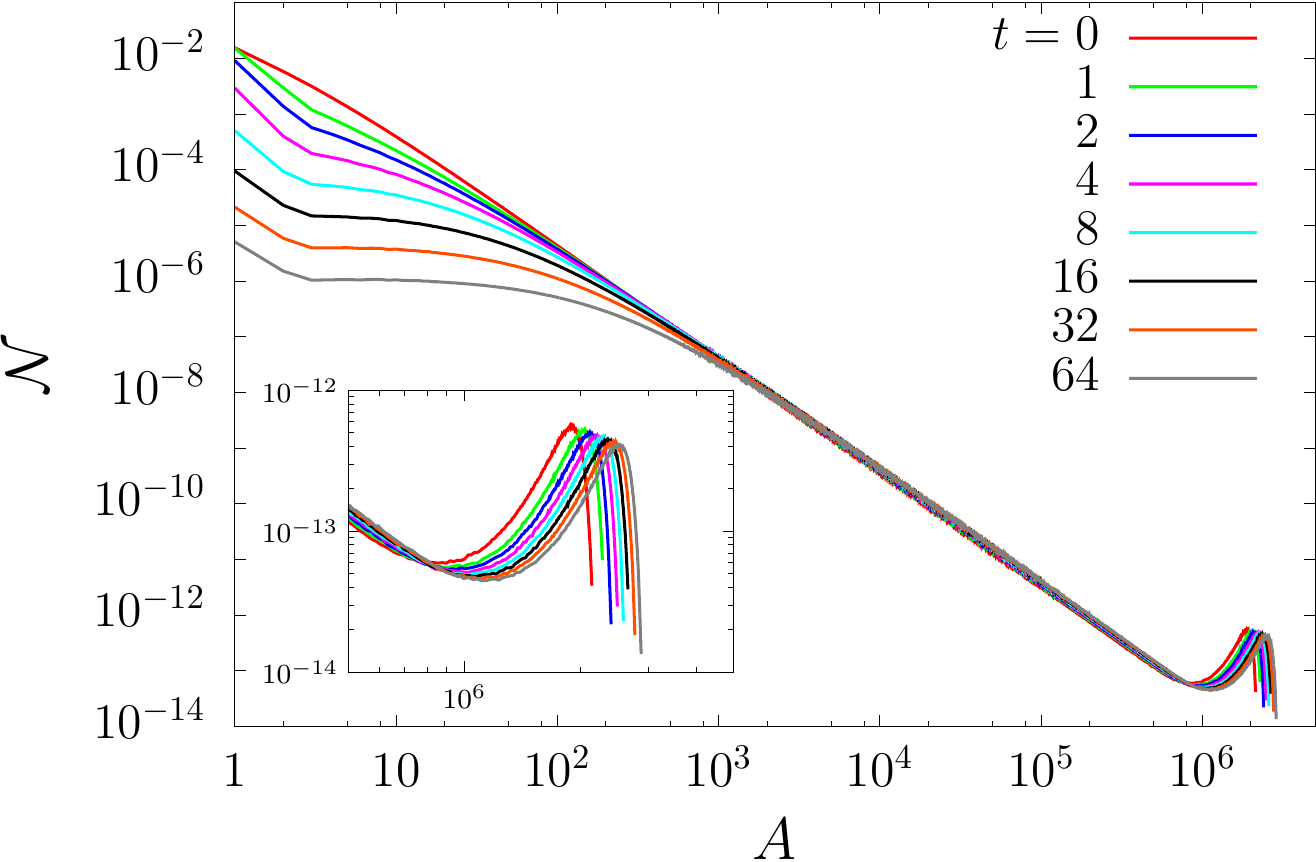}\quad%
\includegraphics[scale=0.54]{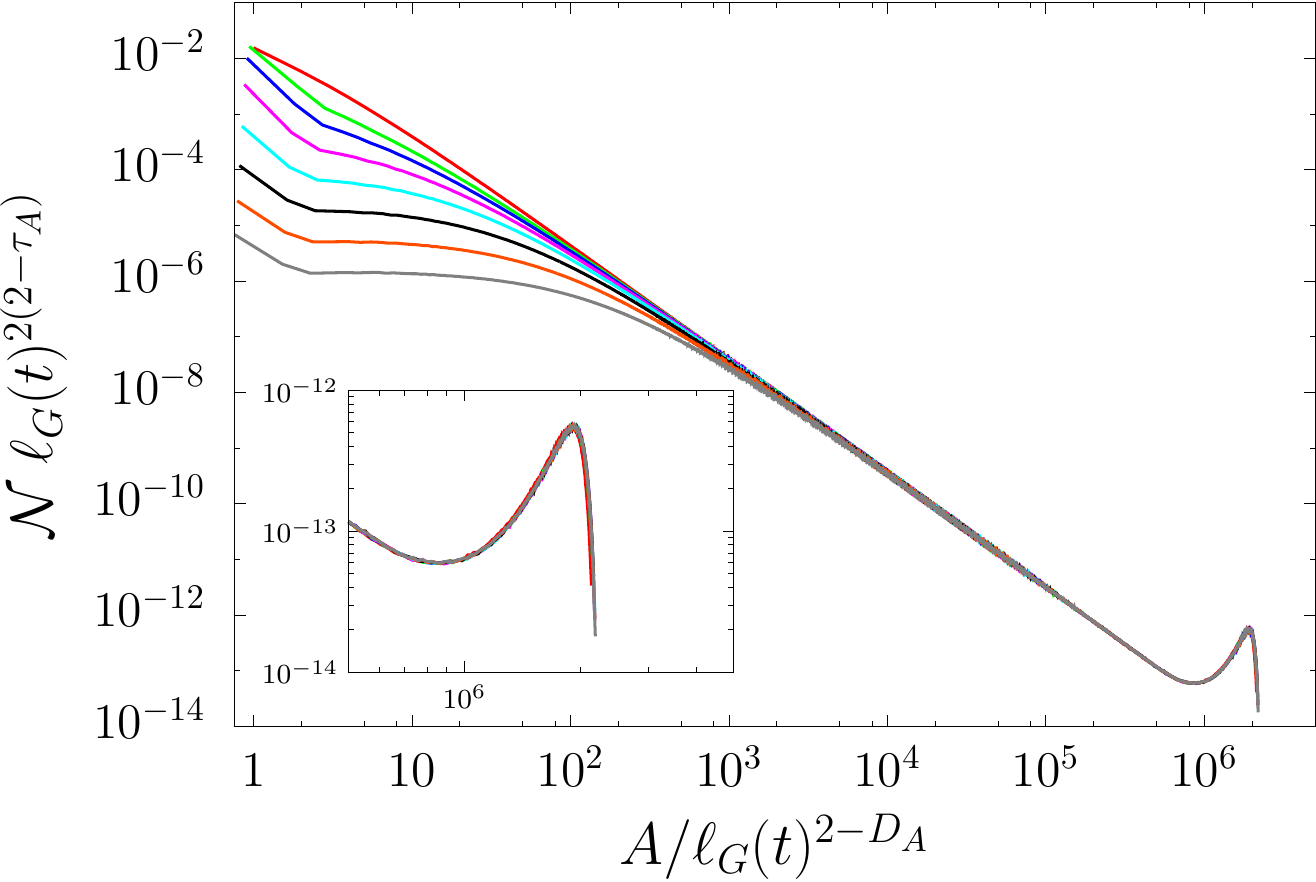}
\end{center}
\caption{\small 
Time evolution of the number density of domain areas for the zero-temperature dynamics on
a triangular lattice of linear size $L=2560$. 
In the left panel we show the bare domain area distribution ${\mathcal N}(A,t)$ vs. 
$A$ at various times given in the key.
In the right panel we present 
${\mathcal N}(A,t) \; \ell_G(t)^{2(2-\tau_A)}$ against the rescaled
area $A/\ell_G(t)^{2-D_A}$, 
with the exponents of critical percolation, $D_A=91/48$ and $\tau_A=187/91$,
and $\ell_G(t)$ the characteristic length scale obtained as the inverse of the excess energy.
In both panels, the insets show a ``zoomed'' view of the \textit{bump}, to
better highlight the difference between the unscaled data and the scaled ones.
The colour code is the same in both panels.
}
\label{NATrIM-2}
\end{figure}

\begin{figure}[h!]
\begin{center}
  \includegraphics[scale=0.7]{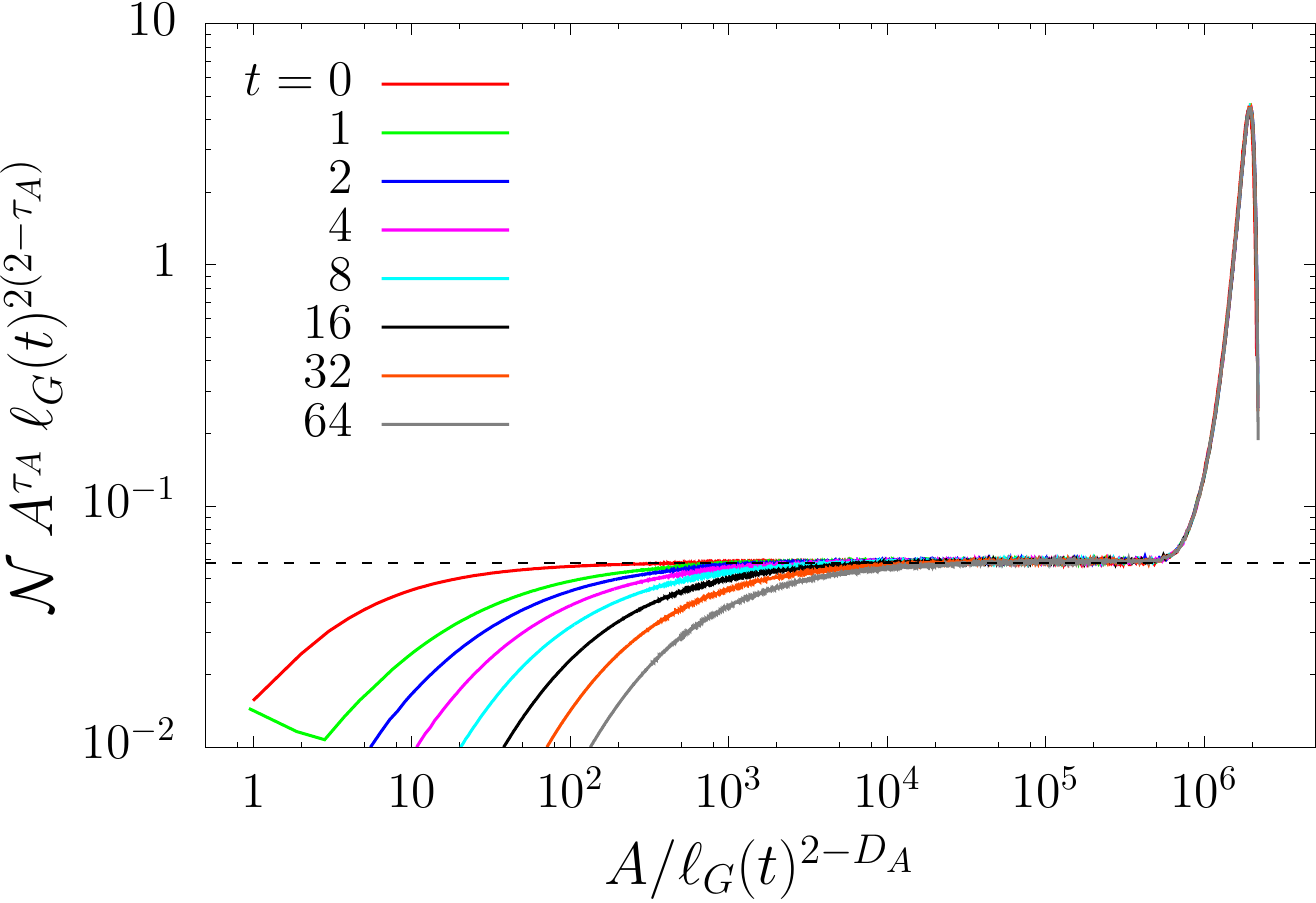}
\end{center}
\caption{\small Time evolution of the number density of domain areas for the zero-temperature dynamics on
a triangular lattice of linear size $L=2560$.
In order to highlight the presence of the algebraic decay $\mathcal{N}(A) \sim N(A) \sim A^{-\tau_A}$ for nonpercolating clusters,
we show  $\mathcal{N}(A,t) \, A^{\tau_A} \, \left[ \ell_G(t) \right]^{2(2-\tau_A)}$ against the rescaled 
area $A/\ell_G(t)^{2-D_A}$, where
$\ell_G(t)$ is the characteristic length scale obtained as the inverse of the excess energy,
$D_A=91/48$ and $\tau_A=187/91$ as in Fig.~\ref{NATrIM-2}. 
The scaling of the area as $A/\ell_G(t)^{2-D_A}$ has been done to collapse the so-called bump, as in the right panel
of Fig.~\ref{NATrIM-2}. The rescaled data presents a \textit{plateau} in the interval 
$[10^4, 5 \times 10^5]$ of the rescaled area, falling approximately onto the expected value for critical percolation, $2 c_d \simeq 0.0579$, 
indicated by the black horizontal line.
}
\label{NATrIM-3}
\end{figure}

In the case of the triangular lattice the initial condition is right at the critical percolation point, thus
$N(A,0) \simeq 2 c_d A^{-\tau_A}$, with $\tau_A=187/91$ and $2 c_d \simeq 0.0579$, if one neglects
effects due to the discreteness of the lattice at very small values of $A$. Added to this finite area weight there is the 
contribution coming from the 
percolating clusters at very large values of $A$, the so-called bump, denoted by $N_p(A,L)$.
The analytic form expressed by Eq.~(\ref{NA-scaling}) should hold for the time evolution of $ \mathcal{N}(A,t)$ 
in the region of sizes $A$ where the aforementioned contribution is negligible.
In order to highlight this last fact, in Fig.~\ref{NATrIM-1} we present the rescaled
domain area distribution for the $T=0$ dynamics on a triangular lattice of linear size $L=2560$:
we plot $\mathcal{N}(A,t) \, \ell_G(t)^4$ against the rescaled area $A/\ell_G(t)^2$, where
$\ell_G(t)$ is the characteristic length scale obtained as the inverse of the excess energy.
As done before for the scaling of other observables, $\ell_G(t)$ is taken as a measure of $\ell_d(t)$,
the dynamical characteristic length for non-conserved order-parameter dynamics, 
which, for sufficiently long time, behaves as $\ell_d(t) \simeq ( \lambda_d t )^{1/2}$.
By so doing, the datasets corresponding to different times should collapse onto the same master curve, which is represented
by $f(x) = 2 \, c_d \, \left( 1 + x \right)^{-\tau_A}$.
The result of the scaling is very  good.
Deviations  from the master curve occur for
very small values of the variable $A/\ell^2_G(t)$, where the scaling is supposed to break, and for very large values corresponding
to the appearance of the bump, which is the contribution $N_p$ of the percolating clusters, as expected.

Let us now turn to the properties of the bump.
Most of the contribution to $N_p$ comes from clusters that are 
either the largest or the second largest ones in the sample (for $A>L^2/2$, only the largest cluster contributes, and the bump is truly 
the size distribution of the largest cluster).
In Sec.~\ref{subsec:largest_cluster_intro} we argued that the fraction of sites belonging to the largest cluster (or to the second largest one), 
$A_c/L^{2}$, should scale dynamically as $\ell_d(t)^{2-D_A}$, and 
the results shown in Fig.~\ref{F2} strongly suggests the validity of this argument.
Accordingly, in order to collapse the bumps at different times $t$ and fixed $L$, 
the area $A$ should be rescaled by $\ell_d(t)^{2-D_A}$.
At the same time, $\mathcal{N}(A,t)$
must be multiplied by $\ell_d(t)^{2(2-\tau_A)}$ to remove the time-dependence of 
the pre-factor $2 c^{\mathrm{eff}}_d(t)$, as explained in the previous Section.
We present the result of this scaling in the right panel of Fig.~\ref{NATrIM-2}, where we plot
$\mathcal{N} \; \ell_G(t)^{2(2-\tau_A)}$ against $A/\ell_G(t)^{2-D_A}$,
using the same data as the ones in Fig.~\ref{NATrIM-1}. 
Again,  $\ell_G(t)$ is taken as a measure of the dynamical characteristic length scale $\ell_d(t)$.
In the left panel of the same figure we show the unscaled distribution against $A$ to let the reader make a comparison.
By looking at the whole distribution one is not able to notice a significant difference between the unscaled and the scaled versions of the data
since both $D_A$ and $\tau_A$ are close to $2$. However, if one focuses only on the bump, 
as done in the insets, it becomes clear that the scaling makes the data collapse in that specific region.

In order to prove that the tail of the finite areas weight fall as $c_d \, A^{-\tau_A}$,
in Fig.~\ref{NATrIM-3} we show 
$\mathcal{N} \, A^{\tau_A} \, \ell_G(t)^{2(2-\tau_A)}$ against the rescaled area $A/\ell_G(t)^{2-D_A}$.
For $A \gg \ell_G(t)^{2}$, the data corresponding to different times should all collapse onto a \textit{plateau}
at the constant $2 c_d$ (up to the point where the contribution due to percolating clusters, $N_p$, starts to be significant).
In fact, the rescaled data present a plateau in the interval 
$[10^4, 5 \times 10^5]$ of the rescaled area, falling approximately onto the expected 
value for critical percolation, $2 \, c_d \simeq 0.0579$ (indicated by a dashed line).
As one can see, the point at which the plateau sets in, that is the point around where there is the crossover between the two different regions
described by Eq.~(\ref{eq:NA-Sicilia}), increases with time. In fact it should go as $\ell_d(t)^2$.
At the same time, by scaling the area as $A/\ell_G(t)^{2-D_A}$, as done in the right panel of Fig.~\ref{NATrIM-2}, it is possible
to collapse the data in the region corresponding to the so-called bump
(the scaling of the horizontal axis is not needed to observe the plateau, but only to enforce the bump to collapse).

\subsubsection{Pre-percolation scaling on the square and honeycomb lattices.}

We adopt a dynamical scaling hypothesis to describe the behaviour of $N(A,t)$ during the approach to 
critical percolation on lattices other than triangular. The argument is the same as the one used in Sec.~\ref{sec:phenomenon}. The area is measured in units 
of the dynamic lattice spacing, $A/\ell^2_d(t)$ and, the (dimensionless) largest cluster area should then be proportional to 
$ \left( \ell_p(t)/\ell_d(t) \right)^{D_A}$ at criticality. We therefore use 
\begin{equation}
\frac{A/\ell_d^2(t)}
{\left(\ell_p(t)/\ell_d(t)\right)^{D_A}} 
\end{equation}
that generalises Eq.~(\ref{eq:rescaling-vertical-axis})
and can also be written as
\begin{equation}
\frac{A/\ell^{2-D_A}_d(t)}{\ell^{D_A}_p(t)}
\label{eq:correct-scaling-variable-pdf}
\end{equation}
as the scaling variable and we suggest that, after some microscopic time-scale, the large-size 
areas (sufficiently large such that $A\gg \ell_d^2(t)$) are distributed according to
\begin{equation}
N(A,t) \, \sim \, 2c_d^{\rm eff}(t) \; A^{-\tau_A} \; \Phi \left( \frac{A/\ell_d^{2-D_A}(t)}{\ell^{D_A}_p(t)} \right)
\qquad\qquad t<t_p
\; . 
 \label{eq:na_dyn_scaling}
\end{equation}
$\Phi$ is a scaling function such that 
\begin{eqnarray}
\Phi(x) \to 
\left\{ 
\begin{array}{ll}
1 \qquad & x \ll 1 
\\
x^a \qquad & x \stackrel{>}{\sim} 1
\end{array}
\right.
\end{eqnarray}
with $a$ an exponent that we study numerically.
$c_d^{\rm eff}(t)$ is defined in Eq.~(\ref{eq:eff-cd}) and $c_d \simeq 0.0289$, see Eq.~(\ref{eq:cd-value}).
These limits imply 
\begin{eqnarray}
N(A,t) \, \to
\left\{ 
\begin{array}{ll}
2 c^{\rm eff}_d(t) \;  A^{-\tau_A} \qquad & x \ll 1 
\\
2 c^{\rm eff}_d(t) \; A^{-\tau_A} \; (A/\ell_p^{D_A}(t))^a \qquad & x \stackrel{>}{\sim} 1 
\end{array}
\right.
\end{eqnarray}
and $x= A/[\ell_d^{2-D_A}(t) \ell^{D_A}_p(t)]$.
In the first line we see that the statistics of the small areas nicely coincide with the one in the second limit in Eq.~(\ref{eq:NA-Sicilia}), 
and the second limit above corresponds to the matching  between the power law tail and the bump 
represented by $N_p$ in Eq.~(\ref{eq:area-number-density}).

\begin{figure}[h!]
\begin{center}
\includegraphics[scale=0.55]{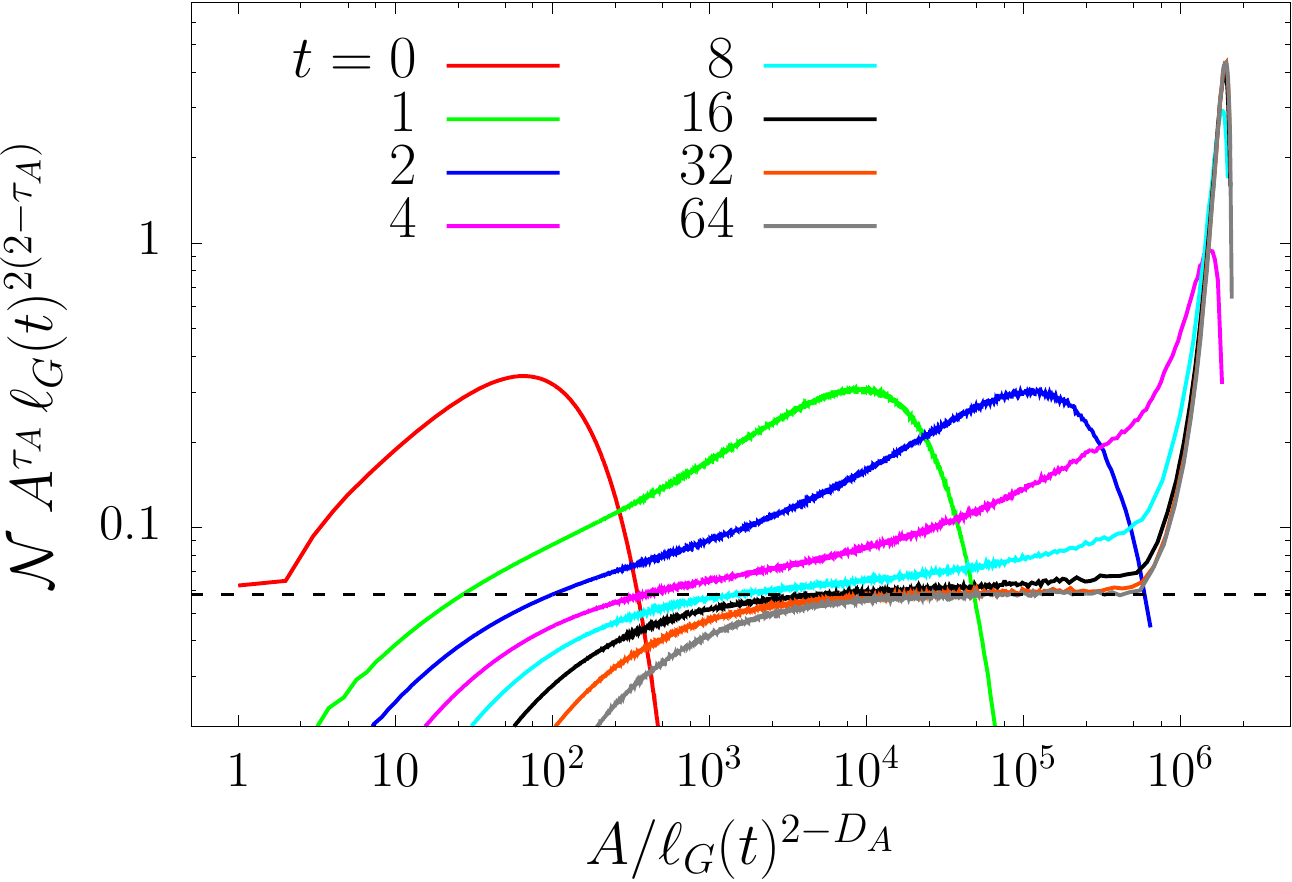}\quad%
\includegraphics[scale=0.54]{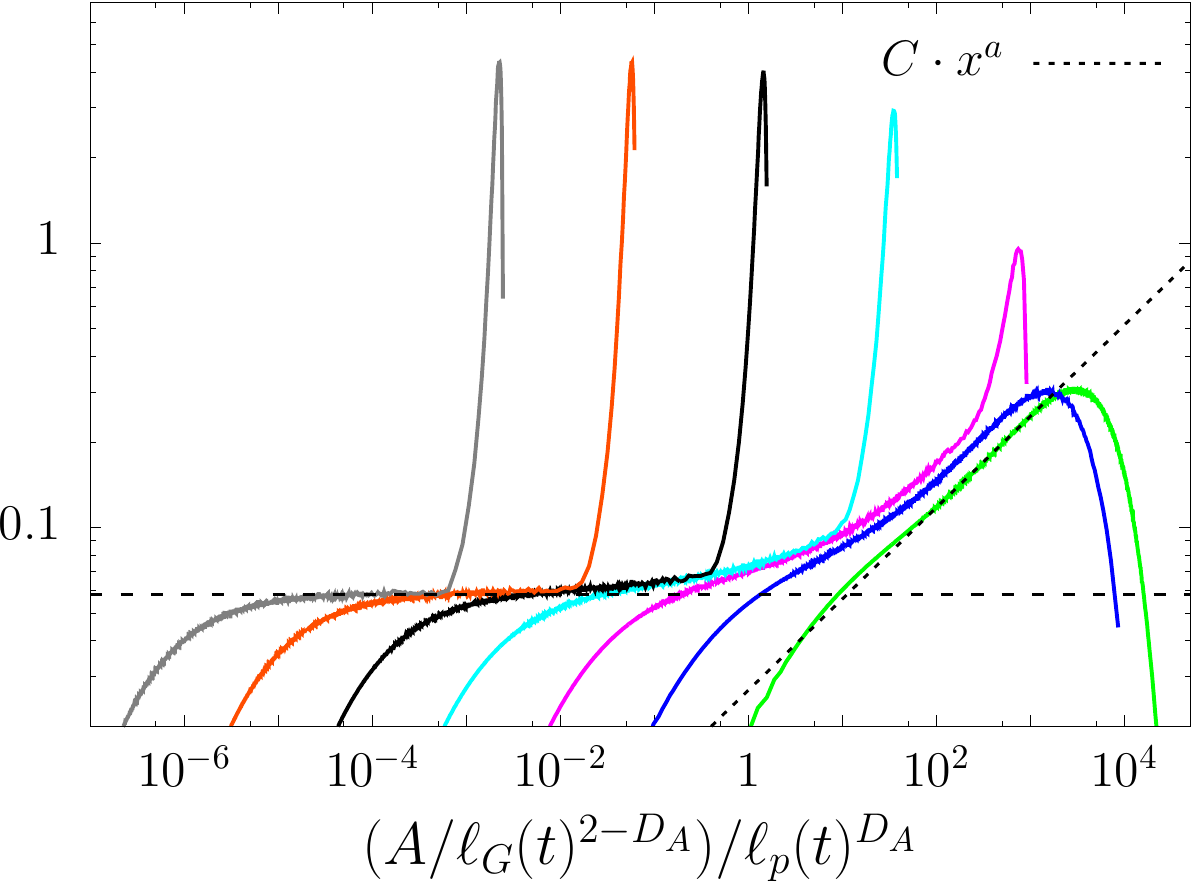}
\end{center}
\caption{\small Dynamics of the square lattice Ising model with $L=2560$ quenched to $T=0$. 
Pre-percolation scaling of the number of cluster areas. The quantity 
$A^{\tau_A} \, \ell_G(t)^{2(2-\tau_A)} \, {\mathcal N}(A,t, L)$,
with $\ell_G(t)$ the characteristic length obtained as the inverse of the excess energy,
is plotted against $A/\ell_G(t)^{2-D_A}$ in the left panel and
against $ \left( A/\ell_G(t)^{2-D_A} \right) /\ell^{D_A}_p(t)$ in the right panel,
where $\ell_p(t) = \ell_G(t) \, t^{1/\zeta}$,  with $\tau_A=187/91$,\, $D_A=91/48$ and $\zeta=0.5$.
The dotted straight line in the right panel corresponds to the power law $\Phi(x) \simeq C \, x^a$, with $a = 0.321(1)$,
which is the best fit to the data at time $t=1$ in the interval $[10, 10^3]$ of the scaling variable.
The black dashed horizontal line corresponds to $2c_d = 0.0579$.
}
\label{NASqIM}
\end{figure}

\begin{figure}[h!]
\begin{center}
\includegraphics[scale=0.55]{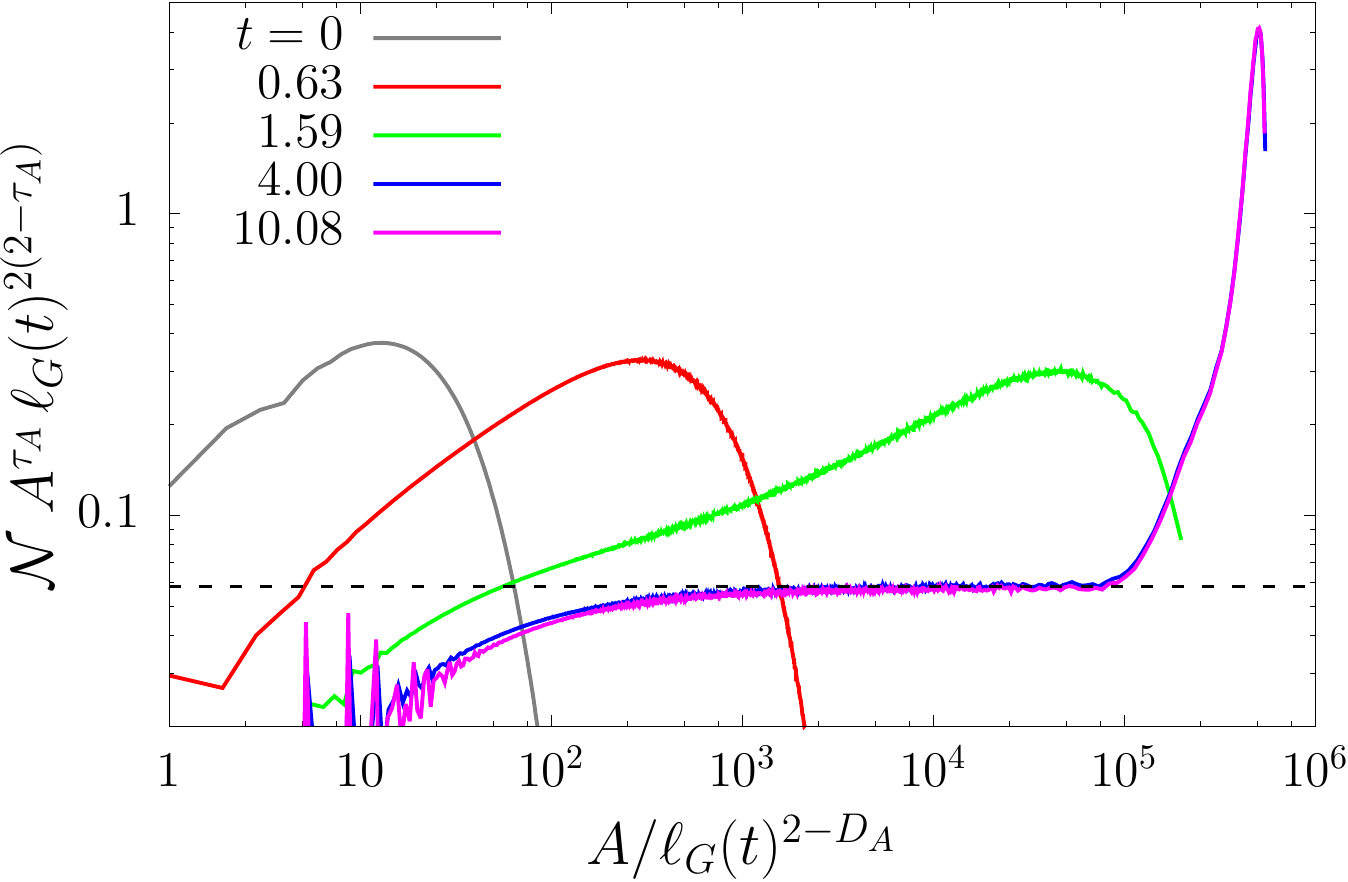} 
\includegraphics[scale=0.55]{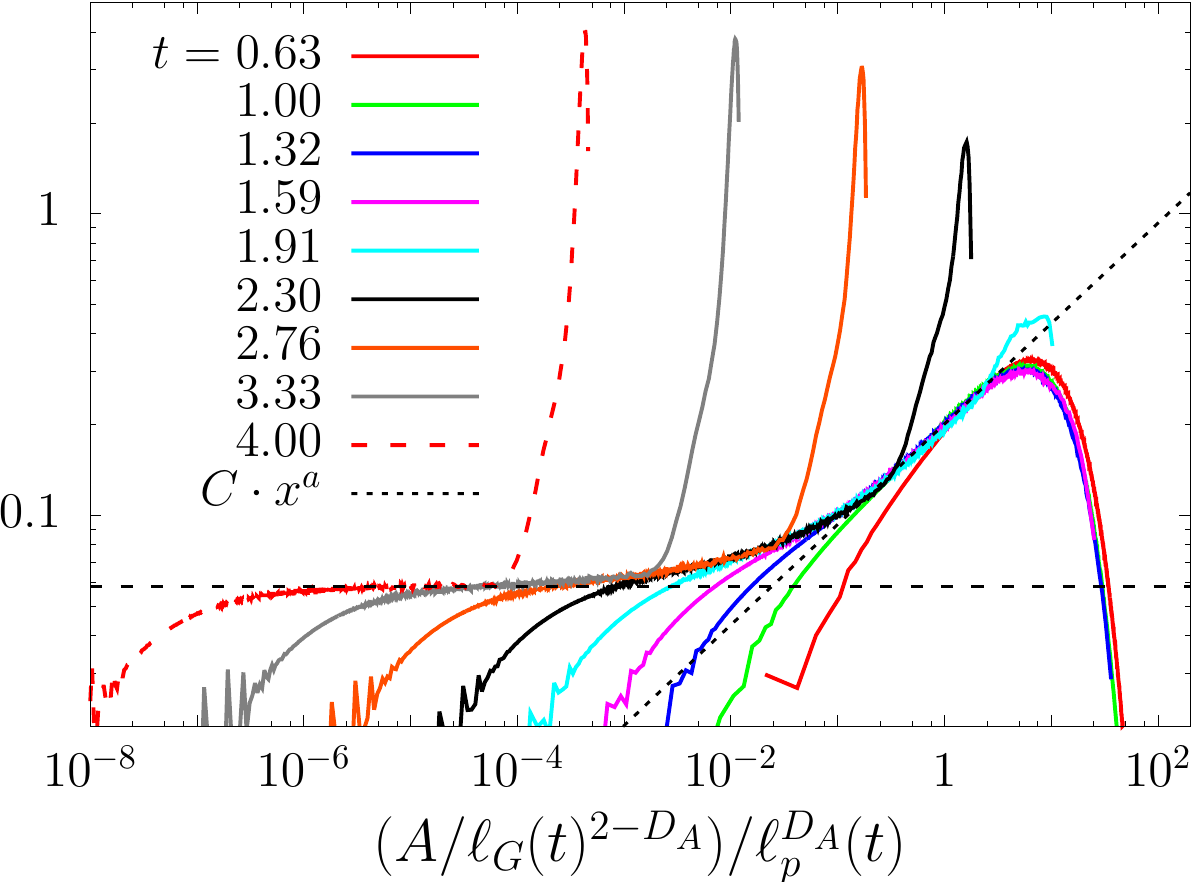}
\end{center}
\caption{\small Dynamics of the honeycomb lattice Ising model with $L=1280$ quenched to $T=0$.
Pre-percolation scaling of the number density of cluster areas. 
The quantity
$A^{\tau_A} \; \ell_G(t)^{2(2- \tau_A)} \; {\mathcal N}(A,t,L)$
is plotted against  $A/\ell_G(t)^{2-D_A}$ in the left panel and against 
$ \left( A/\ell_G(t)^{2-D_A} \right) /\ell^{D_A}_p(t)$ in the right panel,
with $\ell_G(t)$ the characteristic length obtained as the inverse of the excess energy,
and $\ell_p(t) = \ell_G(t) \, \mathrm{e}^{\alpha \, t}$ with $\alpha \simeq 2.6$, as explained
in more detail in the main text.
The constant $2 c_d\simeq 0.0579$ is represented by a horizontal dashed line. 
The function $\Phi(x) = C \, x^a$ has been fitted to the data at time $t=1$
in the region of the scaling variable corresponding to the pre-percolating regime (approximately, the interval
$[0.1,1]$), yielding $a = 0.332(1)$, and it is represented by the dotted line.
}
\label{NAHexIM}
\end{figure}

In Figs.~\ref{NASqIM} and \ref{NAHexIM}
we present our numerical results for the early evolution of the cluster size distribution after $T=0$ quenches of the 
$2d$ Ising model on the square and honeycomb lattices, respectively.
As done before, we use  $\ell_G(t)$, the characteristic length obtained as the inverse of the excess energy, 
as a measure of $\ell_d(t)$.

%%%%%%%%%%%%%%%%%%%% Square %%%%%%%%%%%%%%%%%%%%%

After a time of the order of $t_p$ the 
number density of cluster areas should approach the critical percolation form
and $A^{\tau_A} \ {\mathcal N}(A,t,L) $ should collapse onto a plateau corresponding to
the constant  $2c_d$.
In Fig.~\ref{NASqIM} we show $ A^{\tau_A} \ \ell_G(t)^{2(2-\tau_A)} \, {\mathcal N}(A,t,L) $ against
the rescaled area $A/\ell_G(t)^{2-D_A}$ for the zero-temperature dynamics on a square lattice with $L=2560$.
The factor $\ell_G(t)^{2(2-\tau_A)}$ is necessary to get rid of the time dependence in $c^{\mathrm{eff}}_d(t)$.
Notice that, apart from the 
behaviour at very small areas and the very steep increase at late times (due to the percolating clusters)
a plateau is clearly visible. 
It falls  on top of the expected value, $2c_d  \simeq 0.0579$, indicated by the horizontal dashed line.

\vspace{0.5cm}

In order to highlight the existence of 
the extra growing length $\ell_p(t)$, introduced by the transient between the initial configuration and the
state with a stable pattern of percolating clusters (attained at time $t_p$),
we plot the same quantity
against the rescaled area $ (A/ \ell_G(t)^{2-D_A} )/ \ell^{D_A}_p(t)$ 
where we assume $\ell_p(t) \simeq \ell_G(t) \, t^{1/\zeta}$
in the case of the square lattice, as conjectured in Sec.~\ref{subsec:largest_cluster_intro} and confirmed by the 
scaling of time in the analysis of the largest cluster geometrical properties.
For the square lattice we expect $\zeta=0.5$.
With this choice, we obtain a fairly good collapse, as
seen in the right panel in Fig.~\ref{NASqIM}. The master curve
highlights the presence of two regimes (save the behaviour at very small areas and finite size effects mentioned above): 
the asymptotic one for $A/[\ell_G^{2-D_A}(t)\ell^{D_A}_p(t)] \le 1$, where the rescaled distribution 
is flat and inherits the properties of the critical percolation point, and the ``pre-percolation'' 
one for $A/[\ell_G^{2-D_A}(t)\ell^{D_A}_p(t)] \ge 1$, see Eq.~(\ref{eq:na_dyn_scaling}), where 
the scaling function $\Phi(x)$ is close to a power-law, $\Phi(x)\propto x^a$, with $a = 0.321(1)$.
This curve is shown as an inclined dotted line in
the same figure. The horizontal dashed line corresponds to the constant $2c_d\simeq 0.0579$.

%%%%%%%%%%%%%%%%%%%% Honeycomb %%%%%%%%%%%%%%%%%%%%%

As we have already mentioned when describing  the largest cluster properties, the approach to percolation on a honeycomb lattice 
is much faster, and this is confirmed by the study of ${\mathcal N}(A,t,L)$, see the left panel of Fig.~\ref{NAHexIM}.
One can associate to the characteristic timescale $t_p \sim \ln L$, a characteristic growing length 
$\ell_p(t) \propto \ell_d(t) \,  \mathrm{e}^{\alpha t} $, with $\alpha$ a constant to be determined,
in a way which is similar to what we conjectured for the square lattice, see Eq.~(\ref{eq:ell_p}),
but with a time dependence which is not a simple power law. Again, we take  $\ell_G(t)$ 
as a measure of the characteristic length scale $\ell_d(t)$ associated to coarsening, and we assume that
$\ell_p(t) = \ell_G(t) \, \mathrm{e}^{\alpha t}$. The value of the constant $\alpha$ is not known a priori, but
we can provide a rough estimate of it by looking at the value which yields the best collapse of the data after proper rescaling.
By plotting $A^{\tau_A} \, \ell_G(t)^{2(2- \tau_A)} \ {\mathcal N}(A,t,L)$ against the rescaled
area $ (A /\ell_G(t)^{2-D_A} )/\ell^{D_A}_p(t)$ (right panel in Fig.~\ref{NAHexIM}), 
the data for different times can be collapsed onto a master curve (apart from deviations at small areas and in the region
of the scaling variable where the contribution from percolating clusters become significant),
the shape of which is pretty similar to the one obtained in the case of the square lattice, and the value of $\alpha$ giving the best result
is $\alpha \simeq 2.65(5)$.
As in the case of the data relative to the dynamics on the square lattice, in the region corresponding to the
pre-percolation regime, that is for $A/[\ell_G^{2-D_A}(t)\ell^{D_A}_p(t)] \ge 1$ and before finite-size effects take over, 
the rescaled cluster size distribution can be described by a power law $C \cdot x^{a}$ 
in the scaling variable $x=A/[\ell_G^{2-D_A}(t)\ell^{D_A}_p(t)]$.
The best fit of the function $f(x) = C \cdot x^{a}$ to the rescaled data $A^{\tau_A} \, \ell_G(t)^{2(2- \tau_A)} \ {\mathcal N}$ at time $t=1$
in the interval $[0.1,10]$ of $x=A/[\ell_G^{2-D_A}(t)\ell^{D_A}_p(t)]$
gives $a = 0.332(1)$, a value which is close to the one found in the case of the square lattice.

\vspace{0.5cm}

\subsubsection{Percolating clusters on the square and honeycomb lattices.}
\label{subsubsec:percolating_clusters_distribution}

In~\cite{BlCoCuPi14} we presented a scaling of the so-called
bump, that is the contribution given by the percolating clusters 
(or clusters whose size is comparable with the system size) to the full cluster size distribution $\mathcal{N}$,
for the zero-temperature Glauber dynamics on the square lattice, for different system sizes. 
Here we perform a similar analysis on the three  lattices considered.

As explained in Sec.~\ref{sec:phenomenon}, the very few largest clusters that survive the coarsening process after
a sufficiently long time are the ones that we use to define the characteristic time $t_p$. At the time $t_p$,
these clusters usually span most of the lattice and their geometrical and statistical properties resemble the ones
of the clusters at critical site percolation on the same lattice. Usually, at this time, the largest and second largest clusters (with
opposite spin orientation) are percolating and become ``stable'' with respect to the coarsening dynamics in the sense 
explained in Sec.~\ref{subsec:largest_cluster_intro}.
This is the reason why  $N_p$ that constitutes the contribution given by the percolating clusters to $\mathcal{N}$, 
is mainly due to the two largest clusters.
Then, for all practical purposes,  $N_p$ is just the size distribution of the two largest clusters in the system.

Let us discuss the scaling of $N_p$ in general.
The distribution $N_p(A,t,L)$ satisfies 
\begin{equation}
\int d A \, N_p(A,t,L) = {1 \over L^2}\; .
\end{equation}
The result $1/L^2$ is due to the definition of ${\mathcal N}(A,t,L)$ 
which counts the number of clusters with area $A$ per spin,
and to the fact that we have rescaled  the distribution by a factor $2$ to compare it to the one of percolation for which 
there is only a single percolating cluster.
In site percolation, finite-size scaling implies that the size distribution of the largest cluster $N_p(A,L)$,
for a system of linear size $L$, depend on $A$ and $L$ through the ratio $A/L^{D_A}$ at the threshold occupation probability,
with $D_A=91/48$ the fractal dimension of the critical percolating cluster.
The same should be true for  $N_p(A,t,L)$ in the dynamical problem for $t \ge t_p$.
If we rescale $A$ as $A \rightarrow A/L^{D_A}$, we need also to rescale the measure accordingly, i.e. 
$dA \rightarrow L^{D_A} d A$.
However, in the dynamical problem we need to take into account the effects of coarsening, and
we have seen that the largest cluster size (but also the one of the second largest) scale as $\ell_d(t)^{2-D_A}$.
Thus, the correct quantity to consider is $ L^{D_A}\, \ell_d(t)^{2-D_A}N_p(A,t,L)$ 
as a function of $ A/L^{D_A} \, \ell_d(t)^{2-D_A}$. 

\begin{figure}[h]
\begin{center}
(a) \hspace{6cm} (b)\\
 \includegraphics[scale=0.55]{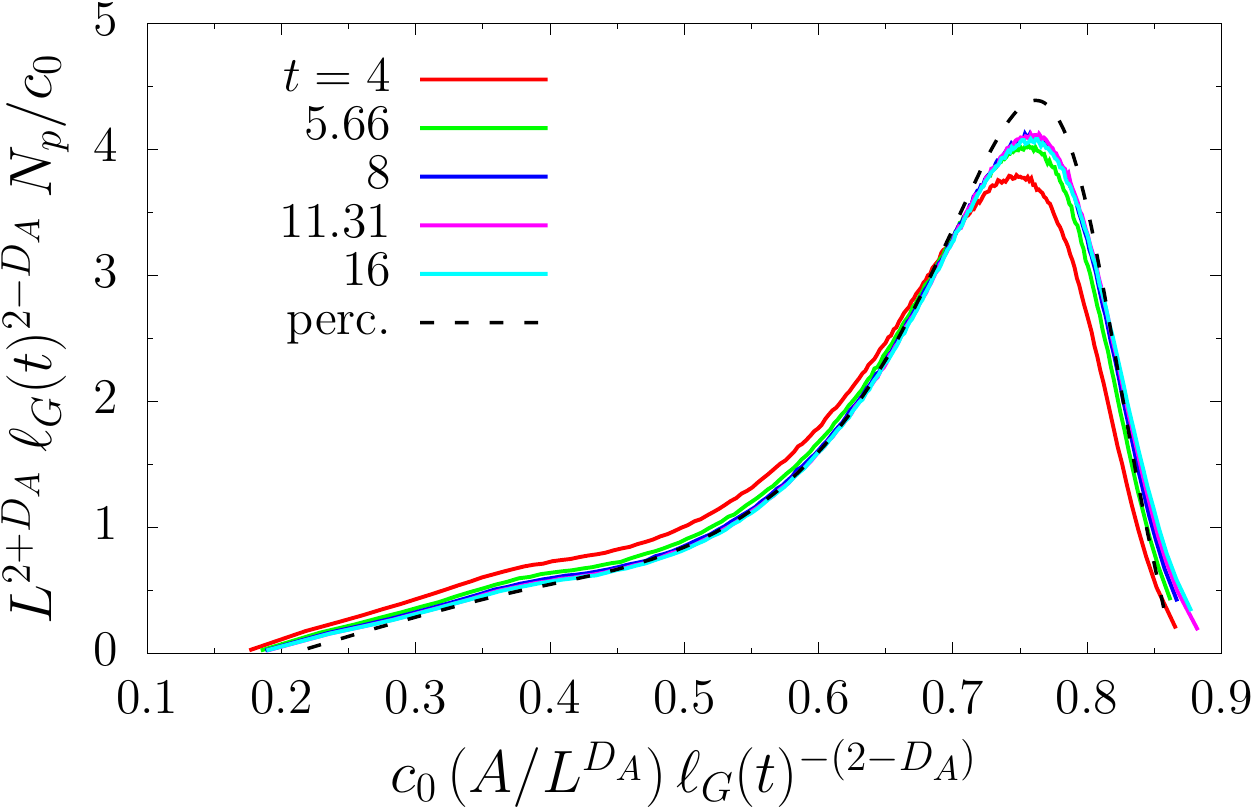}\hspace{1em}%
 \includegraphics[scale=0.55]{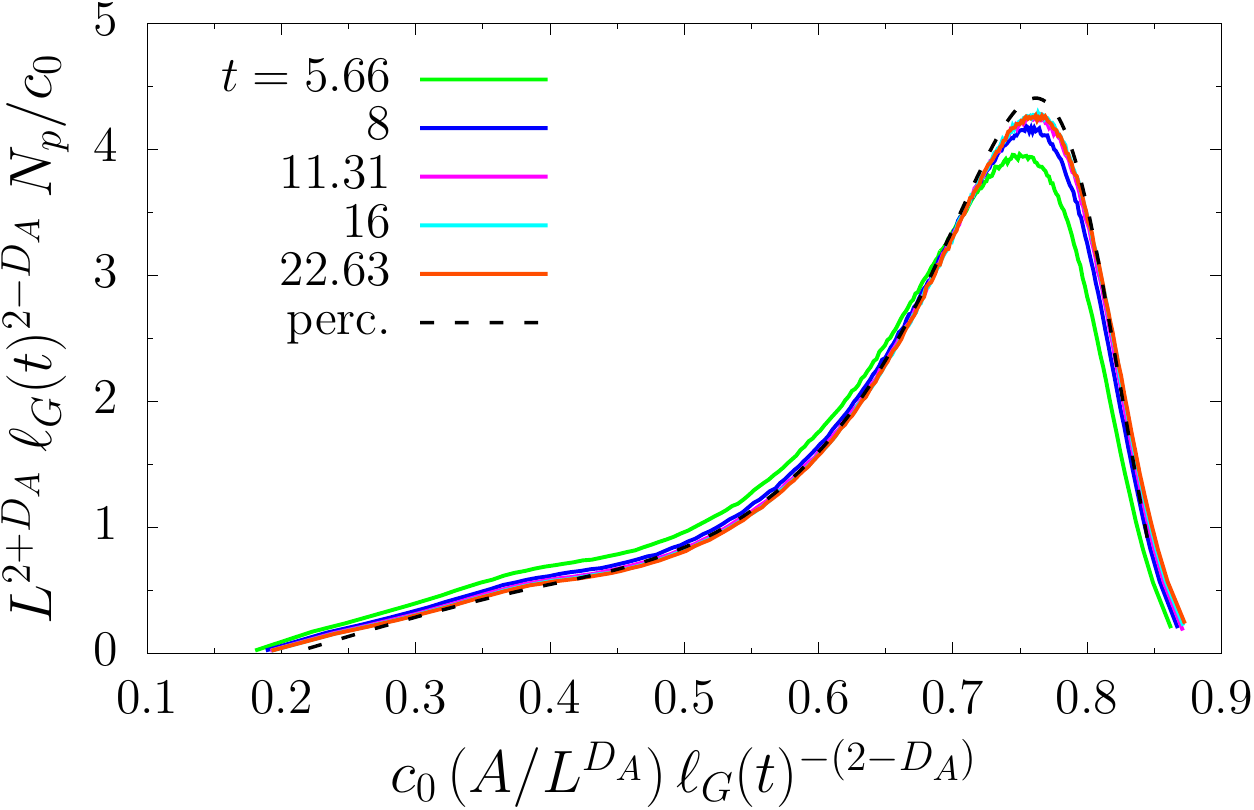}%
\end{center}
\begin{center}
(c) \hspace{6cm} (d)\\
 \includegraphics[scale=0.55]{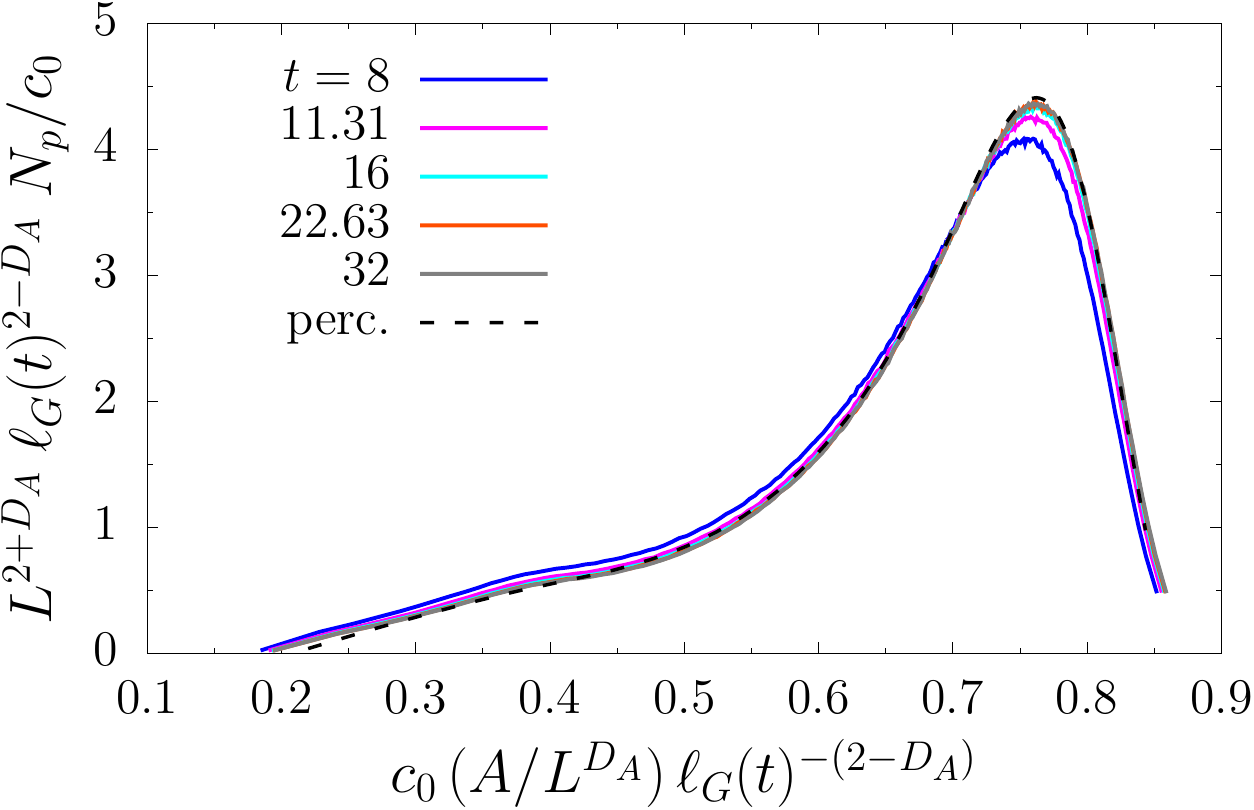}\hspace{1em}%
 \includegraphics[scale=0.55]{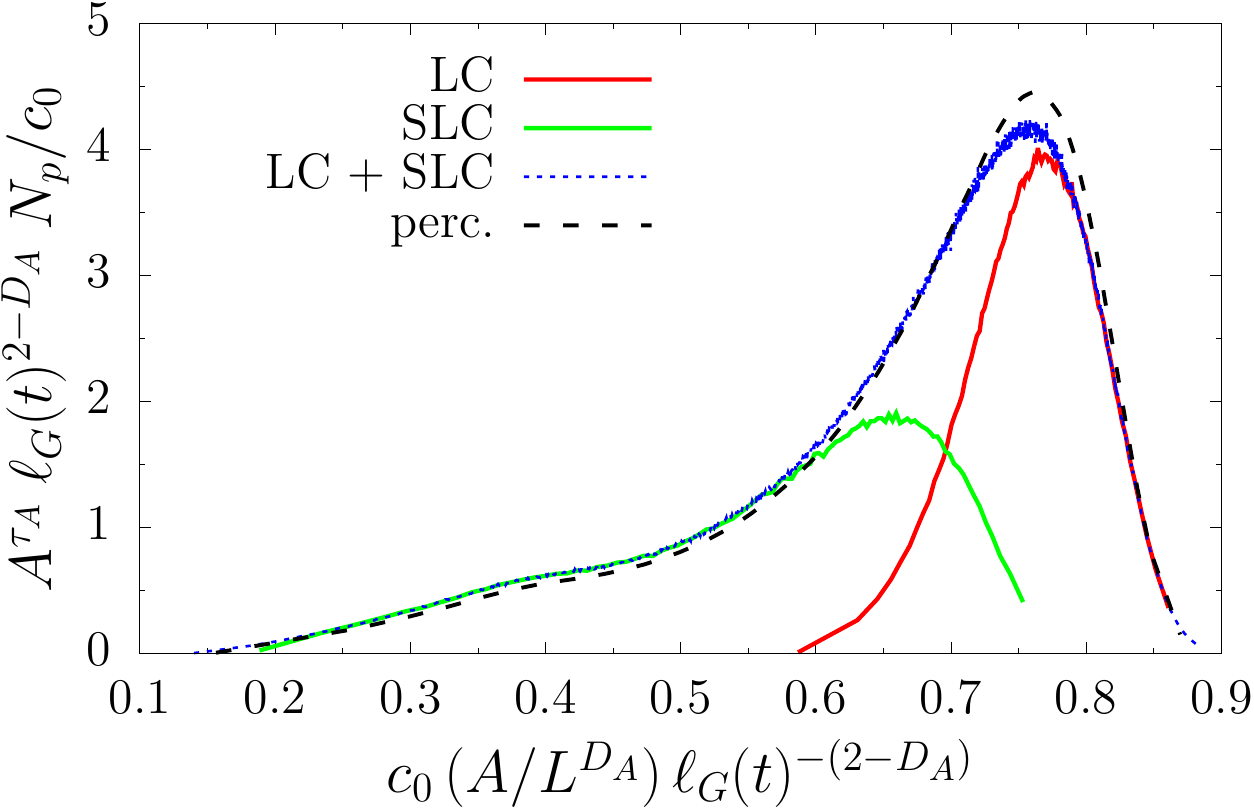}%
\end{center}
\caption{\small
The size distribution of the two largest clusters, $N_p(A,t,L)$,
for the zero-temperature Glauber dynamics on the square lattice, at different times indicated in the key and 
for different values of $L$, $L=160$ (a), $320$ (b) and $640$ (c).
The distribution is rescaled by the factor $ L^{2 + D_A} \, \ell_G(t)^{2-D_A} / c_0$ and plotted against
the rescaled area $ c_0 \, ( A / L^{D_A} ) \, \ell_G(t)^{2-D_A}$, where
$D_A$ is the fractal dimension of the percolating cluster in $2d$ critical percolation, 
$\ell_G(t)$ the characteristic length obtained by the excess energy,
and $c_0 \simeq 1.165$. In panel (d), instead, we show the contributions to $N_p$
coming from the largest (LC) and the  second largest (SLC) clusters, separately,
as well as the whole $N_p$ (LC $+$ SLC), at $t=8$, for the dynamics on a square lattice with
$L=320$. In each panel, the size distribution of the largest cluster for
site percolation, at the threshold occupation probability on the square lattice of corresponding size, is also
shown with a black dashed line, multiplied by $L^{2 + D_A}$ and plotted against $A/L^{D_A}$.
The value of the constant $c_0$ was chosen so that the rescaled distributions for the dynamical problem
coincide with the static one of critical percolation. This values is approximately independent of $L$.
}
\label{DistributionP} 
\end{figure}

We show the data corresponding to the rescaled distribution $N_p(A,t,L)$
in Fig.~\ref{DistributionP} in the case of the zero-temperature Glauber dynamics on
the square lattice, for sizes $L=160$ (a), $320$ (b) and $640$ (c).
Notice that $N_p(A,t,L)$ is multiplied also by a factor $L^2$ to get rid of the
$1/L^2$ present in its definition and make the data fall on a range of values of order $\mathcal{O}(1)$.
For each size, we also show the static size distribution of the largest cluster for site percolation at
threshold occupation probability on the square lattice of same size, rescaled as $N_p(A,L) \, L^{2 + D_A}$ and
plotted against the rescaled size $A/L^{D_A}$.
Our goal is to prove that, with this rescaling, the distribution $N_p(A,t,L)$ for the dynamical problem matches
the static one for critical percolation.
To do so, we need to include an additional scaling factor $c_0$ for the dynamical problem, that is,
we plot $ L^{2 + D_A}\, \ell_d(t)^{2-D_A} \, N_p(A,t,L) / c_0 $ against $ c_0 \, A/L^{D_A} \, \ell_d(t)^{2-D_A}$.
The value of the constant $c_0$ is not known a priori.
The value which gives the best collapse is  $c_0\simeq1.165(5)$, independently of the lattice linear size $L$.

Note that the agreement between the data for the quenched system and 
the critical percolation one becomes much better as we increase the system size. 
For $L=160$, the distributions are too large and not tall enough, the agreement 
is better for $L=320$, and it is nearly  perfect for $L=640$. 

In panel~(d) of Fig.~\ref{DistributionP}, we show the contributions to the size distribution $N_p$
coming from the largest cluster (LC) and the one from the second largest (SLC) separately,
as well as the whole $N_p$ (LC $+$ SLC), at  $t=8$, for the dynamics on a square lattice with
$L=320$. The data is scaled as in the other panels and we have also included  the size distribution of the largest
cluster at critical percolation (dashed line) on the same lattice (properly rescaled as in the other panels).
The whole distribution $N_p$ is  $N_p = \frac{1}{2} ( N_{\mathrm{LC}} + N_{\mathrm{SLC}})$.

\begin{figure}[h!]
\begin{center}
\includegraphics[scale=0.7]{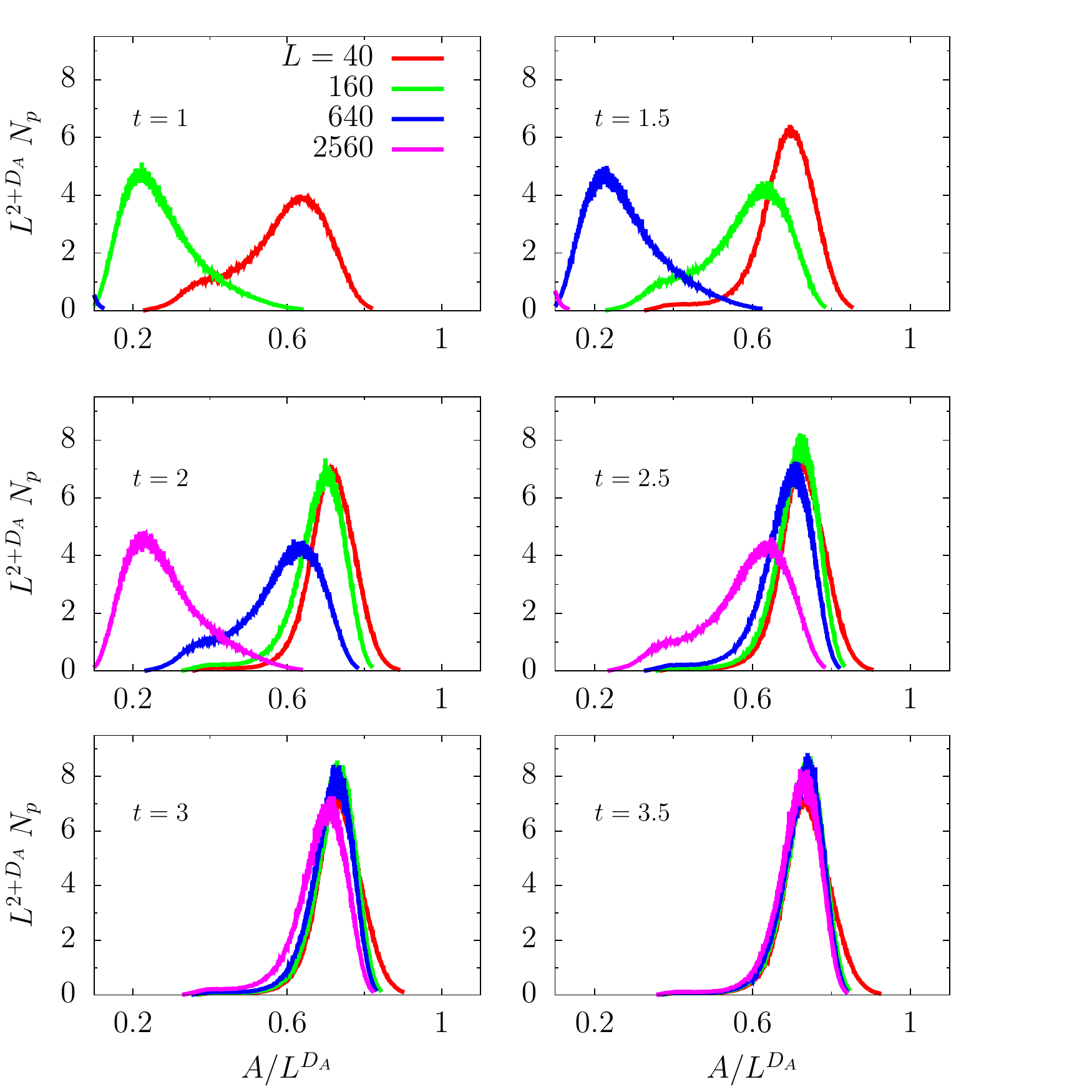}
\end{center}
\caption{\small Rescaling of the bump, $L^{2 +D_A} \, N_p$ vs. $A/L^{D_A}$ for the zero-temperature dynamics on a honeycomb lattice,
with $D_A=91/48$.
The data correspond to systems with linear sizes $L = 40$, $160$, $640$, and $2560$ as indicated in the key 
(the colour code is the same in each panel). The time that the data refers to is written in each panel.
}
\label{QuenchTI_new} 
\end{figure}  

From the time and $L$ dependence of $N_p$ it is also possible to reach an understanding of the dependence of the
characteristic time $t_p$ on $L$, as we show in the following in the case of the dynamics on the honeycomb lattice.
In Fig.~\ref{QuenchTI_new} we show $L^{2+D_A} N_p$ against the rescaled area $A/L^{D_A}$,
in the case of the $T=0$ Glauber dynamics on the honeycomb lattice,
for different values of the linear size $L$ and at different times (given in the key of each panel).  

At short $t$,  the overall shape of the rescaled bump
depends strongly on the size of the lattice and time, 
while at sufficiently long $t$ it seems to approach a stationary form that depends only on $A$ and $L$ through
on $A/L^{D_A}$, so that $N_p(A,t,L) \sim n_p(A/L^{D_A})$ for $t \gg 1$,
with $n_p$ a proper scaling function.
Moreover, we note a very particular scaling behaviour as both $t$ and $L$ vary, in the pre-percolating regime:
the curves for $L=40$ and $L=160$ at $t=1$ are 
replaced by the curves for $L=160$ and $L=640$, respectively, at $t=1.5$.
The same is true when passing from  $t=1.5$ to $t=2$: the
curves for $L=160$, $640$ and $2560$ replace the ones for $L = 40$, $160$ and $640$, in this order, and so on.
At $t=3.5$ all the curves, except for the one relative to $L=40$, have collapsed onto the same master curve.
Notice that this time corresponds approximately to the time at which the excess energy growing length $\ell_G$ saturates, see Fig.~\ref{GL_H}.
From this observation we can deduce that the typical time scale associated to the approach to percolation, $t_p$, roughly satisfies the rule
$t_p(4 L) = t_p(L) + \mathrm{const}$ with $\mathrm{const} \simeq 0.5$.
This result confirms our previous claim that $t_p(L) \propto \ln{L}$ for the zero-temperature dynamics on the honeycomb lattice.
On top, from the above observation, we can infer $t_p(L) \simeq (0.5/\ln 4) \, \ln{L} \simeq 0.36 \, \ln{L}$ which is in reasonable agreement 
with the reverse relation $\ell_p(t) \simeq \exp(2.6 \, t)$ that we found from scaling the full cluster size distribution, 
see Fig.~\ref{NAHexIM}.

\subsubsection{The shape of the bump.}

It is also interesting to study the shape of the probability distribution of the largest cluster. In the context of
percolation, quite a few results have been established both analytically and numerically~\cite{Sen01}.
When $p < p_c$, the distribution of the largest-cluster size was proved to follow a Gumbel distribution~\cite{Bazant00,RedigHofstad},
while for $p  > p_c$ the largest cluster size is distributed approximately as a Gaussian~\cite{Bazant02}.

A simple argument that justifies these observations is the following.
If there were no dependence between the cluster sizes, then $A_{ \rm max}$, the largest cluster size
would be the largest amongst $N_c$ i.i.d random variables, with $N_c$ the number of clusters in which the system is divided.
Then, according to the Extreme Value Theory~\cite{Gnedenko}, as $N_c \rightarrow  \infty$, the random variable 
$A_{ \rm max} = \max_{i=1,...,N_c} \, A_i$, after proper rescaling,
would be distributed as a Gumbel, a Fr\'echet or a Weibull random variable depending on the shape of the tail of the parent
distribution. 

However,  the clusters are correlated for any value of $p \in (0, 1) $ and the argument above is not fully correct.
Nevertheless, assuming that the correlations are weak (and this is the case when the system is sufficiently far away 
from the critical point)  these results can still be established~\cite{Bazant00,Bazant02,HoviAharony-97}.
Instead, when the system approaches the critical point, the largest cluster size (which in the context of percolation can also be 
seen as an
order parameter) experiences large non-Gaussian fluctuations and little is known about its distribution, except for a remarkable exact result
in the mean-field case~\cite{BotetPlos}. Numerical studies~\cite{BotetPOS} suggest that there is a smooth
crossover between the subcritical and the supercritical phase, and that the probability distribution of the order parameter
can be approximated by a weighted sum of a Gumbel and a Gaussian distribution.

We think that this might be exactly what happens in the case of subcritical quench dynamics.
During the time regime in which the system is approaching the critical percolation situation but long before attaining it, 
the distribution of the two largest cluster sizes (after proper rescaling) should be close to a Gumbel, while long
after having trespassed the critical percolation point it should approach a Gaussian. 
%This seems to be confirmed by our numerical simulations
%(see Fig.~\ref{DistributionP}). But we stress the fact that we need to take into account the second largest cluster as well, 
%as it was already explained, and mentioned when showing the snapshots in 
%Fig.~\ref{fig:snapshots-triangular}.
However the system spends most of its time in the vicinity of the critical percolation state (at least in the time
window explored by our numerical simulations), where large fluctuations are present and correlations between cluster sizes cannot be considered
weak. The probability distributions which are shown in Fig.~\ref{DistributionP} clearly are not resembling neither a Gumbel nor a Gaussian distribution.
A possible way to characterize the shape of the so-called ``bump'' would be to consider a mixture of Gumbel and Gaussian probability
distributions. In particular, a linear combination of the two with relative weights measuring the ``distance'' from the the two
extremal situations: long before the critical point and long after it.
We checked this possibility by fitting this trial distribution to the rescaled numerical data as presented in Fig.~\ref{DistributionP}, but
we could not get any satisfactory result, so we decided not to show it. 
Indeed, the fitting requires too many parameters (two parameters for each individual distribution, the Gumbel and the Gaussian,
a relative weight and a global scaling factor) and thus it seems pretty unreliable.
%We think that an other approach should be used instead, but at this stage we have not figure it out.

\subsection{Summary}

In all plots shown the system is initially prepared
at infinite temperature with correlations of the order of the lattice spacing. The same results hold for 
initial states in the high temperature phase, $T>T_c$, where correlations are short-ranged.
After a sudden quench to $T<T_c$ 
the dynamics are characterised by an initial approach to critical percolation lasting up to a time of the order $L^{z_p}$, 
for a system of linear size $L$, when
a stable pattern of percolating domains establishes. After this time, the percolating cluster(s) become fatter and fatter
evolving in a second dynamic regime characterised by the curvature driven 
growing length $\ell_d(t) \simeq t^{1/z_d}$,
where $z_d$ is the usual asymptotic dynamical exponent of the non-conserved order parameter class.
For certain lattice geometries, and depending on temperature being zero or different from zero, the 
system can remain blocked and not reach equilibrium ($T=0$) or it can do ($T\neq 0$) on an 
even longer time-scale $t_{\rm eq}$ that diverges with the system size faster than $L^2$.

The results in this Section confirm that for the triangular and square lattices, as well as for the bow-tie and 
Kagome lattices studied in~\cite{Blanchard14}, 
the growth of $t_p$ with $L$ is algebraic
\begin{equation}
t_p \simeq L^{z_p}
\end{equation}
while for the honeycomb lattice the system size dependent deviates from this form and is instead
\begin{equation}
t_p \simeq\ln L
\; . 
\end{equation}
The values of the exponents $z_p$ depend on the lattice geometry. The more detailed analysis of many 
observables developed in this Section suggests that the values of $z_p$ are 
\begin{eqnarray}
z_p = 
\left\{ 
\begin{array}{ll}
2/5 \qquad\qquad \mbox{square lattice}
\\
1/3 \qquad\qquad \mbox{triangular lattice}
\end{array}
\right.
\end{eqnarray}
The value $2/5$ for the square lattice is slightly different from the one we measured in~\cite{BlCoCuPi14} using the 
overlap function $Q$. The more extended analysis presented in this paper, addressing the scaling 
properties of many other observables, allowed us to measure this exponent with better precision and 
therefore obtain this slightly modified value. As regards the triangular lattice, being the initial condition 
at critical percolation, we did not need to rescale time in the analysis of all these other observables. The 
regime taking from $t_{p_1}=0$ to $t_p$ is one in which the percolating domains are present and, 
although they change shape by eating the small domains within them, they do now change considerably their 
geometric properties. Therefore, the best way to measure $z_p$ remains the one used in~\cite{BlCoCuPi14}
and the value $z_p$ is unchanged with respect to our previous claim.

\section{Metastability}
\label{sec:metastability}

In general, because of the existence of diagonal percolating clusters, and competing domains that wrap simultaneously 
around the system, equilibrium is not always reached at the end of the usual dynamic scaling regime. This means that the 
complete freezing or equilibration times can be notably longer than $L^{z_d}$.  Whether these configurations remain 
stable or decay in an even longer
time-scale depends on the geometry of the lattice, the boundary conditions and temperature. Moreover, 
some lattices allow for finite-size clusters with infinite life time at zero temperature. We discuss some 
of these cases here.

On most regular lattices, finite size clusters are unstable towards single spin flip zero-temperature dynamics.
The honeycomb lattice is special in this respect as finite-size clusters with infinite life-time are possible with local dynamics
that do not conserve the order parameter. 
Only under temperature fluctuations, and hence moves that increase the energy, these clusters acquire a finite but 
very long life-times~\cite{TakanoMiyashita,Cheong04}.

The choice of boundary conditions can have some influence on the final state reached after a quench 
to zero temperature.  More precisely, for all the cases considered, after critical percolation establishes, 
the dynamics at low temperatures are dominated by the coarsening of domains. 
After the characteristic time scale $L^{z_d}$, 
most of the finite domains with linear size much smaller than the lattice linear size $L$
disappeared. For instance,  the arrival 
configuration is either completely magnetised such that all the spins take the same value, or in a striped state 
with interfaces crossing the lattice~\cite{BaKrRe09} (for zero-temperature dynamics on the honeycomb lattice, one can also have
more complex domain patterns).
Next, and depending on the  lattice geometry and the boundary conditions, 
these stripe states can be stable or not. In the latter case, there is some additional evolution on  a much longer time scale.

In short, the stability of the striped states with respect to the zero-temperature dynamics, for the various lattices and 
boundary conditions used, can be classified as follows.

\begin{figure}[h]
\begin{center}
\includegraphics[angle=90,scale=0.65]{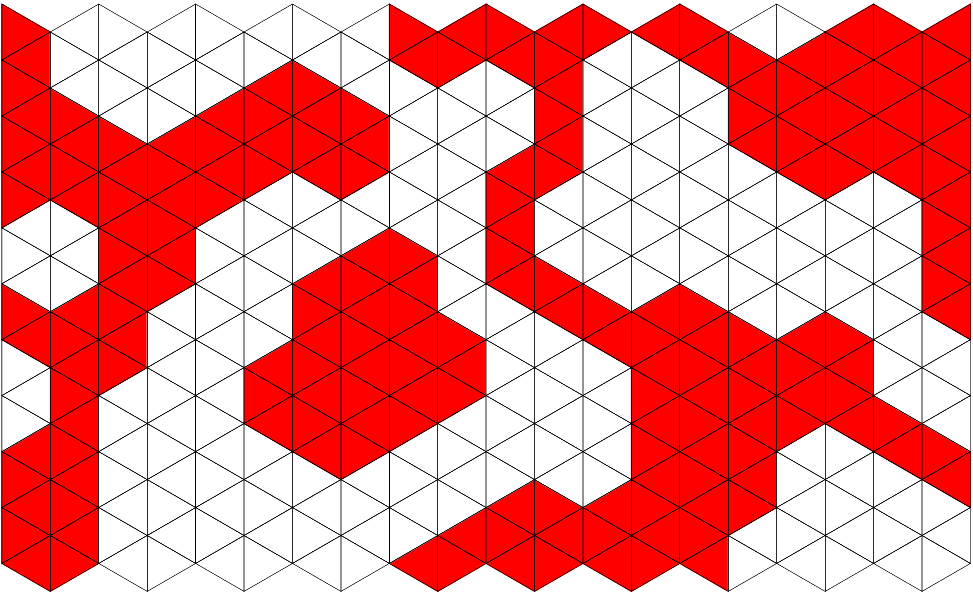}%
\end{center}
\caption{ \small Example of a \textit{frozen} configuration for the zero-temperature dynamics on a honeycomb lattice
of size $20 \times 20$ with PBC. In this picture, each site on the lattice is represented by
a triangular cell, with the color (red or white) indicating the spin orientation.
}
\label{Hon_frozen}
\end{figure}

\begin{figure}[h]
\begin{center}
  \subfloat[$t=1$]{\includegraphics[scale=0.45]{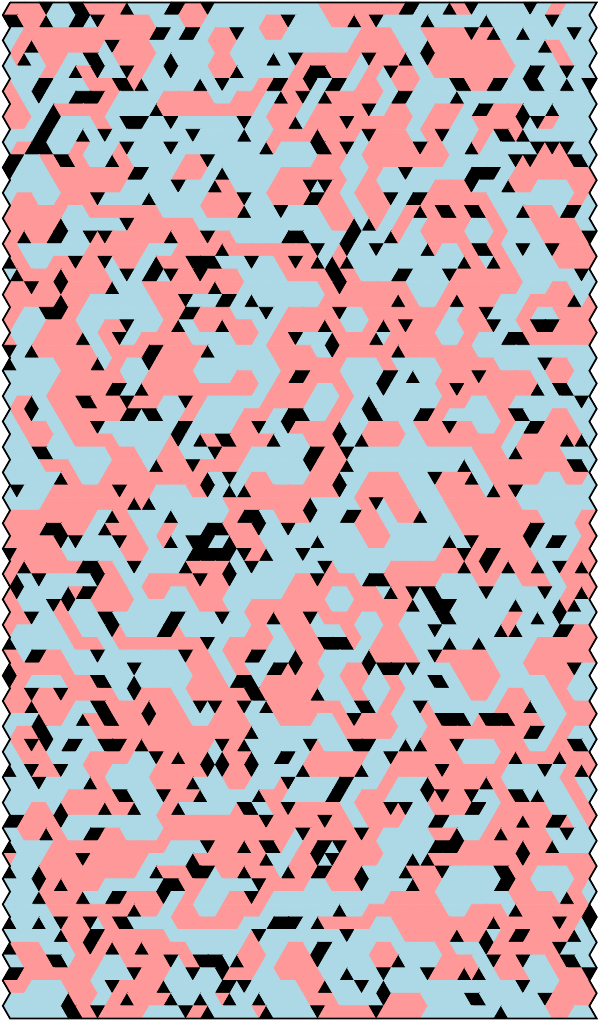}}\quad%
  \subfloat[$t=2$]{\includegraphics[scale=0.45]{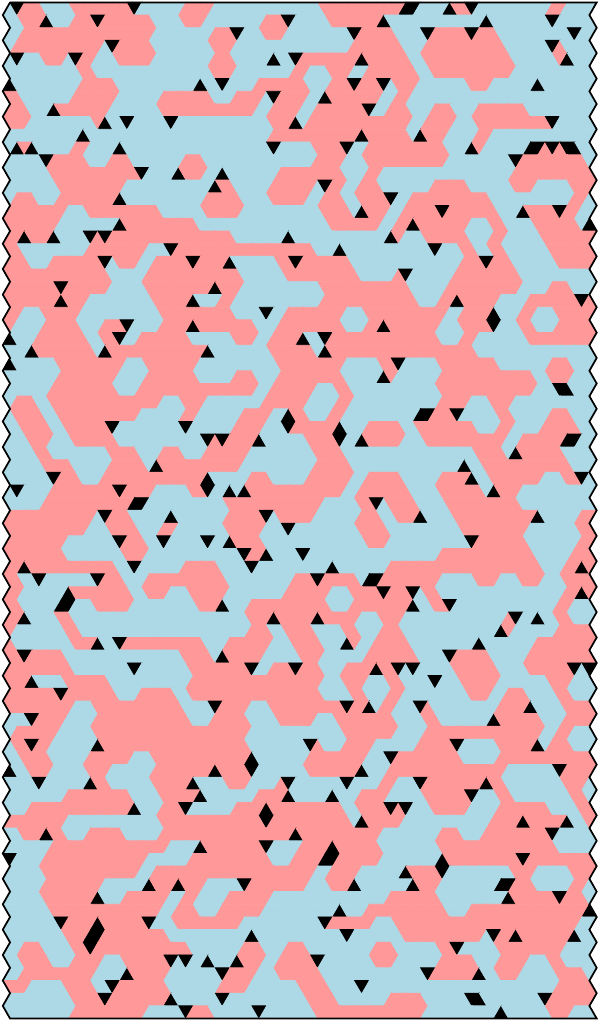}}\quad%
  \subfloat[$t=4$]{\includegraphics[scale=0.45]{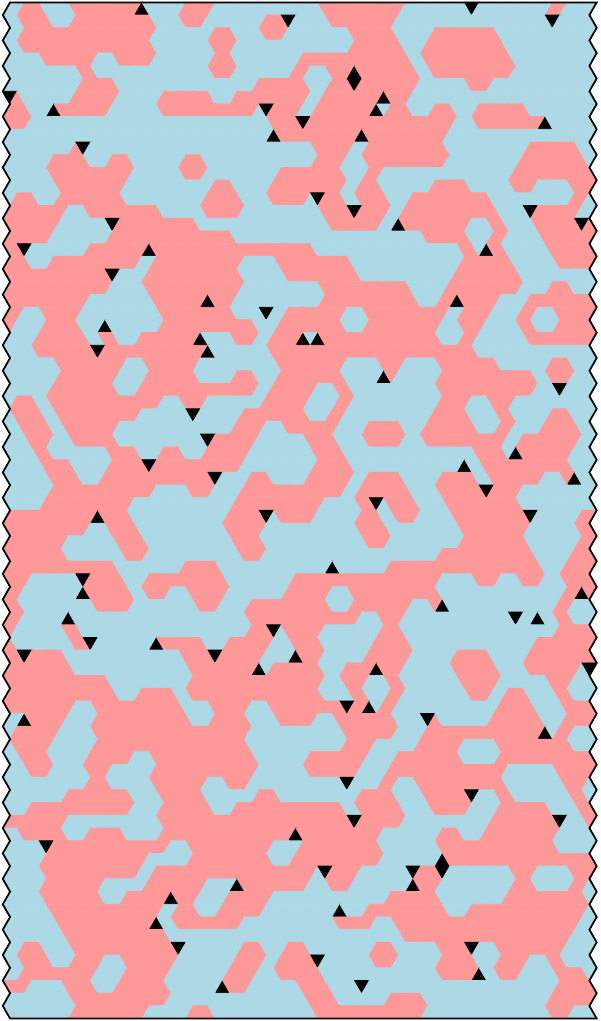}}\quad%

  \subfloat[$t=6$]{\includegraphics[scale=0.45]{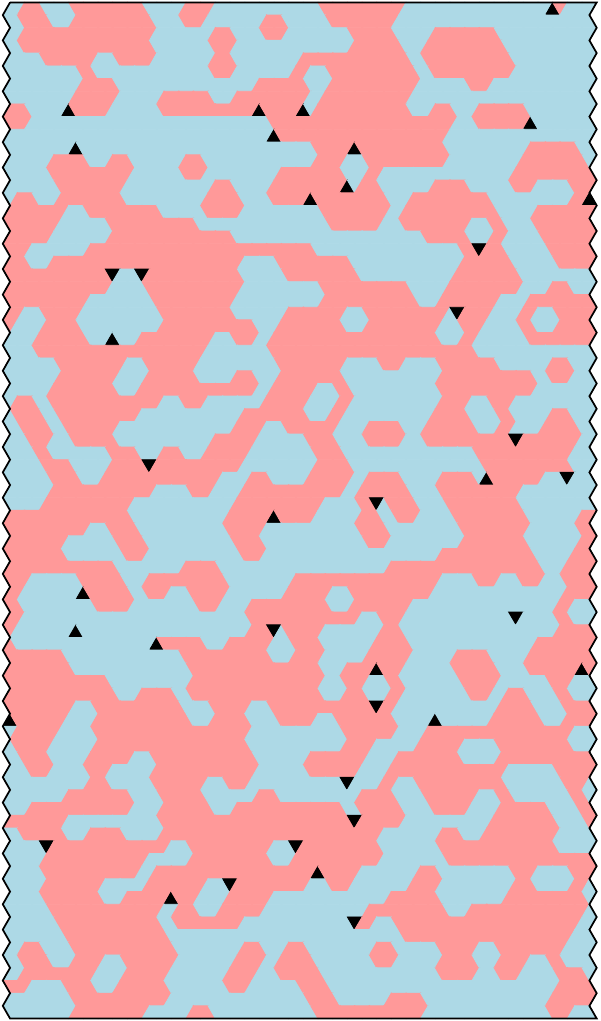}}\quad%
  \subfloat[$t=8$]{\includegraphics[scale=0.45]{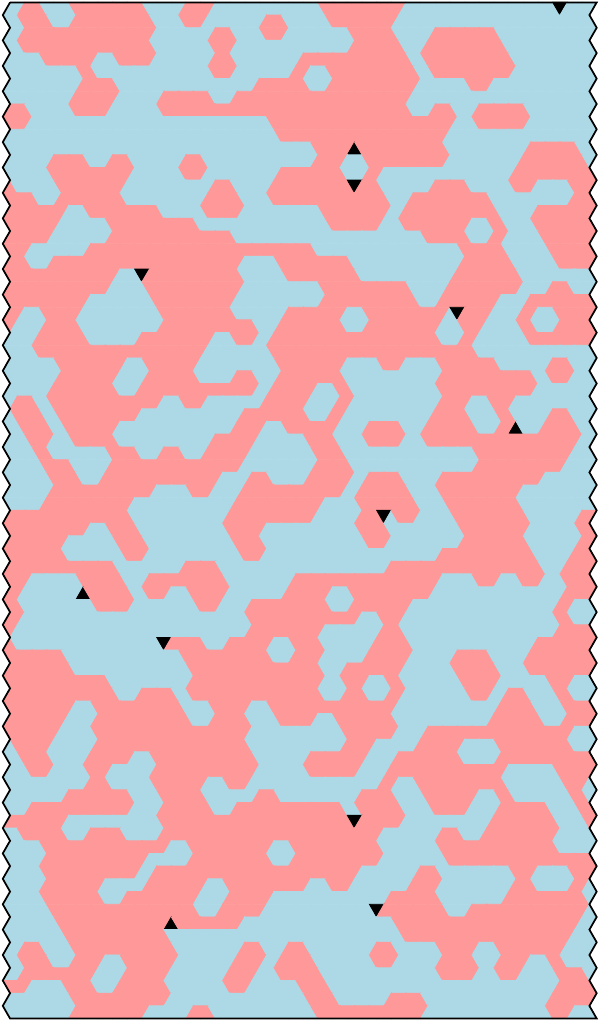}}\quad%
  \subfloat[$t=10$]{\includegraphics[scale=0.45]{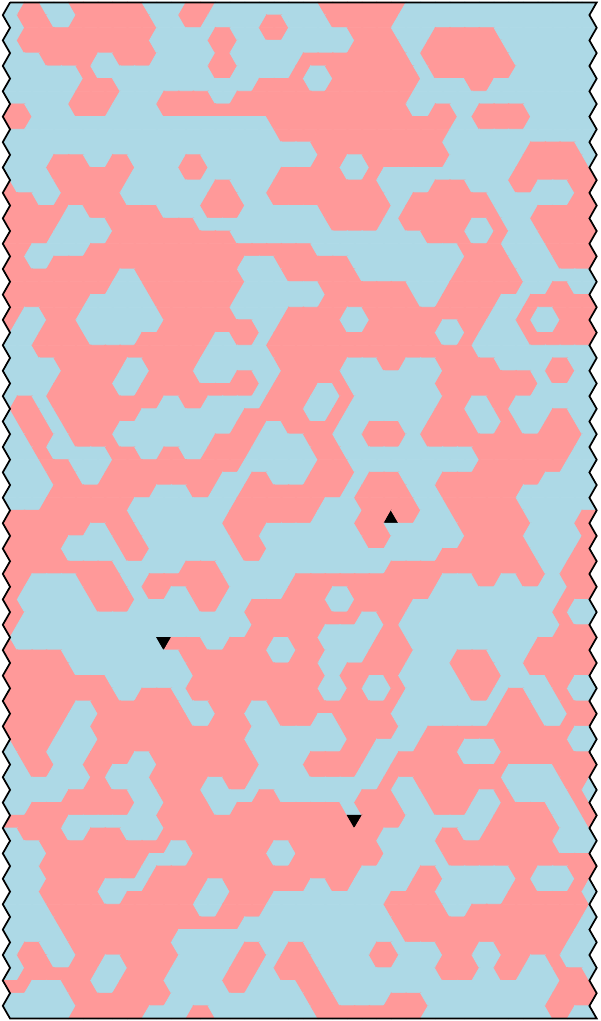}}\quad%
\end{center}
\caption{
\small Some snapshots of the evolution of a spin configuration under zero-temperature dynamics on a honeycomb lattice with linear size $L=80$ and
PBC. 
Each site of the lattice is represented by a triangular cell, as in Fig.~\ref{Hon_frozen}.
The spins that are \textit{frozen} at a given time, that is the ones that cannot be flipped without an energy cost, are represented with light colours,
red for $+1$ spins and blue for $-1$ spins. The black cells represent spins belonging to the two phases and that can still be flipped.
}
\label{fig:Gl-Hon-snapshots}
\end{figure}

\begin{itemize}

\item Square lattice, PBC: 
diagonal striped states are very long-lived but they progressively 
convert into clusters percolating in both directions, see 
Fig.~\ref{number-cluster-percolate}, that in turn
grow to cover the full system
in a characteristic time scale $t_{\rm eq} \simeq L^3$~\cite{SpKrRe02, OlKrRe12}.

\item Square lattice, FBC:  the striped states are stable
and there is no additional time scale beyond $L^{z_d}$.

\item Triangular lattice, PBC: the striped states are stable with no additional time scale. 

\item Triangular lattice, FBC: the horizontal (or vertical) striped states are not stable. 
This is due to our choice of triangular lattice for which a straight interface is not stable on the borders. 
This adds a second characteristic time scale $t_{\rm eq} \simeq L^{3.333}$.

\item Honeycomb lattice: 
this is a particular case since the honeycomb lattice is odd-coordinated:
\textit{frozen} states can have a very complex and varied structure and thus they are very large in number
(see Fig,~\ref{Hon_frozen} for an example of such a frozen state). 
In Fig.~\ref{fig:Gl-Hon-snapshots} we show some snapshots of the evolution of a spin configuration
under zero-temperature dynamics on a honeycomb lattice with linear size $L=80$. 
In each snapshot, the spins that can still be flipped are represented by black cells.
The overall domain structure (number of wrapping domains and their topology) is decided very early in the dynamics and the
later evolution does not change significantly their shape.

\end{itemize}

\subsection{The honeycomb lattice.}
\label{subsubsec:Hon-freeze}

We have already stated that the honeycomb lattice is special due to the 
existence of finite size frozen configurations. The two panels in Fig.~\ref{GL_H} demonstrate that 
the growing length saturates at $\ell_G\simeq 4$ independently of the system 
size while the freezing time scales as $t_{\rm freeze} \simeq \ln L$. These  results seem to be in contradiction. 
We argue now that they are not.

A fit of  the time-dependence of the approach to saturation of the growing length yields $\ell_G(t) \simeq 3.98 \, (1- 0.8 \, e^{-0.4 \, t})$
(not shown).  

On the other hand, the snapshots in Fig.~\ref{fig:Gl-Hon-snapshots} prove that at late times the spins that are 
free to flip are not very numerous and are far apart in the sample.
Let us assume that at time $t$ there are $N_f(t)$ flippable spins the update of which
will lead to an actual decrease in energy, and that they disappear following a
``radioactive'' law, $N_f(t) \simeq N_f(t_0) \, e^{-a(t-t_0)}$, with $N_{f}(t_0) = \rho N$, $\rho$ their density 
at  a reference time $t_0$, and $N=L^2$ the total number of spins in the sample.  

The exponential energy decay, concomitant with the exponential approach of $\ell_G(t)$ towards its asymptote, 
and the exponential decay of $N_f(t)$ imply $a \simeq 0.4$. 

The freezing time can now be associated to the time at which $N_f=1$ implying 
$\ln (\rho N) = 0.4 \, (t_{\rm freeze} -t_0)$ that for $t_{\rm freeze} \gg t_0$ yields 
$2\ln L \simeq 0.4 \, t_{\rm freeze}$ and 
$t_{\rm freeze} \simeq 5 \, \ln L$ as observed numerically in Fig.~\ref{GL_H}-right.

\subsection{Finite temperature quenches}
\label{subsec:finite-temperature}

In this Section we show some measurements relative to finite temperature quenches.
Thermal fluctuations eventually destroy the configurations with stable crossing interfaces and the system 
must asymptotically approach a  magnetised state. 
The magnetisation density and crossing correlations at zero and finite temperature demonstrate that 
a finite working temperature does not destroy the approach to random critical percolation. Moreover, they allow us to 
investigate the very late dynamics with the final approach to a fully blocked state at $T=0$ 
or equilibrium at $T>0$.

\begin{figure}[h]
\begin{center}
\includegraphics[scale=0.6]{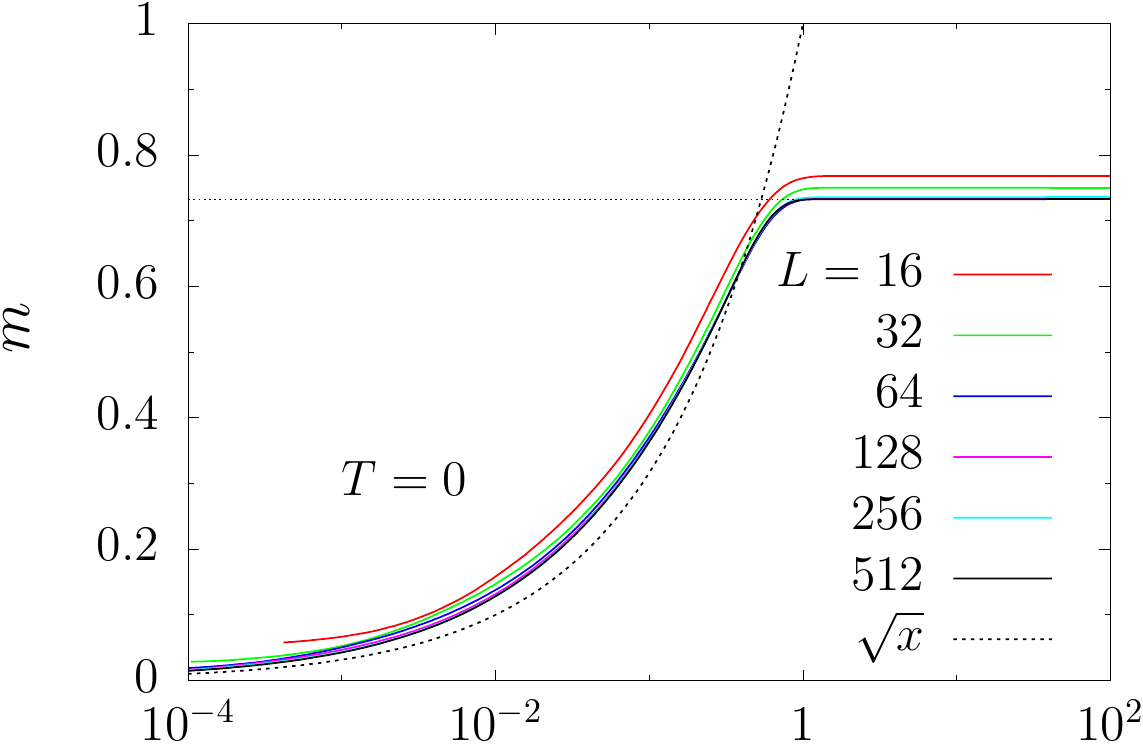}\hspace{1em}%
\includegraphics[scale=0.6]{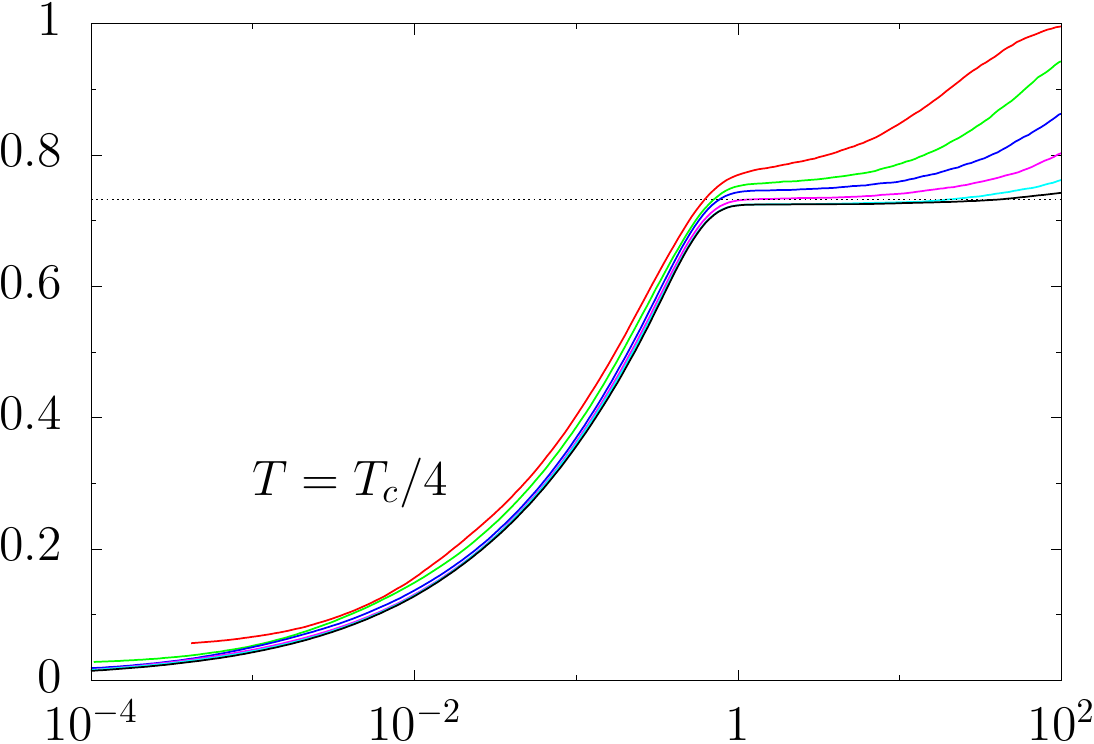}%

\includegraphics[scale=0.6]{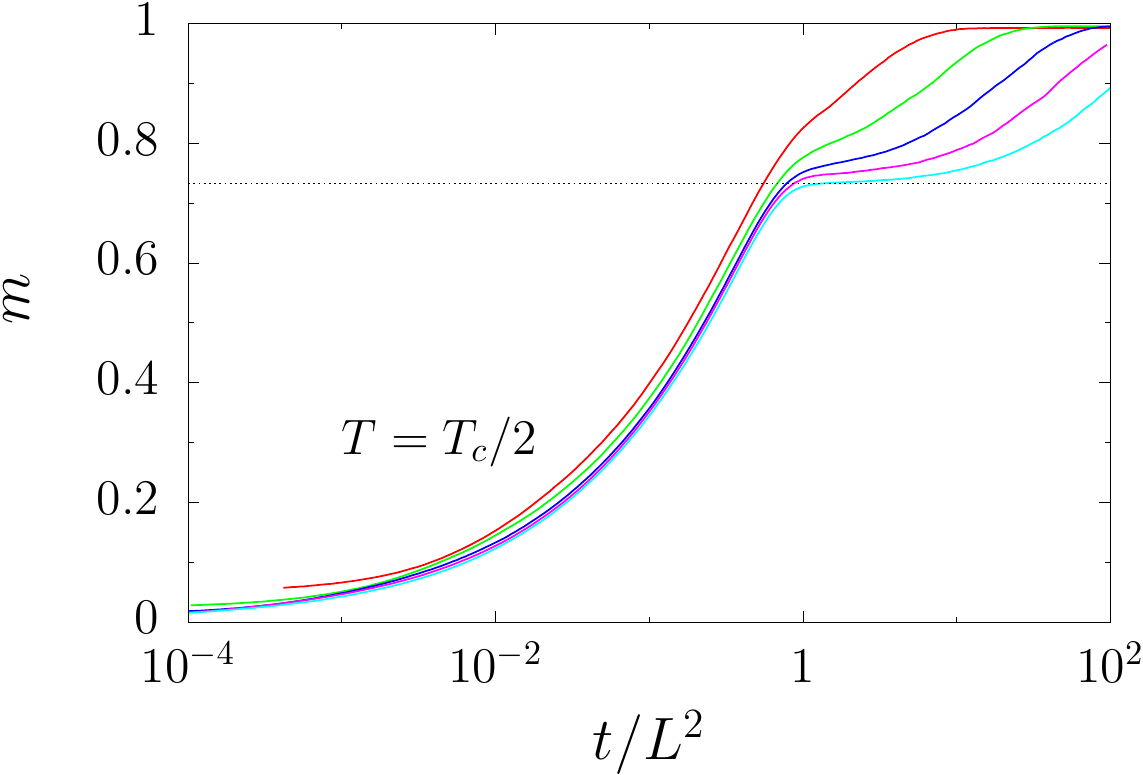}\hspace{1em}%
\includegraphics[scale=0.6]{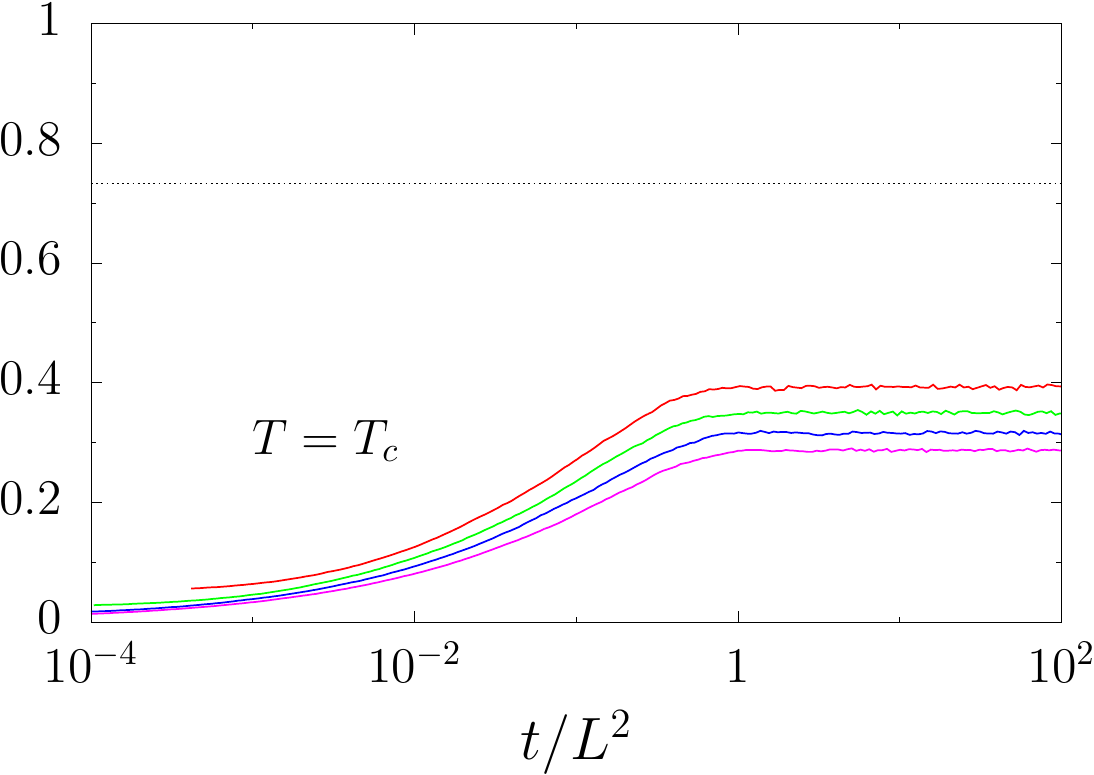}%
\end{center}
\caption{\small Square lattice Ising model with FBC 
evolved with kinetic MC dynamics with non-conserved order parameter. 
Averaged magnetisation density, $m$, vs. $t/L^2$ for various system sizes 
given in the key of the first panel and  various final temperatures $T$ on the different panels. The dotted horizontal  line 
is the infinite time limit of the magnetisation density after a $T = 0$ quench, $m_\infty \simeq 0.7332$.
We notice that at $T=0$ and for $t/L^2<1$, the master curve 
can be roughly approximated by the power law $x^{1/2}$, indicated with a dotted line.
}
\label{FTmag}
\end{figure}

In Fig.~\ref{FTmag}  we show the average magnetisation density 
against $t/L^2$ on the square lattice with FBC and various system sizes $L$ given in the 
key. The working  temperatures are $T=0, \, T_c/4, \, T_c/2, \, T_c$ on the different panels. In the infinite time limit after a $T=0$ quench, 
the magnetisation density converges to $0.7332$ (dotted line). 
This value can be understood by the following simple argument. As shown by Barros et al.~\cite{BaKrRe09}, the probability of 
having a spin configuration with a cluster crossing in both directions, that will evolve to  a state with magnetisation density $1$,
is given by the corresponding probability $ \pi^{\rm FBC}_{\rm hv}   \simeq0.6442$ from $2d$
critical percolation~\cite{Cardy92,Watts96}. 
The complementary probability $\pi^{\rm FBC}_{\rm h} + \pi^{\rm FBC}_{\rm v} = 1- \pi^{\rm FBC}_{\rm hv}$ corresponds to the case with horizontal or vertical stripes 
that will evolve to states with, on average, magnetisation density $1/4$. 
These are the only possibilities for the FBC case as stable diagonal stripes are not allowed in this case. 
Then the magnetisation density in the final state
is expected to be given by $m_\infty =  \pi^{\rm FBC}_{\rm hv} + (1 - \pi^{\rm FBC}_{\rm hv}) /4 \simeq 0.7332$. 

\vspace{0.25cm}

\begin{figure}[h!]
\begin{center}
\includegraphics[scale=0.6]{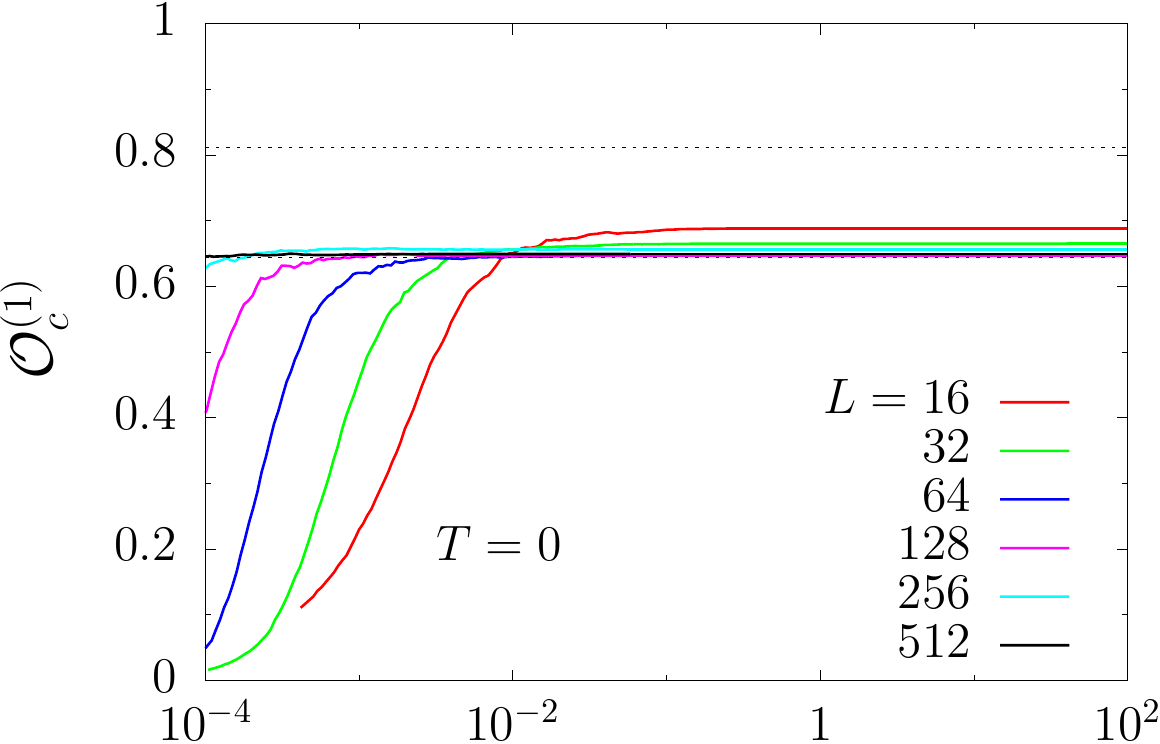}\hspace{1em}%
\includegraphics[scale=0.6]{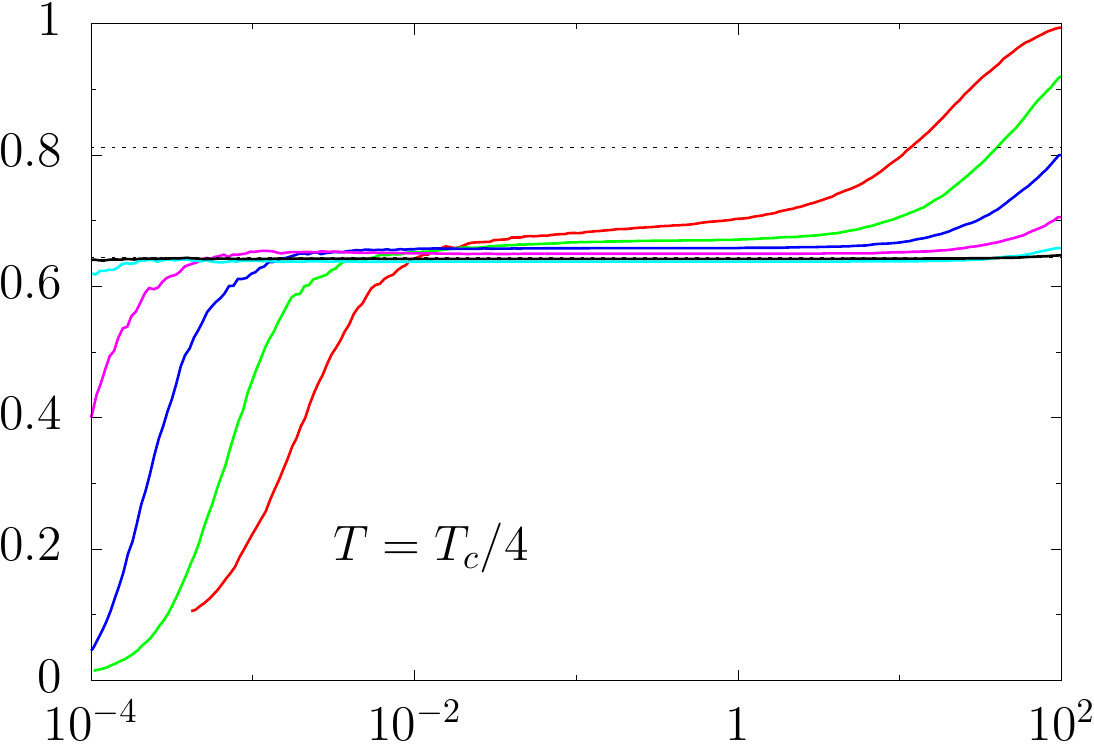}%

\includegraphics[scale=0.6]{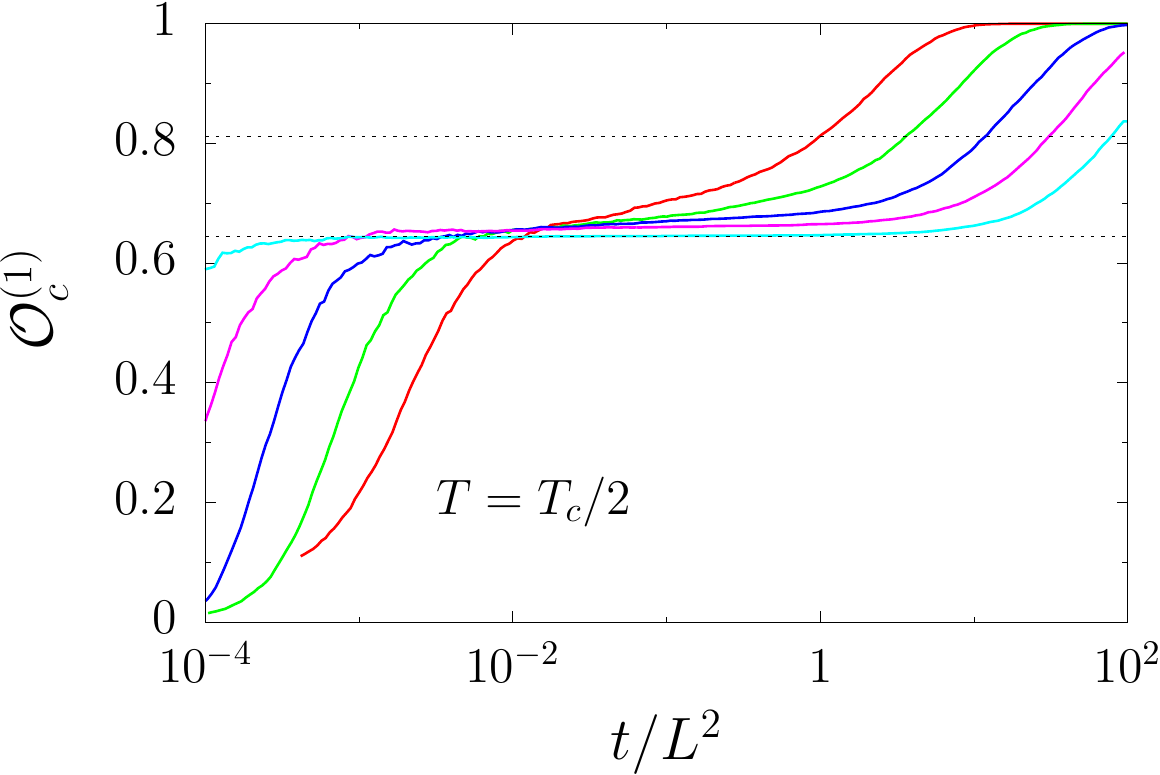}\hspace{1em}%
\includegraphics[scale=0.6]{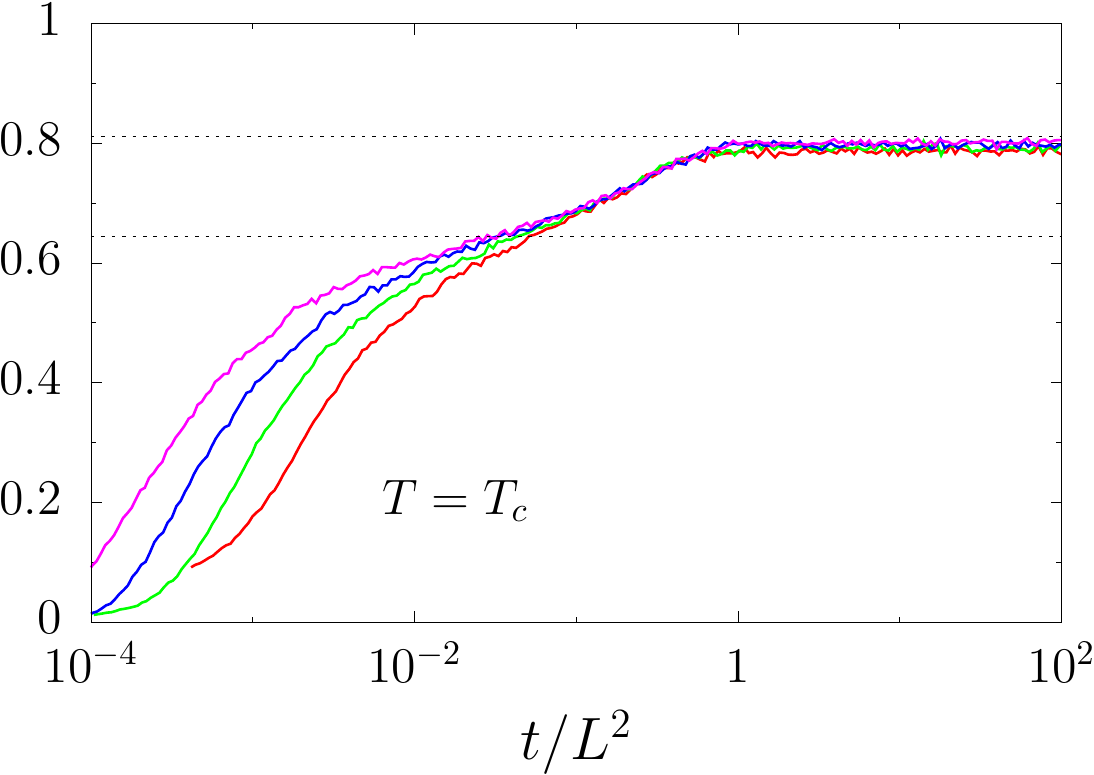}%
\end{center}
\caption{\small Square lattice Ising model with FBCs. 
The correlation between the number of crossings at time $t$ and at the final state, 
${\cal O}^{(1)}_c$, vs. $t/L^{2}$ for different final temperatures $T$ on the various panels. 
The colour code  for the different sizes $L$ is the same as in Fig~\ref{FTmag}.
The dotted horizontal lines are at $0.64$ and $0.81$ the probabilities of 
having a cluster that percolates on both horizontal and vertical directions at critical 
percolation, and at the Ising critical temperature, respectively.
}
\label{FTcr12}
\end{figure}
For  $0 < T < T_c$ we see that the behaviour is similar up to $t/L^2 \simeq 1$. For $t/L^2 > 1$, the 
magnetisation density will eventually approach
$m_{\rm eq}(T)$, the average magnetization density of the Ising model at equilibrium at the temperature $T$,
but after a time that increases with $L$ and the distance of $T$ from $T_c$.
For example, for $T=T_c/2$ the equilibrium magnetisation density is $m_{\rm eq}\simeq0.9980$~\cite{Onsager44,Yang52}.
Instead, for a quench to the critical point we clearly see that the magnetisation reaches a plateau in the
characteristic time $t_L \sim L^2$, but with a value that is decreasing with the system size:
naturally, we expect $ m_{\rm eq }$ to vanish as $L \rightarrow \infty$.
We also notice that at $T=0$ and for $t/L^2<1$, the master curve 
can be roughly approximated by the power law $x^{1/2}$, indicated with a dotted line in the upper left panel in Fig.~\ref{FTmag}.

In Fig.~\ref{FTcr12} we show ${\cal O}^{(1)}_c(t)$, the correlation function of the {\it crossing number},
for the spin configuration at time $t$ and a state with a unique cluster crossing the lattice in both directions (see Sec.~\ref{sec:observables} for more details
on its definition), as a function of the rescaled time $t/L^2$, for the same cases as in Fig.~\ref{FTmag}. 
There is a clear correspondence with the evolution of 
the magnetisation density. The change of behaviour towards a state with magnetisation density
$m_{\rm eq}$ in Fig.~\ref{FTmag} takes place at the same time as the change towards 
${\cal O}^{(1)}_c(t)=1$. Two horizontal dotted lines are also shown, corresponding to $ \pi^{\rm FBC}_{\rm hv} = 0.6442$ and
$ \pi^{\rm FBC}_{\rm hv} |_{T_c} = 0.8113$
which is the probability of having a cluster crossing in both directions at the critical Ising point~\cite{BlPi13} (this state
is reached asymptotically by the dynamics following a quench to $T_c$).

\begin{figure}[h!]
\begin{center}
\includegraphics[scale=0.6]{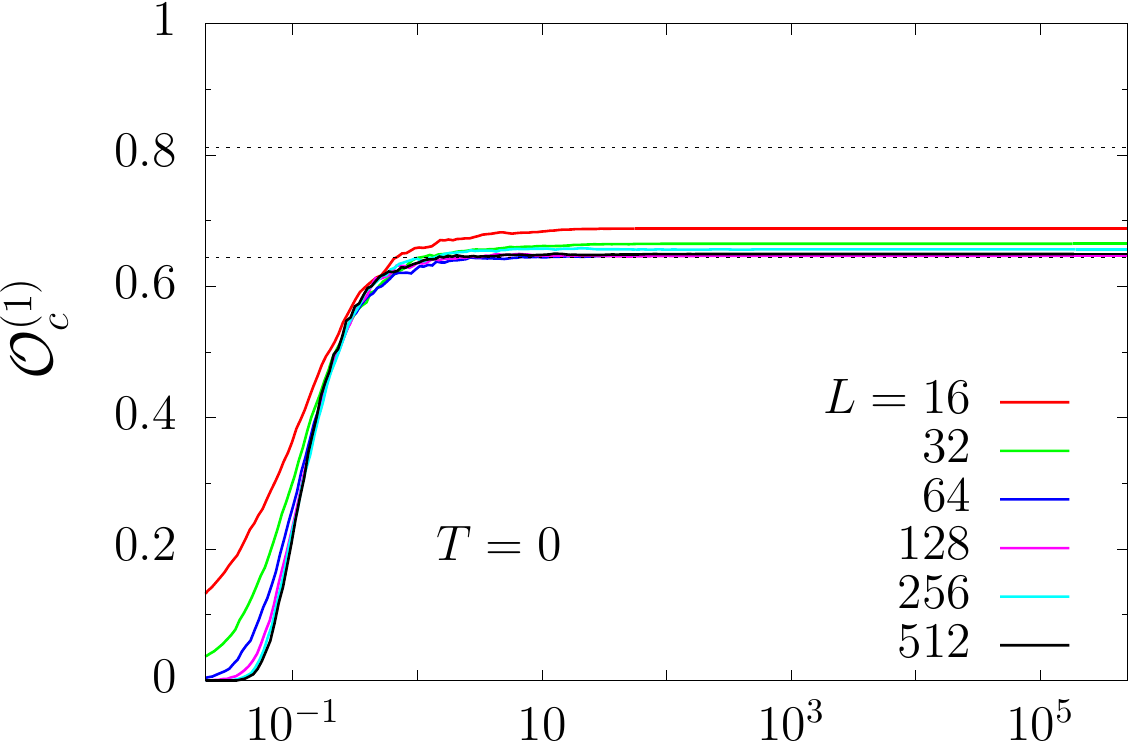}\hspace{1em}%
\includegraphics[scale=0.6]{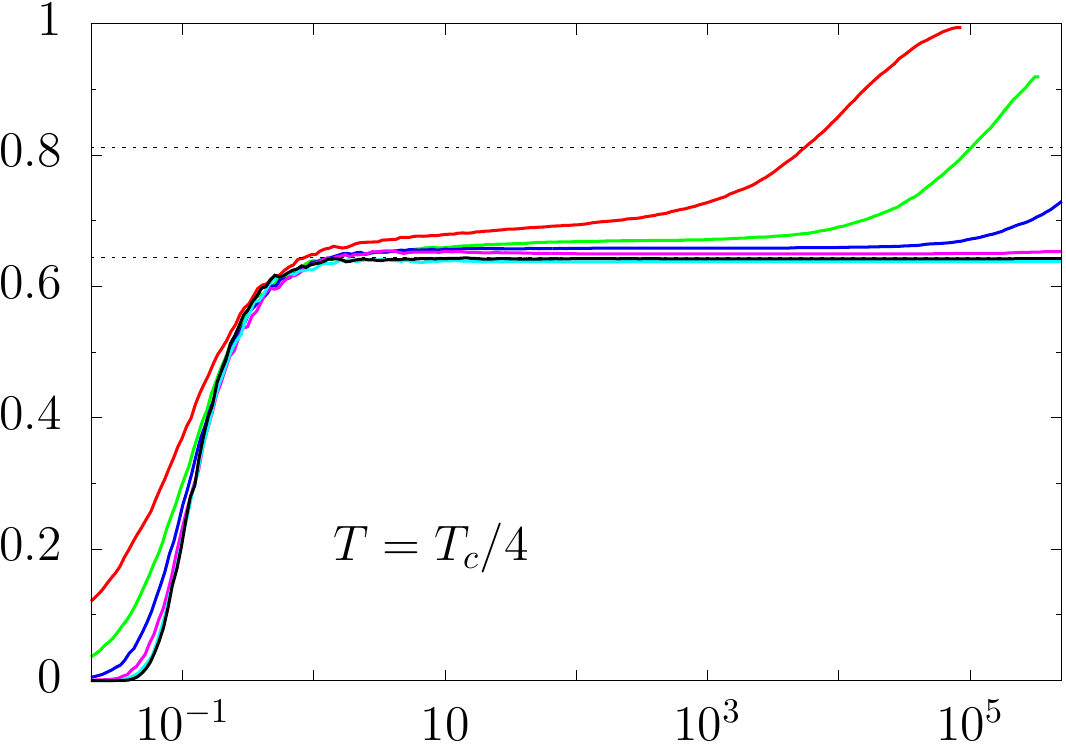}%

\includegraphics[scale=0.6]{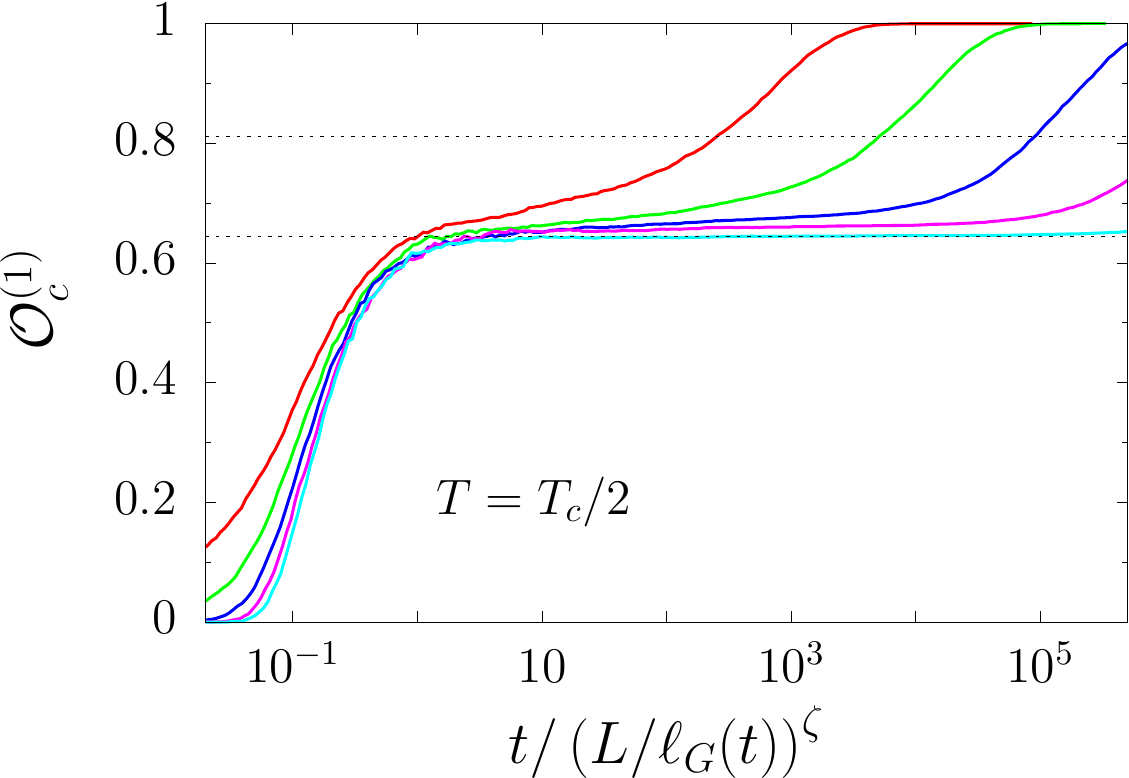}\hspace{1em}%
\includegraphics[scale=0.6]{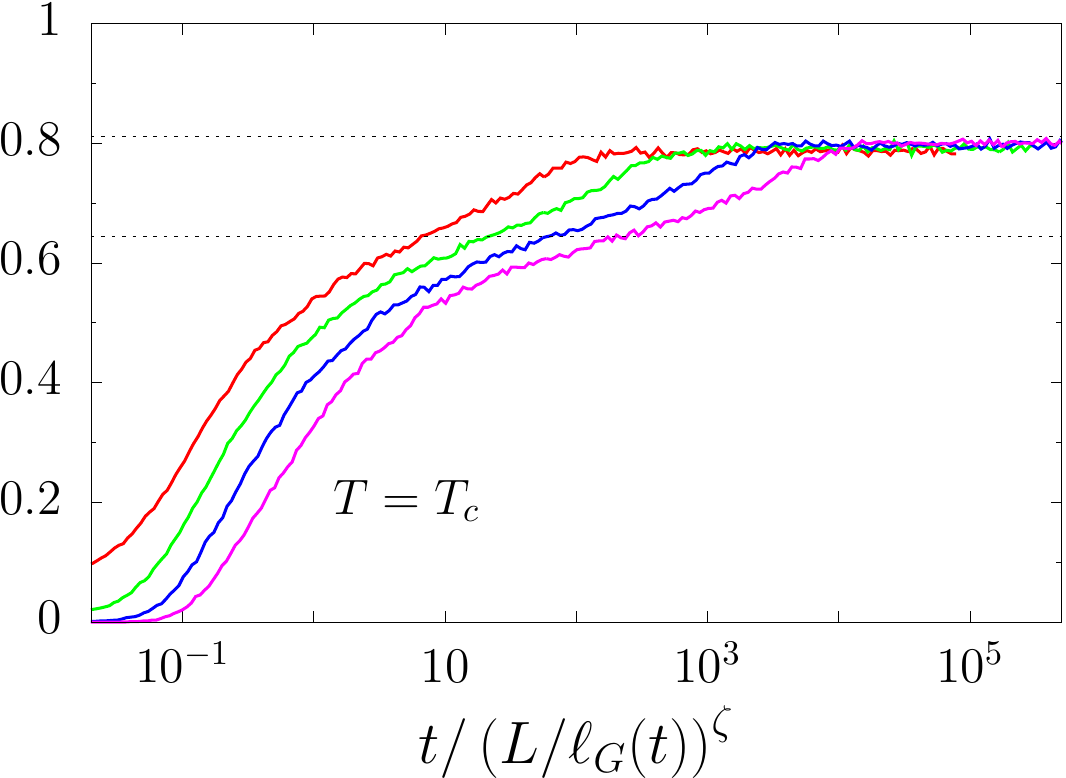}%
\end{center}
\caption{\small
Square lattice Ising model with FBC. 
The correlation between the number of crossings at time $t$ and at the final state,
${\cal O}^{(1)}_c(t)$, against the rescaled time
$t / \left( L / \ell_G(t) \right)^{\zeta}$, with $\zeta=0.5$, for different final temperatures $T$. The characteristic length scale
$\ell_G(t)$ is the numerical value obtained as the inverse of the excess energy, in all cases.
The colour code  for the different sizes $L$ is the same as in Fig.~\ref{FTcr12}.
As in Fig.~\ref{FTcr12}, the dotted horizontal lines are at $0.64$ and $0.81$, the probabilities of 
having a cluster that percolates on both horizontal and vertical directions at critical 
percolation, and at the Ising critical temperature, respectively.
}
\label{FTcr13_new}
\end{figure}

Finally, in Fig.~\ref{FTcr13_new} we show ${\cal O}^{(1)}_c$ against the rescaled time
$t / (L/\ell_G(t))^{\zeta}$ to highlight the region corresponding to the approach to critical percolation, as done for
other observables before. Here we take $\zeta=0.5$ in agreement with the results obtained for the scaling of the largest cluster size
and the wrapping probabilities on the square lattice.
We expect the exponent $\zeta$ to take the same value for any sub-critical quench.
The characteristic length $\ell_G(t)$ derived from the excess energy is again taken as a measure of the usual
dynamical characteristic length $\ell_d(t)$.
We observe that the scaling is good, at least for the largest sizes, up to the point where ${\cal O}^{(1)}_c$ reaches the 
plateau corresponding to $\pi^{\rm FBC}_{\rm hv}$, for the quenches to $T=0, \ T_c/4, T_c/2$, while for the quench to $T_c$ there is no 
collapse for $t / (L/\ell_G(t))^{\zeta} \le 1$, which is the region where the scaling should hold.

\begin{figure}[h]
\begin{center}
\includegraphics[scale=0.52]{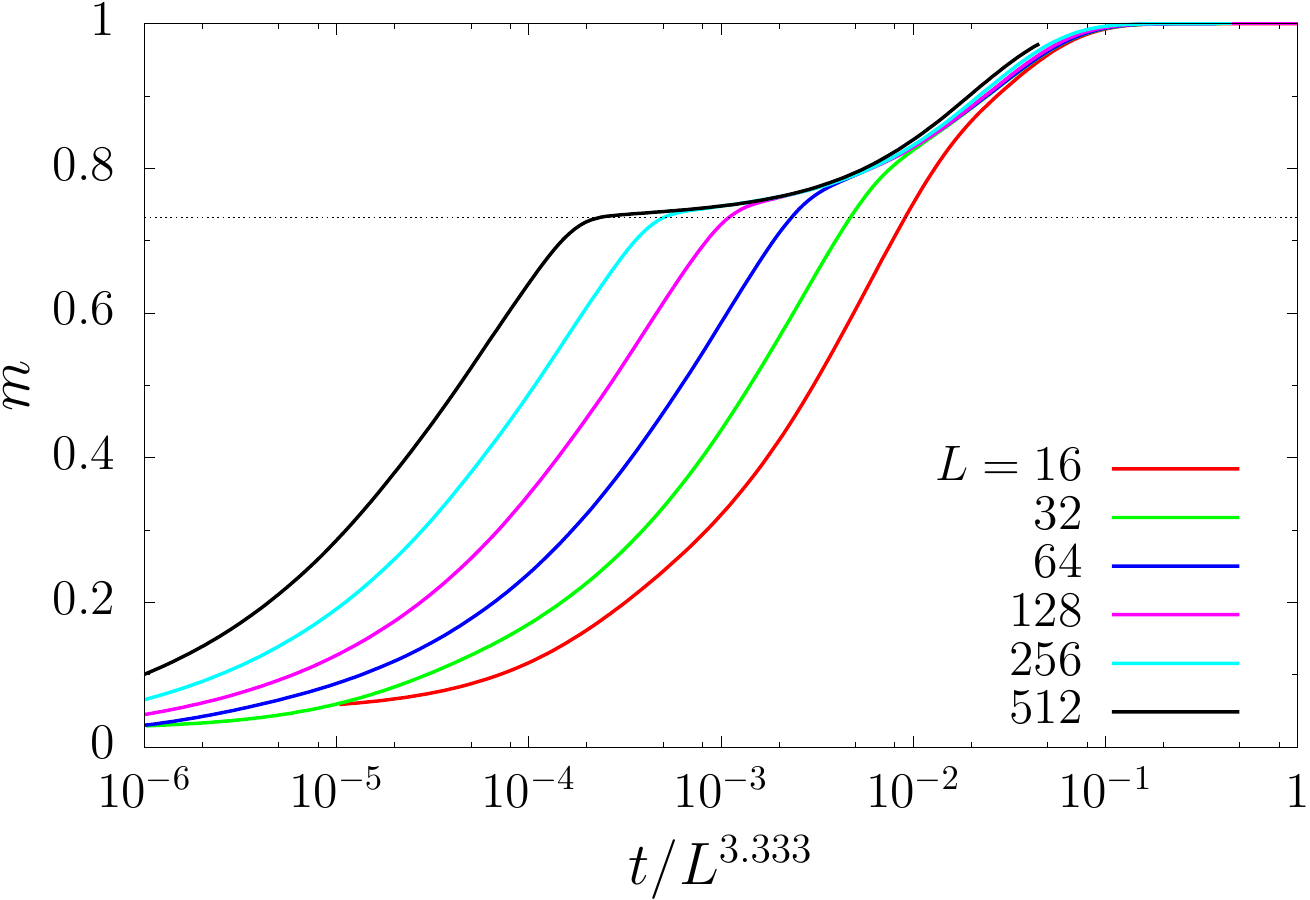} \quad
\includegraphics[scale=0.52]{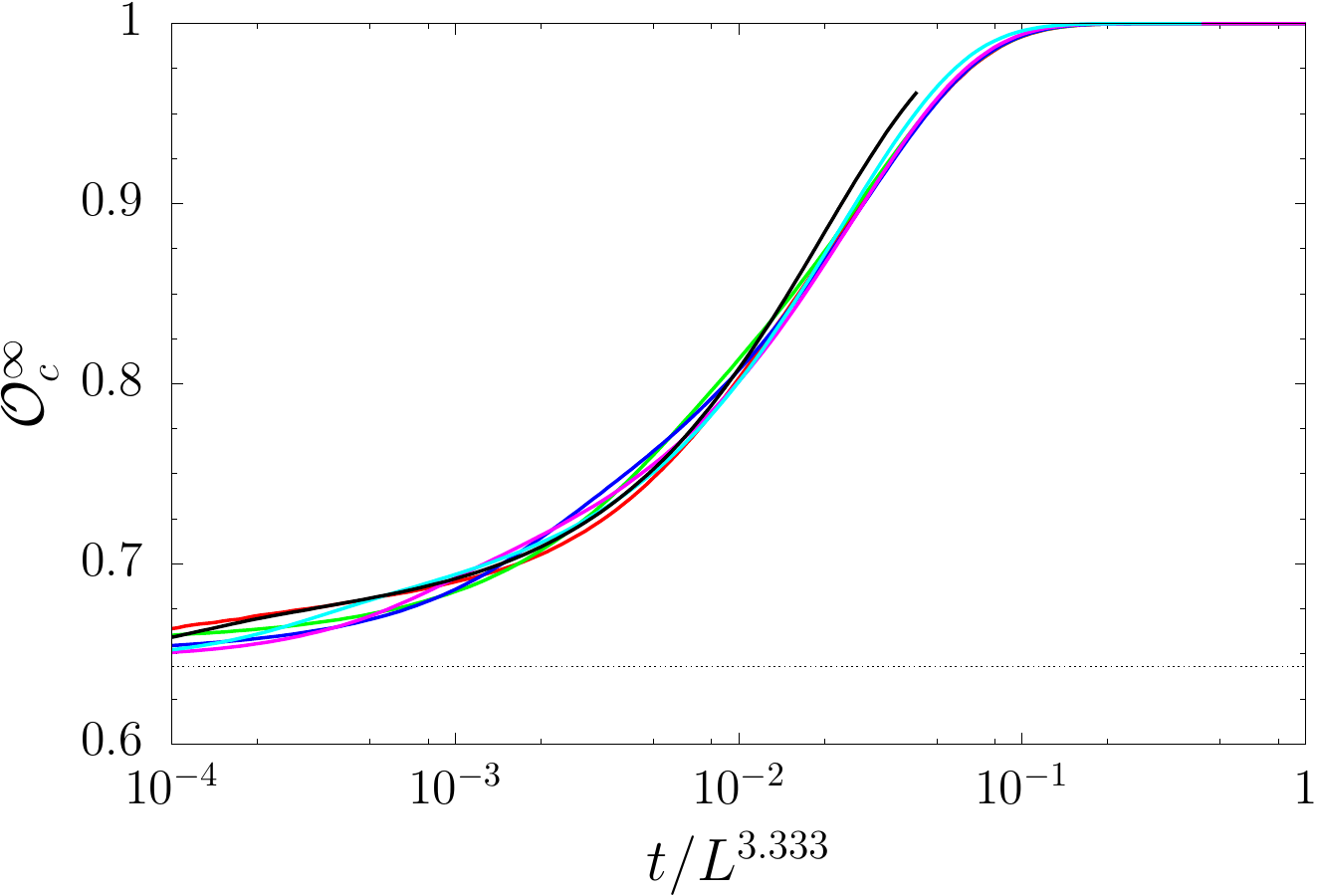}

\includegraphics[scale=0.52]{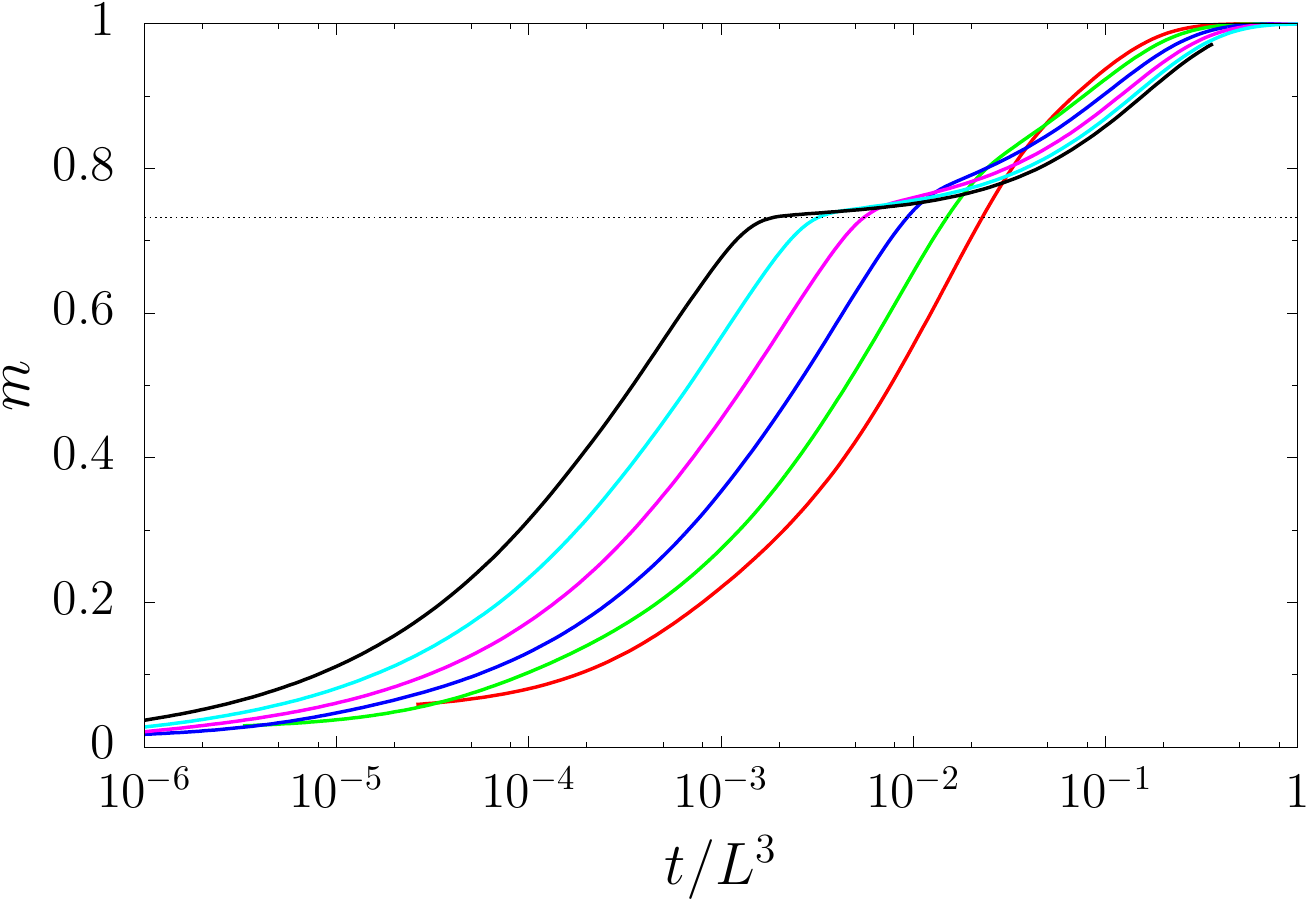} \quad
\includegraphics[scale=0.52]{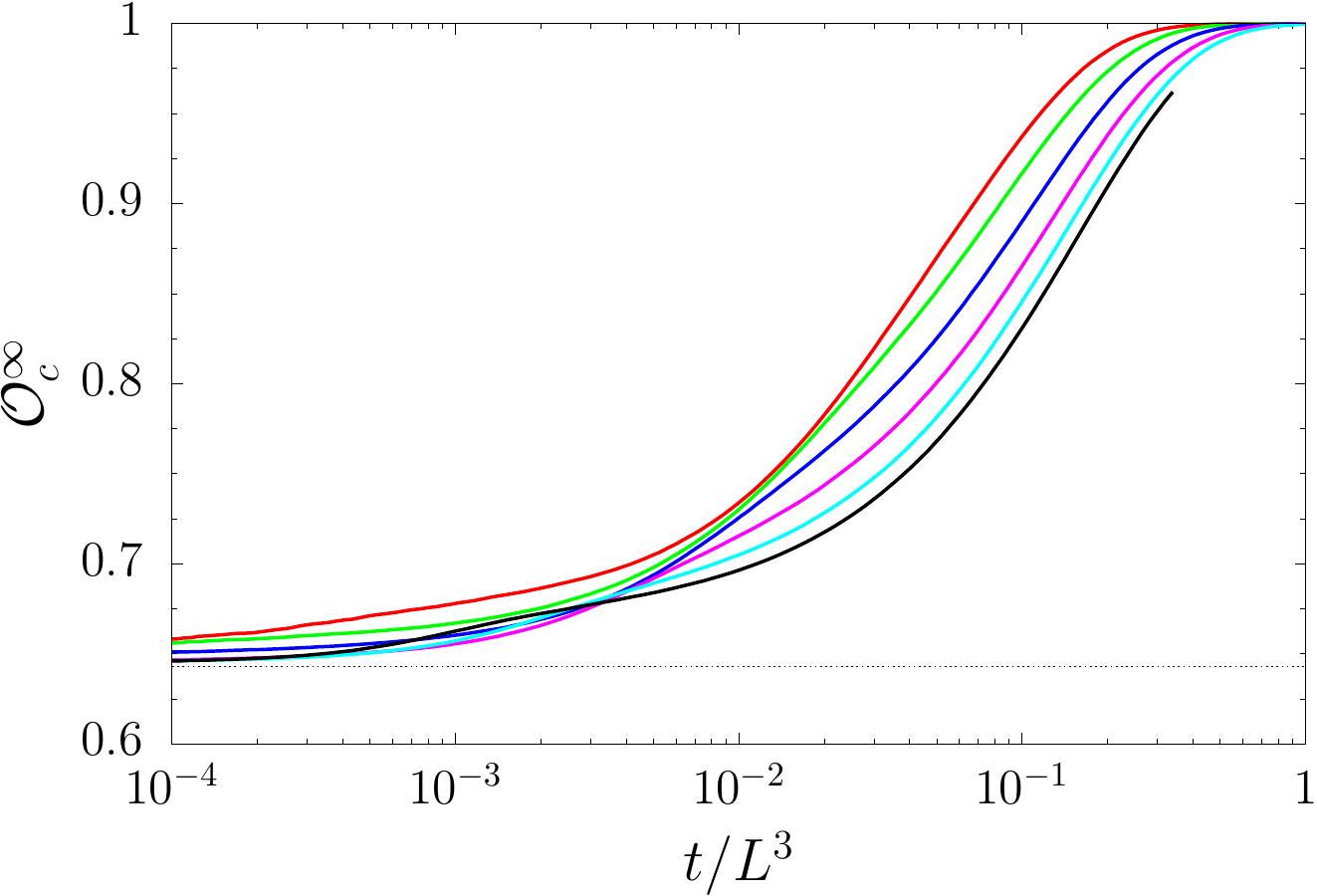}
\end{center}
\caption{\small Dynamics of the Ising model on a triangular lattice with FBC after a quench to $T=0$, for
different values of the lattice linear size $L$.
In the top row, we show the average magnetisation density $m$ (left panel) and the 
crossing correlation ${\cal O}^{\infty}_c$ (right panel), both plotted against the rescaled time $t/L^{3.333}$,
for different values of $L$.
For comparison, in the bottom row, we plot these two observables
against $t/L^3$ to prove that the curves fail to fall on top of each other with this choice of 
scaling variable.
The colour encoding each $L$ is the same in all plots and it is indicated in the key in the first panel.
}
\label{Tr2}
\end{figure}

\vspace{0.5cm}

It is now interesting to compare the behaviour on the square lattice with FBC  
to the $T=0$ dynamics on the triangular lattice also with FBC, a situation 
in which the stripes (of all types) are not stable. 
In Fig.~\ref{Tr2}, we show the magnetisation density $m$ and 
correlation between the number of crossings at time $t$ and at the final state,
${\cal O}^{\infty}_c$, for the latter problem.
The usual scaling against $t/L^2$ describes the data  up to $t/L^2 \simeq 0.1$ with
approximate saturation at $\simeq 0.7332$, a value that coincides with  the
asymptotic one for the square lattice with FBC (not shown). 
However, on the triangular lattice, the evolution goes on and for $t/L^{2} \geq 0.1$,  
the magnetisation enters a new growing regime at the end of which it attains $m_{\rm eq}=1$.
Indeed, the first regime includes the usual coarsening one at $T=0$ while the second one corresponds 
to the disappearance of the vertical or horizontal crossing clusters.
The mechanism through which these clusters disappear is not the same as the one discussed for the
diagonal stripes in the case of the square lattice with PBC.
Because of the particular way in which we constructed the triangular lattice, vertical (or horizontal) crossing clusters first
rotate until they get a diagonal crossing configuration, and then they expand invading the 
non-crossing regions of opposite phase.

The typical time scale for this last process is $t_{\rm eq} \sim L^{3.333}$, as shown in Fig.~\ref{Tr2}
where the rescaled time $t/L^{3.333}$ gave us the best collapse for ${\cal O}^{\infty}_c$.
${\cal O}^{\infty}_c$ is constant up to $t/L^2 \simeq 0.1$  with a value close to $ \pi^{\rm FBC}_{\rm hv} = 0.6442$ (not shown). 
This is due to the fact that on the triangular lattice, 
the infinite temperature initial condition also corresponds to the critical percolation point.
Thus, soon after the quench to $T=0$, ${\rm n}_c(t)$ has a probability $ \pi^{\rm FBC}_{\rm hv}$ of being one and in the final state, 
${\rm n}_c(t)=1$ always. 
Then ${\cal O}^{\infty}_c(0) = \pi^{\rm FBC}_{\rm hv}$ and the value starts to increase for $t/L^{2} \geq 0.1$, indicating 
that the vertical or horizontal crossing clusters are transformed into clusters crossing in both directions
with the mechanism described above, until it reaches $1$.

\section{Conclusions}
\label{sec:conclusions}

The aim of this paper was to quantify, with great precision, the approach to critical percolation 
previously observed in sub-critical quenches of the clean $2d$ ferromagnetic 
Ising model with non-conserved order parameter dynamics~\cite{ArBrCuSi07,SiArBrCu07,BlCoCuPi14}. 
In the analysis we treated zero and finite (though low) temperature
dynamics.  As a by-product we also quantified the latter approach to equilibrium of finite size systems.

The numerical data show that at a time-scale $t_{p_1}$ soon after or right at the quench 
(depending on the lattice geometry)
the configurations have, typically,
two large clusters that almost always are percolating (or at least have linear size comparable with $L$), 
that are also the two largest in the system and have opposite spin orientation.
These two large structures are accompanied by smaller non-percolating ones. 
At $t_{p_1}$ none of the two largest domains are stable against the dynamics: they break,
reconnect and grow by incorporating some smaller domains of opposite orientation surrounded by them 
until a time $t_p$ at which 
at least one of them percolates and remains percolating (and growing) at all subsequent times.
We call this a stable percolating structure.

Globally, we showed that after a subcritical instantaneous quench 
the systems evolve in three time-scales that are well separated and can be 
identified numerically. These are the following.
\begin{itemize}
\item
A short though macroscopic time scale, $t \leq t_p$, satisfying dynamic scaling with respect to the algebraically growing length 
\begin{equation}
\ell_p(t) \simeq t^{1/z_p}
\; 
\end{equation}
on the square and triangular lattices and an exponentially growing length
on the special honeycomb lattice.
From  $\ell_p(t_p) \simeq L$ we identify the characteristic time
$t_p \simeq L^{z_p}$ in the first two cases and a logarithmic dependence in the 
latter. At $ t \simeq t_p$
the morphology and the statistics of the geometric structures are those 
of random critical percolation. The systems are  very far away from equilibrium at $t_p$.
The global pattern is no longer destroyed by the dynamics and the later evolution continues following 
the rules of the next time regime.

\item  
At times $t_p \ll t\ll t_L$ the usual dynamic scaling regime characterised by the 
growing correlation length 
\begin{equation}
\ell_d(t) \simeq t^{1/z_d}
\end{equation}
with $z_d$ the dynamic exponent dictated by the curvature driven dynamics~\cite{AlCa79,ArBrCuSi07,SiArBrCu07}, 
that is to say $z_d=2$, establishes. 
The systems remain very far away from equilibrium with domain growth corresponding to the disappearance of 
small bubbles in favour of their embeding larger neighbouring domains.

\item At time-scales $t \simeq t_L = L^{z_d}$ the systems either reach equilibrium or get blocked in a
metastable state with stripes. In the latter case, depending on the lattice 
geometry, boundary conditions and working temperature, these stripes can eventually disappear leading
the system towards the equilibrium state on a typical time
$t_{\mathrm{eq}}(L) \sim L^{z_{\mathrm{eq}}}$, which defines a new dynamical exponent $z_{\mathrm{eq}}$,
such that $z_{\mathrm{eq}}> z_d$.
During this third regime (when it exists), the relevant lenght scale is given by
\begin{equation}
\ell_{\rm eq}(t) \simeq t^{1/z_{\rm eq}}
\; .
\end{equation}

\end{itemize}

In this paper we focused on the first of these regimes and we spent sometime dealing with the last one.
The intermediate regime is the one that has been mostly addressed in the literature so far. We used three
lattice geometries to test three distinct cases: the square lattice in which $0 \neq t_{p_1} < t_p$, 
the triangular lattice in which $0  = t_{p_1} < t_p$ and the honeycomb lattice 
in which $0 \neq t_{p_1} < t_p$ and, moreover,  there are metastable states at zero temperature.

The main conceptual idea to understand the early approach to critical percolation
in lattices that are not the triangular one is the interpretation of the dynamics in this regime as one 
of percolation with an effective growing lattice spacing~\cite{InCoCuPi16}
\begin{equation}
\ell_p(t) \simeq \ell_d(t) \, (t/t_0)^{1/\zeta}
\end{equation}
that, for an algebraically growing coarsening length, $\ell_d(t) \simeq t^{1/z_d}$, 
leads to 
\begin{equation}
\ell_p(t) \simeq t^{1/z_p}
\; . 
\end{equation}
With massive numerical simulations, and the 
evaluation of many observables that include the percolating probabilities, the winding angles, the geometric 
properties of the largest cluster and the number densities of domain areas,
we studied the dependence of $\ell_p$ on the coordination of the lattice. These studies point 
towards a slight difference in the value of the exponent $z_p$ on the square lattice compared to the one measured in~\cite{BlCoCuPi14}, 
that we here find to be better characterised by $z_p = 2/5$ instead of $z_p=1/2$.
The blocked local configurations on the honeycomb lattice make the scalings be peculiar and, in particular, 
the growing length $\ell_p(t)$ was found to be exponentially growing with time, implying a logarithmic
divergence of $t_p$ with the system size. Finally, the triangular lattice is also special, due to the fact that 
the initial configuration already has a critical percolating cluster in it (since $p_c=1/2$ on this lattice). The regime going from $t_{p_1}=0$ to $t_p>0$ is 
one in which the global characteristics of the structure do not change much. The exponent $z_p$ becomes apparent only 
in the analysis  of the overlap between two replicas of the system studied in~\cite{BlCoCuPi14}.
Finally, 
as already stressed in~\cite{BlCoCuPi14}, we confirm that the dynamic scaling of the correlation functions and other observables 
at times  $t$ of the order of $t_p$ needs the use of the two length scales $\ell_p$ and $\ell_d$.

We also showed that non-zero sub-critical temperatures have no large effect on this initial regime. More details on this issue, as well as 
on the effects of a slow cooling across the critical point~\cite{Biroli10}, 
will be given in~\cite{Ricateau17}.

In a recent paper the effects of weak disorder on the stochastic dynamics 
of the $2d$ Ising model were analysed~\cite{InCoCuPi16}.
The dependence of $t_p$ on $L$ for conserved order-parameter dynamics and the voter model
were studied in~\cite{TaCuPi16} and~\cite{TaCuPi15}, respectively. These two last cases will be 
revisited in view of the detailed analysis performed in this paper. The scaling of $t_p$ with $L$
on generic lattices needs to be rendered  more accurate in these cases and the analysis of 
the large variety of observables used in this paper will allow us to do it.

\appendix

\section{Continuous time Monte Carlo and Glauber dynamics}
\label{app:MC}

The overall stochastic dynamics of the spin variables 
is fully described by a master equation, that is a differential equation for the 
time-dependent probability density function in the state space of the system, with the following form
\begin{equation}
 \frac{\mathrm{d}}{\mathrm{d} t} P ( \mathbf{s}, t) \, = \,
  \sum_{\mathbf{s}'} \big[
  W (\mathbf{s}' \rightarrow \mathbf{s}, t ) P ( \mathbf{s}', t)
 - W (\mathbf{s} \rightarrow \mathbf{s}', t ) P ( \mathbf{s}, t)
  \big]
  \; , 
\label{eq:mastereq}
\end{equation}
where one sums over all possible states of the system and $ W (\mathbf{s} \rightarrow \mathbf{s}', t )$ represents the rate of transition from state $\mathbf{s}$ 
to state $\mathbf{s}' $ at time $t$.
In the case of the Ising model, $\mathbf{s} \in \{-1,+1\}^N$ represents the spin configuration of the system. 
For non-conserved order parameter dynamics, 
the transition rates are chosen such that
$  W (\mathbf{s} \rightarrow \mathbf{s}', t ) \ne 0 $ if and only if the configurations $ \mathbf{s} $ and  $ \mathbf{s}' $ differ 
in the value of a single spin. One then speaks of single spin flip dynamics and the master equation takes the simplified form
\begin{equation}
 \frac{\mathrm{d}}{\mathrm{d} t} P ( \mathbf{s}, t) \, = \,
  \sum_{\mathbf{x}} \big[
  W_{\mathbf{x}}(\mathbf{s}^{\mathbf{x}}, t ) P ( \mathbf{s}^{\mathbf{x}}, t)
 - W_{\mathbf{x}}(\mathbf{s}, t ) P ( \mathbf{s}, t)
  \big]
  \; , 
\label{eq:KIM}
\end{equation}
where the sum now runs over all the sites $\mathbf{x}$ of the lattice, 
$ W_{\mathbf{x}}(\mathbf{s}, t ) $ represents the spin-flip rate for the site
$\mathbf{x}$, given that the system is in the configuration  
$\mathbf{s}$ at time $t$, and $\mathbf{s}^{\mathbf{x}}$ is the configuration
obtained from $\mathbf{s}$ by flipping the spin at site $\mathbf{x}$.

By requiring that the transition rates satisfy the detailed balance condition, 
one makes sure that the dynamics bring the system to a
steady-state. In the particular case of Glauber dynamics, the spin-flip rates are given by 
\begin{equation}
 W_{\mathbf{x}}(\mathbf{s}; \beta) =
  \frac{1}{2 \tau} \left[
  1 - s_{\mathbf{x}} \tanh{ \left( \beta J 
  \sum_{\mathbf{y} \in \mathcal{N} (\mathbf{x}) } s_{\mathbf{y}}  \right) }
  \right]
\label{eq:Glauber_dyn}
\end{equation}
$ s_{\mathbf{x}} $ being the value of the spin at site $\mathbf{x} $ and $  \mathcal{N} (\mathbf{x}) $ 
the set of all its nearest-neighbour sites. The parameter $\tau$ represents the microscopic time scale for the transition processes.
In this form, the spin flip rate describes the relaxational dynamics towards the equilibrium distribution at inverse 
temperature $\beta$. At zero temperature,
\begin{equation}
 W_{\mathbf{x}}(\mathbf{s}; T = 0) \propto
  \left[
  1 -  \mathrm{sign} { \left(
 s_{\mathbf{x}} \sum_{\mathbf{y} \in \mathcal{N} (\mathbf{x}) } s_{\mathbf{y}}
  \right) }
  \right]
  \; , 
\label{eq:zerotemp}
\end{equation}
{\it i.e.} the $\beta \rightarrow \infty $ limit of Eq.~(\ref{eq:Glauber_dyn}).

In the simulations we adopted a heat bath Monte Carlo algorithm (or Metropolis Monte Carlo algorithm), which consists 
in a slightly different expression of the single spin-flip rates, namely
\begin{equation}
 W_{\mathbf{x}}(\mathbf{s}) =
        \left\{
                \begin{array}{l l }
                  \exp{ \left[ -2 \beta J \,  e(\mathbf{s},\mathbf{x}) \right] } & \qquad\qquad \mbox{if} \;\; e(\mathbf{s},\mathbf{x}) > 0  \\
                 \frac{1}{2}  & \qquad\qquad \mbox{if} \;\; e(\mathbf{s},\mathbf{x}) = 0   \\
                  1 & \qquad\qquad \mbox{if} \;\; e(\mathbf{s},\mathbf{x}) < 0 
                \end{array}
              \right.
\label{eq:Metropolis}
\end{equation}
where $e(\mathbf{s},\mathbf{x})= s_{\mathbf{x}} h_{\mathbf{x}} = s_{\mathbf{x}} 
 \left( \sum_{ \mathbf{y} \in \mathcal{N} (\mathbf{x}) } s_{\mathbf{y}} \right)  $,
$ 2 J e(\mathbf{s},\mathbf{x})$ being the energy change caused by flipping the spin at the site $\mathbf{x}$, 
and $h_{\mathbf{x}} $ the local field.

The dynamics are particularly simple at zero temperature. 
After choosing a lattice site at random, one computes the 
local field $h$ produced by its nearest-neighbours.
The spin is flipped with probability $1$ if $e$ is negative, {\it i.e.} \,  if the majority of the nearest-neighbor sites
have antiparallel spin with respect to the chosen site.
If $e$ vanishes, the spin is flipped with probability $ \frac{1}{2} $.
In the remaining case, $e > 0$, the spin is left unchanged.
At exactly zero temperature there is thus no bulk noise, which means that changes occur only at the interface between 
domains of opposite
phase. In the context of a continuum space approximation 
({\it i.e.}~when the lattice spacing becomes infinitesimal) and for long time, the zero temperature
Glauber-Ising dynamics has a very nice description in terms of the motion of the interfaces: 
all the interfaces tend to move with a local velocity that points in the direction that makes the local 
curvature decrease (see~\cite{AlCa79,ArBrCuSi07,SiArBrCu07}).
At the end, the interfaces can only annihilate or become straight and percolate through the system.

Given a $2d$ lattice with linear size $L$, for the usual Monte Carlo method 
$L^2$ spin flip attempts correspond to a single unit of time, namely $\tau \propto L^{-2}$, with
$ \tau$ the microscopic time scale entering in Eq.~(\ref{eq:Glauber_dyn}).
Quite naturally, the number of spins that can be flipped under the rule described by Eq.~(\ref{eq:Metropolis}) decreases in time. 
Therefore, testing all the possible spins in the sample results in a waste of computer time.

It is much faster to consider only the spins that can be actually flipped, namely those that
are characterised by a local field that is opposite to the spin. 
In order to accelerate
our numerical simulations,  we used the Continuous Time Monte Carlo (CTMC) method~\cite{Bortz75}. 
This  algorithm works as follows.
Since $W_{\mathbf{x}} ( \mathbf{s} ) $ depends on the spin configuration $ \mathbf{s} $ and on the lattice site $\mathbf{x} $ only through
the quantity $e(\mathbf{s},\mathbf{x})= s_{\mathbf{x}} \,  \left( \sum_{ \mathbf{y} \in \mathcal{N} (\mathbf{x}) } s_{\mathbf{y}} \right)  $,
we build a list $A_e(t)$ of all the sites that at time $t$ have local field equal to $e$, for each value of $e\in \{-c,-c+2, ..., c-2, c\} $,
with $c$ the coordination number of the lattice.
Before attempting to flip a spin, we compute the associated time increment $\Delta t$ 
by drawing it randomly from an exponential distribution
of parameter $\lambda(t) = N_{{\rm flips}}(t) $, where 
$  N_{\rm flips}(t) $ is the expected number of spins that can be flipped given the configuration at time $t$, 
so that $\langle \Delta t \rangle = 1/N_{\rm flips} $.
Namely, if $n_e(t) = | A_e(t) | $ is the number of sites that have local field $e$ at time $t$, then
$  N_{\rm flips}(t) =  \sum W_e \,  n_e(t) $, where ${W_e}$ are the spin-flip probabilities described by Eq.~(\ref{eq:Metropolis}) 
for each value of $e$.
Then a value $e^*$ is chosen randomly in $\{-c,-c+2, ..., c-2, c\} $ and the site that undergoes a spin flip is chosen randomly amongst 
the ones in $A_{e^*}(t)$.
After the spin has been flipped, one must update the time and all the lists $A_{e}$. This do not represent a great deal, since
the only sites that have a different value of $e$ at time $t'=t+\Delta t$ are the ones which had their spins flipped as well as 
their nearest-neighbours.

For zero-temperature dynamics, this procedure is further simplified: one only needs to keep memory of the list of sites that
have negative local field, $ A_{-}$, the ones that have zero local field, $ A_{0}$, and their respective numbers $n_-$ and $n_0$.
\begin{comment}
An update is done by
choosing a random number $u \in ]0:1]$. If $ u (n_- + \frac{1}{2} n_0 ) < n_-$, we will reverse a spin 
among the ones with negative local field, otherwise a spin with zero local field. Such a spin is chosen at random 
among the $n_-$ or $n_0$ ones.
\end{comment}
We repeat the operation described above until $n_- + n_0 = 0$ at which point we have a stable
configuration. 

It was shown in~\cite{Bortz75} that this algorithm is equivalent to the ordinary heat bath Monte Carlo 
algorithm with discrete time steps
if the time increments $\Delta t$ are drawn from an exponential distribution in the manner explained above.

\vspace{0.75cm}

\noindent
{\bf Acknowledgements.}
L. F. C. is a member of Institut Universitaire de France. We thank H. Ricateau for very useful discussions.

\vspace{1cm}

\bibliographystyle{phaip}
\bibliography{coarsening}

\begin{thebibliography}{10}

\bibitem{Bray94}
A.~J. Bray,
\newblock Adv. Phys. {\bf 43}, 357 (1994).

\bibitem{Puri09-article}
S.~Puri,
\newblock Kinetics of phase transitions,
\newblock in {\em Kinetics of Phase transitions}, edited by S.~Puri and
  V.~Wadhawan, Taylor and Francis, 2009.

\bibitem{CorberiPoliti}
F.~Corberi and P.~Politi,
\newblock Comptes Rendus de Physique {\bf 16}, 255 (2015).

\bibitem{Hohenberg-Halperin}
P.~C. Hohenberg and B.~I. Halperin,
\newblock Rev. Mod. Phys. {\bf 49}, 435 (1977).

\bibitem{ArBrCuSi07}
J.~J. Arenzon, A.~J. Bray, L.~F. Cugliandolo, and A.~Sicilia,
\newblock Phys. Rev. Lett. {\bf 98}, 145701 (2007).

\bibitem{SiArBrCu07}
A.~Sicilia, J.~J. Arenzon, A.~J. Bray, and L.~F. Cugliandolo,
\newblock Phys. Rev. E {\bf 76}, 061116 (2007).

\bibitem{SiArBrCu08}
A.~Sicilia, J.~J. Arenzon, A.~J. Bray, and L.~F. Cugliandolo,
\newblock Europhys. Lett. {\bf 82}, 10001 (2008).

\bibitem{SiSaArBrCu09}
A.~Sicilia, Y.~Sarrazin, J.~J. Arenzon, A.~J. Bray, and L.~F. Cugliandolo,
\newblock Phys. Rev. E {\bf 80}, 031121 (2009).

\bibitem{SpKrRe01}
V.~Spirin, P.~L. Krapivsky, and S.~Redner,
\newblock Phys. Rev. E {\bf 63}, 036118 (2001).

\bibitem{SpKrRe02}
V.~Spirin, P.~Krapivsky, and S.~Redner,
\newblock Phys. Rev. E {\bf 65}, 016119 (2002).

\bibitem{BaKrRe09}
K.~Barros, P.~L. Krapivsky, and S.~Redner,
\newblock Phys. Rev. E {\bf 80}, 040101 (2009).

\bibitem{OlKrRe12}
J.~Olejarz, P.~L. Krapivsky, and S.~Redner,
\newblock Phys. Rev. Lett. {\bf 109}, 195702 (2012).

\bibitem{BlCoCuPi14}
T.~Blanchard, F.~Corberi, L.~F. Cugliandolo, and M.~Picco,
\newblock EPL {\bf 106}, 66001 (2014).

\bibitem{InCoCuPi16}
F.~Corberi, L.~F. Cugliandolo, F.~Insalata, and M.~Picco,
\newblock Phys. Rev. E {\bf 95}, 022101 (2017).

\bibitem{TaCuPi16}
A.~Tartaglia, L.~F. Cugliandolo, and M.~Picco,
\newblock EPL {\bf 116} (2016).

\bibitem{TaCuPi15}
A.~Tartaglia, L.~F. Cugliandolo, and M.~Picco,
\newblock Phys. Rev. E {\bf 92}, 042109 (2015).

\bibitem{TakanoMiyashita}
H.~Takano and S.~Miyashita,
\newblock Phys. Rev. B {\bf 48}, 7221 (1993).

\bibitem{Bortz-etal74}
A.~B. Bortz, M.~H. Kalos, J.~L. Lebowitz, and M.~A. Zendejas,
\newblock Phys. Rev. B {\bf 10}, 535 (1974).

\bibitem{Barkema}
G.~T. Barkema and M.~E.~J. Newman,
\newblock {\em Monte Carlo methods in statistical physics},
\newblock Oxford University Press, Oxford, 1999.

\bibitem{BrayHumayunNewman}
A.~J. Bray, K.~Humayun, and T.~J. Newman,
\newblock Phys. Rev. B {\bf 43}, 3699 (1991).

\bibitem{ChakrabortyDas}
S.~Chakraborty and S.~K. Das,
\newblock Eur. Phys. J. B {\bf 88}, 160 (2015).

\bibitem{Corberi16}
F.~Corberi and R.~Villavicencio-Sanchez,
\newblock Phys. Rev. E {\bf 93}, 052105 (2016).

\bibitem{DBG0}
B.~Derrida, A.~J. Bray, and C.~Godr\`eche,
\newblock J. Phys. A: Math. Gen. {\bf 27}, L357 (1994).

\bibitem{BMS}
A.~J. Bray, S.~N. Majumdar, and G.~Schehr,
\newblock Adv. in Phys. {\bf 62}, 225 (2013).

\bibitem{Pi94}
H.~Pinson,
\newblock J. Stat. Phys. {\bf 75}, 1167 (1994).

\bibitem{PruMol04}
G.~Pruessner and N.~R. Moloney,
\newblock J. Stat. Phys. {\bf 115}, 839 (2004).

\bibitem{Cardy92}
J.~Cardy,
\newblock J. Phys. A {\bf 25}, L201 (1992).

\bibitem{Watts96}
G.~M.~T. Watts,
\newblock J. Phys. A: Math. Gen. {\bf 29}, 363 (1996).

\bibitem{SaDu87}
H.~Saleur and B.~Duplantier,
\newblock Phys. Rev. Lett. {\bf 58}, 2325 (1987).

\bibitem{Smirnov01}
S.~Smirnov,
\newblock C. R. Acad. Sci. Paris I {\bf 333}, 239 (2001).

\bibitem{Stauffer94}
D.~Stauffer and A.~Aharony,
\newblock {\em Introduction To Percolation Theory},
\newblock Taylor and Francis, London, 1994.

\bibitem{CaZi03}
J.~Cardy and R.~M. Ziff,
\newblock J. Stat. Phys. {\bf 110}, 1 (2003).

\bibitem{DuSa88}
B.~Duplantier and H.~Saleur,
\newblock Phys. Rev. Lett. {\bf 60}, 2343 (1988).

\bibitem{WiWi03}
B.~Wieland and D.~B. Wilson,
\newblock Phys. Rev. E {\bf 68}, 056101 (2003).

\bibitem{ChristensenMoloney}
K.~Christensen and N.~R. Moloney,
\newblock {\em Complexity and Criticality},
\newblock Imperial College Press, 2005.

\bibitem{Saberi15}
A.~A. Saberi,
\newblock Phys. Rep. {\bf 578}, 1 (2015).

\bibitem{Arenzon-etal15}
J.~J. Arenzon, L.~F. Cugliandolo, and M.~Picco,
\newblock Phys. Rev. E {\bf 91}, 032142 (2015).

\bibitem{Machta13}
J.~Ye, J.~Machta, C.~M. Newman, and D.~L. Stein,
\newblock Phys. Rev. E {\bf 88}, 040101 (2013).

\bibitem{BlCuPi14}
T.~Blanchard, L.~F. Cugliandolo, and M.~Picco,
\newblock J. Stat. Mech. , P12021 (2014).

\bibitem{ChakrabortyDas16}
S.~Chakraborty and S.~K. Das,
\newblock  {\bf 93}, 032139 (2016).

\bibitem{AlCa79}
S.~M. Allen and J.~W. Cahn,
\newblock Acta Metall. {\bf 27}, 1085 (1979).

\bibitem{Sen01}
P.~Sen,
\newblock J. Phys. A {\bf 34}, 8477 (2001).

\bibitem{Bazant00}
M.~Z. Bazant,
\newblock Phys. Rev. E {\bf 62}, 1660 (2000).

\bibitem{RedigHofstad}
R.~van Der~Hofstad and F.~Redig,
\newblock Journal of Statistical Physics {\bf 122}, 671  (2006).

\bibitem{Bazant02}
M.~Z. Bazant,
\newblock Physica A: Statistical Mechanics and its Applications {\bf 316}, 29
  (2002).

\bibitem{Gnedenko}
B.~V. Gnedenko,
\newblock Annals of Mathematics {\bf 44}, 423 (1943).

\bibitem{HoviAharony-97}
J.~P. Hovi and A.~Aharony,
\newblock Phys. Rev. E {\bf 56}, 172 (1997).

\bibitem{BotetPlos}
R.~Botet and M.~Ploszajczak,
\newblock Phys. Rev. Lett. {\bf 95}, 185702 (2005).

\bibitem{BotetPOS}
R.~Botet,
\newblock Proceedings of Science {\bf 007}, 1 (2012).

\bibitem{Blanchard14}
T.~Blanchard,
\newblock {\em Morphology of domains in and out of equilibrium},
\newblock PhD thesis, Universit\'e Pierre et Marie Curie - Paris VI,
  https://tel.archives-ouvertes.fr/tel-01081275, 2014.

\bibitem{Cheong04}
M.~Cheong and I.~Chang,
\newblock Int. J. Mod. Phys. C {\bf 15}, 835 (2004).

\bibitem{Onsager44}
L.~Onsager,
\newblock Phys. Rev. {\bf 65}, 117 (1944).

\bibitem{Yang52}
C.~N. Yang,
\newblock Phys. Rev. {\bf 85}, 808 (1952).

\bibitem{BlPi13}
T.~Blanchard and M.~Picco,
\newblock Phys. Rev. E {\bf 88}, 032131 (2013).

\bibitem{Biroli10}
G.~Biroli, L.~F. Cugliandolo, and A.~Sicilia,
\newblock Phys. Rev. E {\bf 81}, 050101 (2010).

\bibitem{Ricateau17}
H.~Ricateau, L.~F. Cugliandolo, and M.~Picco,
\newblock to be published, 2017.

\bibitem{Bortz75}
A.~B. Bortz, M.~H. Kalos, and J.~L. Lebowitz,
\newblock J. Comp. Phys. {\bf 17}, 10 (1975).

\end{thebibliography}

\end{document}